%% file: main.tex
\documentclass[lettersize,journal]{IEEEtran}
\ifCLASSINFOpdf
\else
\fi

\usepackage{amsmath,graphicx,mathtools}
\usepackage{amsfonts,amssymb}
\usepackage{siunitx}
\usepackage[bookmarks=false]{hyperref}
\usepackage{subcaption}
\usepackage{psfrag}
\usepackage{multirow}
\usepackage{tikz,pgfplots}
\usepackage{import,subfiles}
\usetikzlibrary{positioning}
\usepackage{colortbl}

\pgfplotsset{
    compat=newest,
    /pgfplots/legend image code/.code={%
        \draw[mark repeat=2,mark phase=2,#1] 
            plot coordinates {
                (0cm,0cm) 
                (0.1cm,0cm)
                (0.1cm,0cm)
                (0.1cm,0cm)
                (0.2cm,0cm)%
            };
    },
}

\def\addlegendimage{\csname pgfplots@addlegendimage\endcsname}

\usepackage{subfiles}
\begin{document}
%\title{Bare Demo of IEEEtran.cls\\ for IEEE Journals}
\title{Wavefront Coding for Accommodation-Invariant Near-Eye Displays}

%\author{Michael~Shell,~\IEEEmembership{Member,~IEEE,}
%        John~Doe,~\IEEEmembership{Fellow,~OSA,}
%        and~Jane~Doe,~\IEEEmembership{Life~Fellow,~IEEE}% <-this % stops a space
%\thanks{M. Shell was with the Department
%of Electrical and Computer Engineering, Georgia Institute of Technology, Atlanta,

\author{Ugur~Akpinar,
        Erdem~Sahin,
        Tina M. Hayward,
        Apratim Majumder,
        Rajesh~Menon,
        and~Atanas~Gotchev%
\thanks{U. Akpinar, E. Sahin, and A. Gotchev are with the Faculty of Information Technology and Communication Sciences, Tampere University, 33720 Tampere, Finland (e-mail: \href{mailto:ugur.akpinar@tuni.fi}{ugur.akpinar@tuni.fi})}%
\thanks{T. M. Hayward, A. Majumder and R. Menon is with the Department of Electrical and Computer Engineering, University of Utah, Salt Lake City, Utah 84102, USA}}
\maketitle


\begin{abstract}
We present a new computational near-eye display method that addresses the vergence-accommodation conflict problem in stereoscopic displays through accommodation-invariance. 
Our system integrates a refractive lens eyepiece with a novel wavefront coding diffractive optical element, operating in tandem with a pre-processing convolutional neural network. 
We employ end-to-end learning to jointly optimize the wavefront-coding optics and the image pre-processing module. 
To implement this approach, we develop a differentiable retinal image formation model that accounts for limiting aperture and chromatic aberrations introduced by the eye optics. We further integrate the neural transfer function and the contrast sensitivity function into the loss model to account for related perceptual effects. To tackle off-axis distortions, we incorporate position dependency into the pre-processing module. 
In addition to conducting rigorous analysis based on simulations, we also fabricate the designed diffractive optical element and build a benchtop setup, demonstrating accommodation-invariance for depth ranges of up to four diopters.
\end{abstract}

\begin{IEEEkeywords}
Near-Eye Displays, Vergence-Accommodation Conflict, Accommodation-Invariance, Diffractive Optics, End-to-end Learning.
\end{IEEEkeywords}

\IEEEpeerreviewmaketitle

\section{Introduction}
\IEEEPARstart{T}{he} simplicity of stereoscopic near-eye display (NED) design has made these systems particularly attractive for virtual reality (VR) and augmented reality (AR) applications. 
However, a major drawback hindering their widespread adoption is the vergence-accommodation conflict (VAC), which is caused by the mismatch between the two visual cues. In natural viewing conditions, vergence and accommodation work in synchrony, but the link between them gets broken in stereoscopic NEDs, resulting in severe visual discomfort \cite{VAC1,VAC2,VAC3}.
Two groups of methods have addressed the VAC. \textit{Accommodation-enabling (AE) displays} have aimed at delivering close-to-natural viewing experience by recreating near-correct retinal blur to drive the accommodation to the vergence distance of the object. We discuss AE display approaches in more details in Sec.~\ref{sec:Related_works}.
Instead of recreating focus cues, \textit{accommodation-invariant (AI) displays} have aimed at coupling vergence with accommodation by removing the retinal defocus blur completely. In general, this can be achieved by extending the display depth of field (DoF) by either delivering images through pinholes \cite{Maxwellian} or by using focus-tunable lenses \cite{KonradAI}.
User studies suggest that display DoF extension leads to a more natural vergence–accommodation interplay, with the potential to mitigate VAC and associated visual discomfort in NEDs \cite{EDOFPerceptualTest}. 

In this paper, we propose to advance AI display development by employing wavefront coding via a passive diffractive optical element (DOE), which works in tandem with a refractive main lens to form the display eyepiece. It is combined with an image pre-processing convolutional neural network (CNN) in a differentiable display model that further incorporates a fully differentiable mathematical model of retinal image formation. The differentiability of the entire pipeline is crucial as it enables joint optimization of both the CNN parameters and the DOE phase profile using stochastic gradient descent over a large dataset of training images. Such end-to-end optimization has proven effective in several image acquisition tasks \cite{Akpinar_TIP, Kim_OJSP, Sahin_ICIP, EndtoEndHaim,EndtoEndHDR,EndtoEndSitzmann}.
We extend our preliminary works on AI display \cite{ICIP2020,MultifocalAI,Makinen_2021}, making several crucial improvements. We consider more realistic viewing conditions through a new retinal image formation model, where the eye pupil is separated from the eyepiece and its size is smaller than the eyepiece. We also build a benchtop setup incorporating the newly designed and fabricated DOE to demonstrate the performance of the proposed method through optical measurements.
The key contributions of the proposed method can be summarized as follows:
\begin{itemize}
    \item We propose the design principles of a novel NED type alleviating the VAC with \textit{static} optics. Our solution is based on the accommodation invariance, where retinal defocus blur is removed from the system and the convergence-accommodation is expected to take effect.
    \item We optimize the proposed display system in an end-to-end manner, where the pre-processing and the display optics are designed jointly.
    \item We incorporate \textit{position dependency} into the pre-processing module, which is instrumental for tackling the off-axis distortions.
    \item We further integrate \textit{perceptual modeling} into the loss function by incorporating both the neural transfer function and the contrast sensitivity function.
    \item We fabricate a custom-designed DOE and implement the proposed method in a benchtop optical setup, validating its effectiveness via optical measurements.
\end{itemize}

\section{Related Works}
\label{sec:Related_works}

\subfile{figures/Literature/figure_Literature.tex}

Fig.~\ref{fig:NED_Literature} illustrates the advanced display architectures aimed at tackling the VAC, organized into the two categories of AE and AI displays. We refer also to the recent surveys \cite{NEDReview1,NEDReview2} for further details.

\subsection{Accommodation-Enabling Near-Eye Displays}

Varifocal and multifocal approaches aim at recreating several focal planes via spatial or temporal multiplexing.
In varifocal displays \cite{VarifocalSugihara,VarifocalLiu,Varifocal3,VarifocalAksit,Varifocal2} the depth of a single virtual image plane is dynamically adjusted to match the vergence distance. 
This adjustment can be done either by mechanically shifting the 2D display plane \cite{VarifocalSugihara,Varifocal3}, or by using tunable optics \cite{VarifocalLiu,Varifocal3,Varifocal2}, or by employing diffractive optics \cite{VarifocalAksit}. Vergence estimation is typically performed via gaze tracking \cite{VarifocalFoveatedAR}.
Varifocal displays do not provide optically accurate retinal defocus blur. Instead, blur is simulated through depth-of-field (DoF) rendering \cite{DOFRendering,DOFRendering2} using the scene depth information.
The gaze tracking requirement and the needed synchronization with the optical setup are the main challenges pertaining to the varifocal approaches.

Multifocal displays \cite{MultifocalRoland,Multifocal,Multifocal2,Multifocal3,FocalSurface,FocalSurfaceAR} approximate volumetric scene representations by means of a dense set of virtual image planes.
Conventional methods rely on spatial multiplexing, where multiple physical displays are stacked together to simultaneously reconstruct the image planes \cite{MultifocalRoland,Multifocal}. This is however challenging for achieving the compact form factor, preferred in modern NEDs.
Adaptive optics has been incorporated to realize multifocal displays via temporal multiplexing \cite{Multifocal2,Multifocal3}. 
The main problems of this approach are the requirements for high display refresh rate and synchronization between the optics and the content.
Alternatively to the above-mentioned methods, fixed focal surfaces have been optimized against the target scene depth by means of spatial light modulators \cite{FocalSurface} or freeform projection surfaces \cite{FocalSurfaceAR}.

More advanced techniques aim at reconstructing the 4D light field (LF) \cite{LF, Bregovic2019}, which is the most rigorous representation of a scene within the constraints of ray optics.
LF NEDs \cite{LFDisplayLuebke,LFDispHua,LFStereoscope,SMVNED,LFNEDPinhole} have demonstrated the ability to deliver near-correct focus cues, effectively mitigating the VAC.
Traditionally, LF NEDs allocate the available pixel budget between angular and spatial information, by means of a 2D display equipped with an array of microlenses or pinhole apertures (Fig.~\ref{fig:NED_Literature}).
In a typical setup, $2 \times 2$ or more views are projected into the eye pupil, aiming to stimulate natural accommodation and monocular parallax \cite{HuangLFSim,Miyanishi2022}.
An evident limitation of this approach is the inherent trade-off between spatial and angular resolutions. 
To overcome it, alternative methods have been proposed including multiplicative \cite{LFStereoscope} or additive \cite{AdditiveLF,AdditiveLF2} compressive LF displays, as well as high-resolution LF retinal projection assisted by gaze tracking \cite{RetinaLF}.
While these techniques can achieve higher spatio-angular resolution, they also face challenges, such as diffraction artifacts, decreased light throughput, or reduced frame rate.

Holographic NEDs \cite{Holography,Holographychen2015,Holographymetalens,HolographySphericalWave,FoveatedHolography2020,TensorHolography,HolographyHogelFree,HolographyOLAS,HolographyHigResHOGD,AccommodativeHolography} aim at recreating the complex hologram of the scene, which provides a virtual experience closest to the natural view in terms of depth perception. 
Typically, a spatial light modulator (SLM) is employed, to modify the phase of the incoming coherent light.
Majority of the applications are proposed for efficient scene hologram generation \cite{FoveatedHolography2020,TensorHolography,HolographyHogelFree,HolographyOLAS,Sahin_CGH}.
While holography is considered the ultimate technology for achieving immersive visual experience, current implementations face several challenges, such as a limited eyebox and the presence of speckle noise.

\subsection{Accommodation-Invariant Near-Eye Displays}

Maxwellian view displays \cite{Maxwellian,Maxwellian2,Maxwellian3,Maxwellian4,Maxwellian5,Maxwellian6} represent one of the most well-established implementations of AI NEDs.
Such displays project the image pixels directly onto the retina through a small aperture (pinhole) at the eye pupil plane.
This approach is analogous to reducing a camera's aperture to achieve extended DoF (EDoF) imaging.
The inherent trade-off in Maxwellian displays is the reduced eyebox size, as light is funneled through a single pinhole.
Recent attempts have addressed this issue \cite{MaxwellianEyebox1,MaxwellianEyebox3,MaxwellianEyebox4}, indicating an increasing interest in AI NEDs for solving the VAC.

Another approach to achieving AI display performance is to modulate the system's point spread function (PSF) to be depth-invariant by using adaptive optics.
One of the earliest demonstrations of this concept was in projectors \cite{CAProjection}, where the display DoF is extended using a coded pattern to the projector's aperture combined with inverse filtering of the input image.
To improve light efficiency, Iwai et al. \cite{IwaiFocusTunable} replaced the coded aperture with a fast focus-tunable lens.
By oscillating the lens' focal length faster than the perceivable temporal resolution, they created an average PSF that remains consistent across a wide depth range.
A similar technique has been adopted by Konrad et al. \cite{KonradAI} specifically for NEDs.
Their work investigates the trade-off between the extended depth range and the spatial resolution, with the aim to optimize the so-called \textit{multi-plane AI mode}.
In this mode, the display backlight and the lens oscillation are synchronized to create discrete virtual image planes at two or three image depths. This approach avoids the spatial resolution loss associated with continuous focal sweeps, which would otherwise increase the effective PSF size. 

In our work, we pursue a streamlined and lightweight display design that effectively alleviates the VAC. To this end, we adopt the AI NED approach and attempt the EDoF by means of static optics, eliminating the need for dynamic adjustment and/or synchronization of optical components. 
The following sections present a formal problem definition based on frequency-domain analysis, followed by a detailed discussion of the proposed implementation.
\begin{figure}[ht]
%\centering
\begin{subfigure}{0.95\columnwidth}
    \includegraphics[width=\columnwidth]{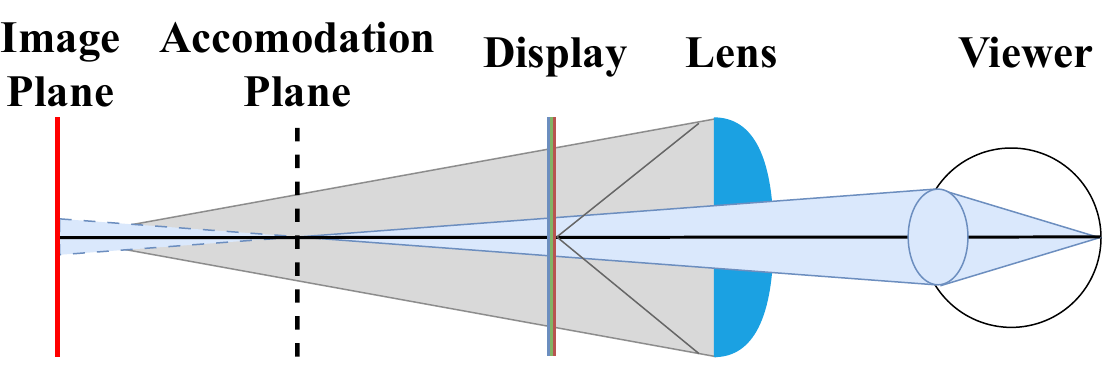}
    \caption{\hspace*{-5em}}
    \label{sfig:NED_conv}
\end{subfigure}

\hspace{-2.5mm}
\begin{subfigure}{0.35\columnwidth}
    \subfile{figures/MTF/MTF_cubic/MTF1D_green_lensonly}
    \caption{\hspace*{-5em}}
    \label{sfig:MTFLens}
\end{subfigure}
\hspace{12mm}
\begin{subfigure}{0.35\columnwidth}
    \subfile{figures/MTF/MTF_cubic/MTF1D_green_lensonly_gradient}
    \caption{\hspace*{-5em}}
    \label{sfig:MTFLensGrad}
\end{subfigure}
\caption{Top: Illustration of a typical NED system, including the viewer's eye. The viewing module consists of a 2D display and a magnifying lens. The lens focuses the display image onto a fixed virtual image plane (red line). Accommodation, on the other hand, is expected to dynamically change with respect to the distance of the virtual object, shown as the dash-lined accommodation plane. Bottom: Frequency analysis through the varying accommodation range of 0-4 diopter (D), illustrated via the MTF (left) as well as the MTF gradient (right). The display is capable of presenting high-frequency information at the virtual image plane (red line), around which the frequency response decreases rapidly.} 
\label{fig:MTFConv}
\end{figure}

\section{Problem Formulation}
\label{sec:Problem}

Understanding the focusing characteristics of a conventional NED is essential for motivating and contextualizing the proposed method. 
Fig.~\ref{fig:MTFConv} illustrates a typical NED setup, comprising a 2D display and a magnifying lens in front of the eye.
The distance between the display and the lens is set to be shorter than the lens' focal length in order to map the display onto a single virtual plane at a fixed distance, referred to as the \textit{image plane}.
The \textit{accommodation plane} refers to the 2D plane within the scene where the eye is focused at a given instant, which is dynamic and expected to follow the intended distance of the virtual object.
To analyze the retinal image quality, we calculate the modulation transfer functions (MTFs) by simulating the system responses at different accommodation states. Specifically, we vary the accommodation plane in Fig.~\ref{fig:MTFConv} (top) over a range of 0-4 D. 
We then stack the 1D cross-sections of the simulated MTFs and plot them as a function of the accommodation state.
The results are given in Fig.~\ref{fig:MTFConv} (bottom left). We calculate MTFs assuming an ideal thin lens with $\SI{30}{\mm}$ focal length and $\SI{10}{\mm}$ diameter of the eyepiece. The eye pupil diameter is set to $\SI{3.5}{\mm}$. 
The MTFs are illustrated up to 16 cycles per degree (cpd), which is the assumed bandwidth of the underlying 2D display.
As can be concluded from the figure, the frequency response is the highest and matches the display bandwidth when the accommodation is in the vicinity of the image plane. The frequency response drops significantly due to the defocus blur when the accommodation is forced to move further away from the image plane.
Defocus blur is the primary cue driving accommodation: the eye tends to accommodate at a distance where the image appears sharpest \cite{ACCAutoFocus1,ACCAutoFocus2}.
Particularly in the NED setup, the blur gradient is expected to drive the accommodation \cite{HuangLFSim}, with the maximum gradient typically occuring near the image plane (Fig.~\ref{fig:MTFConv}, bottom right). Hence, accommodation in conventional NED setups is fixated at or near the virtual image plane, regardless of the vergence distance of the object.
\begin{figure}[ht]
%\centering
\hspace{-2.5mm}
\begin{subfigure}{0.35\columnwidth}
    \subfile{figures/MTF/MTF_cubic/MTF1D_green}
    \caption{\hspace*{-4.2em}}
    \label{sfig:MTFCubic}
\end{subfigure}
\hspace{12mm}
\begin{subfigure}{0.35\columnwidth}
    \subfile{figures/MTF/MTF_cubic/MTF1D_green_gradient}
    \caption{\hspace*{-4.2em}}
    \label{sfig:MTFCubicGrad}
\end{subfigure}

\caption{The frequency analysis illustrating the effect of the wavefront coding to DoF extension in NEDs. Here we assume the cubic phase mask \cite{Dowski} as the underlying phase plate. Left: One-dimensional cross-sections of the frequency responses through the target depth range of 0-4 D. Right: The gradient of the MTFs with respect to changing depth.} 
\label{fig:MTFCubic}
\end{figure}

Defocus blur is not the only accommodation-driving factor. 
Studies have shown that binocular disparity, which is the primary cue for vergence, is also partially responsible for driving accommodation, thereby contributing to the natural coupling between vergence and accommodation in real-world viewing conditions. \cite{Disparity-Accommodation,Disparity-Accommodation2,Disparity-Accommodation3}. 
The core objective of the proposed method is to leverage the relationship between defocus blur and binocular disparity. By eliminating retinal defocus blur from the system, we aim at creating an open-loop condition \cite{KonradAI}, wherein accommodation is primarily dictated by binocular disparity rather than blur cues.
This concept can alternatively be reformulated as an extension of the display DoF. 
We illustrate such a relation in Fig.~\ref{fig:MTFCubic} using one of the well-known methods for EDoF, the wavefront coding with a cubic phase mask \cite{Dowski}.
As shown, wavefront coding enables a relatively uniform frequency response across a wide depth range, in contrast to a conventional lens. This results in a near-zero contrast gradient (Fig.~\ref{fig:MTFCubic}, right), meaning no specific depth plane is favored for accommodation, thus achieving accommodation invariance. However, as we discuss in more detail in the following section, such an approach comes with a trade-off between spatial resolution and extended depth range. Notably, the average frequency response of the wavefront coding system at high frequencies is significantly lower than that of a conventional lens focused at the virtual image plane. To mitigate this loss, wavefront coding is usually accompanied by post-processing in imaging and pre-processing in displays, to partially compensate the resolution-depth trade-off.

\begin{figure}[ht]
\centering
    \includegraphics[width=\columnwidth]{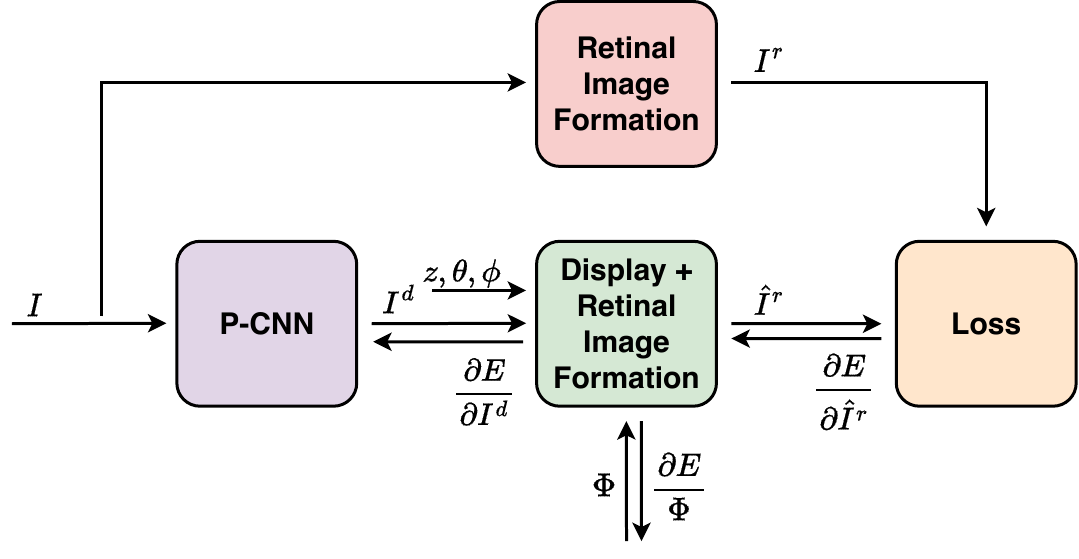}
    %\caption{\hspace*{-5em}}
\caption{The proposed end-to-end learning procedure for AI display optimization.} 
\label{fig:end-to-end}
\end{figure}

\section{Method}

We propose a model that uses an end-to-end learning framework to jointly optimize a pre-processing convolutional neural network (CNN) and a novel display optics. The pre-processing CNN, hereafter referred to as P-CNN, digitally encodes the AI image on the display. Then the novel display optics, comprising a refractive lens and a DOE at the exit pupil, optically reconstructs it. 
Fig.~\ref{fig:end-to-end} illustrates the overall learning procedure.
Assume that a viewer is to perceive a sharp input image $I$ that appears at a certain distance $z$ from the lens plane as illustrated in Fig.~\ref{fig:end-to-end}.
This distance defines the eye accommodation state, i.e. the depth the eye is focused. We use P-CNN to transform $I$ into $I^d$, which is the image that we drive the display with.
We employ a physics-based differentiable simulation model, denoted as \textit{Display + Retinal Image Formation Model} in Fig.~\ref{fig:end-to-end}, to propagate $I^d$ through the display and the viewer optics, and form an image on the retina, denoted as $\hat{I}^r$.
We compute a ground-truth retina image, $I^r$, in a parallel block denoted as \textit{Retinal Image Formation}. This simulates how the original sharp image $I$ at the accommodation distance would appear on the retina without the display. The simulation accounts for diffraction effects due to the finite pupil size and chromatic aberrations caused by the eye optics. 
Finally, in the \textit{Loss} block, we compare $\hat{I}^r$ and $I^r$ using both pixel-to-pixel and structural similarity losses. Additionally, we incorporate neural contrast sensitivity to account for perceptual factors, thus guiding the optimization toward perceptually meaningful improvements.

Our simulation model considers both the display optics and the assumed accommodation state $z$ in each iteration of the training process to preserve image quality across a range of accommodation states. 
Specifically, we search for the optimal phase profile of the DOE, denoted as $\Phi$ in Fig.~\ref{fig:end-to-end}, to enable accommodation invariance.
Upon completion of training, the learned P-CNN weights and optimized DOE parameters together define the characteristics of the proposed computational AI-display. 
This display can then create and show AI images of a 3D scene to the viewer, using the ideal (target) image of the scene as input. 
In the following sections, we detail the end-to-end learning procedure, including the image formation model, the P-CNN architecture and the loss function.

\subsection{Near-Eye Display Image Formation Model}
\label{ssec:Optics}

\begin{figure}[ht]
\centering
\begin{subfigure}{0.8\columnwidth}
    \includegraphics[clip, trim=0 0 0 0, width=\columnwidth]{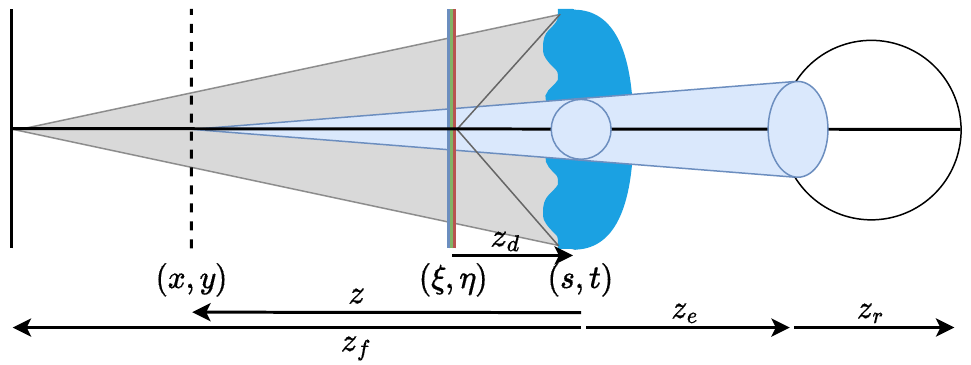}
    %\caption{}
    \label{sfig:Disp_setup}
\end{subfigure}
\begin{subfigure}{0.8\columnwidth}
    \includegraphics[clip, trim=0 0 0 60, width=\columnwidth]{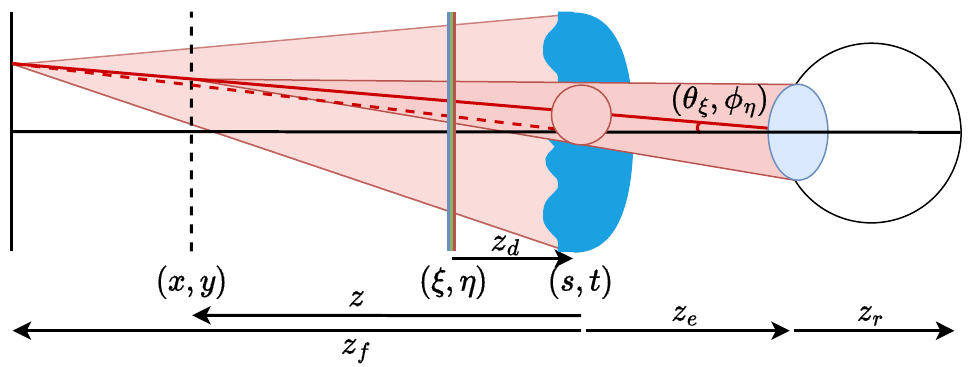}
    %\caption{}
    \label{sfig:Disp_setup_rotated}
\end{subfigure}
\caption{Near-eye display setup including the viewer. For each pixel on the display, only a subsection of the incoming light enters the retina, limited by the eye pupil. The subsection can be introduced via a virtual sub-aperture at the lens plane. The center of the sub-aperture as well as the angle of incidence to the eye, $(\theta,\phi)$, shifts with the pixel location, $(\xi,\eta)$.}
\label{fig:Display}
\end{figure}

The optical setup shown in Fig.~\ref{fig:Display}, comprises a display panel $(\xi,\eta)$ and a refractive lens-DOE pair at the lens plane $(s,t)$; the two planes being at a distance $z_d$ from each other. 
The viewer is located at a viewing distance $z_e$ from the lens plane, and focuses at a distance $z$.
Assuming a thin lens model for the eye and a planar retina, we map the retina to the accommodation plane at distance $z$, referred to as the reference plane $(x,y)$, where we form the equivalent retinal image. 
This mapping simplifies the image formation model by allowing a single wave propagation step while still considering the viewer optics. For the sake of simplicity, we derive the model in one dimension, noting that the extension to 2D is straightforward. 

The perceived image depends on both the eye's accommodation state and the pupil size.
As the pupil is smaller than the main lens, only a portion of the light emanated by a pixel can pass through it. 
We model this effect by introducing a sub-aperture at the lens plane, as illustrated in Fig.~\ref{fig:Display}.
The position of each sub-aperture is related to the pixel position at the display plane.
Specifically, a display pixel at $\xi$ is first imaged by the refractive lens to the virtual image plane at distance $z_f$, and then traced back to the eye pupil.
The incident angle of this pixel at the pupil plane, i.e., eccentricity, $\theta_\xi$, can be found via the geometric relation as
\begin{equation}
    \theta_\xi = \arctan \biggl( \frac{z_f}{z_d (z_e + z_f)} \xi \biggr).
\label{eq:theta_xi}
\end{equation}
The center of the corresponding sub-aperture at the lens plane, $s_\xi$, is then
\begin{equation}
    s_\xi = z_e \tan \theta_\xi = \frac{z_e z_f}{z_d (z_e+z_f)} \xi.
\label{eq:s_xi}
\end{equation}
The sub-aperture $A^e$ is defined as a circular function centered at $s_\xi$, i.e.,
\begin{equation}
    A^e(s;\xi) = \text{circ}\biggl(\frac{s-s_\xi}{a^e} \biggr),
    \label{eq:subaperture}
\end{equation}
with $a^e$ being the sub-aperture diameter. 
Eq.~\ref{eq:subaperture} assumes an ideal, circular-aperture thin-lens, eye model. In reality, an average eye suffers from aberrations, having direct effects on accommodation \cite{DefocusAcc}. We model these effects by defining a complex sub-aperture function $ H_\lambda^e(s;\xi) =  A^e(s;\xi) \exp(j \Phi_\lambda^e(s;\xi))$. Specifically, we incorporate the chromatic aberration in the form of defocus $\Phi^e(s;\xi) = \pi/\lambda D_\lambda (s-s_\xi)^2$, where $D_\lambda$ represents the wavelength-dependent defocus coefficient \cite{DefocusAcc,DefocusAcc2}.

Consider a point source at $\xi$, emitting monochromatic light with wavelength $\lambda$. Within the paraxial optics regime, the resulting wavefront right before the refractive lens is described as \cite{Goodman}
\begin{equation}
    U_\lambda^-(s;\xi) = \exp\biggl(\frac{j\pi}{\lambda z_d} (s-\xi)^2 \biggr).
    \label{eq:Uminus}
\end{equation}
The incoming wave $U_\lambda^-(s;\xi)$ is modified by both the refractive lens and the DOE as
\begin{equation}
\begin{multlined}
    U_\lambda^+(s;\xi) = U_\lambda^-(s;\xi) H_\lambda^e(s;\xi) A(s) \exp(j\Phi_\lambda(s)) \times
    \\  \exp(j\Phi^l_\lambda(s)),
\end{multlined}
\label{eq:UplusWavefront}
\end{equation}
where $U_\lambda^+(s;\xi)$ is the wavefront right after the refractive lens, $\Phi_\lambda(s)$ and $\Phi^l_\lambda(s)$ are the phase delays introduced by the DOE and the refractive lens, correspondingly, and $A(s)$ is the lens aperture function. The final wavefront at the reference plane $(x,y)$ subject to the point source at $\xi$, $U_{z,\lambda}(x;\xi)$, is found using Fresnel propagation as
\begin{equation}
    U_{z,\lambda}(x;\xi) \propto \mathcal{F}\biggl\{U_\lambda^+(s;\xi) \exp\biggl(-\frac{j\pi}{\lambda z}  s^2 \biggr) \biggr\}\bigg|_{\frac{x}{\lambda z}}, 
    \label{eq:coherentPSF}
\end{equation}
where $\mathcal{F}\{.\}$ is the Fourier transform operator. 
The incoherent PSF is defined as the resulting light intensity at $x$,
\begin{equation}
    h_{z,\lambda}(x;\xi) = |U_{z,\lambda}(x;\xi)|^2.
    \label{eq:PSF}
\end{equation}
Under incoherent illumination, the final 2D retinal image $\hat{I}^r_{z,\lambda}(x,y)$ reconstructed at $(x,y)$ is the superposition of the incoherent PSFs for each point source $\xi$. Denoting the image shown at the display as $I^d_\lambda(\xi,\eta)$, that is   
\begin{equation}
    \hat{I}^r_{z,\lambda}(x,y) = \int \int I^d_\lambda(\xi,\eta) h_{z,\lambda}(x,y;\xi,\eta) d \xi d \eta.
\label{eq:ShiftVarCont}
\end{equation}
Note that $h_{z,\lambda}(x,y;\xi,\eta)$ represents a shift-variant response of the retinal image formation process. Each display point is associated with a unique sub-aperture function, $H_\lambda^e(s,t;\xi,\eta)$, resulting in slightly different incoherent PSFs.
We assume that such change is negligible within a local patch around a pixel $(\xi^p,\eta^p)$, $\mathbf{P} = ([\xi^p-\epsilon,\xi^p+\epsilon],[\eta^p-\mu,\eta^p+\mu])$ 
\begin{equation}
    H_\lambda^e(s,t;\xi,\eta) \approx H_\lambda^e(s,t;\xi^p,\eta^p), \forall (\xi,\eta) \in \mathbf{P}.
\end{equation}
Then, following Eq.~\ref{eq:Uminus} through Eq.~\ref{eq:PSF}, one can show that
\begin{equation}
    h_{z,\lambda}(x,y;\xi,\eta) \approx h_{z,\lambda}(x-M\hat{\xi},y-M\hat{\eta};\xi^p,\eta^p),
\label{eq:PSF_Shiftinv}
\end{equation}
where $M=z/z_d$ is the magnification factor and $\hat{\xi} = \xi - \xi^p, \hat{\eta} = \eta - \eta^p$ are centered around $\xi^p,\eta^p$.
Inserting Eq.~\ref{eq:PSF_Shiftinv} into Eq.~\ref{eq:ShiftVarCont}, we get
\begin{equation}
    \hat{I}^r_{z,\mathbf{P},\lambda}(x,y) = \int \int I^d_{\mathbf{P}, \lambda}(\hat{\xi},\hat{\eta}) h_{z,\lambda}(x-M\hat{\xi},y-M\hat{\eta};\xi^p\eta^p) d\hat{\xi} d\hat{\eta}.
    \label{eq:ShiftInvVarConv}
\end{equation}
Let us finally define the geometric (pinhole) mapping of the display image to the reconstruction plane, $\Tilde{I}^d(M\xi,M\eta) = I^d(\xi,\eta)$. 
Replacing $\Tilde{I}^d$ into Eq.~\ref{eq:ShiftInvVarConv}, we obtain the shift-invariant approximation of Eq.~\ref{eq:ShiftVarCont}
\begin{equation}
    \hat{I}^r_{z,\mathbf{P},\lambda}(x,y) = \frac{1}{M} \Tilde{I}^d_{\mathbf{p}, \lambda}(x,y) \ast h_{z,\lambda}(x,y;\xi^p,\eta^p).
    \label{eq:Forwardmodel}
\end{equation}
We use the shift-invariant approximation in the forward training pass to calculate the retinal images for each training input patch.
We arrange the sub-apertures corresponding to different local patches in such a way that they cover the entire main lens aperture. They can be overlapping or non-overlapping (perfectly tiling) subapertures. In any case, we set the sub-aperture size according to the eye pupil size, as shown in Fig.~\ref{fig:Display}. 

We train the network with color (RGB) images. This accounts for three distinct wavelength values for each branch in Fig.~\ref{fig:end-to-end}. The ground-truth data is generated by applying a separate retina model to the input sharp image, which only considers the viewer optics.
Specifically, we calculate the PSF with respect to the viewer, $h^e_{\lambda}(x,y)$, as
\begin{equation}
    h^e_{\lambda}(x,y) = \mathcal{F} \{ H_\lambda^e(s,t;0,0) \},
\end{equation}
wherein we incorporate the chromatic aberrations. Its convolution with the input image results in the ground-truth retinal image,
\begin{equation}
    I^r_{z,\lambda} (x,y) = I_\lambda (x,y) \ast h^e_{z,\lambda}(x,y).
\end{equation}

\subsubsection{DOE Parametrization}
We define the phase transmission function of the DOE, $\Phi(s,t)$, as a set of discrete samples that serve as the optimization parameters of the optical system within the display module.
To ensure that the optimized phase profile can be physically fabricated, we model the DOE using the height map of the material that is used for fabrication, i.e., $d(s,t)$. 
The mathematical relation between the height map, $d(s,t)$, and the wavelength-dependent phase delay, $\Phi_\lambda(s,t)$, is expressed as
\begin{equation}
    \Phi_\lambda(s,t) = \frac{2\pi}{\lambda} (n_\lambda - 1) d(s,t),
    \label{eq:htophi}
\end{equation}
where $n_\lambda$ is the wavelength-dependent refractive index. 
One option is to optimize a single height map and use Eq.~\ref{eq:htophi} to compute the phase delay for each color channel at each iteration. 
Another option is to choose a phase delay parameter for a single color $\Phi_{\lambda_0}$, at a nominal wavelength $\lambda_0$, and then derive the other color channels as
\begin{equation}
    \Phi_\lambda(s,t) = \frac{\lambda_0(n_\lambda - 1)}{\lambda(n_{\lambda_0} - 1)}  \Phi_{\lambda_0}(s,t).
    \label{eq:phi0tophi}
\end{equation}
This approach can improve the numerical stability of the results, since the height map has values in the micrometer range, while the phase mask has values in the $2\pi$ range.
In our model, we optimize $\Phi_{\lambda_0}(s,t)$.
Note that Eq.~\ref{eq:phi0tophi} is a general relation that can also be applied to other phase elements, such as the main refractive lens with chromatic aberrations.

In our previous work, we have proposed an optimal sampling strategy for the DOE that significantly reduces the number of optimization parameters \cite{EDOFAkpinar}.
We use the same formulation here, which can be briefly summarized as follows. According to Eq.~\ref{eq:coherentPSF}, the wavefront after the lens, $U_\lambda^+(s;\xi)$, and the coherent PSF, $U_{z,\lambda}(x;\xi)$, are related by a Fourier transform. To accurately capture this relationship and avoid aliasing, the sampling must satisfy the Nyquist criterion. 
By combining Eq.~\ref{eq:Uminus} and \ref{eq:UplusWavefront} with Eq.~\ref{eq:coherentPSF}, we obtain a second order chirp expression with the aperture functions and the phase terms of the DOE.
The theoretical maximum spatial frequency of this chirp function, $\omega_{\lambda,z}$, is proportional to its instantaneous frequency at the aperture radius $r$ that is given by the first-order derivative of its phase,
\begin{equation}
    \omega_{z,\lambda} = \frac{2 \pi}{\lambda}\biggl(\frac{1}{z_d} - \frac{1}{f_\lambda^l} - \frac{1}{z} + D_\lambda \biggr) r
    \label{eq:omegachirp}
\end{equation}
where $f_\lambda^l$ is the wavelength-dependent focal length of the underlying refractive lens. 
The required minimum sampling rate is then found as $\Delta_s = \pi / 4 \max\limits_{z,\lambda}\{|\omega_{z,\lambda}|\}$ \cite{EDOFAkpinar}.

To further decrease the number of optimization parameters, we model the DOE to be rotationally symmetric, i.e.
\begin{equation}
    \Phi(s,t) = \Phi(\sqrt{s^2+t^2}),
\end{equation}
with $(s,t)$ being the 2D coordinates at the lens plane. This choice is intuitive because the defocus aberration itself is rotationally symmetric.

\subsection{Pre-processing}

The AI display's image quality depends on the interplay between the optics and the pre-processing algorithm. 
This is analogous to the EDoF post-processing in sensing, where the system PSF deblurs the sensor image. 
The goal of the pre-processing in our method is to counteract the blurring optical effects beforehand, so that fine details are preserved after light passes through the optics.
However, in addition to the space-bandwidth limitations, the display pre-processing is limited also by the display dynamic range, which it must fit.
This restricts the available set of solutions.

We employ a standard U-net architecture for the pre-processing stage \cite{Unet}. This encoder-decoder network consists of multiple layers: each encoder layer applies convolution followed by a rectified linear unit (ReLU) activation, and then downsamples the feature map by max pooling. 
In the decoder, each layer upsamples the feature maps by transposed convolution. 
To preserve spatial detail, skip connections concatenate the output of each encoder layer with the corresponding decoder layer.

We modify the standard U-net to account for the variations in the PSFs as the pixel location changes.
Specifically, the input image patch is augmented with the pixel coordinates $(\xi,\eta)$, which change according to the position of the patch on the display plane at each iteration.
This way, the network can process different parts of the display image differently, creating a position-aware pre-processing.
As a result, the input to the modified U-net consists of five channels: the RGB image concatenated with the $\xi$ and $\eta$ coordinate maps. The network outputs three color channels. 
The output of the modified U-net is added to the original image patch at the end to obtain the display image $I^d$.

\begin{figure}[ht]
%\centering
\hspace{-2.5mm}
\begin{subfigure}{0.40\columnwidth}
    \subfile{figures/NTF}
    \caption{\hspace*{-5em}}
    \label{sfig:NCSF}
\end{subfigure}
\hspace{11.5mm}
\begin{subfigure}{0.40\columnwidth}
    \subfile{figures/NTF_1D}
    \caption{\hspace*{-4.2em}}
    \label{sfig:PSF1DCont}
\end{subfigure}
\caption{Neural contrast sensitivity function as adopted from \cite{WatsonNCSF}.} 
\label{fig:NSCF}
\end{figure}

\subsection{Loss Function}

Before applying the loss functions, we process both the target image $I^r$ and the network output $\hat{I}^r$ with the neural contrast sensitivity function (NCSF) to incorporate perceptual factors into training \cite{WatsonNCSF,NCSF}. Fig.~\ref{fig:NSCF} shows the NSCF used in this work, as adopted from \cite{WatsonNCSF}. One benefit of NSCF is that it helps optimize the trade-off between spatial resolution and DoF by emphasizing certain frequencies. Fig.~\ref{fig:NSCF} reveals two main features of NSCF. First, its frequency response peaks around 10 cpd, unlike the low-pass behaviour of a typical MTF. Second, the sensitivity drops for the oblique frequencies, reflecting the orientation-selectivity of the human visual system (HVS). We integrate NCSF into the system by filtering in the frequency domain
\begin{equation}
    \begin{aligned}
        I^N_{z,\lambda} (x,y) &= \mathcal{F}^{-1} \{\mathcal{F} \{I^r_{z,\lambda} (x,y)\} N(\hat{x},\hat{y}) \} \\
        \hat{I}^N_{z,\lambda} (x,y) &= \mathcal{F}^{-1} \{\mathcal{F} \{\hat{I}^r_{z,\lambda} (x,y)\} N(\hat{x},\hat{y}) \},      
    \end{aligned}
\end{equation}
where $I^N_{z,\lambda},\hat{I}^N_{z,\lambda}$ are the target and output neural images, respectively, $N(\hat{x},\hat{y})$ is the NSCF, and $\hat{x},\hat{y}$ are the spatial frequencies. The overall network loss is
\begin{equation}
    E(I^N,\hat{I}^N) = \mathcal{L}_{l_1}(I^N,\hat{I}^N) + \mathcal{L}_{ssim}(I^N,\hat{I}^N),
\label{eq:loss}
\end{equation}
where $\mathcal{L}_{l_1}(I^N,\hat{I}^N)$ is the L1-loss, and $\mathcal{L}_{ssim}(I^N,\hat{I}^N)$ is the SSIM-loss \cite{SSIM}
\begin{equation}
    \mathcal{L}_{ssim}(I^N,\hat{I}^N) = 1 - \text{SSIM}(I^N,\hat{I}^N).
    \label{eq:ssim}
\end{equation}
Since our goal is to provide equally sharp images within the depth range of interest, we use a per-pixel loss that favors sharpness (L1-loss) \cite{L1loss}. Furthermore, we also include the SSIM loss to maintain the perceived structural image quality.

\section{Simulations}
\label{sec:Simulations}

We train the proposed model with the following display parameters. 
We assume a plano-convex refractive lens with a focal length of $f_{\lambda_s} = \SI{30}{\mm}$ for the specification wavelength of $\lambda_s = \SI{587.6}{\nm}$.
We use a single wavelength for each color channel of the display: $\lambda_r = \SI{630}{\nm},\lambda_g = \SI{525}{\nm},\lambda_b = \SI{458}{\nm}$.
The refractive lens is made of silica, with the refractive indices of $n^l_{\lambda_r}=1.457, n^l_{\lambda_g}=1.461, n^l_{\lambda_b}=1.465$.
We include the corresponding color aberration in the system by using the wavelength-dependent refractive indices and Eq.~\ref{eq:phi0tophi}.
We also model the spherical aberration by using the spherical height profile of the lens, which has a central thickness of $\SI{2.90}{\mm}$ and a radius of $\SI{13.75}{\mm}$.
We set the lens-to-display-distance as $z_d = \SI{28.2}{\mm}$, to focus the green channel at $z_f^g = 2$ D away from the lens plane.
The lens aperture is $\SI{10}{\mm}$, with an f-number of 3. 
The 2D display plane has a pixel pitch of $\Delta_\xi = \SI{15}{\micro\metre}$, resulting in a resolution of $\approx 16$ cpd.

As explained in Sec.~\ref{ssec:Optics}, we design the DOE to have rotational symmetry. 
We select the virtual sub-apertures described in Sec.~\ref{ssec:Optics} from a discrete set of non-overlapping sub-apertures. 
To cover the whole lens aperture without any gaps in between, we divide the main lens into hexagonal tiles during training.
The outer diameter of each tile is $a^e =\SI{3.5}{\mm}$, which represents an average eye pupil size. This results in 19 distinct sub-apertures within the main lens.
We use Eq.~\ref{eq:s_xi} to calculate the eccentricity range for each sub-aperture region, which is about $[\SI{- 5}{\degree}, \SI{5}{\degree}]$ for an eye relief of $z_e=\SI{18}{\mm}$.
During testing, we use circular sub-apertures to simulate the perceived images, matching the eye pupil shape. 

We train the network with TAU Agent \cite{TAUDataset}, a stereo RGB-D dataset created from the open-source animated movie Agent 327 in the 3D animation software Blender \cite{Blender}.
The dataset contains 525 high-quality RGB images and their depth maps.
We use synthetic data to control the noise and also due to its suitability for VR.
We divide the images into patches of 256$\times$256 pixels and use a batch size of 3. We reserve $10\%$ of the data for validation.
For each training instance, we randomly sample the accommodation state from a uniform distribution within the scene depth range in diopters. 
We augment the training data by passing each image through all the predefined sub-aperture regions.
We train the network for 8 epochs with Adam optimizer \cite{AdamOptimizer}, setting the learning rate, the first decay rate, the second decay rate, and the weight decay to 1e-3, 0.9, 0.999, and 1e-4, respectively.

\begin{figure}[ht]
%\centering
\hspace{-5mm}
\begin{subfigure}{0.30\columnwidth}
    \subfile{figures/heightmapdraw}
    \caption{\hspace*{-5em}}
    \label{sfig:HmapCont}
\end{subfigure}
\hspace{7mm}
\begin{subfigure}{0.30\columnwidth}
    \subfile{figures/MTF/PSF1D_wo_network}
    \caption{\hspace*{-4.2em}}
    \label{sfig:PSF1D}
\end{subfigure}
\hspace{2.7mm}
\begin{subfigure}{0.30\columnwidth}
    \subfile{figures/MTF/PSF1D_lensonly}
    \caption{\hspace*{-4.2em}}
    \label{sfig:PSF1D_lensonly}
\end{subfigure}
\caption{The optimized height map at the fabrication resolution of \SI{3}{\um} (left), one-dimensional cross-sections of the on-axis PSFs at various depths (middle), one-dimensional PSFs using only refractive lens (right).} 
\label{fig:Hmap}
\end{figure}

We set the sampling rate of the DOE and the lens plane during training to $\Delta_s=\SI{5}{\micro\metre}$ following the optimal sampling requirements in Sec.~\ref{ssec:Optics}.
After training, we upsample the DOE profile to the fabrication resolution of $\Delta_s^f=\SI{3}{\micro\metre}$ using bicubic interpolation.
Fig.~\ref{fig:Hmap} shows the optimized DOE at the fabrication resolution. 
We also present one-dimensional cross section of the optimized PSF within the training depth range of 0-4 D. For comparison, the corresponding PSF produced by the refractive lens alone over the same depth range is shown on the right side of Fig.\ref{fig:Hmap}. The proposed method yields significantly narrower PSF outside the lens DoF, demonstrating the EDoF and, consequently, accommodation-invariance.

\begin{figure}[ht]
%\centering
\hspace{2.5mm}
\begin{subfigure}{0.35\columnwidth}
    \subfile{figures/MTF/MTF_1D_green_lensonly_lines}
    \caption{\hspace*{-5em}}
    \label{sfig:MTFSpherical}
\end{subfigure}
\hspace{2mm}
\begin{subfigure}{0.35\columnwidth}
    \subfile{figures/MTF/MTF_1D_green_lines}
    \caption{\hspace*{-4.2em}}
    \label{sfig:MTFProposed}
\end{subfigure}

\vspace{-2.5mm}
\hspace{4mm}
\begin{subfigure}{0.35\columnwidth}
    \subfile{figures/MTF/MTF_plot_0_5D}
    \caption{\hspace*{-5em}}
    \label{sfig:MTFSpherical_3D}
\end{subfigure}
\hspace{2mm}
\begin{subfigure}{0.35\columnwidth}
    \subfile{figures/MTF/MTF_plot_3D}
    \caption{\hspace*{-4.2em}}
    \label{sfig:MTF_3D}
\end{subfigure}
\caption{Modulation transfer function (MTF) analysis. Top: One-dimensional cross-sections of MTFs throughout the target depth range. The red line represents the display bandwidth, while the green and blue lines illustrate MTFs at 0.5 D and 3 D, respectively. The gray curve maps the cut-off frequencies at each depth for an MTF threshold of 0.1. Bottom: One-dimensional MTF plots for 0.5 D (left) and 3 D (right). The black curve indicates the frequency-dependent contrast threshold map of HVS, calculated as the reciprocal of the CSF. The dashed curves correspond to the conventional display and the solid curves to the proposed method.} 
\label{fig:MTFAnalysis}
\end{figure}

\subsubsection{MTF Analysis} We use frequency analysis to further examine the effectiveness and limitations of our method, considering the AI and spatial resolution.
Fig.~\ref{fig:MTFAnalysis} shows the stack of MTFs for different accommodation distances in the scene depth range, and the one-dimensional plots at two out-of-focus depths (0.5 D and 3 D). The dashed curves are for the conventional approach, and the solid curves are for our method.
We also plot the cut-off frequencies for a contrast threshold of 0.1 (gray curve, Fig.~\ref{fig:MTFAnalysis}, top row).
As expected from the spatio-angular resolution trade-off discussed in Sec.~\ref{sec:Problem}, our method exhibits a more uniform frequency response across depth, albeit with reduced spatial resolution near the focus plane.
Notably, the conventional method's response drops sharply at around 5 cpd for both 0.5 D and 3 D, whereas our method maintains a relatively flat response. 
We also plot the estimated contrast threshold map of the HVS \cite{CAProjection}, which is the minimum contrast needed to detect each frequency component (black plot).
The threshold map is the inverse of the contrast sensitivity function (CSF), which is the overall sensitivity of the HVS to different spatial frequencies \cite{DalyCSF,BartenCSF}.
We use Barten's CSF model \cite{BartenCSF}, with maximum and minimum display luminances of $\SI{200}{\candela/\metre\squared}$ and $\SI{0.04}{\candela/\metre\squared}$, respectively, for a contrast ratio of 5000:1.
Importantly, our method's MTF remains above the threshold map up to the display bandwidth at both tested depths, indicating perceptually sufficient frequency content.
The gray curves in Fig.~\ref{fig:MTFAnalysis} show the display cut-off frequency with respect to the MTF threshold of 0.1, which is a common means for resolution analysis. The cut-off frequency is about 16 cpd for an approximate accommodation range of 0.5 D to 3 D, which matches the display bandwidth. For accommodation states outside the range, the cut-off frequency decreases to a minimum frequency of 8 cpd at 4 D.

\begin{figure}
    \centering
    \hspace{-10mm}
    \begin{subfigure}{0.3\columnwidth}
        \subfile{figures/MTF/Viewing_angle/MTF1D_green_th-10}
        %\caption{\hspace*{-5em}}
        \label{sfig:MTF_Viewing_angle_th-10}
    \end{subfigure}
    %\hspace{12mm}
    \begin{subfigure}{0.3\columnwidth}
        \subfile{figures/MTF/Viewing_angle/MTF1D_green_th0}
        %\caption{\hspace*{-4.2em}}
        \label{sfig:MTF_Viewing_angle_th0}
    \end{subfigure}
    \begin{subfigure}{0.3\columnwidth}
        \subfile{figures/MTF/Viewing_angle/MTF1D_green_th10}
        %\caption{\hspace*{-5em}}
        \label{sfig:MTF_Viewing_angle_lens_th10}
    \end{subfigure}
    \caption{Spatially-variant MTFs for the proposed AI-NED with respect to changing eccentricities, $\theta_\xi$, within the sub-aperture regions of $\SI{\pm 5}{\degree}$ centralized at $\SI{-10}{\degree}$ (left), $\SI{0}{\degree}$ (middle), and $\SI{10}{\degree}$ (right). For each eccentricity value, the MTF is calculated at the virtual image depth of 2 D.}
    \label{fig:MTF_Viewing_angle}
\end{figure}

\subsubsection{Eccentricity-Dependence} As discussed in Section \ref{ssec:Optics}, the proposed NED model is inherently shift-variant, meaning that the PSF changes slightly with lateral shifts in pixel position, due to changes in the corresponding lens sub-apertures. 
We examine how much the spatial variance affects our proposed method and test the validity of the shift-invariant PSF approximation we use for training. 
To do this, we stack the one-dimensional MTF cross-sections for varying pixel positions along the horizontal axis $\xi$. We plot the MTFs against the eccentricity, $\theta_\xi$, which we get from the pixel location using Eq.~\ref{eq:theta_xi}.
Fig.~\ref{fig:MTF_Viewing_angle} shows the results. We choose the green channel at the virtual image depth of 2 D, where the refractive lens is focused and select three consecutive sub-aperture regions along the horizontal axis to cover the full lens aperture.
The total field-of-view (FoV) is about $\SI{30}{\degree}$.
As Fig.~\ref{fig:MTF_Viewing_angle} shows, our method exhibits a fairly flat response in the central sub-aperture, which agrees with the locally shift-invariant PSF approximation.
However, we observe a slight drop in MTFs as the eccentricity moves away from the center.
We also note that our pre-processing depends on the lateral position as well, as the optics and the variations in MTF across the FoV are to be partially compensated by the pre-processing.
Further improvements could be achieved by recalibrating and retraining the pre-processing module using recorded PSFs.

\subfile{figures/figure_compare_soa_Depthim_rect}

\subsubsection{Comparison with the state-of-the-art} Fig.~\ref{fig:compare_soa} compares our method with two alternatives: a conventional stereoscopic display and a state-of-the-art AI-NED that employs focus-tunable lenses, as proposed in \cite{KonradAI}. 
We use a synthetic test image from \cite{TAUDataset} and evaluate performance across multiple accommodation depths. 
The conventional display is simulated using a single refractive lens that has a spherical height profile as the imager.
The method of \cite{KonradAI} is simulated in a discrete mode, where the focus-tunable lens focuses on a discrete set of depth planes at 0, 1, 2, 3, and 4 D to maximize resolution within the target depth range, as suggested in the authors' implementation.
The results are given for five accommodation depths between 0-4 D.
Our model achieves better performance for a larger accommodation depth range than the conventional method, which is especially noticeable at the near and far ends of the target depth range.
Due to the inherent trade-off between resolution and depth, the refractive lens-only setup produces a higher-quality image at the image depth of the main lens.
Overall, the AI display with a focus-tunable lens and the proposed method achieve comparable visual quality at the near and far ends of the target depth range, both successfully extending the DoF. The latter demonstrates noticeably better performance around the central depth of 2 D.
In terms of objective image quality metrics, our method consistently outperforms the approach in \cite{KonradAI} across nearly the entire target depth range, achieving higher values in both PSNR and SSIM.

\subfile{figures/images/PCNN_Output_v2}

\subsubsection{Pre-Processing}

Fig.~\ref{fig:PCNN_Output} qualitatively demonstrates the impact of the pre-processing network using a set of images from various datasets \cite{TAUDataset,MPISintel}. 
The figure compares the results of the end-to-end algorithm with and without the pre-processing module for two different accommodation depths: 2 D and 3 D. 
As shown, the pre-processing module helps to produce sharper retinal images at both depths. The degree of enhancement varies depending on the scene content, particularly in terms of spatial frequency and color composition. For instance, the second input scene in Fig.~\ref{fig:PCNN_Output} exhibits a more noticeable improvement than the first scene.

\subfile{figures/HeightNoise2D/heightnoise}

\subsubsection{Noise Analysis} We also analyze how the DOE fabrication inaccuracies affect the image quality.
We model these inaccuracies by adding a zero-mean i.i.d. Gaussian noise to the DOE height profile at various noise standard deviation levels $\sigma_d$. 
Fig.\ref{fig:Hnoise} shows the results for a test image from \cite{TAUDataset}.
Despite the fact that no fabrication noise is considered during training, our method remains robust to inaccuracies up to $\sigma_d = \SI{40}{\nm}$.
The PSNR drops by 1.7 dB at most, however the perceived image quality and SSIM values do not change much.

\subfile{figures/broadband/figure_broadband}

\subsubsection{Perceptual Comparison} PSNR is a common metric to measure the quality of reconstructed images, however it does not fully reflect how humans perceive them \cite{PerceptualLosses}.
Therefore, we use a more advanced metric, referred to as HDR-VDP-2 \cite{HDRVDP2}, which accounts for both display characteristics, such as dynamic range and spectral emission, and neural factors influencing visual perception.
To adapt this metric to our display system, we modify its initial step, which models the optical and retinal pathways.
Specifically, we replace the intra-ocular light scatter block used in \cite{HDRVDP2} with the MTF derived from our retinal image formation model (see Sec.~\ref{ssec:Optics}. Additionally, we use a dense set of wavelengths to better match the spectral sensitivity of human photoreceptors. 
The resulting multispectral retinal image is obtained from the three-channel display image and the emission spectra of the display's color channels.
Fig.~\ref{fig:HDRVDP} shows the results of this metric for our method and the conventional method. We use the same test image \cite{TAUDataset}, zoomed in on the face of the character. The metric produces a map of the probability of seeing artifacts in each image. The brighter regions indicate higher probability of visible artifacts.
In this context, the dominant artifact is blur, so the map effectively highlights regions where blur is visually detectable.
As shown, the conventional method introduces significantly more blur as the accommodation shifts to 4 D, making it impossible for the eye to accommodate at such distances.
In contrast, our method maintains low visibility of artifacts across a broader depth range that is an indication of the accommodation invariance to the retinal blur.

\section{Experiments}

\begin{figure}
    \centering
    \begin{subfigure}{\columnwidth}
        \includegraphics[clip, trim=0 0 0 0, width=\columnwidth]{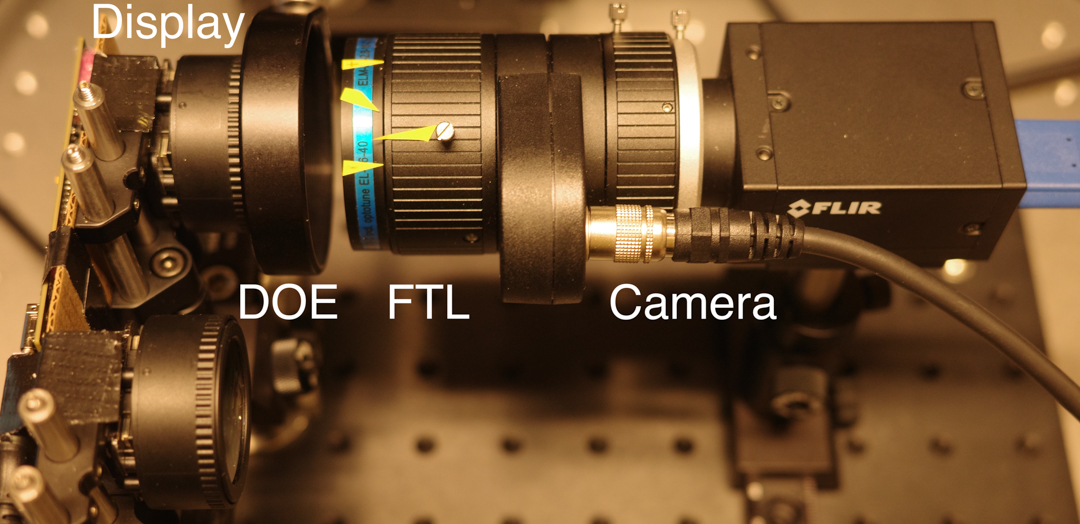}
    \end{subfigure}
    \caption{Experimental setup.}
    \label{fig:Experiment_setup_photo}
\end{figure}

We evaluate the proposed algorithm through a benchtop setup as shown in Fig.~\ref{fig:Experiment_setup_photo}. 
The display module consists of a $2560 \times 2560$ resolution micro-OLED with pixel pitch of $\SI{7.22}{\micro\metre}$ and a lens assembly that supports focal depth adjustment.
To emulate the human eye, we employ a focus-tunable lens from Optotune (ELM-25-2.8-18-C, 25 mm C-mount lens with EL-16-40 integrated) together with an RGB sensor of 5 Megapixels from FLIR (GS3-U3-51S5C-C).
In this setup, we adjust the focal power of the focus-tunable lens to simulate different accommodation states. By varying the lens optical power, we effectively shift the emulated accommodation distance for each experimental condition, enabling a controlled evaluation of the system’s depth-dependent performance. 

\begin{figure}[ht]
%\centering
\hspace{-3.5mm}
\begin{subfigure}{0.44\columnwidth}
    \subfile{figures/HeightMapMeasured}
    %\caption{\hspace*{-5em}}
    \label{sfig:DOEmicrograph}
\end{subfigure}
%\hfill
\begin{subfigure}{0.45\columnwidth}
    \subfile{figures/HeightErrorplot}
    %\caption{\hspace*{-4.2em}}
    \label{sfig:ErrorplotDOE}
\end{subfigure}
\caption{DOE Fabrication. Left: the optical micrograph of the fabricated DOE created by stitching multiple high-resolution images taken with a high-magnification objective in a widefield optical microscope. Right: measured height values (blue) in comparison with the ideal profile (red).} 
\label{fig:DOE_fabrication}
\end{figure}

\subsubsection{DOE Fabrication}

The DOE is fabricated using a grayscale lithography technique. A soda lime glass is spin-coated with Hexamethyldisilazane (HDMS) at 1000 RPM for  $\SI{1}{\minute}$, then coated immediately with photoresist S1813, which is spun on at 800 RPM for  $\SI{1}{\minute}$. The photoresist-coated wafer is then soft-baked on a hotplate. After the sample sits overnight, a laser-pattern generator (Heidelberg DWL66+) writes the design onto the sample with its 256 available gray levels. The exposed sample is baked on a \SI{50}{\celsius} hotplate for  $\SI{1}{\minute}$, developed in an AZ 1:1 solution for  $\SI{1}{\minute}$ and $\SI{12}{\second}$, then rinsed in DI water. A microscope-stitched image of the sample is shown in Fig.~\ref{fig:DOE_fabrication}, left. Before the pattern is generated, a calibration sample (prepared and developed in the same way and at the same time as the sample previously mentioned) is exposed and developed to map the photoresist depths for corresponding laser intensities from the pattern generator. 

The resulting pattern consists of 1,667 concentric rings where the rings are each $\SI{3}{\micro\metre}$ wide and the height of each ring is not uniform around its circumference. The pattern diameter is $\SI{10.002}{\mm}$, and the maximum height is $\SI{1.48}{\micro\metre}$. The ring heights of several of the outermost rings were measured by an Olympus LEXT OLS5000 microscope, and the resulting profile is plotted against a profile cross-section of the ideal heights of the design in Fig.~\ref{fig:DOE_fabrication}, right, where the blue line is the measured height of the features of the device and the red is the ideal height profile. The average and maximum differences between the measured and ideal heights for this profile are $\SI{54}{\nm}$ and $\SI{116}{\nm}$, respectively. The standard deviation of the fabrication inaccuracies is found to be $\SI{30.9}{\nm}$.

\begin{figure}
    \centering
    \hspace{-8mm}
    \begin{subfigure}{0.38\columnwidth}
        \subfile{figures/MTF/Experiments_Slanted_edge/MTF_Lens_SeeYa_v2}
        \caption{\hspace*{-5em}}
        \label{sfig:Exp_MTFSpherical}
    \end{subfigure}
    \hspace{12mm}
    \begin{subfigure}{0.38\columnwidth}
        \subfile{figures/MTF/Experiments_Slanted_edge/MTF_DOE_PCNN_SeeYa_v2}
        \caption{\hspace*{-4.2em}}
        \label{sfig:Exp_MTF_DOE}
    \end{subfigure}

    \caption{MTF plots derived experimentally by the slanted edge method at different depths for the conventional (left) and the proposed (right) methods.}
    \label{fig:Experiment_MTF}
\end{figure}

\subfile{figures/Exp_figure_compare_soa_SeeYa_v2}

\subsubsection{MTF Analysis}

We measure the MTFs of both the proposed AI display and the conventional lens-only display using the slanted edge method \cite{SlantedEdge}. 
The results are shown in Fig.~\ref{fig:Experiment_MTF}.
Due to the limited diopter adjustment range of the focus-tunable lens, we set the virtual image plane of the display at 3 D and perform measurements across the depth range of 1-5 D.
While deriving the MTF plots and conducting the subsequent experiments, we utilize half of the available display bandwidth to match the display resolution to the target resolution used during training. The resulting maximum display resolution is around 13 cpd. 
The proposed method exhibits a relatively consistent spatial frequency response across the target depth range. The cut-off frequencies that can provide a contrast value of 0.1 (gray lines in Fig.~\ref{fig:Experiment_MTF})) are chosen to characterize the available display resolution. The figure reveals that the proposed method achieves an average resolution of around 8 cpd, with a minimum resolution of around 6 cpd for the accommodation depth of 2 D. 
The conventional display delivers the maximum available display resolution of 13 cpd, but only for the accommodation depth of 3 D, and except the accommodation depth of 2 D, all other accommodation depths are supported with much lower resolution (i.e., around 2-4 cpd). Our method is seemingly subject to an average resolution drop of 38.5\% compared to the conventional stereo display case, where, however the viewer is assumed to always accommodate at the virtual image plane (best case in terms of spatial resolution) and experience VAC.

\subsubsection{Qualitative Inspection}

To compare the image quality of the conventional and the proposed method, we conduct another experiment using color images from various datasets. 
Fig.~\ref{fig:Exp_compare_soa} shows the captured images. As the camera focuses away from the virtual image plane of 3 D, the conventional setup causes significant blurring, while the proposed method maintains a more consistent image quality across wider range of focus states.
The conventional display produces a sharper image at the virtual image than our method, which reflects the trade-off discussed earlier. 

We encourage the reader to also view the supplementary video where we exemplify continuous refocusing and demonstrate the accommodation invariance across several scenes. The effect is particularly pronounced in a text-based scene.

\section{Discussion}

In this section, we discuss some limitations of our method and propose directions for future research. 

\subsubsection{DoF-resolution Trade-off} The presented theoretical analysis and experiments confirm the inherent trade-off between DoF and spatial resolution.
This trade-off can be further manipulated through the design of the so-called multifocal-mode AI NED, where the aim is to create distinct focal planes instead of a continuous DoF extension. The HVS can tolerate vergence–accommodation mismatches of up to 0.5 D within the so-called zone of comfort \cite{PercivalZoneofComfort, Shibata2011}. Leveraging this tolerance, multifocal-mode AI NEDs have been demonstrated beneficial for improving the resolution at the dedicated focal depths, for the price of degraded images at intermediate depths \cite{KonradAI,MultifocalAI}.

\subsubsection{Field-of-view} Our current AI NED design facilitates an eyepiece with an aperture diameter of $\SI{10}{\mm}$. This limits the FoV to approximately $\SI{30}{\degree}$. An extension of the current architecture to larger FoV designs would require a more rigorous framework for image formation to account for non-paraxial modeling. The primary challenge is to manage the increased computational complexity associated with such modeling.

\subsubsection{Viewer Optics} The current formulation assumes a fixed eye pupil position located at a fixed viewing distance and aligned with the optical axis of the display. In practice, the eye is subject to rotation and shifts due to the differences between the interpupillary distances of individuals. We plan to explore such changes and their effects on the optimized display in future work. 

\subsubsection{Perceptual Assessment} We demonstrate the effectiveness of our AI NED with simulations and optical measurements. The ultimate way to show how AI displays can overcome VAC is to conduct well-planned and properly executed user tests with human subjects. Future work will include controlled experiments with human participants to measure accommodation responses across different image depths within the targeted depth ranges. In addition to objective accommodation measurements, we also aim to incorporate subjective assessment of visual comfort to comprehensively evaluate the perceptual benefits of our architecture.

\section{Conclusions}

This work has demonstrated the potential of a DOE-based NED architecture to address the VAC inherent in conventional 3D displays. 
We have proposed a novel AI-NED design that aims to eliminate retinal defocus blur and couple accommodation with vergence, relying solely on binocular disparity.
We have shown that this objective can be effectively formulated as a DoF extension problem, which can be addressed by a wavefront coding approach.
The proposed method leverages wavefront coding to co-optimize a novel DOE design for providing accommodation invariance and a pre-processing module to further improve the perceived image quality.
A key advantage of this method is the use and optimization of \textit{static optics}, eliminating the need of complex adaptive optics or gaze tracking.
Through simulations and a benchtop setup, we have demonstrated that the proposed architecture can extend the DoF for up to four diopters. 

At the current stage of deployment, we quantify the image quality provided by the benchtop setup through a focus-tunable camera that emulates the human eye's accommodation response.  
Our next steps include the development of a wearable prototype and the execution of user studies to objectively measure accommodation responses and subjectively assess visual comfort.
Additionally, we plan to investigate how end-to-end optimization and wavefront coding can be extended to address other challenges of existing NEDs, such as achieving a wide field-of-view, integrating a large eyebox, and enabling immersive visualization.

\section*{Acknowledgment}

We thank Lauri Varjo and Johan Rengstedt for their valuable assistance with the benchtop experimental setup and execution, and Mehmet Ugur Gudelek for his insightful discussions on code optimization.
This work is supported in part by the Academy of Finland research project “Modeling and Visualization of Perceivable Light Fields”, under Grant 325530. Funding from Office of Naval Research DURIP Award: N00014-19-1-2458 and N00014-22-1-2014 are acknowledged.


\ifCLASSOPTIONcaptionsoff
  \newpage
\fi



%\bibliographystyle{IEEEtran}
%\begin{thebibliography}{1}


%\end{thebibliography}

\bibliographystyle{IEEEbib}
\bibliography{references}


%\subfile{figures/Authors/AuthorBiography}


%\newpage


%\vfill


\end{document}

%% file: figures/Literature/figure_Literature.tex
\begin{figure*}[ht]
\centering
\begin{subfigure}{0.125\textwidth}
    \includegraphics[width=\textwidth]{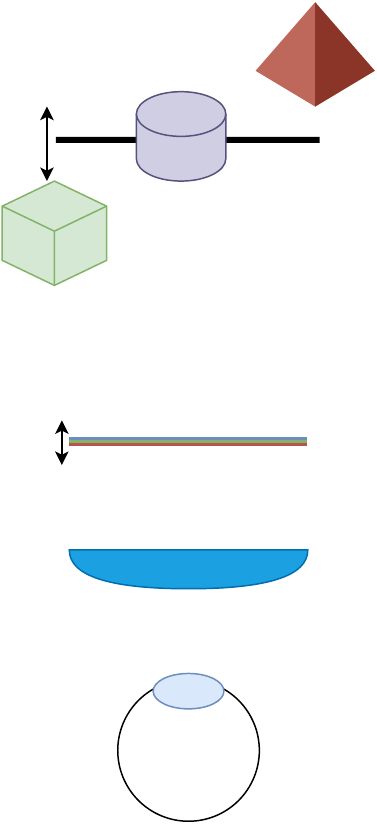}
    \caption{Varifocal}
    \label{sfig:VarifocalNED}
\end{subfigure}
\begin{subfigure}{0.125\textwidth}
    \includegraphics[width=\textwidth]{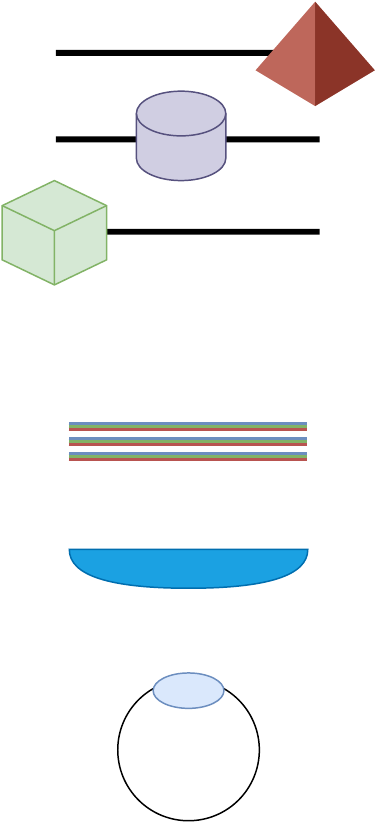}
    \caption{Multifocal}
    \label{sfig:MultifocalfocalNED}
\end{subfigure}
\begin{subfigure}{0.125\textwidth}
    \includegraphics[width=\textwidth]{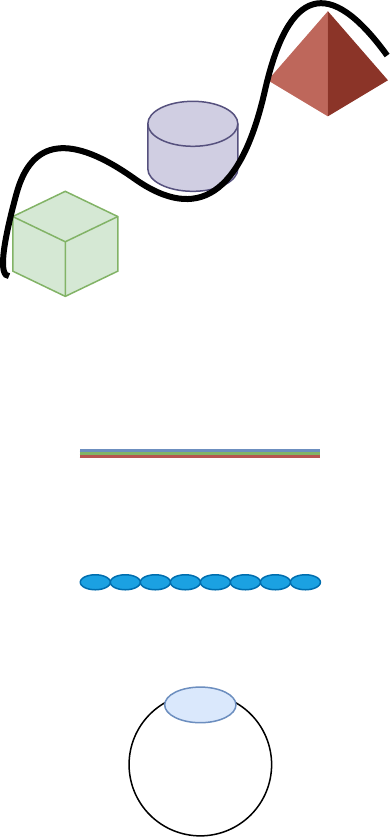}
    \caption{Light Field}
    \label{sfig:LightFieldNED}
\end{subfigure}
\begin{subfigure}{0.125\textwidth}
    \includegraphics[width=\textwidth]{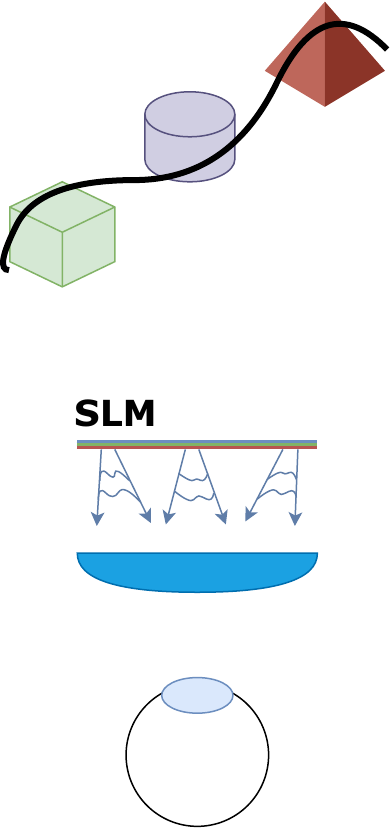}
    \caption{Holographic}
    \label{sfig:HolographicNED}
\end{subfigure}
\begin{subfigure}{0.125\textwidth}
    \includegraphics[width=\textwidth]{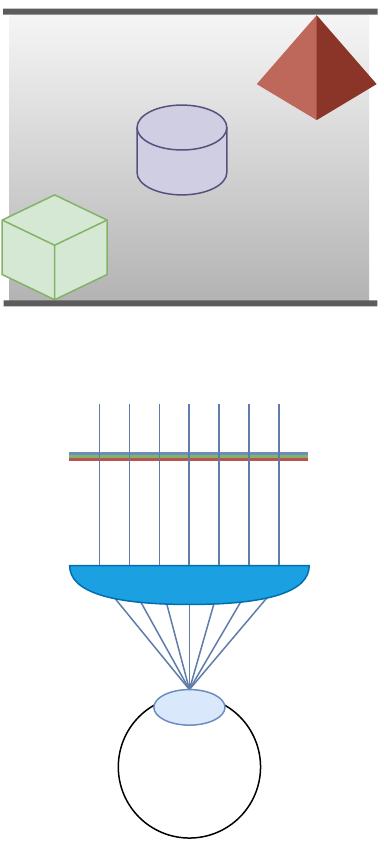}
    \caption{Maxwellian}
    \label{sfig:MaxwellianNED}
\end{subfigure}
\begin{subfigure}{0.125\textwidth}
    \includegraphics[width=\textwidth]{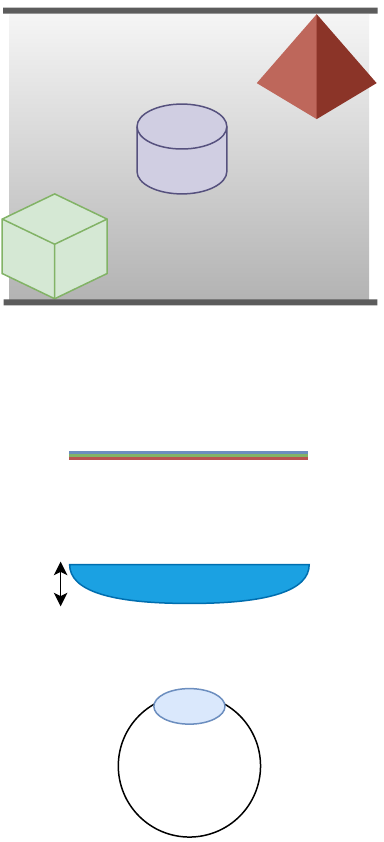}
    \caption{Focus Tunable}
    \label{sfig:FocusTunableNED}
\end{subfigure}
\begin{subfigure}{0.125\textwidth}
    \includegraphics[width=\textwidth]{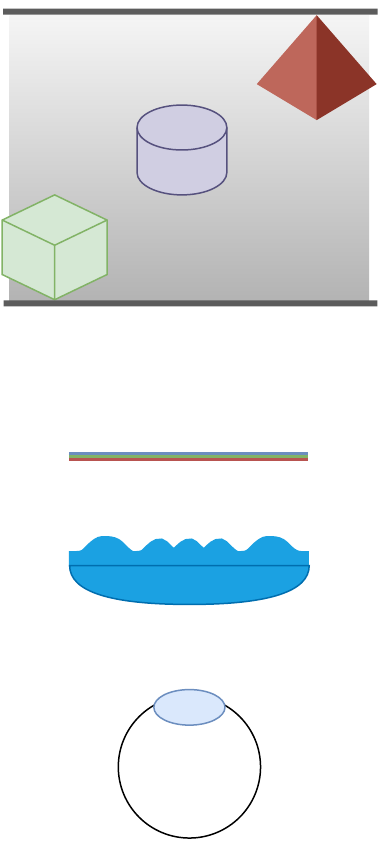}
    \caption{Proposed}
    \label{sfig:ProposedNED}
\end{subfigure}

\caption{Existing near-eye display architectures to address the VAC. %The varifocal, multifocal, light field, and holographic displays belong to the class of accommodation-enabling displays as they are capable of recreating the retinal defocus blur, whereas the Maxwellian, focus-tunable, and the proposed methods are referred as accommodation-invariant displays as they remove the blur altogether from the system.
Each method incorporates one or more display planes as well as a light modulator such as a refractive lens or a microlens array.
Depending on the architecture, the focal surfaces with varying numbers and shapes can be created, shown as the solid black lines within the scene.
Some parts of the display can also be dynamically adjusted as illustrated with arrows, in order to manipulate the focusing mechanism.
Please, note that the drawing is not to scale and some elements are exaggerated in size to illustrate the underlying principles.
} 
\label{fig:NED_Literature}
\end{figure*}

%% file: figures/MTF/MTF_cubic/MTF1D_green_lensonly.tex
\begin{tikzpicture}

\begin{axis}[%
width=0.85\columnwidth,
height=0.75\columnwidth,
at={(0\columnwidth,0\columnwidth)},
scale only axis,
point meta min=0,
point meta max=1,
axis on top,
xmin=-0.0125786163522013,
xmax=4.0125786163522,
xlabel style={font=\color{white!15!black},yshift=3pt},
xlabel={D},
xtick={0,1,2,3,4},
ymin=-0.0692598827793667,
ymax=16.4145922187099,
ylabel style={font=\color{white!15!black},yshift=-5pt},
ylabel={Frequency (cpd)},
ytick={0,15},
axis background/.style={fill=white},
title style={font=\bfseries},
title={MTF},
colormap={mymap}{[1pt] rgb(0pt)=(0.2422,0.1504,0.6603); rgb(1pt)=(0.25039,0.164995,0.707614); rgb(2pt)=(0.257771,0.181781,0.751138); rgb(3pt)=(0.264729,0.197757,0.795214); rgb(4pt)=(0.270648,0.214676,0.836371); rgb(5pt)=(0.275114,0.234238,0.870986); rgb(6pt)=(0.2783,0.255871,0.899071); rgb(7pt)=(0.280333,0.278233,0.9221); rgb(8pt)=(0.281338,0.300595,0.941376); rgb(9pt)=(0.281014,0.322757,0.957886); rgb(10pt)=(0.279467,0.344671,0.971676); rgb(11pt)=(0.275971,0.366681,0.982905); rgb(12pt)=(0.269914,0.3892,0.9906); rgb(13pt)=(0.260243,0.412329,0.995157); rgb(14pt)=(0.244033,0.435833,0.998833); rgb(15pt)=(0.220643,0.460257,0.997286); rgb(16pt)=(0.196333,0.484719,0.989152); rgb(17pt)=(0.183405,0.507371,0.979795); rgb(18pt)=(0.178643,0.528857,0.968157); rgb(19pt)=(0.176438,0.549905,0.952019); rgb(20pt)=(0.168743,0.570262,0.935871); rgb(21pt)=(0.154,0.5902,0.9218); rgb(22pt)=(0.146029,0.609119,0.907857); rgb(23pt)=(0.138024,0.627629,0.89729); rgb(24pt)=(0.124814,0.645929,0.888343); rgb(25pt)=(0.111252,0.6635,0.876314); rgb(26pt)=(0.0952095,0.679829,0.859781); rgb(27pt)=(0.0688714,0.694771,0.839357); rgb(28pt)=(0.0296667,0.708167,0.816333); rgb(29pt)=(0.00357143,0.720267,0.7917); rgb(30pt)=(0.00665714,0.731214,0.766014); rgb(31pt)=(0.0433286,0.741095,0.73941); rgb(32pt)=(0.0963952,0.75,0.712038); rgb(33pt)=(0.140771,0.7584,0.684157); rgb(34pt)=(0.1717,0.766962,0.655443); rgb(35pt)=(0.193767,0.775767,0.6251); rgb(36pt)=(0.216086,0.7843,0.5923); rgb(37pt)=(0.246957,0.791795,0.556743); rgb(38pt)=(0.290614,0.79729,0.518829); rgb(39pt)=(0.340643,0.8008,0.478857); rgb(40pt)=(0.3909,0.802871,0.435448); rgb(41pt)=(0.445629,0.802419,0.390919); rgb(42pt)=(0.5044,0.7993,0.348); rgb(43pt)=(0.561562,0.794233,0.304481); rgb(44pt)=(0.617395,0.787619,0.261238); rgb(45pt)=(0.671986,0.779271,0.2227); rgb(46pt)=(0.7242,0.769843,0.191029); rgb(47pt)=(0.773833,0.759805,0.16461); rgb(48pt)=(0.820314,0.749814,0.153529); rgb(49pt)=(0.863433,0.7406,0.159633); rgb(50pt)=(0.903543,0.733029,0.177414); rgb(51pt)=(0.939257,0.728786,0.209957); rgb(52pt)=(0.972757,0.729771,0.239443); rgb(53pt)=(0.995648,0.743371,0.237148); rgb(54pt)=(0.996986,0.765857,0.219943); rgb(55pt)=(0.995205,0.789252,0.202762); rgb(56pt)=(0.9892,0.813567,0.188533); rgb(57pt)=(0.978629,0.838629,0.176557); rgb(58pt)=(0.967648,0.8639,0.16429); rgb(59pt)=(0.96101,0.889019,0.153676); rgb(60pt)=(0.959671,0.913457,0.142257); rgb(61pt)=(0.962795,0.937338,0.12651); rgb(62pt)=(0.969114,0.960629,0.106362); rgb(63pt)=(0.9769,0.9839,0.0805)},
colorbar,
colorbar style={
            at={(0.85\columnwidth,0.75\columnwidth)},
            anchor=north west,
            align=left,
            ytick = {0,1},
            },
colorbar/width=0.05\columnwidth
]
\addplot [forget plot] graphics [xmin=-0.0125786163522013, xmax=4.0125786163522, ymin=-0.0692598827793667, ymax=16.4145922187099] {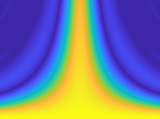};
\addplot [color=red, line width=1.4pt, forget plot]
  table[row sep=crcr]{%
2	0\\
2	0.13851976555874\\
2	0.277039531117467\\
2	0.415559296676193\\
2	0.554079062234933\\
2	0.69259882779366\\
2	0.8311185933524\\
2	0.969638358911141\\
2	1.10815812446987\\
2	1.24667789002859\\
2	1.38519765558733\\
2	1.52371742114607\\
2	1.6622371867048\\
2	1.80075695226354\\
2	1.93927671782227\\
2	2.07779648338099\\
2	2.21631624893973\\
2	2.35483601449847\\
2	2.4933557800572\\
2	2.63187554561594\\
2	2.77039531117467\\
2	2.90891507673339\\
2	3.04743484229213\\
2	3.18595460785087\\
2	3.3244743734096\\
2	3.46299413896834\\
2	3.60151390452707\\
2	3.74003367008579\\
2	3.87855343564453\\
2	4.01707320120327\\
2	4.155592966762\\
2	4.29411273232074\\
2	4.43263249787947\\
2	4.57115226343821\\
2	4.70967202899693\\
2	4.84819179455566\\
2	4.9867115601144\\
2	5.12523132567314\\
2	5.26375109123187\\
2	5.40227085679061\\
2	5.54079062234933\\
2	5.67931038790806\\
2	5.8178301534668\\
2	5.95634991902553\\
2	6.09486968458427\\
2	6.23338945014301\\
2	6.37190921570173\\
2	6.51042898126046\\
2	6.6489487468192\\
2	6.78746851237793\\
2	6.92598827793667\\
2	7.06450804349541\\
2	7.20302780905413\\
2	7.34154757461286\\
2	7.4800673401716\\
2	7.61858710573033\\
2	7.75710687128907\\
2	7.89562663684781\\
2	8.03414640240653\\
2	8.17266616796526\\
2	8.311185933524\\
2	8.44970569908273\\
2	8.58822546464147\\
2	8.72674523020021\\
2	8.86526499575893\\
2	9.00378476131766\\
2	9.1423045268764\\
2	9.28082429243513\\
2	9.41934405799387\\
2	9.55786382355261\\
2	9.69638358911133\\
2	9.83490335467006\\
2	9.9734231202288\\
2	10.1119428857875\\
2	10.2504626513463\\
2	10.388982416905\\
2	10.5275021824637\\
2	10.6660219480225\\
2	10.8045417135812\\
2	10.9430614791399\\
2	11.0815812446987\\
2	11.2201010102574\\
2	11.3586207758161\\
2	11.4971405413749\\
2	11.6356603069336\\
2	11.7741800724923\\
2	11.9126998380511\\
2	12.0512196036098\\
2	12.1897393691685\\
2	12.3282591347273\\
2	12.466778900286\\
2	12.6052986658447\\
2	12.7438184314035\\
2	12.8823381969622\\
2	13.0208579625209\\
2	13.1593777280797\\
2	13.2978974936384\\
2	13.4364172591971\\
2	13.5749370247559\\
2	13.7134567903146\\
2	13.8519765558733\\
2	13.9904963214321\\
2	14.1290160869908\\
2	14.2675358525495\\
2	14.4060556181083\\
2	14.544575383667\\
2	14.6830951492257\\
2	14.8216149147845\\
2	14.9601346803432\\
2	15.0986544459019\\
2	15.2371742114607\\
2	15.3756939770194\\
2	15.5142137425781\\
2	15.6527335081369\\
2	15.7912532736956\\
2	15.9297730392543\\
2	16.0682928048131\\
2	16.2068125703718\\
2	16.3453323359305\\
2	16.4838521014893\\
2	16.622371867048\\
2	16.7608916326067\\
2	16.8994113981655\\
2	17.0379311637242\\
2	17.1764509292829\\
2	17.3149706948417\\
2	17.4534904604004\\
2	17.5920102259591\\
2	17.7305299915179\\
2	17.8690497570766\\
2	18.0075695226353\\
2	18.1460892881941\\
2	18.2846090537528\\
2	18.4231288193115\\
2	18.5616485848703\\
2	18.700168350429\\
2	18.8386881159877\\
2	18.9772078815465\\
2	19.1157276471052\\
2	19.2542474126639\\
2	19.3927671782227\\
2	19.5312869437814\\
2	19.6698067093401\\
2	19.8083264748989\\
2	19.9468462404576\\
2	20.0853660060163\\
2	20.2238857715751\\
2	20.3624055371338\\
2	20.5009253026925\\
2	20.6394450682513\\
2	20.77796483381\\
2	20.9164845993687\\
2	21.0550043649275\\
2	21.1935241304862\\
2	21.3320438960449\\
2	21.4705636616037\\
2	21.6090834271624\\
2	21.7476031927211\\
2	21.8861229582799\\
2	22.0246427238386\\
2	22.1631624893973\\
2	22.3016822549561\\
2	22.4402020205148\\
2	22.5787217860735\\
2	22.7172415516323\\
2	22.855761317191\\
2	22.9942810827497\\
2	23.1328008483085\\
2	23.2713206138672\\
2	23.4098403794259\\
2	23.5483601449847\\
2	23.6868799105434\\
2	23.8253996761021\\
2	23.9639194416609\\
2	24.1024392072196\\
2	24.2409589727783\\
2	24.3794787383371\\
2	24.5179985038958\\
2	24.6565182694545\\
2	24.7950380350133\\
2	24.933557800572\\
2	25.0720775661307\\
2	25.2105973316895\\
2	25.3491170972482\\
2	25.4876368628069\\
2	25.6261566283657\\
2	25.7646763939244\\
2	25.9031961594831\\
2	26.0417159250419\\
2	26.1802356906006\\
2	26.3187554561593\\
2	26.4572752217181\\
2	26.5957949872768\\
2	26.7343147528355\\
2	26.8728345183943\\
2	27.011354283953\\
2	27.1498740495117\\
2	27.2883938150705\\
2	27.4269135806292\\
2	27.5654333461879\\
2	27.7039531117467\\
2	27.8424728773054\\
2	27.9809926428641\\
2	28.1195124084229\\
2	28.2580321739816\\
2	28.3965519395403\\
2	28.5350717050991\\
2	28.6735914706578\\
2	28.8121112362165\\
2	28.9506310017753\\
2	29.089150767334\\
2	29.2276705328927\\
2	29.3661902984515\\
2	29.5047100640102\\
2	29.6432298295689\\
2	29.7817495951277\\
2	29.9202693606864\\
2	30.0587891262451\\
2	30.1973088918039\\
2	30.3358286573626\\
2	30.4743484229213\\
2	30.6128681884801\\
2	30.7513879540388\\
2	30.8899077195975\\
2	31.0284274851563\\
2	31.166947250715\\
2	31.3054670162737\\
2	31.4439867818325\\
2	31.5825065473912\\
2	31.7210263129499\\
2	31.8595460785087\\
2	31.9980658440674\\
2	32.1365856096261\\
2	32.2751053751849\\
2	32.4136251407436\\
2	32.5521449063023\\
2	32.6906646718611\\
2	32.8291844374198\\
2	32.9677042029786\\
2	33.1062239685373\\
2	33.244743734096\\
2	33.3832634996547\\
2	33.5217832652135\\
2	33.6603030307722\\
2	33.798822796331\\
2	33.9373425618897\\
2	34.0758623274484\\
2	34.2143820930071\\
2	34.3529018585659\\
2	34.4914216241246\\
2	34.6299413896833\\
2	34.7684611552421\\
2	34.9069809208008\\
2	35.0455006863595\\
2	35.1840204519183\\
2	35.322540217477\\
2	35.4610599830357\\
2	35.5995797485945\\
2	35.7380995141532\\
2	35.8766192797119\\
2	36.0151390452707\\
2	36.1536588108294\\
2	36.2921785763881\\
2	36.4306983419469\\
2	36.5692181075056\\
2	36.7077378730643\\
2	36.8462576386231\\
2	36.9847774041818\\
2	37.1232971697405\\
2	37.2618169352993\\
2	37.400336700858\\
2	37.5388564664167\\
2	37.6773762319755\\
2	37.8158959975342\\
2	37.9544157630929\\
2	38.0929355286517\\
2	38.2314552942104\\
2	38.3699750597691\\
2	38.5084948253279\\
2	38.6470145908866\\
2	38.7855343564453\\
2	38.9240541220041\\
2	39.0625738875628\\
2	39.2010936531215\\
2	39.3396134186803\\
2	39.478133184239\\
2	39.6166529497977\\
2	39.7551727153565\\
2	39.8936924809152\\
2	40.0322122464739\\
2	40.1707320120327\\
2	40.3092517775914\\
2	40.4477715431501\\
2	40.5862913087089\\
2	40.7248110742676\\
2	40.8633308398263\\
2	41.0018506053851\\
2	41.1403703709438\\
2	41.2788901365025\\
2	41.4174099020613\\
2	41.55592966762\\
2	41.6944494331787\\
2	41.8329691987375\\
2	41.9714889642962\\
2	42.1100087298549\\
2	42.2485284954137\\
2	42.3870482609724\\
2	42.5255680265311\\
2	42.6640877920899\\
2	42.8026075576486\\
2	42.9411273232073\\
2	43.0796470887661\\
2	43.2181668543248\\
2	43.3566866198835\\
2	43.4952063854423\\
2	43.633726151001\\
2	43.7722459165597\\
2	43.9107656821185\\
2	44.0492854476772\\
2	44.1878052132359\\
2	44.3263249787947\\
2	44.4648447443534\\
2	44.6033645099121\\
2	44.7418842754709\\
2	44.8804040410296\\
2	45.0189238065883\\
2	45.1574435721471\\
2	45.2959633377058\\
2	45.4344831032645\\
2	45.5730028688233\\
2	45.711522634382\\
2	45.8500423999407\\
2	45.9885621654995\\
2	46.1270819310582\\
2	46.2656016966169\\
2	46.4041214621757\\
2	46.5426412277344\\
2	46.6811609932931\\
2	46.8196807588519\\
2	46.9582005244106\\
2	47.0967202899693\\
2	47.2352400555281\\
2	47.3737598210868\\
2	47.5122795866455\\
2	47.6507993522043\\
2	47.789319117763\\
2	47.9278388833217\\
2	48.0663586488805\\
2	48.2048784144392\\
2	48.3433981799979\\
2	48.4819179455567\\
2	48.6204377111154\\
2	48.7589574766741\\
2	48.8974772422329\\
2	49.0359970077916\\
2	49.1745167733503\\
2	49.3130365389091\\
2	49.4515563044678\\
2	49.5900760700265\\
2	49.7285958355853\\
2	49.867115601144\\
2	50.0056353667027\\
2	50.1441551322615\\
2	50.2826748978202\\
2	50.4211946633789\\
2	50.5597144289377\\
2	50.6982341944964\\
2	50.8367539600551\\
2	50.9752737256139\\
2	51.1137934911726\\
2	51.2523132567313\\
2	51.3908330222901\\
2	51.5293527878488\\
2	51.6678725534075\\
2	51.8063923189663\\
2	51.944912084525\\
2	52.0834318500837\\
2	52.2219516156425\\
2	52.3604713812012\\
2	52.4989911467599\\
2	52.6375109123187\\
2	52.7760306778774\\
2	52.9145504434361\\
2	53.0530702089949\\
2	53.1915899745536\\
2	53.3301097401123\\
2	53.4686295056711\\
2	53.6071492712298\\
2	53.7456690367885\\
2	53.8841888023473\\
2	54.022708567906\\
2	54.1612283334647\\
2	54.2997480990235\\
2	54.4382678645822\\
2	54.5767876301409\\
2	54.7153073956997\\
2	54.8538271612584\\
2	54.9923469268171\\
2	55.1308666923759\\
2	55.2693864579346\\
2	55.4079062234933\\
2	55.5464259890521\\
2	55.6849457546108\\
2	55.8234655201695\\
2	55.9619852857283\\
2	56.100505051287\\
2	56.2390248168457\\
2	56.3775445824045\\
2	56.5160643479632\\
2	56.6545841135219\\
2	56.7931038790807\\
2	56.9316236446394\\
2	57.0701434101981\\
2	57.2086631757569\\
2	57.3471829413156\\
2	57.4857027068743\\
2	57.6242224724331\\
2	57.7627422379918\\
2	57.9012620035505\\
2	58.0397817691093\\
2	58.178301534668\\
2	58.3168213002267\\
2	58.4553410657855\\
2	58.5938608313442\\
2	58.7323805969029\\
2	58.8709003624617\\
2	59.0094201280204\\
2	59.1479398935791\\
2	59.2864596591379\\
2	59.4249794246966\\
2	59.5634991902553\\
2	59.7020189558141\\
2	59.8405387213728\\
2	59.9790584869315\\
2	60.1175782524903\\
2	60.256098018049\\
2	60.3946177836077\\
2	60.5331375491665\\
2	60.6716573147252\\
2	60.8101770802839\\
2	60.9486968458427\\
2	61.0872166114014\\
2	61.2257363769601\\
2	61.3642561425189\\
2	61.5027759080776\\
2	61.6412956736363\\
2	61.7798154391951\\
2	61.9183352047538\\
2	62.0568549703125\\
2	62.1953747358713\\
2	62.33389450143\\
2	62.4724142669887\\
2	62.6109340325475\\
2	62.7494537981062\\
2	62.8879735636649\\
2	63.0264933292237\\
2	63.1650130947824\\
2	63.3035328603411\\
2	63.4420526258999\\
2	63.5805723914586\\
2	63.7190921570173\\
2	63.8576119225761\\
2	63.9961316881348\\
2	64.1346514536936\\
2	64.2731712192523\\
2	64.411690984811\\
2	64.5502107503697\\
2	64.6887305159285\\
2	64.8272502814872\\
2	64.9657700470459\\
2	65.1042898126047\\
2	65.2428095781634\\
2	65.3813293437221\\
};
\end{axis}
\end{tikzpicture}%

%% file: figures/MTF/MTF_cubic/MTF1D_green_lensonly_gradient.tex
\begin{tikzpicture}

\begin{axis}[%
width=0.85\columnwidth,
height=0.75\columnwidth,
at={(0\columnwidth,0\columnwidth)},
scale only axis,
point meta min=0,
point meta max=0.05,
axis on top,
xmin=-0.0125786163522013,
xmax=4.0125786163522,
xlabel style={font=\color{white!15!black},yshift=3pt},
xlabel={D},
xtick={0,1,2,3,4},
ymin=-0.0692598827793667,
ymax=16.4145922187099,
ylabel style={font=\color{white!15!black},yshift=-5pt},
ylabel={Frequency (cpd)},
ytick={0,15},
axis background/.style={fill=white},
title style={font=\bfseries},
title={MTF Gradient},
colormap={mymap}{[1pt] rgb(0pt)=(0.2422,0.1504,0.6603); rgb(1pt)=(0.25039,0.164995,0.707614); rgb(2pt)=(0.257771,0.181781,0.751138); rgb(3pt)=(0.264729,0.197757,0.795214); rgb(4pt)=(0.270648,0.214676,0.836371); rgb(5pt)=(0.275114,0.234238,0.870986); rgb(6pt)=(0.2783,0.255871,0.899071); rgb(7pt)=(0.280333,0.278233,0.9221); rgb(8pt)=(0.281338,0.300595,0.941376); rgb(9pt)=(0.281014,0.322757,0.957886); rgb(10pt)=(0.279467,0.344671,0.971676); rgb(11pt)=(0.275971,0.366681,0.982905); rgb(12pt)=(0.269914,0.3892,0.9906); rgb(13pt)=(0.260243,0.412329,0.995157); rgb(14pt)=(0.244033,0.435833,0.998833); rgb(15pt)=(0.220643,0.460257,0.997286); rgb(16pt)=(0.196333,0.484719,0.989152); rgb(17pt)=(0.183405,0.507371,0.979795); rgb(18pt)=(0.178643,0.528857,0.968157); rgb(19pt)=(0.176438,0.549905,0.952019); rgb(20pt)=(0.168743,0.570262,0.935871); rgb(21pt)=(0.154,0.5902,0.9218); rgb(22pt)=(0.146029,0.609119,0.907857); rgb(23pt)=(0.138024,0.627629,0.89729); rgb(24pt)=(0.124814,0.645929,0.888343); rgb(25pt)=(0.111252,0.6635,0.876314); rgb(26pt)=(0.0952095,0.679829,0.859781); rgb(27pt)=(0.0688714,0.694771,0.839357); rgb(28pt)=(0.0296667,0.708167,0.816333); rgb(29pt)=(0.00357143,0.720267,0.7917); rgb(30pt)=(0.00665714,0.731214,0.766014); rgb(31pt)=(0.0433286,0.741095,0.73941); rgb(32pt)=(0.0963952,0.75,0.712038); rgb(33pt)=(0.140771,0.7584,0.684157); rgb(34pt)=(0.1717,0.766962,0.655443); rgb(35pt)=(0.193767,0.775767,0.6251); rgb(36pt)=(0.216086,0.7843,0.5923); rgb(37pt)=(0.246957,0.791795,0.556743); rgb(38pt)=(0.290614,0.79729,0.518829); rgb(39pt)=(0.340643,0.8008,0.478857); rgb(40pt)=(0.3909,0.802871,0.435448); rgb(41pt)=(0.445629,0.802419,0.390919); rgb(42pt)=(0.5044,0.7993,0.348); rgb(43pt)=(0.561562,0.794233,0.304481); rgb(44pt)=(0.617395,0.787619,0.261238); rgb(45pt)=(0.671986,0.779271,0.2227); rgb(46pt)=(0.7242,0.769843,0.191029); rgb(47pt)=(0.773833,0.759805,0.16461); rgb(48pt)=(0.820314,0.749814,0.153529); rgb(49pt)=(0.863433,0.7406,0.159633); rgb(50pt)=(0.903543,0.733029,0.177414); rgb(51pt)=(0.939257,0.728786,0.209957); rgb(52pt)=(0.972757,0.729771,0.239443); rgb(53pt)=(0.995648,0.743371,0.237148); rgb(54pt)=(0.996986,0.765857,0.219943); rgb(55pt)=(0.995205,0.789252,0.202762); rgb(56pt)=(0.9892,0.813567,0.188533); rgb(57pt)=(0.978629,0.838629,0.176557); rgb(58pt)=(0.967648,0.8639,0.16429); rgb(59pt)=(0.96101,0.889019,0.153676); rgb(60pt)=(0.959671,0.913457,0.142257); rgb(61pt)=(0.962795,0.937338,0.12651); rgb(62pt)=(0.969114,0.960629,0.106362); rgb(63pt)=(0.9769,0.9839,0.0805)},
colorbar,
colorbar style={
            at={(0.85\columnwidth,0.75\columnwidth)},
            anchor=north west,
            align=left,
            ytick = {0,0.05},
            yticklabel style={
            /pgf/number format/fixed,
            scaled ticks=false,
              },
            },
colorbar/width=0.05\columnwidth
]
\addplot [forget plot] graphics [xmin=-0.0125786163522013, xmax=4.0125786163522, ymin=-0.0692598827793667, ymax=16.4145922187099] {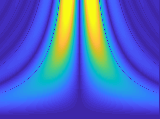};
\addplot [color=red, line width=1.4pt, forget plot]
  table[row sep=crcr]{%
2	0\\
2	0.13851976555874\\
2	0.277039531117467\\
2	0.415559296676193\\
2	0.554079062234933\\
2	0.69259882779366\\
2	0.8311185933524\\
2	0.969638358911141\\
2	1.10815812446987\\
2	1.24667789002859\\
2	1.38519765558733\\
2	1.52371742114607\\
2	1.6622371867048\\
2	1.80075695226354\\
2	1.93927671782227\\
2	2.07779648338099\\
2	2.21631624893973\\
2	2.35483601449847\\
2	2.4933557800572\\
2	2.63187554561594\\
2	2.77039531117467\\
2	2.90891507673339\\
2	3.04743484229213\\
2	3.18595460785087\\
2	3.3244743734096\\
2	3.46299413896834\\
2	3.60151390452707\\
2	3.74003367008579\\
2	3.87855343564453\\
2	4.01707320120327\\
2	4.155592966762\\
2	4.29411273232074\\
2	4.43263249787947\\
2	4.57115226343821\\
2	4.70967202899693\\
2	4.84819179455566\\
2	4.9867115601144\\
2	5.12523132567314\\
2	5.26375109123187\\
2	5.40227085679061\\
2	5.54079062234933\\
2	5.67931038790806\\
2	5.8178301534668\\
2	5.95634991902553\\
2	6.09486968458427\\
2	6.23338945014301\\
2	6.37190921570173\\
2	6.51042898126046\\
2	6.6489487468192\\
2	6.78746851237793\\
2	6.92598827793667\\
2	7.06450804349541\\
2	7.20302780905413\\
2	7.34154757461286\\
2	7.4800673401716\\
2	7.61858710573033\\
2	7.75710687128907\\
2	7.89562663684781\\
2	8.03414640240653\\
2	8.17266616796526\\
2	8.311185933524\\
2	8.44970569908273\\
2	8.58822546464147\\
2	8.72674523020021\\
2	8.86526499575893\\
2	9.00378476131766\\
2	9.1423045268764\\
2	9.28082429243513\\
2	9.41934405799387\\
2	9.55786382355261\\
2	9.69638358911133\\
2	9.83490335467006\\
2	9.9734231202288\\
2	10.1119428857875\\
2	10.2504626513463\\
2	10.388982416905\\
2	10.5275021824637\\
2	10.6660219480225\\
2	10.8045417135812\\
2	10.9430614791399\\
2	11.0815812446987\\
2	11.2201010102574\\
2	11.3586207758161\\
2	11.4971405413749\\
2	11.6356603069336\\
2	11.7741800724923\\
2	11.9126998380511\\
2	12.0512196036098\\
2	12.1897393691685\\
2	12.3282591347273\\
2	12.466778900286\\
2	12.6052986658447\\
2	12.7438184314035\\
2	12.8823381969622\\
2	13.0208579625209\\
2	13.1593777280797\\
2	13.2978974936384\\
2	13.4364172591971\\
2	13.5749370247559\\
2	13.7134567903146\\
2	13.8519765558733\\
2	13.9904963214321\\
2	14.1290160869908\\
2	14.2675358525495\\
2	14.4060556181083\\
2	14.544575383667\\
2	14.6830951492257\\
2	14.8216149147845\\
2	14.9601346803432\\
2	15.0986544459019\\
2	15.2371742114607\\
2	15.3756939770194\\
2	15.5142137425781\\
2	15.6527335081369\\
2	15.7912532736956\\
2	15.9297730392543\\
2	16.0682928048131\\
2	16.2068125703718\\
2	16.3453323359305\\
2	16.4838521014893\\
2	16.622371867048\\
2	16.7608916326067\\
2	16.8994113981655\\
2	17.0379311637242\\
2	17.1764509292829\\
2	17.3149706948417\\
2	17.4534904604004\\
2	17.5920102259591\\
2	17.7305299915179\\
2	17.8690497570766\\
2	18.0075695226353\\
2	18.1460892881941\\
2	18.2846090537528\\
2	18.4231288193115\\
2	18.5616485848703\\
2	18.700168350429\\
2	18.8386881159877\\
2	18.9772078815465\\
2	19.1157276471052\\
2	19.2542474126639\\
2	19.3927671782227\\
2	19.5312869437814\\
2	19.6698067093401\\
2	19.8083264748989\\
2	19.9468462404576\\
2	20.0853660060163\\
2	20.2238857715751\\
2	20.3624055371338\\
2	20.5009253026925\\
2	20.6394450682513\\
2	20.77796483381\\
2	20.9164845993687\\
2	21.0550043649275\\
2	21.1935241304862\\
2	21.3320438960449\\
2	21.4705636616037\\
2	21.6090834271624\\
2	21.7476031927211\\
2	21.8861229582799\\
2	22.0246427238386\\
2	22.1631624893973\\
2	22.3016822549561\\
2	22.4402020205148\\
2	22.5787217860735\\
2	22.7172415516323\\
2	22.855761317191\\
2	22.9942810827497\\
2	23.1328008483085\\
2	23.2713206138672\\
2	23.4098403794259\\
2	23.5483601449847\\
2	23.6868799105434\\
2	23.8253996761021\\
2	23.9639194416609\\
2	24.1024392072196\\
2	24.2409589727783\\
2	24.3794787383371\\
2	24.5179985038958\\
2	24.6565182694545\\
2	24.7950380350133\\
2	24.933557800572\\
2	25.0720775661307\\
2	25.2105973316895\\
2	25.3491170972482\\
2	25.4876368628069\\
2	25.6261566283657\\
2	25.7646763939244\\
2	25.9031961594831\\
2	26.0417159250419\\
2	26.1802356906006\\
2	26.3187554561593\\
2	26.4572752217181\\
2	26.5957949872768\\
2	26.7343147528355\\
2	26.8728345183943\\
2	27.011354283953\\
2	27.1498740495117\\
2	27.2883938150705\\
2	27.4269135806292\\
2	27.5654333461879\\
2	27.7039531117467\\
2	27.8424728773054\\
2	27.9809926428641\\
2	28.1195124084229\\
2	28.2580321739816\\
2	28.3965519395403\\
2	28.5350717050991\\
2	28.6735914706578\\
2	28.8121112362165\\
2	28.9506310017753\\
2	29.089150767334\\
2	29.2276705328927\\
2	29.3661902984515\\
2	29.5047100640102\\
2	29.6432298295689\\
2	29.7817495951277\\
2	29.9202693606864\\
2	30.0587891262451\\
2	30.1973088918039\\
2	30.3358286573626\\
2	30.4743484229213\\
2	30.6128681884801\\
2	30.7513879540388\\
2	30.8899077195975\\
2	31.0284274851563\\
2	31.166947250715\\
2	31.3054670162737\\
2	31.4439867818325\\
2	31.5825065473912\\
2	31.7210263129499\\
2	31.8595460785087\\
2	31.9980658440674\\
2	32.1365856096261\\
2	32.2751053751849\\
2	32.4136251407436\\
2	32.5521449063023\\
2	32.6906646718611\\
2	32.8291844374198\\
2	32.9677042029786\\
2	33.1062239685373\\
2	33.244743734096\\
2	33.3832634996547\\
2	33.5217832652135\\
2	33.6603030307722\\
2	33.798822796331\\
2	33.9373425618897\\
2	34.0758623274484\\
2	34.2143820930071\\
2	34.3529018585659\\
2	34.4914216241246\\
2	34.6299413896833\\
2	34.7684611552421\\
2	34.9069809208008\\
2	35.0455006863595\\
2	35.1840204519183\\
2	35.322540217477\\
2	35.4610599830357\\
2	35.5995797485945\\
2	35.7380995141532\\
2	35.8766192797119\\
2	36.0151390452707\\
2	36.1536588108294\\
2	36.2921785763881\\
2	36.4306983419469\\
2	36.5692181075056\\
2	36.7077378730643\\
2	36.8462576386231\\
2	36.9847774041818\\
2	37.1232971697405\\
2	37.2618169352993\\
2	37.400336700858\\
2	37.5388564664167\\
2	37.6773762319755\\
2	37.8158959975342\\
2	37.9544157630929\\
2	38.0929355286517\\
2	38.2314552942104\\
2	38.3699750597691\\
2	38.5084948253279\\
2	38.6470145908866\\
2	38.7855343564453\\
2	38.9240541220041\\
2	39.0625738875628\\
2	39.2010936531215\\
2	39.3396134186803\\
2	39.478133184239\\
2	39.6166529497977\\
2	39.7551727153565\\
2	39.8936924809152\\
2	40.0322122464739\\
2	40.1707320120327\\
2	40.3092517775914\\
2	40.4477715431501\\
2	40.5862913087089\\
2	40.7248110742676\\
2	40.8633308398263\\
2	41.0018506053851\\
2	41.1403703709438\\
2	41.2788901365025\\
2	41.4174099020613\\
2	41.55592966762\\
2	41.6944494331787\\
2	41.8329691987375\\
2	41.9714889642962\\
2	42.1100087298549\\
2	42.2485284954137\\
2	42.3870482609724\\
2	42.5255680265311\\
2	42.6640877920899\\
2	42.8026075576486\\
2	42.9411273232073\\
2	43.0796470887661\\
2	43.2181668543248\\
2	43.3566866198835\\
2	43.4952063854423\\
2	43.633726151001\\
2	43.7722459165597\\
2	43.9107656821185\\
2	44.0492854476772\\
2	44.1878052132359\\
2	44.3263249787947\\
2	44.4648447443534\\
2	44.6033645099121\\
2	44.7418842754709\\
2	44.8804040410296\\
2	45.0189238065883\\
2	45.1574435721471\\
2	45.2959633377058\\
2	45.4344831032645\\
2	45.5730028688233\\
2	45.711522634382\\
2	45.8500423999407\\
2	45.9885621654995\\
2	46.1270819310582\\
2	46.2656016966169\\
2	46.4041214621757\\
2	46.5426412277344\\
2	46.6811609932931\\
2	46.8196807588519\\
2	46.9582005244106\\
2	47.0967202899693\\
2	47.2352400555281\\
2	47.3737598210868\\
2	47.5122795866455\\
2	47.6507993522043\\
2	47.789319117763\\
2	47.9278388833217\\
2	48.0663586488805\\
2	48.2048784144392\\
2	48.3433981799979\\
2	48.4819179455567\\
2	48.6204377111154\\
2	48.7589574766741\\
2	48.8974772422329\\
2	49.0359970077916\\
2	49.1745167733503\\
2	49.3130365389091\\
2	49.4515563044678\\
2	49.5900760700265\\
2	49.7285958355853\\
2	49.867115601144\\
2	50.0056353667027\\
2	50.1441551322615\\
2	50.2826748978202\\
2	50.4211946633789\\
2	50.5597144289377\\
2	50.6982341944964\\
2	50.8367539600551\\
2	50.9752737256139\\
2	51.1137934911726\\
2	51.2523132567313\\
2	51.3908330222901\\
2	51.5293527878488\\
2	51.6678725534075\\
2	51.8063923189663\\
2	51.944912084525\\
2	52.0834318500837\\
2	52.2219516156425\\
2	52.3604713812012\\
2	52.4989911467599\\
2	52.6375109123187\\
2	52.7760306778774\\
2	52.9145504434361\\
2	53.0530702089949\\
2	53.1915899745536\\
2	53.3301097401123\\
2	53.4686295056711\\
2	53.6071492712298\\
2	53.7456690367885\\
2	53.8841888023473\\
2	54.022708567906\\
2	54.1612283334647\\
2	54.2997480990235\\
2	54.4382678645822\\
2	54.5767876301409\\
2	54.7153073956997\\
2	54.8538271612584\\
2	54.9923469268171\\
2	55.1308666923759\\
2	55.2693864579346\\
2	55.4079062234933\\
2	55.5464259890521\\
2	55.6849457546108\\
2	55.8234655201695\\
2	55.9619852857283\\
2	56.100505051287\\
2	56.2390248168457\\
2	56.3775445824045\\
2	56.5160643479632\\
2	56.6545841135219\\
2	56.7931038790807\\
2	56.9316236446394\\
2	57.0701434101981\\
2	57.2086631757569\\
2	57.3471829413156\\
2	57.4857027068743\\
2	57.6242224724331\\
2	57.7627422379918\\
2	57.9012620035505\\
2	58.0397817691093\\
2	58.178301534668\\
2	58.3168213002267\\
2	58.4553410657855\\
2	58.5938608313442\\
2	58.7323805969029\\
2	58.8709003624617\\
2	59.0094201280204\\
2	59.1479398935791\\
2	59.2864596591379\\
2	59.4249794246966\\
2	59.5634991902553\\
2	59.7020189558141\\
2	59.8405387213728\\
2	59.9790584869315\\
2	60.1175782524903\\
2	60.256098018049\\
2	60.3946177836077\\
2	60.5331375491665\\
2	60.6716573147252\\
2	60.8101770802839\\
2	60.9486968458427\\
2	61.0872166114014\\
2	61.2257363769601\\
2	61.3642561425189\\
2	61.5027759080776\\
2	61.6412956736363\\
2	61.7798154391951\\
2	61.9183352047538\\
2	62.0568549703125\\
2	62.1953747358713\\
2	62.33389450143\\
2	62.4724142669887\\
2	62.6109340325475\\
2	62.7494537981062\\
2	62.8879735636649\\
2	63.0264933292237\\
2	63.1650130947824\\
2	63.3035328603411\\
2	63.4420526258999\\
2	63.5805723914586\\
2	63.7190921570173\\
2	63.8576119225761\\
2	63.9961316881348\\
2	64.1346514536936\\
2	64.2731712192523\\
2	64.411690984811\\
2	64.5502107503697\\
2	64.6887305159285\\
2	64.8272502814872\\
2	64.9657700470459\\
2	65.1042898126047\\
2	65.2428095781634\\
2	65.3813293437221\\
};
\end{axis}
\end{tikzpicture}%

%% file: figures/MTF/MTF_cubic/MTF1D_green.tex
\begin{tikzpicture}

\begin{axis}[%
width=0.85\columnwidth,
height=0.75\columnwidth,
at={(0\columnwidth,0\columnwidth)},
scale only axis,
point meta min=0,
point meta max=1,
axis on top,
xmin=-0.0125786163522013,
xmax=4.0125786163522,
xlabel style={font=\color{white!15!black},yshift=3pt},
xlabel={D},
xtick={0,1,2,3,4},
ymin=-0.0692598827793667,
ymax=16.4145922187099,
ylabel style={font=\color{white!15!black},yshift=-5pt},
ylabel={Frequency (cpd)},
ytick={0,15},
axis background/.style={fill=white},
title style={font=\bfseries},
title={MTF},
colormap={mymap}{[1pt] rgb(0pt)=(0.2422,0.1504,0.6603); rgb(1pt)=(0.25039,0.164995,0.707614); rgb(2pt)=(0.257771,0.181781,0.751138); rgb(3pt)=(0.264729,0.197757,0.795214); rgb(4pt)=(0.270648,0.214676,0.836371); rgb(5pt)=(0.275114,0.234238,0.870986); rgb(6pt)=(0.2783,0.255871,0.899071); rgb(7pt)=(0.280333,0.278233,0.9221); rgb(8pt)=(0.281338,0.300595,0.941376); rgb(9pt)=(0.281014,0.322757,0.957886); rgb(10pt)=(0.279467,0.344671,0.971676); rgb(11pt)=(0.275971,0.366681,0.982905); rgb(12pt)=(0.269914,0.3892,0.9906); rgb(13pt)=(0.260243,0.412329,0.995157); rgb(14pt)=(0.244033,0.435833,0.998833); rgb(15pt)=(0.220643,0.460257,0.997286); rgb(16pt)=(0.196333,0.484719,0.989152); rgb(17pt)=(0.183405,0.507371,0.979795); rgb(18pt)=(0.178643,0.528857,0.968157); rgb(19pt)=(0.176438,0.549905,0.952019); rgb(20pt)=(0.168743,0.570262,0.935871); rgb(21pt)=(0.154,0.5902,0.9218); rgb(22pt)=(0.146029,0.609119,0.907857); rgb(23pt)=(0.138024,0.627629,0.89729); rgb(24pt)=(0.124814,0.645929,0.888343); rgb(25pt)=(0.111252,0.6635,0.876314); rgb(26pt)=(0.0952095,0.679829,0.859781); rgb(27pt)=(0.0688714,0.694771,0.839357); rgb(28pt)=(0.0296667,0.708167,0.816333); rgb(29pt)=(0.00357143,0.720267,0.7917); rgb(30pt)=(0.00665714,0.731214,0.766014); rgb(31pt)=(0.0433286,0.741095,0.73941); rgb(32pt)=(0.0963952,0.75,0.712038); rgb(33pt)=(0.140771,0.7584,0.684157); rgb(34pt)=(0.1717,0.766962,0.655443); rgb(35pt)=(0.193767,0.775767,0.6251); rgb(36pt)=(0.216086,0.7843,0.5923); rgb(37pt)=(0.246957,0.791795,0.556743); rgb(38pt)=(0.290614,0.79729,0.518829); rgb(39pt)=(0.340643,0.8008,0.478857); rgb(40pt)=(0.3909,0.802871,0.435448); rgb(41pt)=(0.445629,0.802419,0.390919); rgb(42pt)=(0.5044,0.7993,0.348); rgb(43pt)=(0.561562,0.794233,0.304481); rgb(44pt)=(0.617395,0.787619,0.261238); rgb(45pt)=(0.671986,0.779271,0.2227); rgb(46pt)=(0.7242,0.769843,0.191029); rgb(47pt)=(0.773833,0.759805,0.16461); rgb(48pt)=(0.820314,0.749814,0.153529); rgb(49pt)=(0.863433,0.7406,0.159633); rgb(50pt)=(0.903543,0.733029,0.177414); rgb(51pt)=(0.939257,0.728786,0.209957); rgb(52pt)=(0.972757,0.729771,0.239443); rgb(53pt)=(0.995648,0.743371,0.237148); rgb(54pt)=(0.996986,0.765857,0.219943); rgb(55pt)=(0.995205,0.789252,0.202762); rgb(56pt)=(0.9892,0.813567,0.188533); rgb(57pt)=(0.978629,0.838629,0.176557); rgb(58pt)=(0.967648,0.8639,0.16429); rgb(59pt)=(0.96101,0.889019,0.153676); rgb(60pt)=(0.959671,0.913457,0.142257); rgb(61pt)=(0.962795,0.937338,0.12651); rgb(62pt)=(0.969114,0.960629,0.106362); rgb(63pt)=(0.9769,0.9839,0.0805)},
colorbar,
colorbar style={
            at={(0.85\columnwidth,0.75\columnwidth)},
            anchor=north west,
            align=left,
            ytick = {0,1},
            },
colorbar/width=0.05\columnwidth
]
\addplot [forget plot] graphics [xmin=-0.0125786163522013, xmax=4.0125786163522, ymin=-0.0692598827793667, ymax=16.4145922187099] {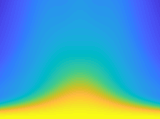};
\end{axis}
\end{tikzpicture}%

%% file: figures/MTF/MTF_cubic/MTF1D_green_gradient.tex
\begin{tikzpicture}

\begin{axis}[%
width=0.85\columnwidth,
height=0.75\columnwidth,
at={(0\columnwidth,0\columnwidth)},
scale only axis,
point meta min=0,
point meta max=0.05,
axis on top,
xmin=-0.0125786163522013,
xmax=4.0125786163522,
xlabel style={font=\color{white!15!black},yshift=3pt},
xlabel={D},
xtick={0,1,2,3,4},
ymin=-0.0692598827793667,
ymax=16.4145922187099,
ylabel style={font=\color{white!15!black},yshift=-5pt},
ylabel={Frequency (cpd)},
ytick={0,15},
axis background/.style={fill=white},
title style={font=\bfseries},
title={MTF Gradient},
colormap={mymap}{[1pt] rgb(0pt)=(0.2422,0.1504,0.6603); rgb(1pt)=(0.25039,0.164995,0.707614); rgb(2pt)=(0.257771,0.181781,0.751138); rgb(3pt)=(0.264729,0.197757,0.795214); rgb(4pt)=(0.270648,0.214676,0.836371); rgb(5pt)=(0.275114,0.234238,0.870986); rgb(6pt)=(0.2783,0.255871,0.899071); rgb(7pt)=(0.280333,0.278233,0.9221); rgb(8pt)=(0.281338,0.300595,0.941376); rgb(9pt)=(0.281014,0.322757,0.957886); rgb(10pt)=(0.279467,0.344671,0.971676); rgb(11pt)=(0.275971,0.366681,0.982905); rgb(12pt)=(0.269914,0.3892,0.9906); rgb(13pt)=(0.260243,0.412329,0.995157); rgb(14pt)=(0.244033,0.435833,0.998833); rgb(15pt)=(0.220643,0.460257,0.997286); rgb(16pt)=(0.196333,0.484719,0.989152); rgb(17pt)=(0.183405,0.507371,0.979795); rgb(18pt)=(0.178643,0.528857,0.968157); rgb(19pt)=(0.176438,0.549905,0.952019); rgb(20pt)=(0.168743,0.570262,0.935871); rgb(21pt)=(0.154,0.5902,0.9218); rgb(22pt)=(0.146029,0.609119,0.907857); rgb(23pt)=(0.138024,0.627629,0.89729); rgb(24pt)=(0.124814,0.645929,0.888343); rgb(25pt)=(0.111252,0.6635,0.876314); rgb(26pt)=(0.0952095,0.679829,0.859781); rgb(27pt)=(0.0688714,0.694771,0.839357); rgb(28pt)=(0.0296667,0.708167,0.816333); rgb(29pt)=(0.00357143,0.720267,0.7917); rgb(30pt)=(0.00665714,0.731214,0.766014); rgb(31pt)=(0.0433286,0.741095,0.73941); rgb(32pt)=(0.0963952,0.75,0.712038); rgb(33pt)=(0.140771,0.7584,0.684157); rgb(34pt)=(0.1717,0.766962,0.655443); rgb(35pt)=(0.193767,0.775767,0.6251); rgb(36pt)=(0.216086,0.7843,0.5923); rgb(37pt)=(0.246957,0.791795,0.556743); rgb(38pt)=(0.290614,0.79729,0.518829); rgb(39pt)=(0.340643,0.8008,0.478857); rgb(40pt)=(0.3909,0.802871,0.435448); rgb(41pt)=(0.445629,0.802419,0.390919); rgb(42pt)=(0.5044,0.7993,0.348); rgb(43pt)=(0.561562,0.794233,0.304481); rgb(44pt)=(0.617395,0.787619,0.261238); rgb(45pt)=(0.671986,0.779271,0.2227); rgb(46pt)=(0.7242,0.769843,0.191029); rgb(47pt)=(0.773833,0.759805,0.16461); rgb(48pt)=(0.820314,0.749814,0.153529); rgb(49pt)=(0.863433,0.7406,0.159633); rgb(50pt)=(0.903543,0.733029,0.177414); rgb(51pt)=(0.939257,0.728786,0.209957); rgb(52pt)=(0.972757,0.729771,0.239443); rgb(53pt)=(0.995648,0.743371,0.237148); rgb(54pt)=(0.996986,0.765857,0.219943); rgb(55pt)=(0.995205,0.789252,0.202762); rgb(56pt)=(0.9892,0.813567,0.188533); rgb(57pt)=(0.978629,0.838629,0.176557); rgb(58pt)=(0.967648,0.8639,0.16429); rgb(59pt)=(0.96101,0.889019,0.153676); rgb(60pt)=(0.959671,0.913457,0.142257); rgb(61pt)=(0.962795,0.937338,0.12651); rgb(62pt)=(0.969114,0.960629,0.106362); rgb(63pt)=(0.9769,0.9839,0.0805)},
colorbar,
colorbar style={
            at={(0.85\columnwidth,0.75\columnwidth)},
            anchor=north west,
            align=left,
            ytick = {0,0.05},
            yticklabel style={
            /pgf/number format/fixed,
            scaled ticks=false,
              },
            },
colorbar/width=0.05\columnwidth
]
\addplot [forget plot] graphics [xmin=-0.0125786163522013, xmax=4.0125786163522, ymin=-0.0692598827793667, ymax=16.4145922187099] {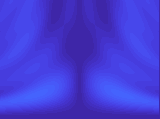};
\end{axis}
\end{tikzpicture}%

%% file: figures/NTF.tex
\begin{tikzpicture}

\begin{axis}[%
width=0.75\columnwidth,
height=0.75\columnwidth,
at={(0\columnwidth,0\columnwidth)},
scale only axis,
axis on top,
xmin=-60,
xmax=60,
xtick={-60,  60},
xlabel style={font=\color{white!15!black},yshift=5pt},
xlabel={$\hat{x}$ (cpd)},
y dir=reverse,
ymin=-60,
ymax=60,
ytick={-60,  60},
ylabel style={font=\color{white!15!black},yshift=-12pt},
ylabel={$\hat{y}$ (cpd)},
axis background/.style={fill=white},
title style={font=\bfseries},
title={NCSF}
]
\addplot [forget plot] graphics [xmin=-67.8068104111632, xmax=67.8068104111632, ymin=-67.8068104111632, ymax=67.8068104111632] {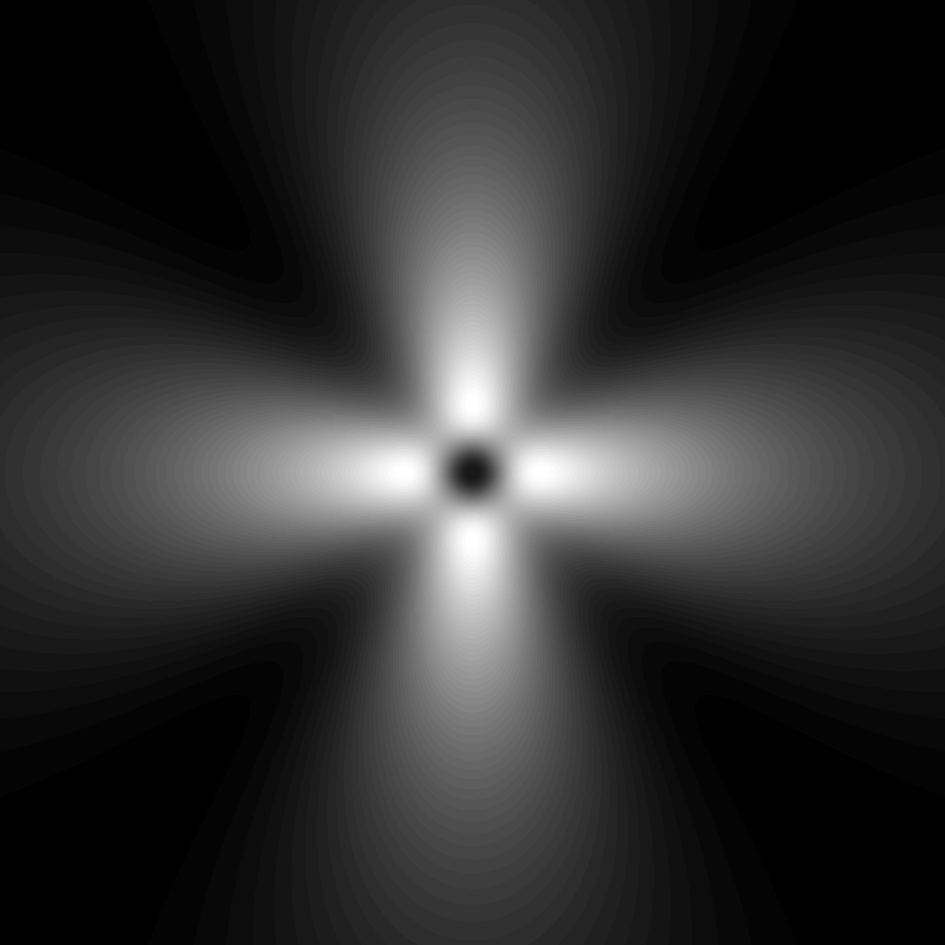};
\end{axis}
\end{tikzpicture}%

%% file: figures/NTF_1D.tex
\definecolor{myred}{rgb}{0.8078,0.1255,0.1608}%
\begin{tikzpicture}

\begin{axis}[%
width=\columnwidth,
height=0.75\columnwidth,
at={(0\columnwidth,0\columnwidth)},
scale only axis,
xmin=0,
xmax=70,
xlabel style={font=\color{white!15!black}},
xlabel={$\hat{x}$ (cpd)},
xmin = 0,
xmax = 60,
ymin=0,
ymax=1,
ytick={0, 1},
axis background/.style={fill=white},
title style={font=\bfseries},
title={NCSF (1D)}
]
\addplot [color=myred, line width=1.4pt, forget plot]
  table[row sep=crcr]{%
0	0.10184453728531\\
0.143506477060669	0.100277181776487\\
0.287012954121323	0.0981136985530638\\
0.430519431181992	0.0971747864100016\\
0.574025908242646	0.0978621749989155\\
0.717532385303315	0.100266613022617\\
0.861038862363969	0.104397244691323\\
1.00454533942464	0.110231918814424\\
1.14805181648531	0.11773198749601\\
1.29155829354598	0.126847834705613\\
1.43506477060663	0.137521418906092\\
1.5785712476673	0.149687701446273\\
1.72207772472795	0.163275615259053\\
1.86558420178862	0.178208831440815\\
2.00909067884929	0.194406433979324\\
2.15259715590994	0.211783552344059\\
2.2961036329706	0.230251974613666\\
2.43961011003127	0.249720750900006\\
2.58311658709194	0.270096790321345\\
2.72662306415259	0.291285451371898\\
2.87012954121326	0.313191123721742\\
3.01363601827393	0.335717798531798\\
3.1571424953346	0.358769623914184\\
3.30064897239524	0.382251442012373\\
3.44415544945591	0.406069304209576\\
3.58766192651657	0.430108700966106\\
3.73116840357724	0.454296632127896\\
3.8746748806379	0.478552323476146\\
4.01818135769857	0.502791194892714\\
4.16168783475922	0.526932029185187\\
4.30519431181989	0.550897292775995\\
4.44870078888054	0.574613418946668\\
4.59220726594121	0.598011052549896\\
4.73571374300188	0.621025255582028\\
4.87922022006255	0.643595673457158\\
5.02272669712319	0.665666662242505\\
5.16623317418386	0.687187377503658\\
5.30973965124453	0.708111825766216\\
5.4532461283052	0.728398879926648\\
5.59675260536585	0.748012260237771\\
5.74025908242652	0.766920482752666\\
5.88376555948719	0.78509677733363\\
6.02727203654783	0.802518977519697\\
6.1707785136085	0.819169384697001\\
6.31428499066917	0.835034609131272\\
6.45779146772983	0.850105390501552\\
6.60129794479049	0.864376400620077\\
6.74480442185116	0.877846031036562\\
6.88831089891181	0.890516168207703\\
7.03181737597248	0.902391958866741\\
7.17532385303313	0.913481568155695\\
7.3188303300938	0.923795932987052\\
7.46233680715447	0.933348512984923\\
7.60584328421514	0.942155041220878\\
7.74934976127578	0.950233276809667\\
7.89285623833645	0.957602761267747\\
8.03636271539712	0.96428458036585\\
8.17986919245779	0.97030113302852\\
8.32337566951844	0.975675908651299\\
8.46688214657911	0.980433274022555\\
8.61038862363978	0.984598270854279\\
8.75389510070043	0.988196424746591\\
8.89740157776109	0.991253566236239\\
9.04090805482176	0.993795664411728\\
9.18441453188242	0.995848673418422\\
9.32792100894309	0.997438392027292\\
9.47142748600375	0.99859033630191\\
9.6149339630644	0.999329625270737\\
9.75844044012507	0.999680879396148\\
9.90194691718574	0.999668131528502\\
10.0454533942464	0.999314749942923\\
10.1889598713071	0.998643372978408\\
10.3324663483677	0.997675854733141\\
10.4759728254284	0.996433221216208\\
10.619479302489	0.994935636313715\\
10.7629857795497	0.993202376896153\\
10.9064922566104	0.991251816372928\\
11.049998733671	0.989101415988614\\
11.1935052107317	0.986767723152876\\
11.3370116877924	0.984266376101283\\
11.480518164853	0.981612114196559\\
11.6240246419137	0.97881879319827\\
11.7675311189743	0.975899404852768\\
11.911037596035	0.972866100183443\\
12.0545440730957	0.969730215893202\\
12.1980505501564	0.966502303325865\\
12.341557027217	0.96319215947006\\
12.4850635042777	0.959808859527522\\
12.6285699813383	0.95636079060688\\
12.772076458399	0.952855686143468\\
12.9155829354597	0.949300660684865\\
13.0590894125203	0.945702244720401\\
13.202595889581	0.942066419270315\\
13.3461023666416	0.938398649986283\\
13.4896088437023	0.93470392054944\\
13.633115320763	0.930986765184559\\
13.7766217978236	0.92725130013949\\
13.9201282748843	0.923501254007349\\
14.063634751945	0.919739996795013\\
14.2071412290056	0.915970567665366\\
14.3506477060663	0.9121957013023\\
14.494154183127	0.908417852866883\\
14.6376606601876	0.904639221530257\\
14.7811671372483	0.90086177258396\\
14.9246736143089	0.89708725814146\\
15.0681800913696	0.893317236455881\\
15.2116865684303	0.889553089888372\\
15.3551930454909	0.885796041569327\\
15.4986995225516	0.88204717080098\\
15.6422059996123	0.878307427254756\\
15.7857124766729	0.874577644020483\\
15.9292189537336	0.870858549567019\\
16.0727254307942	0.867150778675481\\
16.2162319078549	0.863454882406903\\
16.3597383849156	0.85977133716609\\
16.5032448619762	0.85610055292274\\
16.6467513390369	0.852442880649668\\
16.7902578160976	0.848798619036254\\
16.9337642931582	0.845168020533218\\
17.0772707702189	0.841551296782475\\
17.2207772472795	0.837948623483303\\
17.3642837243402	0.83436014474336\\
17.5077902014009	0.830785976960301\\
17.6512966784615	0.827226212276931\\
17.7948031555222	0.82368092164997\\
17.9383096325829	0.820150157569739\\
18.0818161096435	0.816633956465269\\
18.2253225867042	0.813132340826719\\
18.3688290637648	0.809645321074378\\
18.5123355408255	0.806172897201082\\
18.6558420178862	0.802715060212523\\
18.7993484949468	0.799271793387759\\
18.9428549720075	0.795843073380095\\
19.0863614490682	0.792428871176635\\
19.2298679261288	0.789029152932972\\
19.3733744031895	0.785643880697794\\
19.5168808802502	0.782273013040713\\
19.6603873573108	0.778916505595142\\
19.8038938343715	0.77557431152681\\
19.9474003114321	0.772246381937299\\
20.0909067884928	0.76893266621095\\
20.2344132655535	0.765633112312483\\
20.3779197426141	0.762347667041856\\
20.5214262196748	0.759076276252068\\
20.6649326967354	0.755818885034931\\
20.8084391737961	0.752575437879206\\
20.9519456508568	0.749345878804961\\
21.0954521279174	0.746130151477474\\
21.2389586049781	0.742928199303616\\
21.3824650820388	0.739739965513236\\
21.5259715590994	0.736565393227727\\
21.6694780361601	0.733404425517681\\
21.8129845132208	0.730257005451242\\
21.9564909902814	0.72712307613458\\
22.0999974673421	0.72400258074567\\
22.2435039444027	0.720895462562422\\
22.3870104214634	0.71780166498603\\
22.5305168985241	0.714721131560302\\
22.6740233755847	0.711653805987604\\
22.8175298526454	0.708599632141957\\
22.9610363297061	0.705558554079764\\
23.1045428067667	0.702530516048536\\
23.2480492838274	0.699515462493958\\
23.391555760888	0.696513338065571\\
23.5350622379487	0.693524087621299\\
23.6785687150094	0.690547656231025\\
23.82207519207	0.687583989179364\\
23.9655816691307	0.684633031967785\\
24.1090881461914	0.6816947303162\\
24.252594623252	0.678769030164087\\
24.3961011003127	0.675855877671267\\
24.5396075773733	0.672955219218366\\
24.683114054434	0.67006700140703\\
24.8266205314947	0.667191171059949\\
24.9701270085553	0.664327675220705\\
25.113633485616	0.661476461153488\\
25.2571399626767	0.658637476342709\\
25.4006464397373	0.655810668492517\\
25.544152916798	0.652995985526248\\
25.6876593938587	0.650193375585817\\
25.8311658709193	0.647402787031062\\
25.97467234798	0.644624168439051\\
26.1181788250406	0.641857468603362\\
26.2616853021013	0.639102636533334\\
26.405191779162	0.636359621453305\\
26.5486982562226	0.633628372801836\\
26.6922047332833	0.630908840230915\\
26.835711210344	0.628200973605162\\
26.9792176874046	0.625504723001025\\
27.1227241644653	0.622820038705967\\
27.2662306415259	0.620146871217654\\
27.4097371185866	0.61748517124314\\
27.5532435956473	0.614834889698047\\
27.6967500727079	0.61219597770575\\
27.8402565497686	0.609568386596557\\
27.9837630268293	0.606952067906894\\
28.1272695038899	0.60434697337848\\
28.2707759809506	0.601753054957521\\
28.4142824580112	0.59917026479389\\
28.5577889350719	0.596598555240311\\
28.7012954121326	0.594037878851551\\
28.8448018891932	0.59148818838361\\
28.9883083662539	0.588949436792907\\
29.1318148433146	0.586421577235482\\
29.2753213203752	0.583904563066184\\
29.4188277974359	0.581398347837873\\
29.5623342744966	0.578902885300616\\
29.7058407515572	0.576418129400895\\
29.8493472286179	0.573944034280803\\
29.9928537056785	0.571480554277257\\
30.1363601827392	0.569027643921201\\
30.2798666597999	0.566585257936821\\
30.4233731368605	0.564153351240759\\
30.5668796139212	0.561731878941324\\
30.7103860909818	0.55932079633771\\
30.8538925680425	0.556920058919221\\
30.9973990451032	0.554529622364492\\
31.1409055221638	0.552149442540708\\
31.2844119992245	0.549779475502841\\
31.4279184762852	0.547419677492873\\
31.5714249533458	0.545070004939029\\
31.7149314304065	0.542730414455016\\
31.8584379074671	0.540400862839256\\
32.0019443845278	0.538081307074126\\
32.1454508615885	0.535771704325204\\
32.2889573386491	0.533472011940508\\
32.4324638157098	0.531182187449748\\
32.5759702927705	0.528902188563575\\
32.7194767698311	0.52663197317283\\
32.8629832468918	0.524371499347803\\
33.0064897239525	0.522120725337487\\
33.1499962010131	0.51987960956884\\
33.2935026780738	0.517648110646048\\
33.4370091551344	0.515426187349788\\
33.5805156321951	0.513213798636495\\
33.7240221092558	0.511010903637635\\
33.8675285863164	0.508817461658976\\
34.0110350633771	0.506633432179864\\
34.1545415404378	0.504458774852498\\
34.2980480174984	0.502293449501217\\
34.4415544945591	0.500137416121773\\
34.5850609716197	0.497990634880628\\
34.7285674486804	0.495853066114232\\
34.8720739257411	0.493724670328321\\
35.0155804028017	0.491605408197205\\
35.1590868798624	0.489495240563069\\
35.3025933569231	0.487394128435267\\
35.4460998339837	0.485302032989625\\
35.5896063110444	0.483218915567746\\
35.733112788105	0.481144737676313\\
35.8766192651657	0.4790794609864\\
36.0201257422264	0.477023047332783\\
36.163632219287	0.474975458713255\\
36.3071386963477	0.47293665728794\\
36.4506451734084	0.470906605378613\\
36.594151650469	0.468885265468024\\
36.7376581275297	0.466872600199219\\
36.8811646045904	0.464868572374871\\
37.024671081651	0.462873144956601\\
37.1681775587117	0.460886281064322\\
37.3116840357723	0.458907943975562\\
37.455190512833	0.45693809712481\\
37.5986969898937	0.454976704102848\\
37.7422034669543	0.4530237286561\\
37.885709944015	0.451079134685972\\
38.0292164210756	0.4491428862482\\
38.1727228981363	0.447214947552202\\
38.316229375197	0.445295282960427\\
38.4597358522577	0.443383856987714\\
38.6032423293183	0.441480634300645\\
38.746748806379	0.439585579716906\\
38.8902552834396	0.437698658204652\\
39.0337617605003	0.43581983488187\\
39.177268237561	0.433949075015743\\
39.3207747146216	0.432086344022025\\
39.4642811916823	0.430231607464413\\
39.607787668743	0.428384831053918\\
39.7512941458036	0.426545980648246\\
39.8948006228643	0.424715022251173\\
40.0383070999249	0.422891922011935\\
40.1818135769856	0.421076646224603\\
40.3253200540463	0.419269161327481\\
40.4688265311069	0.417469433902485\\
40.6123330081676	0.415677430674541\\
40.7558394852283	0.41389311851098\\
40.8993459622889	0.412116464420931\\
41.0428524393496	0.410347435554724\\
41.1863589164102	0.408585999203287\\
41.3298653934709	0.406832122797558\\
41.4733718705316	0.405085773907883\\
41.6168783475922	0.40334692024343\\
41.7603848246529	0.401615529651599\\
41.9038913017136	0.399891570117435\\
42.0473977787742	0.398175009763045\\
42.1909042558349	0.396465816847016\\
42.3344107328955	0.394763959763836\\
42.4779172099562	0.393069407043317\\
42.6214236870169	0.39138212735002\\
42.7649301640775	0.389702089482683\\
42.9084366411382	0.388029262373654\\
43.0519431181988	0.386363615088317\\
43.1954495952595	0.384705116824533\\
43.3389560723202	0.383053736912074\\
43.4824625493808	0.381409444812063\\
43.6259690264415	0.379772210116417\\
43.7694755035022	0.378142002547287\\
43.9129819805628	0.376518791956511\\
44.0564884576235	0.374902548325056\\
44.1999949346842	0.37329324176247\\
44.3435014117448	0.37169084250634\\
44.4870078888055	0.370095320921741\\
44.6305143658661	0.368506647500698\\
44.7740208429268	0.366924792861642\\
44.9175273199875	0.365349727748874\\
45.0610337970481	0.363781423032033\\
45.2045402741088	0.362219849705553\\
45.3480467511694	0.360664978888142\\
45.4915532282301	0.359116781822245\\
45.6350597052908	0.357575229873522\\
45.7785661823515	0.35604029453032\\
45.9220726594121	0.354511947403153\\
46.0655791364728	0.35299016022418\\
46.2090856135334	0.351474904846685\\
46.3525920905941	0.349966153244565\\
46.4960985676548	0.348463877511813\\
46.6396050447154	0.346968049862006\\
46.7831115217761	0.345478642627798\\
46.9266179988368	0.343995628260409\\
47.0701244758974	0.342518979329122\\
47.2136309529581	0.341048668520778\\
47.3571374300187	0.339584668639275\\
47.5006439070794	0.33812695260507\\
47.6441503841401	0.33667549345468\\
47.7876568612007	0.335230264340189\\
47.9311633382614	0.333791238528751\\
48.0746698153221	0.332358389402107\\
48.2181762923827	0.330931690456085\\
48.3616827694434	0.329511115300125\\
48.505189246504	0.328096637656783\\
48.6486957235647	0.326688231361257\\
48.7922022006254	0.325285870360902\\
48.935708677686	0.323889528714753\\
49.0792151547467	0.322499180593044\\
49.2227216318074	0.32111480027674\\
49.366228108868	0.31973636215706\\
49.5097345859287	0.318363840735006\\
49.6532410629894	0.316997210620895\\
49.79674754005	0.315636446533894\\
49.9402540171107	0.314281523301553\\
50.0837604941713	0.312932415859342\\
50.227266971232	0.31158909925019\\
50.3707734482927	0.310251548624029\\
50.5142799253533	0.308919739237331\\
50.657786402414	0.307593646452659\\
50.8012928794746	0.306273245738211\\
50.9447993565353	0.304958512667366\\
51.088305833596	0.303649422918239\\
51.2318123106566	0.302345952273233\\
51.3753187877173	0.301048076618589\\
51.518825264778	0.299755771943947\\
51.6623317418386	0.298469014341901\\
51.8058382188993	0.29718778000756\\
51.94934469596	0.295912045238112\\
52.0928511730206	0.294641786432382\\
52.2363576500813	0.293376980090404\\
52.3798641271419	0.292117602812984\\
52.5233706042026	0.29086363130127\\
52.6668770812633	0.289615042356323\\
52.8103835583239	0.288371812878692\\
52.9538900353846	0.287133919867984\\
53.0973965124453	0.285901340422444\\
53.2409029895059	0.284674051738534\\
53.3844094665666	0.283452031110509\\
53.5279159436272	0.28223525593\\
53.6714224206879	0.2810237036856\\
53.8149288977486	0.279817351962447\\
53.9584353748092	0.278616178441813\\
54.1019418518699	0.277420160900687\\
54.2454483289306	0.276229277211373\\
54.3889548059912	0.27504350534108\\
54.5324612830519	0.273862823351512\\
54.6759677601125	0.272687209398467\\
54.8194742371732	0.271516641731436\\
54.9629807142339	0.270351098693198\\
55.1064871912945	0.269190558719425\\
55.2499936683552	0.268035000338278\\
55.3935001454159	0.266884402170021\\
55.5370066224765	0.265738742926616\\
55.6805130995372	0.26459800141134\\
55.8240195765978	0.263462156518387\\
55.9675260536585	0.262331187232484\\
56.1110325307192	0.2612050726285\\
56.2545390077798	0.260083791871062\\
56.3980454848405	0.25896732421417\\
56.5415519619012	0.257855649000813\\
56.6850584389618	0.256748745662593\\
56.8285649160225	0.25564659371934\\
56.9720713930832	0.254549172778736\\
57.1155778701438	0.253456462535941\\
57.2590843472045	0.252368442773217\\
57.4025908242651	0.251285093359552\\
57.5460973013258	0.250206394250294\\
57.6896037783865	0.249132325486779\\
57.8331102554471	0.248062867195962\\
57.9766167325078	0.246997999590049\\
58.1201232095684	0.245937702966137\\
58.2636296866291	0.244881957705844\\
58.4071361636898	0.243830744274953\\
58.5506426407504	0.242784043223047\\
58.6941491178111	0.241741835183153\\
58.8376555948718	0.240704100871382\\
58.9811620719324	0.239670821086576\\
59.1246685489931	0.238641976709952\\
59.2681750260538	0.237617548704749\\
59.4116815031144	0.236597518115878\\
59.5551879801751	0.235581866069568\\
59.6986944572357	0.234570573773026\\
59.8422009342964	0.233563622514081\\
59.9857074113571	0.232560993660843\\
60.1292138884177	0.231562668661361\\
60.2727203654784	0.230568629043276\\
60.416226842539	0.229578856413482\\
60.5597333195997	0.228593332457789\\
60.7032397966604	0.22761203894058\\
60.8467462737211	0.226634957704477\\
60.9902527507817	0.225662070670008\\
61.1337592278424	0.224693359835268\\
61.277265704903	0.223728807275593\\
61.4207721819637	0.22276839514322\\
61.5642786590244	0.221812105666968\\
61.707785136085	0.220859921151902\\
61.8512916131457	0.219911823979011\\
61.9947980902063	0.218967796604879\\
62.138304567267	0.218027821561364\\
62.2818110443277	0.217091881455273\\
62.4253175213883	0.216159958968044\\
62.568823998449	0.215232036855421\\
62.7123304755097	0.214308097947141\\
62.8558369525703	0.213388125146614\\
62.999343429631	0.212472101430607\\
63.1428499066916	0.211560009848932\\
63.2863563837523	0.21065183352413\\
63.429862860813	0.209747555651162\\
63.5733693378736	0.208847159497094\\
63.7168758149343	0.207950628400795\\
63.860382291995	0.207057945772622\\
64.0038887690556	0.20616909509412\\
64.1473952461163	0.205284059917712\\
64.290901723177	0.204402823866397\\
64.4344082002376	0.203525370633446\\
64.5779146772983	0.202651683982106\\
64.7214211543589	0.201781747745293\\
64.8649276314196	0.200915545825297\\
65.0084341084803	0.200053062193485\\
65.1519405855409	0.199194280890003\\
65.2954470626016	0.198339186023484\\
65.4389535396623	0.19748776177075\\
65.5824600167229	0.196639992376523\\
65.7259664937836	0.195795862153135\\
65.8694729708442	0.194955355480234\\
66.0129794479049	0.194118456804498\\
66.1564859249656	0.193285150639348\\
66.2999924020262	0.192455421564659\\
66.4434988790869	0.191629254226482\\
66.5870053561476	0.190806633336749\\
66.7305118332082	0.189987543673001\\
66.8740183102689	0.189171970078102\\
67.0175247873295	0.18835989745996\\
67.1610312643902	0.187551310791245\\
67.3045377414509	0.186746195109117\\
67.4480442185116	0.185944535514944\\
67.5915506955722	0.185146317174032\\
67.7350571726329	0.184351525315345\\
};
\end{axis}
\end{tikzpicture}%

%% file: figures/heightmapdraw.tex
\begin{tikzpicture}

\begin{axis}[%
width=0.75\columnwidth,
height=0.75\columnwidth,
at={(0\columnwidth,0\columnwidth)},
scale only axis,
point meta min=0,
point meta max=1.48183733017504, %e-06,
axis on top,
xmin=-5.01,
xmax=5.01,
xtick={-5,  5},
xlabel style={font=\color{white!15!black},yshift=5pt},
xlabel={$s (mm)$},
y dir=reverse,
ymin=-5.01,
ymax=5.01,
ytick={-5,  5},
ylabel style={font=\color{white!15!black},yshift=-14pt},
ylabel={$t (mm)$},
axis background/.style={fill=white},
title style={font=\bfseries},
title={Height map},
colormap={mymap}{[1pt] rgb(0pt)=(0.2422,0.1504,0.6603); rgb(1pt)=(0.25039,0.164995,0.707614); rgb(2pt)=(0.257771,0.181781,0.751138); rgb(3pt)=(0.264729,0.197757,0.795214); rgb(4pt)=(0.270648,0.214676,0.836371); rgb(5pt)=(0.275114,0.234238,0.870986); rgb(6pt)=(0.2783,0.255871,0.899071); rgb(7pt)=(0.280333,0.278233,0.9221); rgb(8pt)=(0.281338,0.300595,0.941376); rgb(9pt)=(0.281014,0.322757,0.957886); rgb(10pt)=(0.279467,0.344671,0.971676); rgb(11pt)=(0.275971,0.366681,0.982905); rgb(12pt)=(0.269914,0.3892,0.9906); rgb(13pt)=(0.260243,0.412329,0.995157); rgb(14pt)=(0.244033,0.435833,0.998833); rgb(15pt)=(0.220643,0.460257,0.997286); rgb(16pt)=(0.196333,0.484719,0.989152); rgb(17pt)=(0.183405,0.507371,0.979795); rgb(18pt)=(0.178643,0.528857,0.968157); rgb(19pt)=(0.176438,0.549905,0.952019); rgb(20pt)=(0.168743,0.570262,0.935871); rgb(21pt)=(0.154,0.5902,0.9218); rgb(22pt)=(0.146029,0.609119,0.907857); rgb(23pt)=(0.138024,0.627629,0.89729); rgb(24pt)=(0.124814,0.645929,0.888343); rgb(25pt)=(0.111252,0.6635,0.876314); rgb(26pt)=(0.0952095,0.679829,0.859781); rgb(27pt)=(0.0688714,0.694771,0.839357); rgb(28pt)=(0.0296667,0.708167,0.816333); rgb(29pt)=(0.00357143,0.720267,0.7917); rgb(30pt)=(0.00665714,0.731214,0.766014); rgb(31pt)=(0.0433286,0.741095,0.73941); rgb(32pt)=(0.0963952,0.75,0.712038); rgb(33pt)=(0.140771,0.7584,0.684157); rgb(34pt)=(0.1717,0.766962,0.655443); rgb(35pt)=(0.193767,0.775767,0.6251); rgb(36pt)=(0.216086,0.7843,0.5923); rgb(37pt)=(0.246957,0.791795,0.556743); rgb(38pt)=(0.290614,0.79729,0.518829); rgb(39pt)=(0.340643,0.8008,0.478857); rgb(40pt)=(0.3909,0.802871,0.435448); rgb(41pt)=(0.445629,0.802419,0.390919); rgb(42pt)=(0.5044,0.7993,0.348); rgb(43pt)=(0.561562,0.794233,0.304481); rgb(44pt)=(0.617395,0.787619,0.261238); rgb(45pt)=(0.671986,0.779271,0.2227); rgb(46pt)=(0.7242,0.769843,0.191029); rgb(47pt)=(0.773833,0.759805,0.16461); rgb(48pt)=(0.820314,0.749814,0.153529); rgb(49pt)=(0.863433,0.7406,0.159633); rgb(50pt)=(0.903543,0.733029,0.177414); rgb(51pt)=(0.939257,0.728786,0.209957); rgb(52pt)=(0.972757,0.729771,0.239443); rgb(53pt)=(0.995648,0.743371,0.237148); rgb(54pt)=(0.996986,0.765857,0.219943); rgb(55pt)=(0.995205,0.789252,0.202762); rgb(56pt)=(0.9892,0.813567,0.188533); rgb(57pt)=(0.978629,0.838629,0.176557); rgb(58pt)=(0.967648,0.8639,0.16429); rgb(59pt)=(0.96101,0.889019,0.153676); rgb(60pt)=(0.959671,0.913457,0.142257); rgb(61pt)=(0.962795,0.937338,0.12651); rgb(62pt)=(0.969114,0.960629,0.106362); rgb(63pt)=(0.9769,0.9839,0.0805)},
colorbar,
colorbar style={
            at={(0.70\columnwidth,0.75\columnwidth)},
            anchor=north west,
            align=left,
            ytick={0,1.4},
            title style = {at = {(0.1\columnwidth,0.7\columnwidth)}},
            title={$\mu m$}},
colorbar/width=0.05\columnwidth
]
\addplot [forget plot] graphics [xmin=-5.01, xmax=5.01, ymin=-5.01, ymax=5.01] {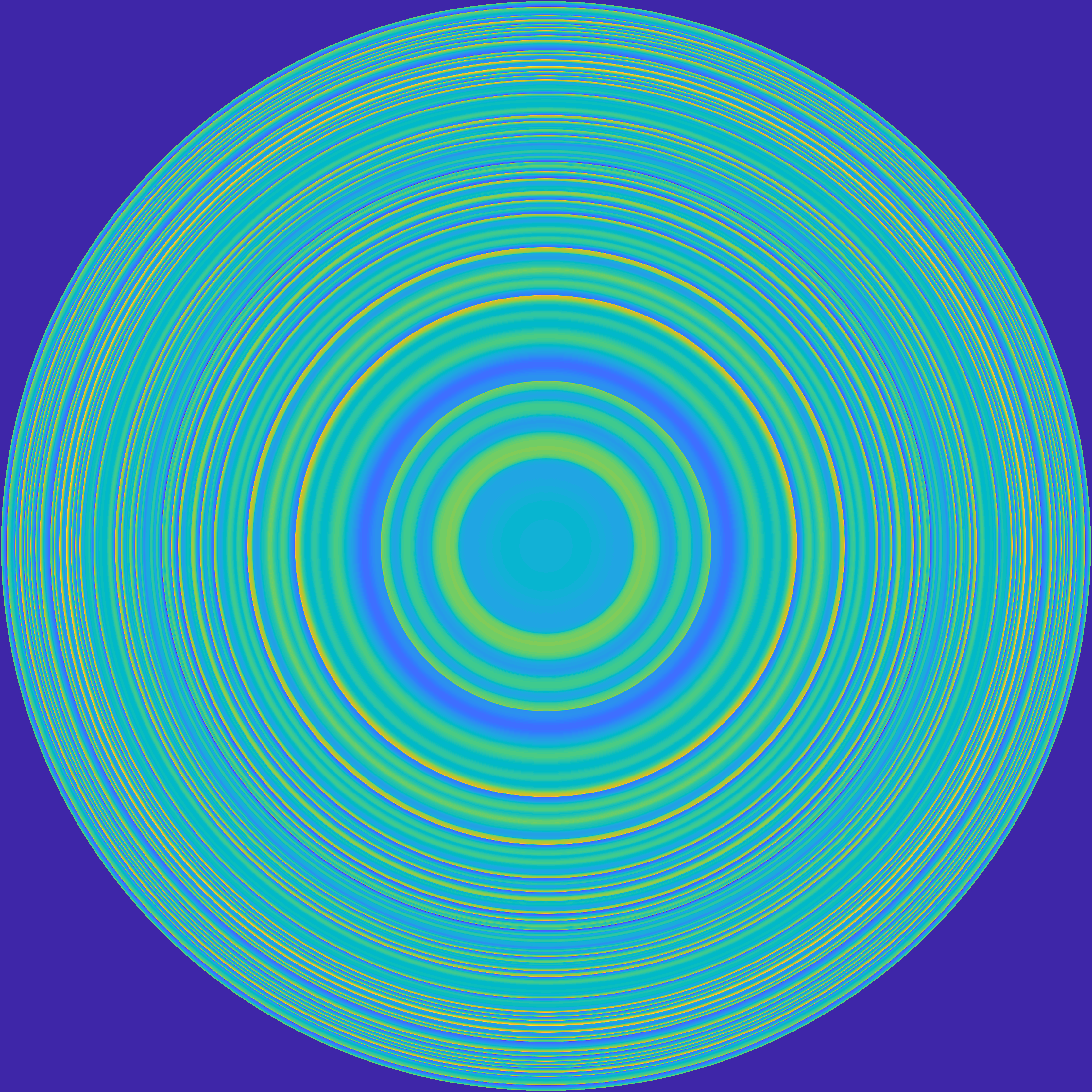};
\end{axis}

\end{tikzpicture}%

%% file: figures/MTF/PSF1D_wo_network.tex
\begin{tikzpicture}

\begin{axis}[%
width=0.75\columnwidth,
height=0.75\columnwidth,
at={(0\columnwidth,0\columnwidth)},
scale only axis,
axis on top,
xmin=-0.0125786163522013,
xmax=4.0125786163522,
xlabel style={font=\color{white!15!black},yshift=5pt},
xlabel={D},
y dir=reverse,
ymin=-27.7602186773872,
ymax=27.7602186773872,
ylabel style={font=\color{white!15!black},yshift=-12pt},
ylabel={arcmin},
ytick={-20,20},
axis background/.style={fill=white},
title style={font=\bfseries},
title={Proposed}
]
\addplot [forget plot] graphics [xmin=-0.0125786163522013, xmax=4.0125786163522, ymin=-27.7602186773872, ymax=27.7602186773872] {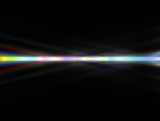};
\end{axis}
\end{tikzpicture}%

%% file: figures/MTF/PSF1D_lensonly.tex
\begin{tikzpicture}

\begin{axis}[%
width=0.75\columnwidth,
height=0.75\columnwidth,
at={(0\columnwidth,0\columnwidth)},
scale only axis,
axis on top,
xmin=-0.0125786163522013,
xmax=4.0125786163522,
xlabel style={font=\color{white!15!black},yshift=5pt},
xlabel={D},
y dir=reverse,
ymin=-27.7602186773872,
ymax=27.7602186773872,
ylabel style={font=\color{white!15!black},yshift=-12pt},
ymajorticks=false,
axis background/.style={fill=white},
title style={font=\bfseries},
title={Conventional}
]
\addplot [forget plot] graphics [xmin=-0.0125786163522013, xmax=4.0125786163522, ymin=-27.7602186773872, ymax=27.7602186773872] {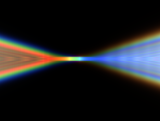};
\end{axis}
\end{tikzpicture}%

%% file: figures/MTF/MTF_1D_green_lensonly_lines.tex
\definecolor{myred}{rgb}{0.8078,0.1255,0.1608}%
\definecolor{mygreen}{rgb}{0.1608,0.8078,0.1255}%
\definecolor{myblue}{rgb}{0.1255,0.1608,0.8078}%

\begin{tikzpicture}

\begin{axis}[%
width=0.75\columnwidth,
height=0.75\columnwidth,
at={(0\columnwidth,0\columnwidth)},
scale only axis,
point meta min=0,
point meta max=1,
axis on top,
xmin=-0.0125786163522013,
xmax=4.0125786163522,
xlabel style={font=\color{white!15!black},yshift=3pt},
xlabel={D},
xtick={0,4},
ymin=-0.0692598827793667,
ymax=30.1280490090245,
ylabel style={font=\color{white!15!black},yshift=-5pt},
ylabel={Frequency (cpd)},
ytick={0,30},
axis background/.style={fill=white},
title style={font=\bfseries},
title={Spherical},
colormap={mymap}{[1pt] rgb(0pt)=(0.2422,0.1504,0.6603); rgb(1pt)=(0.25039,0.164995,0.707614); rgb(2pt)=(0.257771,0.181781,0.751138); rgb(3pt)=(0.264729,0.197757,0.795214); rgb(4pt)=(0.270648,0.214676,0.836371); rgb(5pt)=(0.275114,0.234238,0.870986); rgb(6pt)=(0.2783,0.255871,0.899071); rgb(7pt)=(0.280333,0.278233,0.9221); rgb(8pt)=(0.281338,0.300595,0.941376); rgb(9pt)=(0.281014,0.322757,0.957886); rgb(10pt)=(0.279467,0.344671,0.971676); rgb(11pt)=(0.275971,0.366681,0.982905); rgb(12pt)=(0.269914,0.3892,0.9906); rgb(13pt)=(0.260243,0.412329,0.995157); rgb(14pt)=(0.244033,0.435833,0.998833); rgb(15pt)=(0.220643,0.460257,0.997286); rgb(16pt)=(0.196333,0.484719,0.989152); rgb(17pt)=(0.183405,0.507371,0.979795); rgb(18pt)=(0.178643,0.528857,0.968157); rgb(19pt)=(0.176438,0.549905,0.952019); rgb(20pt)=(0.168743,0.570262,0.935871); rgb(21pt)=(0.154,0.5902,0.9218); rgb(22pt)=(0.146029,0.609119,0.907857); rgb(23pt)=(0.138024,0.627629,0.89729); rgb(24pt)=(0.124814,0.645929,0.888343); rgb(25pt)=(0.111252,0.6635,0.876314); rgb(26pt)=(0.0952095,0.679829,0.859781); rgb(27pt)=(0.0688714,0.694771,0.839357); rgb(28pt)=(0.0296667,0.708167,0.816333); rgb(29pt)=(0.00357143,0.720267,0.7917); rgb(30pt)=(0.00665714,0.731214,0.766014); rgb(31pt)=(0.0433286,0.741095,0.73941); rgb(32pt)=(0.0963952,0.75,0.712038); rgb(33pt)=(0.140771,0.7584,0.684157); rgb(34pt)=(0.1717,0.766962,0.655443); rgb(35pt)=(0.193767,0.775767,0.6251); rgb(36pt)=(0.216086,0.7843,0.5923); rgb(37pt)=(0.246957,0.791795,0.556743); rgb(38pt)=(0.290614,0.79729,0.518829); rgb(39pt)=(0.340643,0.8008,0.478857); rgb(40pt)=(0.3909,0.802871,0.435448); rgb(41pt)=(0.445629,0.802419,0.390919); rgb(42pt)=(0.5044,0.7993,0.348); rgb(43pt)=(0.561562,0.794233,0.304481); rgb(44pt)=(0.617395,0.787619,0.261238); rgb(45pt)=(0.671986,0.779271,0.2227); rgb(46pt)=(0.7242,0.769843,0.191029); rgb(47pt)=(0.773833,0.759805,0.16461); rgb(48pt)=(0.820314,0.749814,0.153529); rgb(49pt)=(0.863433,0.7406,0.159633); rgb(50pt)=(0.903543,0.733029,0.177414); rgb(51pt)=(0.939257,0.728786,0.209957); rgb(52pt)=(0.972757,0.729771,0.239443); rgb(53pt)=(0.995648,0.743371,0.237148); rgb(54pt)=(0.996986,0.765857,0.219943); rgb(55pt)=(0.995205,0.789252,0.202762); rgb(56pt)=(0.9892,0.813567,0.188533); rgb(57pt)=(0.978629,0.838629,0.176557); rgb(58pt)=(0.967648,0.8639,0.16429); rgb(59pt)=(0.96101,0.889019,0.153676); rgb(60pt)=(0.959671,0.913457,0.142257); rgb(61pt)=(0.962795,0.937338,0.12651); rgb(62pt)=(0.969114,0.960629,0.106362); rgb(63pt)=(0.9769,0.9839,0.0805)},
%colorbar,
]
\addplot [forget plot] graphics [xmin=-0.0125786163522013, xmax=4.0125786163522, ymin=-0.0692598827793667, ymax=30.1280490090245] {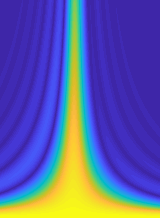};
\addplot [color=mygreen, dashed, line width=1.4pt, forget plot]
  table[row sep=crcr]{%
0.5	0\\
0.5	0.13851976555874\\
0.5	0.277039531117467\\
0.5	0.415559296676193\\
0.5	0.554079062234933\\
0.5	0.69259882779366\\
0.5	0.8311185933524\\
0.5	0.969638358911141\\
0.5	1.10815812446987\\
0.5	1.24667789002859\\
0.5	1.38519765558733\\
0.5	1.52371742114607\\
0.5	1.6622371867048\\
0.5	1.80075695226354\\
0.5	1.93927671782227\\
0.5	2.07779648338099\\
0.5	2.21631624893973\\
0.5	2.35483601449847\\
0.5	2.4933557800572\\
0.5	2.63187554561594\\
0.5	2.77039531117467\\
0.5	2.90891507673339\\
0.5	3.04743484229213\\
0.5	3.18595460785087\\
0.5	3.3244743734096\\
0.5	3.46299413896834\\
0.5	3.60151390452707\\
0.5	3.74003367008579\\
0.5	3.87855343564453\\
0.5	4.01707320120327\\
0.5	4.155592966762\\
0.5	4.29411273232074\\
0.5	4.43263249787947\\
0.5	4.57115226343821\\
0.5	4.70967202899693\\
0.5	4.84819179455566\\
0.5	4.9867115601144\\
0.5	5.12523132567314\\
0.5	5.26375109123187\\
0.5	5.40227085679061\\
0.5	5.54079062234933\\
0.5	5.67931038790806\\
0.5	5.8178301534668\\
0.5	5.95634991902553\\
0.5	6.09486968458427\\
0.5	6.23338945014301\\
0.5	6.37190921570173\\
0.5	6.51042898126046\\
0.5	6.6489487468192\\
0.5	6.78746851237793\\
0.5	6.92598827793667\\
0.5	7.06450804349541\\
0.5	7.20302780905413\\
0.5	7.34154757461286\\
0.5	7.4800673401716\\
0.5	7.61858710573033\\
0.5	7.75710687128907\\
0.5	7.89562663684781\\
0.5	8.03414640240653\\
0.5	8.17266616796526\\
0.5	8.311185933524\\
0.5	8.44970569908273\\
0.5	8.58822546464147\\
0.5	8.72674523020021\\
0.5	8.86526499575893\\
0.5	9.00378476131766\\
0.5	9.1423045268764\\
0.5	9.28082429243513\\
0.5	9.41934405799387\\
0.5	9.55786382355261\\
0.5	9.69638358911133\\
0.5	9.83490335467006\\
0.5	9.9734231202288\\
0.5	10.1119428857875\\
0.5	10.2504626513463\\
0.5	10.388982416905\\
0.5	10.5275021824637\\
0.5	10.6660219480225\\
0.5	10.8045417135812\\
0.5	10.9430614791399\\
0.5	11.0815812446987\\
0.5	11.2201010102574\\
0.5	11.3586207758161\\
0.5	11.4971405413749\\
0.5	11.6356603069336\\
0.5	11.7741800724923\\
0.5	11.9126998380511\\
0.5	12.0512196036098\\
0.5	12.1897393691685\\
0.5	12.3282591347273\\
0.5	12.466778900286\\
0.5	12.6052986658447\\
0.5	12.7438184314035\\
0.5	12.8823381969622\\
0.5	13.0208579625209\\
0.5	13.1593777280797\\
0.5	13.2978974936384\\
0.5	13.4364172591971\\
0.5	13.5749370247559\\
0.5	13.7134567903146\\
0.5	13.8519765558733\\
0.5	13.9904963214321\\
0.5	14.1290160869908\\
0.5	14.2675358525495\\
0.5	14.4060556181083\\
0.5	14.544575383667\\
0.5	14.6830951492257\\
0.5	14.8216149147845\\
0.5	14.9601346803432\\
0.5	15.0986544459019\\
0.5	15.2371742114607\\
0.5	15.3756939770194\\
0.5	15.5142137425781\\
0.5	15.6527335081369\\
0.5	15.7912532736956\\
0.5	15.9297730392543\\
0.5	16.0682928048131\\
};
\addplot [color=myblue, dashed, line width=1.4pt, forget plot]
  table[row sep=crcr]{%
3	0\\
3	0.13851976555874\\
3	0.277039531117467\\
3	0.415559296676193\\
3	0.554079062234933\\
3	0.69259882779366\\
3	0.8311185933524\\
3	0.969638358911141\\
3	1.10815812446987\\
3	1.24667789002859\\
3	1.38519765558733\\
3	1.52371742114607\\
3	1.6622371867048\\
3	1.80075695226354\\
3	1.93927671782227\\
3	2.07779648338099\\
3	2.21631624893973\\
3	2.35483601449847\\
3	2.4933557800572\\
3	2.63187554561594\\
3	2.77039531117467\\
3	2.90891507673339\\
3	3.04743484229213\\
3	3.18595460785087\\
3	3.3244743734096\\
3	3.46299413896834\\
3	3.60151390452707\\
3	3.74003367008579\\
3	3.87855343564453\\
3	4.01707320120327\\
3	4.155592966762\\
3	4.29411273232074\\
3	4.43263249787947\\
3	4.57115226343821\\
3	4.70967202899693\\
3	4.84819179455566\\
3	4.9867115601144\\
3	5.12523132567314\\
3	5.26375109123187\\
3	5.40227085679061\\
3	5.54079062234933\\
3	5.67931038790806\\
3	5.8178301534668\\
3	5.95634991902553\\
3	6.09486968458427\\
3	6.23338945014301\\
3	6.37190921570173\\
3	6.51042898126046\\
3	6.6489487468192\\
3	6.78746851237793\\
3	6.92598827793667\\
3	7.06450804349541\\
3	7.20302780905413\\
3	7.34154757461286\\
3	7.4800673401716\\
3	7.61858710573033\\
3	7.75710687128907\\
3	7.89562663684781\\
3	8.03414640240653\\
3	8.17266616796526\\
3	8.311185933524\\
3	8.44970569908273\\
3	8.58822546464147\\
3	8.72674523020021\\
3	8.86526499575893\\
3	9.00378476131766\\
3	9.1423045268764\\
3	9.28082429243513\\
3	9.41934405799387\\
3	9.55786382355261\\
3	9.69638358911133\\
3	9.83490335467006\\
3	9.9734231202288\\
3	10.1119428857875\\
3	10.2504626513463\\
3	10.388982416905\\
3	10.5275021824637\\
3	10.6660219480225\\
3	10.8045417135812\\
3	10.9430614791399\\
3	11.0815812446987\\
3	11.2201010102574\\
3	11.3586207758161\\
3	11.4971405413749\\
3	11.6356603069336\\
3	11.7741800724923\\
3	11.9126998380511\\
3	12.0512196036098\\
3	12.1897393691685\\
3	12.3282591347273\\
3	12.466778900286\\
3	12.6052986658447\\
3	12.7438184314035\\
3	12.8823381969622\\
3	13.0208579625209\\
3	13.1593777280797\\
3	13.2978974936384\\
3	13.4364172591971\\
3	13.5749370247559\\
3	13.7134567903146\\
3	13.8519765558733\\
3	13.9904963214321\\
3	14.1290160869908\\
3	14.2675358525495\\
3	14.4060556181083\\
3	14.544575383667\\
3	14.6830951492257\\
3	14.8216149147845\\
3	14.9601346803432\\
3	15.0986544459019\\
3	15.2371742114607\\
3	15.3756939770194\\
3	15.5142137425781\\
3	15.6527335081369\\
3	15.7912532736956\\
3	15.9297730392543\\
3	16.0682928048131\\
};
\addplot [color=myred, line width=1.4pt, forget plot]
  table[row sep=crcr]{%
0	16\\
0.0251572327044025	16\\
0.050314465408805	16\\
0.0754716981132075	16\\
0.10062893081761	16\\
0.125786163522013	16\\
0.150943396226415	16\\
0.176100628930818	16\\
0.20125786163522	16\\
0.226415094339623	16\\
0.251572327044025	16\\
0.276729559748428	16\\
0.30188679245283	16\\
0.327044025157233	16\\
0.352201257861635	16\\
0.377358490566038	16\\
0.40251572327044	16\\
0.427672955974843	16\\
0.452830188679245	16\\
0.477987421383648	16\\
0.50314465408805	16\\
0.528301886792453	16\\
0.553459119496855	16\\
0.578616352201258	16\\
0.60377358490566	16\\
0.628930817610063	16\\
0.654088050314465	16\\
0.679245283018868	16\\
0.70440251572327	16\\
0.729559748427673	16\\
0.754716981132076	16\\
0.779874213836478	16\\
0.805031446540881	16\\
0.830188679245283	16\\
0.855345911949686	16\\
0.880503144654088	16\\
0.905660377358491	16\\
0.930817610062893	16\\
0.955974842767296	16\\
0.981132075471698	16\\
1.0062893081761	16\\
1.0314465408805	16\\
1.05660377358491	16\\
1.08176100628931	16\\
1.10691823899371	16\\
1.13207547169811	16\\
1.15723270440252	16\\
1.18238993710692	16\\
1.20754716981132	16\\
1.23270440251572	16\\
1.25786163522013	16\\
1.28301886792453	16\\
1.30817610062893	16\\
1.33333333333333	16\\
1.35849056603774	16\\
1.38364779874214	16\\
1.40880503144654	16\\
1.43396226415094	16\\
1.45911949685535	16\\
1.48427672955975	16\\
1.50943396226415	16\\
1.53459119496855	16\\
1.55974842767296	16\\
1.58490566037736	16\\
1.61006289308176	16\\
1.63522012578616	16\\
1.66037735849057	16\\
1.68553459119497	16\\
1.71069182389937	16\\
1.73584905660377	16\\
1.76100628930818	16\\
1.78616352201258	16\\
1.81132075471698	16\\
1.83647798742138	16\\
1.86163522012579	16\\
1.88679245283019	16\\
1.91194968553459	16\\
1.93710691823899	16\\
1.9622641509434	16\\
1.9874213836478	16\\
2.0125786163522	16\\
2.0377358490566	16\\
2.06289308176101	16\\
2.08805031446541	16\\
2.11320754716981	16\\
2.13836477987421	16\\
2.16352201257862	16\\
2.18867924528302	16\\
2.21383647798742	16\\
2.23899371069182	16\\
2.26415094339623	16\\
2.28930817610063	16\\
2.31446540880503	16\\
2.33962264150943	16\\
2.36477987421384	16\\
2.38993710691824	16\\
2.41509433962264	16\\
2.44025157232704	16\\
2.46540880503145	16\\
2.49056603773585	16\\
2.51572327044025	16\\
2.54088050314465	16\\
2.56603773584906	16\\
2.59119496855346	16\\
2.61635220125786	16\\
2.64150943396226	16\\
2.66666666666667	16\\
2.69182389937107	16\\
2.71698113207547	16\\
2.74213836477987	16\\
2.76729559748428	16\\
2.79245283018868	16\\
2.81761006289308	16\\
2.84276729559748	16\\
2.86792452830189	16\\
2.89308176100629	16\\
2.91823899371069	16\\
2.94339622641509	16\\
2.9685534591195	16\\
2.9937106918239	16\\
3.0188679245283	16\\
3.0440251572327	16\\
3.06918238993711	16\\
3.09433962264151	16\\
3.11949685534591	16\\
3.14465408805031	16\\
3.16981132075472	16\\
3.19496855345912	16\\
3.22012578616352	16\\
3.24528301886792	16\\
3.27044025157233	16\\
3.29559748427673	16\\
3.32075471698113	16\\
3.34591194968553	16\\
3.37106918238994	16\\
3.39622641509434	16\\
3.42138364779874	16\\
3.44654088050314	16\\
3.47169811320755	16\\
3.49685534591195	16\\
3.52201257861635	16\\
3.54716981132075	16\\
3.57232704402516	16\\
3.59748427672956	16\\
3.62264150943396	16\\
3.64779874213836	16\\
3.67295597484277	16\\
3.69811320754717	16\\
3.72327044025157	16\\
3.74842767295597	16\\
3.77358490566038	16\\
3.79874213836478	16\\
3.82389937106918	16\\
3.84905660377358	16\\
3.87421383647799	16\\
3.89937106918239	16\\
3.92452830188679	16\\
3.9496855345912	16\\
3.9748427672956	16\\
4	16\\
};
\addplot [color=gray, line width=1.4pt, forget plot]
  table[row sep=crcr]{%
0	3.04743484229213\\
0.0251572327044025	3.04743484229213\\
0.050314465408805	3.18595460785087\\
0.0754716981132075	3.18595460785087\\
0.10062893081761	3.18595460785087\\
0.125786163522013	3.3244743734096\\
0.150943396226415	3.3244743734096\\
0.176100628930818	3.3244743734096\\
0.20125786163522	3.46299413896834\\
0.226415094339623	3.46299413896834\\
0.251572327044025	3.60151390452707\\
0.276729559748428	3.60151390452707\\
0.30188679245283	3.60151390452707\\
0.327044025157233	3.74003367008579\\
0.352201257861635	3.74003367008579\\
0.377358490566038	3.87855343564453\\
0.40251572327044	3.87855343564453\\
0.427672955974843	4.01707320120327\\
0.452830188679245	4.01707320120327\\
0.477987421383648	4.155592966762\\
0.50314465408805	4.155592966762\\
0.528301886792453	4.29411273232074\\
0.553459119496855	4.43263249787947\\
0.578616352201258	4.43263249787947\\
0.60377358490566	4.57115226343821\\
0.628930817610063	4.70967202899693\\
0.654088050314465	4.70967202899693\\
0.679245283018868	4.84819179455566\\
0.70440251572327	4.9867115601144\\
0.729559748427673	5.12523132567314\\
0.754716981132076	5.26375109123187\\
0.779874213836478	5.26375109123187\\
0.805031446540881	5.40227085679061\\
0.830188679245283	5.54079062234933\\
0.855345911949686	5.67931038790806\\
0.880503144654088	5.95634991902553\\
0.905660377358491	6.09486968458427\\
0.930817610062893	6.23338945014301\\
0.955974842767296	6.37190921570173\\
0.981132075471698	6.6489487468192\\
1.0062893081761	6.78746851237793\\
1.0314465408805	7.06450804349541\\
1.05660377358491	7.20302780905413\\
1.08176100628931	7.4800673401716\\
1.10691823899371	7.75710687128907\\
1.13207547169811	8.03414640240653\\
1.15723270440252	8.311185933524\\
1.18238993710692	8.58822546464147\\
1.20754716981132	9.00378476131766\\
1.23270440251572	9.41934405799387\\
1.25786163522013	9.83490335467006\\
1.28301886792453	10.2504626513463\\
1.30817610062893	10.8045417135812\\
1.33333333333333	11.2201010102574\\
1.35849056603774	11.9126998380511\\
1.38364779874214	12.6052986658447\\
1.40880503144654	13.2978974936384\\
1.43396226415094	14.1290160869908\\
1.45911949685535	15.0986544459019\\
1.48427672955975	16.2068125703718\\
1.50943396226415	17.4534904604004\\
1.53459119496855	18.9772078815465\\
1.55974842767296	20.6394450682513\\
1.58490566037736	22.7172415516323\\
1.61006289308176	25.3491170972482\\
1.63522012578616	28.6735914706578\\
1.66037735849057	32.6906646718611\\
1.68553459119497	32.6906646718611\\
1.71069182389937	32.6906646718611\\
1.73584905660377	32.6906646718611\\
1.76100628930818	32.6906646718611\\
1.78616352201258	32.6906646718611\\
1.81132075471698	32.6906646718611\\
1.83647798742138	32.6906646718611\\
1.86163522012579	32.6906646718611\\
1.88679245283019	32.6906646718611\\
1.91194968553459	32.6906646718611\\
1.93710691823899	32.6906646718611\\
1.9622641509434	32.6906646718611\\
1.9874213836478	32.6906646718611\\
2.0125786163522	32.6906646718611\\
2.0377358490566	32.6906646718611\\
2.06289308176101	32.6906646718611\\
2.08805031446541	32.6906646718611\\
2.11320754716981	32.6906646718611\\
2.13836477987421	28.1195124084229\\
2.16352201257862	24.6565182694545\\
2.18867924528302	22.0246427238386\\
2.21383647798742	19.8083264748989\\
2.23899371069182	18.0075695226353\\
2.26415094339623	16.622371867048\\
2.28930817610063	15.3756939770194\\
2.31446540880503	14.2675358525495\\
2.33962264150943	13.2978974936384\\
2.36477987421384	12.6052986658447\\
2.38993710691824	11.7741800724923\\
2.41509433962264	11.2201010102574\\
2.44025157232704	10.6660219480225\\
2.46540880503145	10.1119428857875\\
2.49056603773585	9.69638358911133\\
2.51572327044025	9.28082429243513\\
2.54088050314465	8.86526499575893\\
2.56603773584906	8.44970569908273\\
2.59119496855346	8.17266616796526\\
2.61635220125786	7.89562663684781\\
2.64150943396226	7.61858710573033\\
2.66666666666667	7.34154757461286\\
2.69182389937107	7.06450804349541\\
2.71698113207547	6.92598827793667\\
2.74213836477987	6.6489487468192\\
2.76729559748428	6.51042898126046\\
2.79245283018868	6.23338945014301\\
2.81761006289308	6.09486968458427\\
2.84276729559748	5.95634991902553\\
2.86792452830189	5.8178301534668\\
2.89308176100629	5.67931038790806\\
2.91823899371069	5.54079062234933\\
2.94339622641509	5.40227085679061\\
2.9685534591195	5.26375109123187\\
2.9937106918239	5.12523132567314\\
3.0188679245283	4.9867115601144\\
3.0440251572327	4.84819179455566\\
3.06918238993711	4.70967202899693\\
3.09433962264151	4.70967202899693\\
3.11949685534591	4.57115226343821\\
3.14465408805031	4.43263249787947\\
3.16981132075472	4.43263249787947\\
3.19496855345912	4.29411273232074\\
3.22012578616352	4.155592966762\\
3.24528301886792	4.155592966762\\
3.27044025157233	4.01707320120327\\
3.29559748427673	4.01707320120327\\
3.32075471698113	3.87855343564453\\
3.34591194968553	3.87855343564453\\
3.37106918238994	3.74003367008579\\
3.39622641509434	3.74003367008579\\
3.42138364779874	3.60151390452707\\
3.44654088050314	3.60151390452707\\
3.47169811320755	3.60151390452707\\
3.49685534591195	3.46299413896834\\
3.52201257861635	3.46299413896834\\
3.54716981132075	3.3244743734096\\
3.57232704402516	3.3244743734096\\
3.59748427672956	3.3244743734096\\
3.62264150943396	3.18595460785087\\
3.64779874213836	3.18595460785087\\
3.67295597484277	3.18595460785087\\
3.69811320754717	3.04743484229213\\
3.72327044025157	3.04743484229213\\
3.74842767295597	3.04743484229213\\
3.77358490566038	3.04743484229213\\
3.79874213836478	2.90891507673339\\
3.82389937106918	2.90891507673339\\
3.84905660377358	2.90891507673339\\
3.87421383647799	2.77039531117467\\
3.89937106918239	2.77039531117467\\
3.92452830188679	2.77039531117467\\
3.9496855345912	2.77039531117467\\
3.9748427672956	2.63187554561594\\
4	2.63187554561594\\
};
\end{axis}
\end{tikzpicture}%

%% file: figures/MTF/MTF_1D_green_lines.tex
\definecolor{myred}{rgb}{0.8078,0.1255,0.1608}%
\definecolor{mygreen}{rgb}{0.1608,0.8078,0.1255}%
\definecolor{myblue}{rgb}{0.1255,0.1608,0.8078}%
\begin{tikzpicture}

\begin{axis}[%
width=0.75\columnwidth,
height=0.75\columnwidth,
at={(0\columnwidth,0\columnwidth)},
scale only axis,
point meta min=0,
point meta max=1,
axis on top,
xmin=-0.0125786163522013,
xmax=4.0125786163522,
xlabel style={font=\color{white!15!black},yshift=3pt},
xlabel={D},
xtick={0,4},
ymin=-0.0692598827793667,
ymax=30.1280490090245,
ylabel style={font=\color{white!15!black},yshift=-5pt},
ylabel={Frequency (cpd)},
ytick={0,30},
axis background/.style={fill=white},
title style={font=\bfseries},
title={Proposed},
colormap={mymap}{[1pt] rgb(0pt)=(0.2422,0.1504,0.6603); rgb(1pt)=(0.25039,0.164995,0.707614); rgb(2pt)=(0.257771,0.181781,0.751138); rgb(3pt)=(0.264729,0.197757,0.795214); rgb(4pt)=(0.270648,0.214676,0.836371); rgb(5pt)=(0.275114,0.234238,0.870986); rgb(6pt)=(0.2783,0.255871,0.899071); rgb(7pt)=(0.280333,0.278233,0.9221); rgb(8pt)=(0.281338,0.300595,0.941376); rgb(9pt)=(0.281014,0.322757,0.957886); rgb(10pt)=(0.279467,0.344671,0.971676); rgb(11pt)=(0.275971,0.366681,0.982905); rgb(12pt)=(0.269914,0.3892,0.9906); rgb(13pt)=(0.260243,0.412329,0.995157); rgb(14pt)=(0.244033,0.435833,0.998833); rgb(15pt)=(0.220643,0.460257,0.997286); rgb(16pt)=(0.196333,0.484719,0.989152); rgb(17pt)=(0.183405,0.507371,0.979795); rgb(18pt)=(0.178643,0.528857,0.968157); rgb(19pt)=(0.176438,0.549905,0.952019); rgb(20pt)=(0.168743,0.570262,0.935871); rgb(21pt)=(0.154,0.5902,0.9218); rgb(22pt)=(0.146029,0.609119,0.907857); rgb(23pt)=(0.138024,0.627629,0.89729); rgb(24pt)=(0.124814,0.645929,0.888343); rgb(25pt)=(0.111252,0.6635,0.876314); rgb(26pt)=(0.0952095,0.679829,0.859781); rgb(27pt)=(0.0688714,0.694771,0.839357); rgb(28pt)=(0.0296667,0.708167,0.816333); rgb(29pt)=(0.00357143,0.720267,0.7917); rgb(30pt)=(0.00665714,0.731214,0.766014); rgb(31pt)=(0.0433286,0.741095,0.73941); rgb(32pt)=(0.0963952,0.75,0.712038); rgb(33pt)=(0.140771,0.7584,0.684157); rgb(34pt)=(0.1717,0.766962,0.655443); rgb(35pt)=(0.193767,0.775767,0.6251); rgb(36pt)=(0.216086,0.7843,0.5923); rgb(37pt)=(0.246957,0.791795,0.556743); rgb(38pt)=(0.290614,0.79729,0.518829); rgb(39pt)=(0.340643,0.8008,0.478857); rgb(40pt)=(0.3909,0.802871,0.435448); rgb(41pt)=(0.445629,0.802419,0.390919); rgb(42pt)=(0.5044,0.7993,0.348); rgb(43pt)=(0.561562,0.794233,0.304481); rgb(44pt)=(0.617395,0.787619,0.261238); rgb(45pt)=(0.671986,0.779271,0.2227); rgb(46pt)=(0.7242,0.769843,0.191029); rgb(47pt)=(0.773833,0.759805,0.16461); rgb(48pt)=(0.820314,0.749814,0.153529); rgb(49pt)=(0.863433,0.7406,0.159633); rgb(50pt)=(0.903543,0.733029,0.177414); rgb(51pt)=(0.939257,0.728786,0.209957); rgb(52pt)=(0.972757,0.729771,0.239443); rgb(53pt)=(0.995648,0.743371,0.237148); rgb(54pt)=(0.996986,0.765857,0.219943); rgb(55pt)=(0.995205,0.789252,0.202762); rgb(56pt)=(0.9892,0.813567,0.188533); rgb(57pt)=(0.978629,0.838629,0.176557); rgb(58pt)=(0.967648,0.8639,0.16429); rgb(59pt)=(0.96101,0.889019,0.153676); rgb(60pt)=(0.959671,0.913457,0.142257); rgb(61pt)=(0.962795,0.937338,0.12651); rgb(62pt)=(0.969114,0.960629,0.106362); rgb(63pt)=(0.9769,0.9839,0.0805)},
colorbar,
colorbar style={
            at={(0.85\columnwidth,0.75\columnwidth)},
            anchor=north west,
            align=left,
            ytick = {0,1},
            yticklabel style={
            /pgf/number format/fixed,
            scaled ticks=false,
              },
            },
colorbar/width=0.05\columnwidth
]
\addplot [forget plot] graphics [xmin=-0.0125786163522013, xmax=4.0125786163522, ymin=-0.0692598827793667, ymax=30.1280490090245] {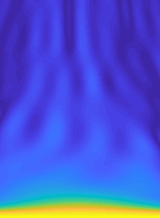};
\addplot [color=mygreen, line width=1.4pt, forget plot]
  table[row sep=crcr]{%
0.5	0\\
0.5	0.13851976555874\\
0.5	0.277039531117467\\
0.5	0.415559296676193\\
0.5	0.554079062234933\\
0.5	0.69259882779366\\
0.5	0.8311185933524\\
0.5	0.969638358911141\\
0.5	1.10815812446987\\
0.5	1.24667789002859\\
0.5	1.38519765558733\\
0.5	1.52371742114607\\
0.5	1.6622371867048\\
0.5	1.80075695226354\\
0.5	1.93927671782227\\
0.5	2.07779648338099\\
0.5	2.21631624893973\\
0.5	2.35483601449847\\
0.5	2.4933557800572\\
0.5	2.63187554561594\\
0.5	2.77039531117467\\
0.5	2.90891507673339\\
0.5	3.04743484229213\\
0.5	3.18595460785087\\
0.5	3.3244743734096\\
0.5	3.46299413896834\\
0.5	3.60151390452707\\
0.5	3.74003367008579\\
0.5	3.87855343564453\\
0.5	4.01707320120327\\
0.5	4.155592966762\\
0.5	4.29411273232074\\
0.5	4.43263249787947\\
0.5	4.57115226343821\\
0.5	4.70967202899693\\
0.5	4.84819179455566\\
0.5	4.9867115601144\\
0.5	5.12523132567314\\
0.5	5.26375109123187\\
0.5	5.40227085679061\\
0.5	5.54079062234933\\
0.5	5.67931038790806\\
0.5	5.8178301534668\\
0.5	5.95634991902553\\
0.5	6.09486968458427\\
0.5	6.23338945014301\\
0.5	6.37190921570173\\
0.5	6.51042898126046\\
0.5	6.6489487468192\\
0.5	6.78746851237793\\
0.5	6.92598827793667\\
0.5	7.06450804349541\\
0.5	7.20302780905413\\
0.5	7.34154757461286\\
0.5	7.4800673401716\\
0.5	7.61858710573033\\
0.5	7.75710687128907\\
0.5	7.89562663684781\\
0.5	8.03414640240653\\
0.5	8.17266616796526\\
0.5	8.311185933524\\
0.5	8.44970569908273\\
0.5	8.58822546464147\\
0.5	8.72674523020021\\
0.5	8.86526499575893\\
0.5	9.00378476131766\\
0.5	9.1423045268764\\
0.5	9.28082429243513\\
0.5	9.41934405799387\\
0.5	9.55786382355261\\
0.5	9.69638358911133\\
0.5	9.83490335467006\\
0.5	9.9734231202288\\
0.5	10.1119428857875\\
0.5	10.2504626513463\\
0.5	10.388982416905\\
0.5	10.5275021824637\\
0.5	10.6660219480225\\
0.5	10.8045417135812\\
0.5	10.9430614791399\\
0.5	11.0815812446987\\
0.5	11.2201010102574\\
0.5	11.3586207758161\\
0.5	11.4971405413749\\
0.5	11.6356603069336\\
0.5	11.7741800724923\\
0.5	11.9126998380511\\
0.5	12.0512196036098\\
0.5	12.1897393691685\\
0.5	12.3282591347273\\
0.5	12.466778900286\\
0.5	12.6052986658447\\
0.5	12.7438184314035\\
0.5	12.8823381969622\\
0.5	13.0208579625209\\
0.5	13.1593777280797\\
0.5	13.2978974936384\\
0.5	13.4364172591971\\
0.5	13.5749370247559\\
0.5	13.7134567903146\\
0.5	13.8519765558733\\
0.5	13.9904963214321\\
0.5	14.1290160869908\\
0.5	14.2675358525495\\
0.5	14.4060556181083\\
0.5	14.544575383667\\
0.5	14.6830951492257\\
0.5	14.8216149147845\\
0.5	14.9601346803432\\
0.5	15.0986544459019\\
0.5	15.2371742114607\\
0.5	15.3756939770194\\
0.5	15.5142137425781\\
0.5	15.6527335081369\\
0.5	15.7912532736956\\
0.5	15.9297730392543\\
0.5	16.0682928048131\\
};
\addplot [color=myblue, line width=1.4pt, forget plot]
  table[row sep=crcr]{%
3	0\\
3	0.13851976555874\\
3	0.277039531117467\\
3	0.415559296676193\\
3	0.554079062234933\\
3	0.69259882779366\\
3	0.8311185933524\\
3	0.969638358911141\\
3	1.10815812446987\\
3	1.24667789002859\\
3	1.38519765558733\\
3	1.52371742114607\\
3	1.6622371867048\\
3	1.80075695226354\\
3	1.93927671782227\\
3	2.07779648338099\\
3	2.21631624893973\\
3	2.35483601449847\\
3	2.4933557800572\\
3	2.63187554561594\\
3	2.77039531117467\\
3	2.90891507673339\\
3	3.04743484229213\\
3	3.18595460785087\\
3	3.3244743734096\\
3	3.46299413896834\\
3	3.60151390452707\\
3	3.74003367008579\\
3	3.87855343564453\\
3	4.01707320120327\\
3	4.155592966762\\
3	4.29411273232074\\
3	4.43263249787947\\
3	4.57115226343821\\
3	4.70967202899693\\
3	4.84819179455566\\
3	4.9867115601144\\
3	5.12523132567314\\
3	5.26375109123187\\
3	5.40227085679061\\
3	5.54079062234933\\
3	5.67931038790806\\
3	5.8178301534668\\
3	5.95634991902553\\
3	6.09486968458427\\
3	6.23338945014301\\
3	6.37190921570173\\
3	6.51042898126046\\
3	6.6489487468192\\
3	6.78746851237793\\
3	6.92598827793667\\
3	7.06450804349541\\
3	7.20302780905413\\
3	7.34154757461286\\
3	7.4800673401716\\
3	7.61858710573033\\
3	7.75710687128907\\
3	7.89562663684781\\
3	8.03414640240653\\
3	8.17266616796526\\
3	8.311185933524\\
3	8.44970569908273\\
3	8.58822546464147\\
3	8.72674523020021\\
3	8.86526499575893\\
3	9.00378476131766\\
3	9.1423045268764\\
3	9.28082429243513\\
3	9.41934405799387\\
3	9.55786382355261\\
3	9.69638358911133\\
3	9.83490335467006\\
3	9.9734231202288\\
3	10.1119428857875\\
3	10.2504626513463\\
3	10.388982416905\\
3	10.5275021824637\\
3	10.6660219480225\\
3	10.8045417135812\\
3	10.9430614791399\\
3	11.0815812446987\\
3	11.2201010102574\\
3	11.3586207758161\\
3	11.4971405413749\\
3	11.6356603069336\\
3	11.7741800724923\\
3	11.9126998380511\\
3	12.0512196036098\\
3	12.1897393691685\\
3	12.3282591347273\\
3	12.466778900286\\
3	12.6052986658447\\
3	12.7438184314035\\
3	12.8823381969622\\
3	13.0208579625209\\
3	13.1593777280797\\
3	13.2978974936384\\
3	13.4364172591971\\
3	13.5749370247559\\
3	13.7134567903146\\
3	13.8519765558733\\
3	13.9904963214321\\
3	14.1290160869908\\
3	14.2675358525495\\
3	14.4060556181083\\
3	14.544575383667\\
3	14.6830951492257\\
3	14.8216149147845\\
3	14.9601346803432\\
3	15.0986544459019\\
3	15.2371742114607\\
3	15.3756939770194\\
3	15.5142137425781\\
3	15.6527335081369\\
3	15.7912532736956\\
3	15.9297730392543\\
3	16.0682928048131\\
};
\addplot [color=myred, line width=1.4pt, forget plot]
  table[row sep=crcr]{%
0	16\\
0.0251572327044025	16\\
0.050314465408805	16\\
0.0754716981132075	16\\
0.10062893081761	16\\
0.125786163522013	16\\
0.150943396226415	16\\
0.176100628930818	16\\
0.20125786163522	16\\
0.226415094339623	16\\
0.251572327044025	16\\
0.276729559748428	16\\
0.30188679245283	16\\
0.327044025157233	16\\
0.352201257861635	16\\
0.377358490566038	16\\
0.40251572327044	16\\
0.427672955974843	16\\
0.452830188679245	16\\
0.477987421383648	16\\
0.50314465408805	16\\
0.528301886792453	16\\
0.553459119496855	16\\
0.578616352201258	16\\
0.60377358490566	16\\
0.628930817610063	16\\
0.654088050314465	16\\
0.679245283018868	16\\
0.70440251572327	16\\
0.729559748427673	16\\
0.754716981132076	16\\
0.779874213836478	16\\
0.805031446540881	16\\
0.830188679245283	16\\
0.855345911949686	16\\
0.880503144654088	16\\
0.905660377358491	16\\
0.930817610062893	16\\
0.955974842767296	16\\
0.981132075471698	16\\
1.0062893081761	16\\
1.0314465408805	16\\
1.05660377358491	16\\
1.08176100628931	16\\
1.10691823899371	16\\
1.13207547169811	16\\
1.15723270440252	16\\
1.18238993710692	16\\
1.20754716981132	16\\
1.23270440251572	16\\
1.25786163522013	16\\
1.28301886792453	16\\
1.30817610062893	16\\
1.33333333333333	16\\
1.35849056603774	16\\
1.38364779874214	16\\
1.40880503144654	16\\
1.43396226415094	16\\
1.45911949685535	16\\
1.48427672955975	16\\
1.50943396226415	16\\
1.53459119496855	16\\
1.55974842767296	16\\
1.58490566037736	16\\
1.61006289308176	16\\
1.63522012578616	16\\
1.66037735849057	16\\
1.68553459119497	16\\
1.71069182389937	16\\
1.73584905660377	16\\
1.76100628930818	16\\
1.78616352201258	16\\
1.81132075471698	16\\
1.83647798742138	16\\
1.86163522012579	16\\
1.88679245283019	16\\
1.91194968553459	16\\
1.93710691823899	16\\
1.9622641509434	16\\
1.9874213836478	16\\
2.0125786163522	16\\
2.0377358490566	16\\
2.06289308176101	16\\
2.08805031446541	16\\
2.11320754716981	16\\
2.13836477987421	16\\
2.16352201257862	16\\
2.18867924528302	16\\
2.21383647798742	16\\
2.23899371069182	16\\
2.26415094339623	16\\
2.28930817610063	16\\
2.31446540880503	16\\
2.33962264150943	16\\
2.36477987421384	16\\
2.38993710691824	16\\
2.41509433962264	16\\
2.44025157232704	16\\
2.46540880503145	16\\
2.49056603773585	16\\
2.51572327044025	16\\
2.54088050314465	16\\
2.56603773584906	16\\
2.59119496855346	16\\
2.61635220125786	16\\
2.64150943396226	16\\
2.66666666666667	16\\
2.69182389937107	16\\
2.71698113207547	16\\
2.74213836477987	16\\
2.76729559748428	16\\
2.79245283018868	16\\
2.81761006289308	16\\
2.84276729559748	16\\
2.86792452830189	16\\
2.89308176100629	16\\
2.91823899371069	16\\
2.94339622641509	16\\
2.9685534591195	16\\
2.9937106918239	16\\
3.0188679245283	16\\
3.0440251572327	16\\
3.06918238993711	16\\
3.09433962264151	16\\
3.11949685534591	16\\
3.14465408805031	16\\
3.16981132075472	16\\
3.19496855345912	16\\
3.22012578616352	16\\
3.24528301886792	16\\
3.27044025157233	16\\
3.29559748427673	16\\
3.32075471698113	16\\
3.34591194968553	16\\
3.37106918238994	16\\
3.39622641509434	16\\
3.42138364779874	16\\
3.44654088050314	16\\
3.47169811320755	16\\
3.49685534591195	16\\
3.52201257861635	16\\
3.54716981132075	16\\
3.57232704402516	16\\
3.59748427672956	16\\
3.62264150943396	16\\
3.64779874213836	16\\
3.67295597484277	16\\
3.69811320754717	16\\
3.72327044025157	16\\
3.74842767295597	16\\
3.77358490566038	16\\
3.79874213836478	16\\
3.82389937106918	16\\
3.84905660377358	16\\
3.87421383647799	16\\
3.89937106918239	16\\
3.92452830188679	16\\
3.9496855345912	16\\
3.9748427672956	16\\
4	16\\
};
\addplot [color=gray, line width=1.4pt, forget plot]
  table[row sep=crcr]{%
0	11.4971405413749\\
0.0251572327044025	11.9126998380511\\
0.050314465408805	12.3282591347273\\
0.0754716981132075	12.6052986658447\\
0.10062893081761	12.8823381969622\\
0.125786163522013	13.1593777280797\\
0.150943396226415	13.4364172591971\\
0.176100628930818	13.5749370247559\\
0.20125786163522	13.8519765558733\\
0.226415094339623	14.1290160869908\\
0.251572327044025	14.2675358525495\\
0.276729559748428	14.544575383667\\
0.30188679245283	14.6830951492257\\
0.327044025157233	14.9601346803432\\
0.352201257861635	15.0986544459019\\
0.377358490566038	15.2371742114607\\
0.40251572327044	15.3756939770194\\
0.427672955974843	15.5142137425781\\
0.452830188679245	15.6527335081369\\
0.477987421383648	15.7912532736956\\
0.50314465408805	16.0682928048131\\
0.528301886792453	16.2068125703718\\
0.553459119496855	16.3453323359305\\
0.578616352201258	16.622371867048\\
0.60377358490566	16.7608916326067\\
0.628930817610063	17.0379311637242\\
0.654088050314465	17.3149706948417\\
0.679245283018868	17.7305299915179\\
0.70440251572327	18.1460892881941\\
0.729559748427673	18.5616485848703\\
0.754716981132076	19.2542474126639\\
0.779874213836478	19.9468462404576\\
0.805031446540881	24.5179985038958\\
0.830188679245283	25.4876368628069\\
0.855345911949686	26.1802356906006\\
0.880503144654088	26.7343147528355\\
0.905660377358491	27.011354283953\\
0.930817610062893	27.1498740495117\\
0.955974842767296	27.011354283953\\
0.981132075471698	26.3187554561593\\
1.0062893081761	21.8861229582799\\
1.0314465408805	21.8861229582799\\
1.05660377358491	21.8861229582799\\
1.08176100628931	21.8861229582799\\
1.10691823899371	21.7476031927211\\
1.13207547169811	21.7476031927211\\
1.15723270440252	21.7476031927211\\
1.18238993710692	15.6527335081369\\
1.20754716981132	32.4136251407436\\
1.23270440251572	32.6906646718611\\
1.25786163522013	32.6906646718611\\
1.28301886792453	32.6906646718611\\
1.30817610062893	32.6906646718611\\
1.33333333333333	32.6906646718611\\
1.35849056603774	32.6906646718611\\
1.38364779874214	29.6432298295689\\
1.40880503144654	29.3661902984515\\
1.43396226415094	29.089150767334\\
1.45911949685535	28.6735914706578\\
1.48427672955975	27.8424728773054\\
1.50943396226415	26.1802356906006\\
1.53459119496855	20.9164845993687\\
1.55974842767296	20.9164845993687\\
1.58490566037736	20.77796483381\\
1.61006289308176	20.6394450682513\\
1.63522012578616	20.5009253026925\\
1.66037735849057	20.3624055371338\\
1.68553459119497	20.0853660060163\\
1.71069182389937	19.6698067093401\\
1.73584905660377	18.8386881159877\\
1.76100628930818	17.0379311637242\\
1.78616352201258	16.0682928048131\\
1.81132075471698	15.2371742114607\\
1.83647798742138	10.5275021824637\\
1.86163522012579	10.388982416905\\
1.88679245283019	10.388982416905\\
1.91194968553459	10.388982416905\\
1.93710691823899	28.1195124084229\\
1.9622641509434	29.6432298295689\\
1.9874213836478	30.3358286573626\\
2.0125786163522	31.0284274851563\\
2.0377358490566	31.4439867818325\\
2.06289308176101	31.9980658440674\\
2.08805031446541	32.6906646718611\\
2.11320754716981	32.6906646718611\\
2.13836477987421	32.6906646718611\\
2.16352201257862	26.7343147528355\\
2.18867924528302	25.4876368628069\\
2.21383647798742	24.3794787383371\\
2.23899371069182	23.2713206138672\\
2.26415094339623	22.7172415516323\\
2.28930817610063	22.3016822549561\\
2.31446540880503	21.8861229582799\\
2.33962264150943	21.6090834271624\\
2.36477987421384	21.4705636616037\\
2.38993710691824	21.1935241304862\\
2.41509433962264	20.77796483381\\
2.44025157232704	20.5009253026925\\
2.46540880503145	19.9468462404576\\
2.49056603773585	19.2542474126639\\
2.51572327044025	18.5616485848703\\
2.54088050314465	17.8690497570766\\
2.56603773584906	17.3149706948417\\
2.59119496855346	16.8994113981655\\
2.61635220125786	16.4838521014893\\
2.64150943396226	32.6906646718611\\
2.66666666666667	32.6906646718611\\
2.69182389937107	32.6906646718611\\
2.71698113207547	15.3756939770194\\
2.74213836477987	15.2371742114607\\
2.76729559748428	14.9601346803432\\
2.79245283018868	14.8216149147845\\
2.81761006289308	14.6830951492257\\
2.84276729559748	14.544575383667\\
2.86792452830189	14.4060556181083\\
2.89308176100629	14.4060556181083\\
2.91823899371069	14.2675358525495\\
2.94339622641509	14.1290160869908\\
2.9685534591195	14.1290160869908\\
2.9937106918239	13.9904963214321\\
3.0188679245283	13.9904963214321\\
3.0440251572327	13.8519765558733\\
3.06918238993711	13.8519765558733\\
3.09433962264151	13.8519765558733\\
3.11949685534591	13.7134567903146\\
3.14465408805031	13.7134567903146\\
3.16981132075472	13.5749370247559\\
3.19496855345912	13.5749370247559\\
3.22012578616352	13.5749370247559\\
3.24528301886792	13.4364172591971\\
3.27044025157233	13.4364172591971\\
3.29559748427673	13.2978974936384\\
3.32075471698113	13.2978974936384\\
3.34591194968553	13.1593777280797\\
3.37106918238994	13.1593777280797\\
3.39622641509434	13.0208579625209\\
3.42138364779874	12.8823381969622\\
3.44654088050314	12.8823381969622\\
3.47169811320755	12.7438184314035\\
3.49685534591195	12.6052986658447\\
3.52201257861635	12.466778900286\\
3.54716981132075	12.3282591347273\\
3.57232704402516	12.1897393691685\\
3.59748427672956	12.0512196036098\\
3.62264150943396	11.7741800724923\\
3.64779874213836	11.6356603069336\\
3.67295597484277	11.4971405413749\\
3.69811320754717	11.2201010102574\\
3.72327044025157	11.0815812446987\\
3.74842767295597	10.8045417135812\\
3.77358490566038	10.6660219480225\\
3.79874213836478	10.388982416905\\
3.82389937106918	10.2504626513463\\
3.84905660377358	9.9734231202288\\
3.87421383647799	9.83490335467006\\
3.89937106918239	9.55786382355261\\
3.92452830188679	9.28082429243513\\
3.9496855345912	9.1423045268764\\
3.9748427672956	8.72674523020021\\
4	8.44970569908273\\
};
\end{axis}
\end{tikzpicture}%

%% file: figures/MTF/MTF_plot_0_5D.tex
\definecolor{myred}{rgb}{0.8078,0.1255,0.1608}%
\definecolor{mygreen}{rgb}{0.1608,0.8078,0.1255}%
\definecolor{myblue}{rgb}{0.1255,0.1608,0.8078}%
\begin{tikzpicture}

\begin{axis}[%
width=\columnwidth,
height=0.75\columnwidth,
at={(0\columnwidth,0\columnwidth)},
scale only axis,
xmin=0,
xmax=16,
xlabel={$\hat{x}$ (cpd)},
ylabel style={font=\color{white!15!black},yshift=-12pt},
ylabel={Contrast},
ymin=0,
ymax=1,
ytick={0, 1},
axis background/.style={fill=white},
title style={font=\bfseries},
title={MTF (0.5 D)},
legend style={legend cell align=left, align=left, draw=white!15!black}
]
\addplot [color=mygreen, line width=1.4pt, forget plot]
  table[row sep=crcr]{%
0	1\\
0.138519763946533	0.987499535083771\\
0.277039527893066	0.968913614749908\\
0.4155592918396	0.945008754730225\\
0.554079055786133	0.913376986980438\\
0.692598819732666	0.877700686454773\\
0.831118583679199	0.83842658996582\\
0.969638347625732	0.796847760677338\\
1.10815811157227	0.75409072637558\\
1.2466778755188	0.711033582687378\\
1.38519763946533	0.66855388879776\\
1.52371740341187	0.627312958240509\\
1.6622371673584	0.587951719760895\\
1.80075693130493	0.550916910171509\\
1.93927669525146	0.516649842262268\\
2.077796459198	0.485399752855301\\
2.21631622314453	0.457445651292801\\
2.35483598709106	0.432822734117508\\
2.4933557510376	0.411585986614227\\
2.63187551498413	0.393533051013947\\
2.77039527893066	0.37847051024437\\
2.9089150428772	0.365800708532333\\
3.04743480682373	0.355126589536667\\
3.18595457077026	0.345917165279388\\
3.3244743347168	0.337832093238831\\
3.46299409866333	0.330452084541321\\
3.60151386260986	0.323630899190903\\
3.7400336265564	0.317013382911682\\
3.87855339050293	0.310579776763916\\
4.01707315444946	0.304148942232132\\
4.155592918396	0.297737687826157\\
4.29411268234253	0.29122006893158\\
4.43263244628906	0.284710079431534\\
4.5711522102356	0.278106033802032\\
4.70967197418213	0.271580636501312\\
4.84819173812866	0.265067428350449\\
4.9867115020752	0.258774906396866\\
5.12523126602173	0.252683371305466\\
5.26375102996826	0.246910363435745\\
5.40227079391479	0.241508290171623\\
5.54079055786133	0.236618131399155\\
5.67931032180786	0.232140153646469\\
5.81783008575439	0.228328287601471\\
5.95634984970093	0.225114151835442\\
6.09486961364746	0.222595036029816\\
6.23338937759399	0.220627263188362\\
6.37190914154053	0.219395071268082\\
6.51042890548706	0.21874538064003\\
6.64894866943359	0.218602329492569\\
6.78746843338013	0.218751475214958\\
6.92598819732666	0.219155967235565\\
7.06450796127319	0.219520136713982\\
7.20302772521973	0.219833329319954\\
7.34154748916626	0.219786122441292\\
7.48006725311279	0.219400808215141\\
7.61858701705933	0.218284174799919\\
7.75710678100586	0.216866925358772\\
7.89562654495239	0.215039372444153\\
8.03414630889893	0.212509498000145\\
8.17266654968262	0.209372520446777\\
8.31118583679199	0.205638706684113\\
8.44970607757568	0.201263636350632\\
8.58822536468506	0.196349635720253\\
8.72674560546875	0.190962463617325\\
8.86526489257813	0.185285791754723\\
9.00378513336182	0.17934462428093\\
9.14230442047119	0.173402309417725\\
9.28082466125488	0.167466178536415\\
9.41934394836426	0.161760538816452\\
9.55786418914795	0.156296849250793\\
9.69638347625732	0.151242643594742\\
9.83490371704102	0.146572962403297\\
9.97342300415039	0.142442956566811\\
10.1119432449341	0.138759851455688\\
10.2504625320435	0.135603323578835\\
10.3889827728271	0.132917284965515\\
10.5275020599365	0.130724608898163\\
10.6660223007202	0.128861129283905\\
10.8045415878296	0.127511635422707\\
10.9430618286133	0.126556515693665\\
11.0815811157227	0.126106098294258\\
11.2201013565063	0.126082569360733\\
11.3586206436157	0.126497134566307\\
11.4971408843994	0.127265304327011\\
11.6356601715088	0.128468006849289\\
11.7741804122925	0.129987746477127\\
11.9126996994019	0.131912991404533\\
12.0512199401855	0.134129971265793\\
12.1897392272949	0.136738464236259\\
12.3282594680786	0.139617502689362\\
12.466778755188	0.142828941345215\\
12.6052989959717	0.146227329969406\\
12.7438182830811	0.149847283959389\\
12.8823385238647	0.153501987457275\\
13.0208578109741	0.157156974077225\\
13.1593780517578	0.160589620471001\\
13.2978973388672	0.163756400346756\\
13.4364175796509	0.166416168212891\\
13.5749368667603	0.168569758534431\\
13.7134571075439	0.170068264007568\\
13.8519763946533	0.170961394906044\\
13.990496635437	0.17108242213726\\
14.1290159225464	0.170471146702766\\
14.2675361633301	0.169071555137634\\
14.4060554504395	0.167016983032227\\
14.5445756912231	0.164184972643852\\
14.6830949783325	0.160742998123169\\
14.8216152191162	0.156623274087906\\
14.9601345062256	0.151946738362312\\
15.0986547470093	0.146694839000702\\
15.2371740341187	0.140996545553207\\
15.3756942749023	0.134809777140617\\
15.5142135620117	0.128336116671562\\
15.6527338027954	0.121531344950199\\
15.7912530899048	0.11459456384182\\
15.9297733306885	0.10752410441637\\
16.0682926177979	0.100506395101547\\
};

\addplot [color=mygreen, dashed, line width=1.4pt, forget plot]
  table[row sep=crcr]{%
0	1\\
0.138519763946533	0.997146844863892\\
0.277039527893066	0.990727484226227\\
0.4155592918396	0.981157839298248\\
0.554079055786133	0.968479633331299\\
0.692598819732666	0.952807068824768\\
0.831118583679199	0.93425464630127\\
0.969638347625732	0.91295450925827\\
1.10815811157227	0.889061152935028\\
1.2466778755188	0.862725973129272\\
1.38519763946533	0.834134817123413\\
1.52371740341187	0.803460299968719\\
1.6622371673584	0.770907878875732\\
1.80075693130493	0.736666202545166\\
1.93927669525146	0.700954735279083\\
2.077796459198	0.663973271846771\\
2.21631622314453	0.625951111316681\\
2.35483598709106	0.587089419364929\\
2.4933557510376	0.547618687152863\\
2.63187551498413	0.507739841938019\\
2.77039527893066	0.467680215835571\\
2.9089150428772	0.427631825208664\\
3.04743480682373	0.387811154127121\\
3.18595457077026	0.348399490118027\\
3.3244743347168	0.309599190950394\\
3.46299409866333	0.271573185920715\\
3.60151386260986	0.234505102038383\\
3.7400336265564	0.198536962270737\\
3.87855339050293	0.163830071687698\\
4.01707315444946	0.1305031478405\\
4.155592918396	0.0986925438046455\\
4.29411268234253	0.0684903115034103\\
4.43263244628906	0.0400065034627914\\
4.5711522102356	0.0133057124912739\\
4.70967197418213	0.011529746465385\\
4.84819173812866	0.0344629175961018\\
4.9867115020752	0.0554391257464886\\
5.12523126602173	0.0744486227631569\\
5.26375102996826	0.0914636179804802\\
5.40227079391479	0.106500573456287\\
5.54079055786133	0.119556993246078\\
5.67931032180786	0.130673825740814\\
5.81783008575439	0.139872416853905\\
5.95634984970093	0.147215515375137\\
6.09486961364746	0.152744516730309\\
6.23338937759399	0.156541004776955\\
6.37190914154053	0.158662810921669\\
6.51042890548706	0.159210205078125\\
6.64894866943359	0.158255815505981\\
6.78746843338013	0.15591099858284\\
6.92598819732666	0.152259290218353\\
7.06450796127319	0.147420853376389\\
7.20302772521973	0.1414864808321\\
7.34154748916626	0.134581834077835\\
7.48006725311279	0.126801699399948\\
7.61858701705933	0.118273884057999\\
7.75710678100586	0.109092973172665\\
7.89562654495239	0.0993850529193878\\
8.03414630889893	0.0892437919974327\\
8.17266654968262	0.0787912011146545\\
8.31118583679199	0.0681148618459702\\
8.44970607757568	0.0573299527168274\\
8.58822536468506	0.0465164817869663\\
8.72674560546875	0.0357808582484722\\
8.86526489257813	0.0251937266439199\\
9.00378513336182	0.0148509526625276\\
9.14230442047119	0.00481056049466133\\
9.28082466125488	0.0048442161642015\\
9.41934394836426	0.0140669932588935\\
9.55786418914795	0.0227849893271923\\
9.69638347625732	0.0309660192579031\\
9.83490371704102	0.0385515354573727\\
9.97342300415039	0.0455216653645039\\
10.1119432449341	0.0518310442566872\\
10.2504625320435	0.0574732460081577\\
10.3889827728271	0.0624167062342167\\
10.5275020599365	0.0666666254401207\\
10.6660223007202	0.0702025517821312\\
10.8045415878296	0.073042243719101\\
10.9430618286133	0.0751764103770256\\
11.0815811157227	0.0766331776976585\\
11.2201013565063	0.0774134024977684\\
11.3586206436157	0.0775537863373756\\
11.4971408843994	0.0770649164915085\\
11.6356601715088	0.0759920999407768\\
11.7741804122925	0.0743506848812103\\
11.9126996994019	0.0721924155950546\\
12.0512199401855	0.0695388987660408\\
12.1897392272949	0.0664468035101891\\
12.3282594680786	0.0629406869411469\\
12.466778755188	0.0590809062123299\\
12.6052989959717	0.0548947229981422\\
12.7438182830811	0.0504422783851624\\
12.8823385238647	0.0457525365054607\\
13.0208578109741	0.0408857315778732\\
13.1593780517578	0.0358699820935726\\
13.2978973388672	0.0307635981589556\\
13.4364175796509	0.0255945511162281\\
13.5749368667603	0.0204202514141798\\
13.7134571075439	0.0152625851333141\\
13.8519763946533	0.0101757487282157\\
13.990496635437	0.00517915235832334\\
14.1290159225464	0.000322210980812088\\
14.2675361633301	0.00437924871221185\\
14.4060554504395	0.00888074468821287\\
14.5445756912231	0.0131732244044542\\
14.6830949783325	0.0172164849936962\\
14.8216152191162	0.0210062563419342\\
14.9601345062256	0.024508697912097\\
15.0986547470093	0.0277230106294155\\
15.2371740341187	0.0306211356073618\\
15.3756942749023	0.0332100987434387\\
15.5142135620117	0.0354659669101238\\
15.6527338027954	0.0374003797769547\\
15.7912530899048	0.0389935784041882\\
15.9297733306885	0.0402627401053905\\
16.0682926177979	0.0411949940025806\\
};
\addplot [color=black, line width=1.4pt, forget plot]
  table[row sep=crcr]{%
0	inf\\
0.13851976555874	0.258979187505763\\
0.277039531117467	0.13055820444367\\
0.415559296676193	0.088217745659358\\
0.554079062234933	0.0673884935421609\\
0.69259882779366	0.0551545028519081\\
0.8311185933524	0.0472092468908607\\
0.969638358911141	0.0417064912126435\\
1.10815812446987	0.0377229793497158\\
1.24667789002859	0.0347459415295017\\
1.38519765558733	0.032468040432637\\
1.52371742114607	0.0306941120779764\\
1.6622371867048	0.0292944931950505\\
1.80075695226354	0.0281798573173818\\
1.93927671782227	0.0272868125448338\\
2.07779648338099	0.0265692451835633\\
2.21631624893973	0.0259929004196091\\
2.35483601449847	0.0255318707906698\\
2.4933557800572	0.0251662530789471\\
2.63187554561594	0.0248805448310009\\
2.77039531117467	0.0246625226778583\\
2.90891507673339	0.0245024424562471\\
3.04743484229213	0.0243924590457509\\
3.18595460785087	0.0243261991692569\\
3.3244743734096	0.0242984425449721\\
3.46299413896834	0.0243048809887485\\
3.60151390452707	0.024341934383963\\
3.74003367008579	0.0244066086661579\\
3.87855343564453	0.0244963852082038\\
4.01707320120327	0.0246091339215869\\
4.155592966762	0.0247430444441152\\
4.29411273232074	0.0248965712444581\\
4.43263249787947	0.0250683895242065\\
4.57115226343821	0.0252573595620825\\
4.70967202899693	0.0254624977063809\\
4.84819179455566	0.0256829526383226\\
4.9867115601144	0.0259179858408891\\
5.12523132567314	0.0261669554431565\\
5.26375109123187	0.0264293027892769\\
5.40227085679061	0.0267045412185436\\
5.54079062234933	0.0269922466489144\\
5.67931038790806	0.0272920496386614\\
5.8178301534668	0.0276036286651129\\
5.95634991902553	0.0279267044100102\\
6.09486968458427	0.0282610348809487\\
6.23338945014301	0.0286064112301206\\
6.37190921570173	0.0289626541569257\\
6.51042898126046	0.0293296108013546\\
6.6489487468192	0.0297071520514427\\
6.78746851237793	0.0300951702013641\\
6.92598827793667	0.0304935769075249\\
7.06450804349541	0.0309023013988173\\
7.20302780905413	0.0313212889044146\\
7.34154757461286	0.031750499268419\\
7.4800673401716	0.0321899057255718\\
7.61858710573033	0.0326394938162908\\
7.75710687128907	0.0330992604226741\\
7.89562663684781	0.0335692129099125\\
8.03414640240653	0.0340493683599095\\
8.17266616796526	0.0345397528858717\\
8.311185933524	0.0350404010182937\\
8.44970569908273	0.035551355154156\\
8.58822546464147	0.0360726650623372\\
8.72674523020021	0.0366043874392386\\
8.86526499575893	0.0371465855094718\\
9.00378476131766	0.0376993286671803\\
9.1423045268764	0.0382626921541785\\
9.28082429243513	0.0388367567716182\\
9.41934405799387	0.0394216086223402\\
9.55786382355261	0.0400173388814438\\
9.69638358911133	0.0406240435929402\\
9.83490335467006	0.0412418234906281\\
9.9734231202288	0.0418707838415648\\
10.1119428857875	0.0425110343107102\\
10.2504626513463	0.0431626888454877\\
10.388982416905	0.0438258655791499\\
10.5275021824637	0.0445006867519565\\
10.6660219480225	0.0451872786492722\\
10.8045417135812	0.0458857715557749\\
10.9430614791399	0.0465962997250322\\
11.0815812446987	0.0473190013637598\\
11.2201010102574	0.0480540186301169\\
11.3586207758161	0.048801497645432\\
11.4971405413749	0.0495615885187764\\
11.6356603069336	0.0503344453838216\\
11.7741800724923	0.0511202264474358\\
11.9126998380511	0.0519190940494799\\
12.0512196036098	0.0527312147332735\\
12.1897393691685	0.0535567593262053\\
12.3282591347273	0.0543959030299643\\
12.466778900286	0.0552488255198715\\
12.6052986658447	0.0561157110527914\\
12.7438184314035	0.0569967485831091\\
12.8823381969622	0.0578921318862556\\
13.0208579625209	0.0588020596892736\\
13.1593777280797	0.0597267358079205\\
13.2978974936384	0.0606663692898103\\
13.4364172591971	0.0616211745631098\\
13.5749370247559	0.062591371590317\\
13.7134567903146	0.0635771860266603\\
13.8519765558733	0.0645788493826777\\
13.9904963214321	0.0655965991905575\\
14.1290160869908	0.0666306791738354\\
14.2675358525495	0.0676813394200777\\
14.4060556181083	0.0687488365561996\\
14.544575383667	0.0698334339260987\\
14.6830951492257	0.0709354017703149\\
14.8216149147845	0.0720550174074567\\
14.9601346803432	0.0731925654171699\\
15.0986544459019	0.0743483378244556\\
15.2371742114607	0.0755226342851821\\
15.3756939770194	0.0767157622726649\\
15.5142137425781	0.0779280372652306\\
15.6527335081369	0.0791597829347084\\
15.7912532736956	0.08041133133583\\
15.9297730392543	0.081683023096554\\
16.0682928048131	0.0829752076093578\\
};
\end{axis}
\end{tikzpicture}%

%% file: figures/MTF/MTF_plot_3D.tex
\definecolor{myred}{rgb}{0.8078,0.1255,0.1608}%
\definecolor{mygreen}{rgb}{0.1608,0.8078,0.1255}%
\definecolor{myblue}{rgb}{0.1255,0.1608,0.8078}%
\begin{tikzpicture}

\begin{axis}[%
width=\columnwidth,
height=0.75\columnwidth,
at={(0\columnwidth,0\columnwidth)},
scale only axis,
xmin=0,
xmax=16,
xlabel={$\hat{x}$ (cpd)},
ylabel style={font=\color{white!15!black},yshift=-12pt},
ymin=0,
ymax=1,
ytick={0, 1},
axis background/.style={fill=white},
title style={font=\bfseries},
title={MTF (3 D)}
]
\addplot [color=myblue, line width=1.4pt, forget plot]
  table[row sep=crcr]{%
0	0.99999988079071\\
0.138519763946533	0.988345324993134\\
0.277039527893066	0.973236739635468\\
0.4155592918396	0.954385280609131\\
0.554079055786133	0.929415225982666\\
0.692598819732666	0.901288509368896\\
0.831118583679199	0.869796931743622\\
0.969638347625732	0.835950911045074\\
1.10815811157227	0.800385117530823\\
1.2466778755188	0.76372504234314\\
1.38519763946533	0.726588070392609\\
1.52371740341187	0.689459979534149\\
1.6622371673584	0.652870714664459\\
1.80075693130493	0.617168545722961\\
1.93927669525146	0.582791030406952\\
2.077796459198	0.549959421157837\\
2.21631622314453	0.51899254322052\\
2.35483598709106	0.489976346492767\\
2.4933557510376	0.463136613368988\\
2.63187551498413	0.438424736261368\\
2.77039527893066	0.415937125682831\\
2.9089150428772	0.395613849163055\\
3.04743480682373	0.377454191446304\\
3.18595457077026	0.36122590303421\\
3.3244743347168	0.346959590911865\\
3.46299409866333	0.334355115890503\\
3.60151386260986	0.323345839977264\\
3.7400336265564	0.313667356967926\\
3.87855339050293	0.305230379104614\\
4.01707315444946	0.29773011803627\\
4.155592918396	0.291103780269623\\
4.29411268234253	0.285091727972031\\
4.43263244628906	0.279670119285584\\
4.5711522102356	0.274561733007431\\
4.70967197418213	0.269769370555878\\
4.84819173812866	0.26506832242012\\
4.9867115020752	0.26051989197731\\
5.12523126602173	0.255916506052017\\
5.26375102996826	0.251287907361984\\
5.40227079391479	0.246592104434967\\
5.54079055786133	0.241915509104729\\
5.67931032180786	0.237250715494156\\
5.81783008575439	0.232872530817986\\
5.95634984970093	0.228675976395607\\
6.09486961364746	0.224787354469299\\
6.23338937759399	0.221193507313728\\
6.37190914154053	0.217994078993797\\
6.51042890548706	0.214822337031364\\
6.64894866943359	0.211755454540253\\
6.78746843338013	0.208536043763161\\
6.92598819732666	0.205151066184044\\
7.06450796127319	0.201418325304985\\
7.20302772521973	0.197388917207718\\
7.34154748916626	0.192867010831833\\
7.48006725311279	0.187929168343544\\
7.61858701705933	0.182479545474052\\
7.75710678100586	0.176646262407303\\
7.89562654495239	0.170815691351891\\
8.03414630889893	0.164983347058296\\
8.17266654968262	0.158984124660492\\
8.31118583679199	0.152926072478294\\
8.44970607757568	0.14667409658432\\
8.58822536468506	0.140457317233086\\
8.72674560546875	0.134204119443893\\
8.86526489257813	0.128142565488815\\
9.00378513336182	0.122236348688602\\
9.14230442047119	0.116695657372475\\
9.28082466125488	0.111493088304996\\
9.41934394836426	0.106811136007309\\
9.55786418914795	0.102563850581646\\
9.69638347625732	0.0989023074507713\\
9.83490371704102	0.0956627577543259\\
9.97342300415039	0.0929737910628319\\
10.1119432449341	0.0906756669282913\\
10.2504625320435	0.0888222306966782\\
10.3889827728271	0.0873710736632347\\
10.5275020599365	0.0863451883196831\\
10.6660223007202	0.0856031700968742\\
10.8045415878296	0.085284598171711\\
10.9430618286133	0.0851764008402824\\
11.0815811157227	0.0854094922542572\\
11.2201013565063	0.0858241021633148\\
11.3586206436157	0.0864756852388382\\
11.4971408843994	0.0872467234730721\\
11.6356601715088	0.0882330536842346\\
11.7741804122925	0.0892617329955101\\
11.9126996994019	0.0904525443911552\\
12.0512199401855	0.0916595533490181\\
12.1897392272949	0.0929770022630692\\
12.3282594680786	0.0942668169736862\\
12.466778755188	0.0956287980079651\\
12.6052989959717	0.0969170182943344\\
12.7438182830811	0.0982265919446945\\
12.8823385238647	0.0994028374552727\\
13.0208578109741	0.100522883236408\\
13.1593780517578	0.101411998271942\\
13.2978973388672	0.10212666541338\\
13.4364175796509	0.102482758462429\\
13.5749368667603	0.102551579475403\\
13.7134571075439	0.102179110050201\\
13.8519763946533	0.101454392075539\\
13.990496635437	0.100227557122707\\
14.1290159225464	0.0986026600003242\\
14.2675361633301	0.0965123251080513\\
14.4060554504395	0.0941271334886551\\
14.5445756912231	0.0913519263267517\\
14.6830949783325	0.0883752927184105\\
14.8216152191162	0.0851041972637177\\
14.9601345062256	0.0817128270864487\\
15.0986547470093	0.0781404450535774\\
15.2371740341187	0.0745604634284973\\
15.3756942749023	0.0709150061011314\\
15.5142135620117	0.0674246773123741\\
15.6527338027954	0.0640594884753227\\
15.7912530899048	0.0610637366771698\\
15.9297733306885	0.0584388449788094\\
16.0682926177979	0.0563972592353821\\
};
%\addlegendentry{Proposed}
\addplot [color=myblue, dashed, line width=1.4pt, forget plot]
  table[row sep=crcr]{%
0	1\\
0.138519763946533	0.997417211532593\\
0.277039527893066	0.991809844970703\\
0.4155592918396	0.983591437339783\\
0.554079055786133	0.972797632217407\\
0.692598819732666	0.959531486034393\\
0.831118583679199	0.943893015384674\\
0.969638347625732	0.925994753837585\\
1.10815811157227	0.905967950820923\\
1.2466778755188	0.883936643600464\\
1.38519763946533	0.860055387020111\\
1.52371740341187	0.834462285041809\\
1.6622371673584	0.807324230670929\\
1.80075693130493	0.778789162635803\\
1.93927669525146	0.749032020568848\\
2.077796459198	0.718206882476807\\
2.21631622314453	0.686494410037994\\
2.35483598709106	0.654045939445496\\
2.4933557510376	0.621041119098663\\
2.63187551498413	0.587628781795502\\
2.77039527893066	0.553984403610229\\
2.9089150428772	0.520247995853424\\
3.04743480682373	0.48658487200737\\
3.18595457077026	0.453125894069672\\
3.3244743347168	0.42002472281456\\
3.46299409866333	0.387397348880768\\
3.60151386260986	0.355382978916168\\
3.7400336265564	0.324081569910049\\
3.87855339050293	0.29361554980278\\
4.01707315444946	0.264068126678467\\
4.155592918396	0.235543787479401\\
4.29411268234253	0.20810654759407\\
4.43263244628906	0.18184258043766\\
4.5711522102356	0.156796902418137\\
4.70967197418213	0.133036687970161\\
4.84819173812866	0.110588319599628\\
4.9867115020752	0.0895003601908684\\
5.12523126602173	0.0697812139987946\\
5.26375102996826	0.0514618381857872\\
5.40227079391479	0.0345334783196449\\
5.54079055786133	0.0190107207745314\\
5.67931032180786	0.00486900471150875\\
5.81783008575439	0.00789252109825611\\
5.95634984970093	0.0193122569471598\\
6.09486961364746	0.0294035617262125\\
6.23338937759399	0.0382166802883148\\
6.37190914154053	0.0457749851047993\\
6.51042890548706	0.052141509950161\\
6.64894866943359	0.0573492981493473\\
6.78746843338013	0.0614679902791977\\
6.92598819732666	0.0645377635955811\\
7.06450796127319	0.066633976995945\\
7.20302772521973	0.067801833152771\\
7.34154748916626	0.0681206658482552\\
7.48006725311279	0.0676388442516327\\
7.61858701705933	0.0664379000663757\\
7.75710678100586	0.0645666494965553\\
7.89562654495239	0.0621063075959682\\
8.03414630889893	0.0591069236397743\\
8.17266654968262	0.0556483902037144\\
8.31118583679199	0.0517780296504498\\
8.44970607757568	0.0475728586316109\\
8.58822536468506	0.0430771037936211\\
8.72674560546875	0.0383640080690384\\
8.86526489257813	0.0334738045930862\\
9.00378513336182	0.0284749772399664\\
9.14230442047119	0.0234012641012669\\
9.28082466125488	0.0183148514479399\\
9.41934394836426	0.013244410045445\\
9.55786418914795	0.00824839156121016\\
9.69638347625732	0.00334808905608952\\
9.83490371704102	0.00140546797774732\\
9.97342300415039	0.00599649641662836\\
10.1119432449341	0.0103803556412458\\
10.2504625320435	0.0145481415092945\\
10.3889827728271	0.0184625964611769\\
10.5275020599365	0.02212018892169\\
10.6660223007202	0.0254888795316219\\
10.8045415878296	0.028572078794241\\
10.9430618286133	0.0313435681164265\\
11.0815811157227	0.0338122546672821\\
11.2201013565063	0.0359575562179089\\
11.3586206436157	0.0377928614616394\\
11.4971408843994	0.039303719997406\\
11.6356601715088	0.0405089110136032\\
11.7741804122925	0.0413960665464401\\
11.9126996994019	0.0419882349669933\\
12.0512199401855	0.0422775074839592\\
12.1897392272949	0.0422906279563904\\
12.3282594680786	0.0420219115912914\\
12.466778755188	0.0415014438331127\\
12.6052989959717	0.0407263971865177\\
12.7438182830811	0.0397272631525993\\
12.8823385238647	0.0385039187967777\\
13.0208578109741	0.0370883569121361\\
13.1593780517578	0.0354814454913139\\
13.2978973388672	0.0337153822183609\\
13.4364175796509	0.0317933820188046\\
13.5749368667603	0.0297495182603598\\
13.7134571075439	0.0275839474052191\\
13.8519763946533	0.0253308154642582\\
13.990496635437	0.0229913201183081\\
14.1290159225464	0.0205984469503164\\
14.2675361633301	0.0181535836309195\\
14.4060554504395	0.0156887210905552\\
14.5445756912231	0.0132024148479104\\
14.6830949783325	0.0107263922691345\\
14.8216152191162	0.00825845543295145\\
14.9601345062256	0.00582796987146139\\
15.0986547470093	0.00343322614207864\\
15.2371740341187	0.00110167637467384\\
15.3756942749023	0.0011724253417924\\
15.5142135620117	0.00336209475062788\\
15.6527338027954	0.00547418370842934\\
15.7912530899048	0.00748254684731364\\
15.9297733306885	0.00939639657735825\\
16.0682926177979	0.0111935203894973\\
};
%\addlegendentry{Spherical}
\addplot [color=black, line width=1.4pt, forget plot]
  table[row sep=crcr]{%
0	inf\\
0.13851976555874	0.258979187505763\\
0.277039531117467	0.13055820444367\\
0.415559296676193	0.088217745659358\\
0.554079062234933	0.0673884935421609\\
0.69259882779366	0.0551545028519081\\
0.8311185933524	0.0472092468908607\\
0.969638358911141	0.0417064912126435\\
1.10815812446987	0.0377229793497158\\
1.24667789002859	0.0347459415295017\\
1.38519765558733	0.032468040432637\\
1.52371742114607	0.0306941120779764\\
1.6622371867048	0.0292944931950505\\
1.80075695226354	0.0281798573173818\\
1.93927671782227	0.0272868125448338\\
2.07779648338099	0.0265692451835633\\
2.21631624893973	0.0259929004196091\\
2.35483601449847	0.0255318707906698\\
2.4933557800572	0.0251662530789471\\
2.63187554561594	0.0248805448310009\\
2.77039531117467	0.0246625226778583\\
2.90891507673339	0.0245024424562471\\
3.04743484229213	0.0243924590457509\\
3.18595460785087	0.0243261991692569\\
3.3244743734096	0.0242984425449721\\
3.46299413896834	0.0243048809887485\\
3.60151390452707	0.024341934383963\\
3.74003367008579	0.0244066086661579\\
3.87855343564453	0.0244963852082038\\
4.01707320120327	0.0246091339215869\\
4.155592966762	0.0247430444441152\\
4.29411273232074	0.0248965712444581\\
4.43263249787947	0.0250683895242065\\
4.57115226343821	0.0252573595620825\\
4.70967202899693	0.0254624977063809\\
4.84819179455566	0.0256829526383226\\
4.9867115601144	0.0259179858408891\\
5.12523132567314	0.0261669554431565\\
5.26375109123187	0.0264293027892769\\
5.40227085679061	0.0267045412185436\\
5.54079062234933	0.0269922466489144\\
5.67931038790806	0.0272920496386614\\
5.8178301534668	0.0276036286651129\\
5.95634991902553	0.0279267044100102\\
6.09486968458427	0.0282610348809487\\
6.23338945014301	0.0286064112301206\\
6.37190921570173	0.0289626541569257\\
6.51042898126046	0.0293296108013546\\
6.6489487468192	0.0297071520514427\\
6.78746851237793	0.0300951702013641\\
6.92598827793667	0.0304935769075249\\
7.06450804349541	0.0309023013988173\\
7.20302780905413	0.0313212889044146\\
7.34154757461286	0.031750499268419\\
7.4800673401716	0.0321899057255718\\
7.61858710573033	0.0326394938162908\\
7.75710687128907	0.0330992604226741\\
7.89562663684781	0.0335692129099125\\
8.03414640240653	0.0340493683599095\\
8.17266616796526	0.0345397528858717\\
8.311185933524	0.0350404010182937\\
8.44970569908273	0.035551355154156\\
8.58822546464147	0.0360726650623372\\
8.72674523020021	0.0366043874392386\\
8.86526499575893	0.0371465855094718\\
9.00378476131766	0.0376993286671803\\
9.1423045268764	0.0382626921541785\\
9.28082429243513	0.0388367567716182\\
9.41934405799387	0.0394216086223402\\
9.55786382355261	0.0400173388814438\\
9.69638358911133	0.0406240435929402\\
9.83490335467006	0.0412418234906281\\
9.9734231202288	0.0418707838415648\\
10.1119428857875	0.0425110343107102\\
10.2504626513463	0.0431626888454877\\
10.388982416905	0.0438258655791499\\
10.5275021824637	0.0445006867519565\\
10.6660219480225	0.0451872786492722\\
10.8045417135812	0.0458857715557749\\
10.9430614791399	0.0465962997250322\\
11.0815812446987	0.0473190013637598\\
11.2201010102574	0.0480540186301169\\
11.3586207758161	0.048801497645432\\
11.4971405413749	0.0495615885187764\\
11.6356603069336	0.0503344453838216\\
11.7741800724923	0.0511202264474358\\
11.9126998380511	0.0519190940494799\\
12.0512196036098	0.0527312147332735\\
12.1897393691685	0.0535567593262053\\
12.3282591347273	0.0543959030299643\\
12.466778900286	0.0552488255198715\\
12.6052986658447	0.0561157110527914\\
12.7438184314035	0.0569967485831091\\
12.8823381969622	0.0578921318862556\\
13.0208579625209	0.0588020596892736\\
13.1593777280797	0.0597267358079205\\
13.2978974936384	0.0606663692898103\\
13.4364172591971	0.0616211745631098\\
13.5749370247559	0.062591371590317\\
13.7134567903146	0.0635771860266603\\
13.8519765558733	0.0645788493826777\\
13.9904963214321	0.0655965991905575\\
14.1290160869908	0.0666306791738354\\
14.2675358525495	0.0676813394200777\\
14.4060556181083	0.0687488365561996\\
14.544575383667	0.0698334339260987\\
14.6830951492257	0.0709354017703149\\
14.8216149147845	0.0720550174074567\\
14.9601346803432	0.0731925654171699\\
15.0986544459019	0.0743483378244556\\
15.2371742114607	0.0755226342851821\\
15.3756939770194	0.0767157622726649\\
15.5142137425781	0.0779280372652306\\
15.6527335081369	0.0791597829347084\\
15.7912532736956	0.08041133133583\\
15.9297730392543	0.081683023096554\\
16.0682928048131	0.0829752076093578\\
};
%\addlegendentry{$\Delta L$}

\end{axis}
\end{tikzpicture}%

%% file: figures/MTF/Viewing_angle/MTF1D_green_th-10.tex
\begin{tikzpicture}

\begin{axis}[%
width=0.75\columnwidth,
height=0.75\columnwidth,
at={(0\columnwidth,0\columnwidth)},
scale only axis,
point meta min=0.0418069772422314,
point meta max=1.00000011920929,
axis on top,
xmin=-15, %4.42890999476988,
xmax=-5, %15.7187636459119,
xtick={-15,-5},
xlabel style={font=\color{white!15!black},yshift=3pt},
xlabel={$\theta (^\circ)$},
ymin=-0.0692598827793667,
ymax=16.4145922187099,
ytick={0,16},
ylabel style={font=\color{white!15!black},yshift=-14pt},
ylabel={cpd},
axis background/.style={fill=white},
title style={font=\bfseries},
title={$\SI{-10}{\degree}$},
colormap={mymap}{[1pt] rgb(0pt)=(0.2422,0.1504,0.6603); rgb(1pt)=(0.25039,0.164995,0.707614); rgb(2pt)=(0.257771,0.181781,0.751138); rgb(3pt)=(0.264729,0.197757,0.795214); rgb(4pt)=(0.270648,0.214676,0.836371); rgb(5pt)=(0.275114,0.234238,0.870986); rgb(6pt)=(0.2783,0.255871,0.899071); rgb(7pt)=(0.280333,0.278233,0.9221); rgb(8pt)=(0.281338,0.300595,0.941376); rgb(9pt)=(0.281014,0.322757,0.957886); rgb(10pt)=(0.279467,0.344671,0.971676); rgb(11pt)=(0.275971,0.366681,0.982905); rgb(12pt)=(0.269914,0.3892,0.9906); rgb(13pt)=(0.260243,0.412329,0.995157); rgb(14pt)=(0.244033,0.435833,0.998833); rgb(15pt)=(0.220643,0.460257,0.997286); rgb(16pt)=(0.196333,0.484719,0.989152); rgb(17pt)=(0.183405,0.507371,0.979795); rgb(18pt)=(0.178643,0.528857,0.968157); rgb(19pt)=(0.176438,0.549905,0.952019); rgb(20pt)=(0.168743,0.570262,0.935871); rgb(21pt)=(0.154,0.5902,0.9218); rgb(22pt)=(0.146029,0.609119,0.907857); rgb(23pt)=(0.138024,0.627629,0.89729); rgb(24pt)=(0.124814,0.645929,0.888343); rgb(25pt)=(0.111252,0.6635,0.876314); rgb(26pt)=(0.0952095,0.679829,0.859781); rgb(27pt)=(0.0688714,0.694771,0.839357); rgb(28pt)=(0.0296667,0.708167,0.816333); rgb(29pt)=(0.00357143,0.720267,0.7917); rgb(30pt)=(0.00665714,0.731214,0.766014); rgb(31pt)=(0.0433286,0.741095,0.73941); rgb(32pt)=(0.0963952,0.75,0.712038); rgb(33pt)=(0.140771,0.7584,0.684157); rgb(34pt)=(0.1717,0.766962,0.655443); rgb(35pt)=(0.193767,0.775767,0.6251); rgb(36pt)=(0.216086,0.7843,0.5923); rgb(37pt)=(0.246957,0.791795,0.556743); rgb(38pt)=(0.290614,0.79729,0.518829); rgb(39pt)=(0.340643,0.8008,0.478857); rgb(40pt)=(0.3909,0.802871,0.435448); rgb(41pt)=(0.445629,0.802419,0.390919); rgb(42pt)=(0.5044,0.7993,0.348); rgb(43pt)=(0.561562,0.794233,0.304481); rgb(44pt)=(0.617395,0.787619,0.261238); rgb(45pt)=(0.671986,0.779271,0.2227); rgb(46pt)=(0.7242,0.769843,0.191029); rgb(47pt)=(0.773833,0.759805,0.16461); rgb(48pt)=(0.820314,0.749814,0.153529); rgb(49pt)=(0.863433,0.7406,0.159633); rgb(50pt)=(0.903543,0.733029,0.177414); rgb(51pt)=(0.939257,0.728786,0.209957); rgb(52pt)=(0.972757,0.729771,0.239443); rgb(53pt)=(0.995648,0.743371,0.237148); rgb(54pt)=(0.996986,0.765857,0.219943); rgb(55pt)=(0.995205,0.789252,0.202762); rgb(56pt)=(0.9892,0.813567,0.188533); rgb(57pt)=(0.978629,0.838629,0.176557); rgb(58pt)=(0.967648,0.8639,0.16429); rgb(59pt)=(0.96101,0.889019,0.153676); rgb(60pt)=(0.959671,0.913457,0.142257); rgb(61pt)=(0.962795,0.937338,0.12651); rgb(62pt)=(0.969114,0.960629,0.106362); rgb(63pt)=(0.9769,0.9839,0.0805)},
%colorbar
]
\addplot [forget plot] graphics [xmin=-15.7187636459119, xmax=-4.42890999476988, ymin=-0.0692598827793667, ymax=16.4145922187099] {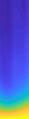};
\end{axis}
\end{tikzpicture}%

%% file: figures/MTF/Viewing_angle/MTF1D_green_th0.tex
\begin{tikzpicture}

\begin{axis}[%
width=0.75\columnwidth,
height=0.75\columnwidth,
at={(0\columnwidth,0\columnwidth)},
scale only axis,
point meta min=0.0418069772422314,
point meta max=1.00000011920929,
axis on top,
xmin=-5, %4.42890999476988,
xmax=5, %15.7187636459119,
xtick={-5,5},
xlabel style={font=\color{white!15!black},yshift=3pt},
xlabel={$\theta (^\circ)$},
ymin=-0.0692598827793667,
ymax=16.4145922187099,
ytick={0,16},
ylabel style={font=\color{white!15!black},yshift=-14pt},
axis background/.style={fill=white},
title style={font=\bfseries},
title={$\SI{0}{\degree}$},
colormap={mymap}{[1pt] rgb(0pt)=(0.2422,0.1504,0.6603); rgb(1pt)=(0.25039,0.164995,0.707614); rgb(2pt)=(0.257771,0.181781,0.751138); rgb(3pt)=(0.264729,0.197757,0.795214); rgb(4pt)=(0.270648,0.214676,0.836371); rgb(5pt)=(0.275114,0.234238,0.870986); rgb(6pt)=(0.2783,0.255871,0.899071); rgb(7pt)=(0.280333,0.278233,0.9221); rgb(8pt)=(0.281338,0.300595,0.941376); rgb(9pt)=(0.281014,0.322757,0.957886); rgb(10pt)=(0.279467,0.344671,0.971676); rgb(11pt)=(0.275971,0.366681,0.982905); rgb(12pt)=(0.269914,0.3892,0.9906); rgb(13pt)=(0.260243,0.412329,0.995157); rgb(14pt)=(0.244033,0.435833,0.998833); rgb(15pt)=(0.220643,0.460257,0.997286); rgb(16pt)=(0.196333,0.484719,0.989152); rgb(17pt)=(0.183405,0.507371,0.979795); rgb(18pt)=(0.178643,0.528857,0.968157); rgb(19pt)=(0.176438,0.549905,0.952019); rgb(20pt)=(0.168743,0.570262,0.935871); rgb(21pt)=(0.154,0.5902,0.9218); rgb(22pt)=(0.146029,0.609119,0.907857); rgb(23pt)=(0.138024,0.627629,0.89729); rgb(24pt)=(0.124814,0.645929,0.888343); rgb(25pt)=(0.111252,0.6635,0.876314); rgb(26pt)=(0.0952095,0.679829,0.859781); rgb(27pt)=(0.0688714,0.694771,0.839357); rgb(28pt)=(0.0296667,0.708167,0.816333); rgb(29pt)=(0.00357143,0.720267,0.7917); rgb(30pt)=(0.00665714,0.731214,0.766014); rgb(31pt)=(0.0433286,0.741095,0.73941); rgb(32pt)=(0.0963952,0.75,0.712038); rgb(33pt)=(0.140771,0.7584,0.684157); rgb(34pt)=(0.1717,0.766962,0.655443); rgb(35pt)=(0.193767,0.775767,0.6251); rgb(36pt)=(0.216086,0.7843,0.5923); rgb(37pt)=(0.246957,0.791795,0.556743); rgb(38pt)=(0.290614,0.79729,0.518829); rgb(39pt)=(0.340643,0.8008,0.478857); rgb(40pt)=(0.3909,0.802871,0.435448); rgb(41pt)=(0.445629,0.802419,0.390919); rgb(42pt)=(0.5044,0.7993,0.348); rgb(43pt)=(0.561562,0.794233,0.304481); rgb(44pt)=(0.617395,0.787619,0.261238); rgb(45pt)=(0.671986,0.779271,0.2227); rgb(46pt)=(0.7242,0.769843,0.191029); rgb(47pt)=(0.773833,0.759805,0.16461); rgb(48pt)=(0.820314,0.749814,0.153529); rgb(49pt)=(0.863433,0.7406,0.159633); rgb(50pt)=(0.903543,0.733029,0.177414); rgb(51pt)=(0.939257,0.728786,0.209957); rgb(52pt)=(0.972757,0.729771,0.239443); rgb(53pt)=(0.995648,0.743371,0.237148); rgb(54pt)=(0.996986,0.765857,0.219943); rgb(55pt)=(0.995205,0.789252,0.202762); rgb(56pt)=(0.9892,0.813567,0.188533); rgb(57pt)=(0.978629,0.838629,0.176557); rgb(58pt)=(0.967648,0.8639,0.16429); rgb(59pt)=(0.96101,0.889019,0.153676); rgb(60pt)=(0.959671,0.913457,0.142257); rgb(61pt)=(0.962795,0.937338,0.12651); rgb(62pt)=(0.969114,0.960629,0.106362); rgb(63pt)=(0.9769,0.9839,0.0805)},
%colorbar
]
\addplot [forget plot] graphics [xmin=-5.22458131568492, xmax=5.22458131568492, ymin=-0.0692598827793667, ymax=16.4145922187099] {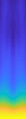};
\end{axis}
\end{tikzpicture}%

%% file: figures/MTF/Viewing_angle/MTF1D_green_th10.tex
\begin{tikzpicture}

\begin{axis}[%
width=0.75\columnwidth,
height=0.75\columnwidth,
at={(0\columnwidth,0\columnwidth)},
scale only axis,
point meta min=0.0418069772422314,
point meta max=1.00000011920929,
axis on top,
xmin=5, %4.42890999476988,
xmax=15, %15.7187636459119,
xtick={5,15},
xlabel style={font=\color{white!15!black},yshift=3pt},
xlabel={$\theta (^\circ)$},
ymin=-0.0692598827793667,
ymax=16.4145922187099,
ytick={0,16},
ylabel style={font=\color{white!15!black},yshift=-14pt},
axis background/.style={fill=white},
title style={font=\bfseries},
title={$\SI{10}{\degree}$},
colormap={mymap}{[1pt] rgb(0pt)=(0.2422,0.1504,0.6603); rgb(1pt)=(0.25039,0.164995,0.707614); rgb(2pt)=(0.257771,0.181781,0.751138); rgb(3pt)=(0.264729,0.197757,0.795214); rgb(4pt)=(0.270648,0.214676,0.836371); rgb(5pt)=(0.275114,0.234238,0.870986); rgb(6pt)=(0.2783,0.255871,0.899071); rgb(7pt)=(0.280333,0.278233,0.9221); rgb(8pt)=(0.281338,0.300595,0.941376); rgb(9pt)=(0.281014,0.322757,0.957886); rgb(10pt)=(0.279467,0.344671,0.971676); rgb(11pt)=(0.275971,0.366681,0.982905); rgb(12pt)=(0.269914,0.3892,0.9906); rgb(13pt)=(0.260243,0.412329,0.995157); rgb(14pt)=(0.244033,0.435833,0.998833); rgb(15pt)=(0.220643,0.460257,0.997286); rgb(16pt)=(0.196333,0.484719,0.989152); rgb(17pt)=(0.183405,0.507371,0.979795); rgb(18pt)=(0.178643,0.528857,0.968157); rgb(19pt)=(0.176438,0.549905,0.952019); rgb(20pt)=(0.168743,0.570262,0.935871); rgb(21pt)=(0.154,0.5902,0.9218); rgb(22pt)=(0.146029,0.609119,0.907857); rgb(23pt)=(0.138024,0.627629,0.89729); rgb(24pt)=(0.124814,0.645929,0.888343); rgb(25pt)=(0.111252,0.6635,0.876314); rgb(26pt)=(0.0952095,0.679829,0.859781); rgb(27pt)=(0.0688714,0.694771,0.839357); rgb(28pt)=(0.0296667,0.708167,0.816333); rgb(29pt)=(0.00357143,0.720267,0.7917); rgb(30pt)=(0.00665714,0.731214,0.766014); rgb(31pt)=(0.0433286,0.741095,0.73941); rgb(32pt)=(0.0963952,0.75,0.712038); rgb(33pt)=(0.140771,0.7584,0.684157); rgb(34pt)=(0.1717,0.766962,0.655443); rgb(35pt)=(0.193767,0.775767,0.6251); rgb(36pt)=(0.216086,0.7843,0.5923); rgb(37pt)=(0.246957,0.791795,0.556743); rgb(38pt)=(0.290614,0.79729,0.518829); rgb(39pt)=(0.340643,0.8008,0.478857); rgb(40pt)=(0.3909,0.802871,0.435448); rgb(41pt)=(0.445629,0.802419,0.390919); rgb(42pt)=(0.5044,0.7993,0.348); rgb(43pt)=(0.561562,0.794233,0.304481); rgb(44pt)=(0.617395,0.787619,0.261238); rgb(45pt)=(0.671986,0.779271,0.2227); rgb(46pt)=(0.7242,0.769843,0.191029); rgb(47pt)=(0.773833,0.759805,0.16461); rgb(48pt)=(0.820314,0.749814,0.153529); rgb(49pt)=(0.863433,0.7406,0.159633); rgb(50pt)=(0.903543,0.733029,0.177414); rgb(51pt)=(0.939257,0.728786,0.209957); rgb(52pt)=(0.972757,0.729771,0.239443); rgb(53pt)=(0.995648,0.743371,0.237148); rgb(54pt)=(0.996986,0.765857,0.219943); rgb(55pt)=(0.995205,0.789252,0.202762); rgb(56pt)=(0.9892,0.813567,0.188533); rgb(57pt)=(0.978629,0.838629,0.176557); rgb(58pt)=(0.967648,0.8639,0.16429); rgb(59pt)=(0.96101,0.889019,0.153676); rgb(60pt)=(0.959671,0.913457,0.142257); rgb(61pt)=(0.962795,0.937338,0.12651); rgb(62pt)=(0.969114,0.960629,0.106362); rgb(63pt)=(0.9769,0.9839,0.0805)},
colorbar,
colorbar style={
            at={(0.75\columnwidth,0.75\columnwidth)},
            anchor=north west,
            align=left,
            ytick = {0,1},
            },
colorbar/width=0.05\columnwidth
]
\addplot [forget plot] graphics [xmin=4.42890999476988, xmax=15.7187636459119, ymin=-0.0692598827793667, ymax=16.4145922187099] {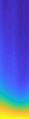};
\end{axis}
\end{tikzpicture}%

%% file: figures/figure_compare_soa_Depthim_rect.tex
\input{figures/images/DepthIm/X0Y0/Spherical/PSNR}
\input{figures/images/DepthIm/X0Y0/Spherical/SSIM}
\input{figures/images/DepthIm/X0Y0/Konrad5/PSNR}
\input{figures/images/DepthIm/X0Y0/Konrad5/SSIM}
\input{figures/images/DepthIm/X0Y0/Our/PSNR}
\input{figures/images/DepthIm/X0Y0/Our/SSIM}

\begin{figure}[t!]
    \centering
    \begin{subfigure}[t]{0.32\columnwidth}
		\centering
		\caption{\textbf{Conventional}}
		\includegraphics[width=\columnwidth]{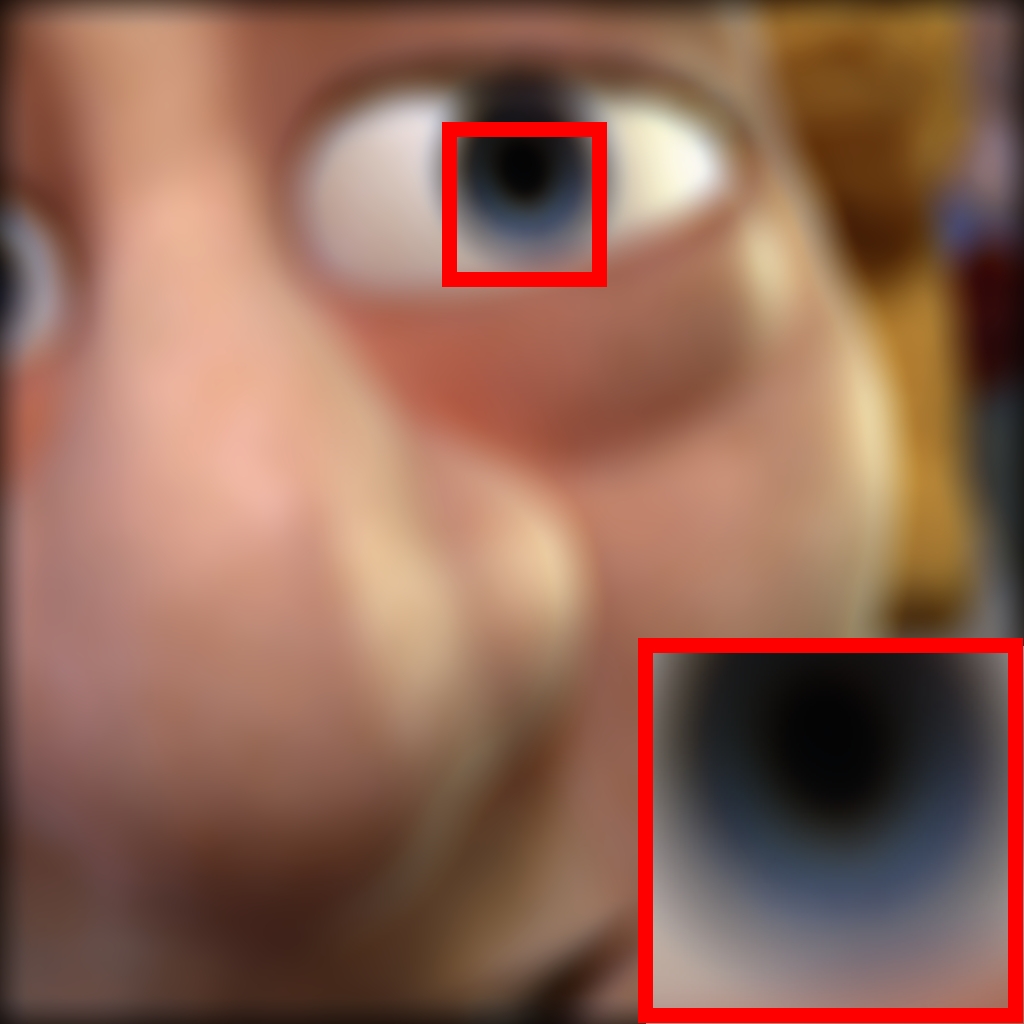}
		\put(-63,-9){\TLPSNRCSFtwentyfive \text{ / }\TLSSIMCSFtwentyfive}
		\put(-90,25){\rotatebox{90}{4 D}}
	\end{subfigure}
	\begin{subfigure}[t]{0.32\columnwidth}
		\centering
		\caption{\textbf{AI-NED \cite{KonradAI}}}
		\includegraphics[width=\columnwidth]{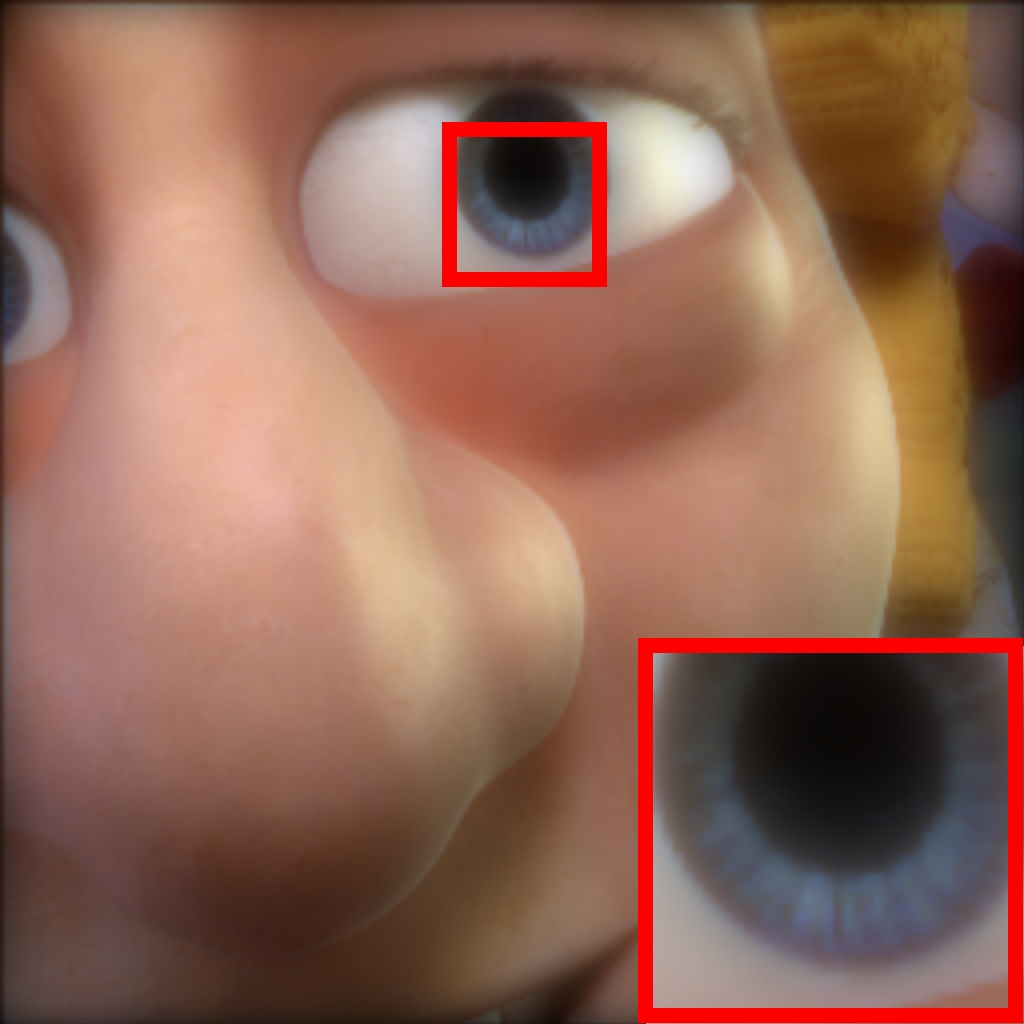}
		\put(-63,-9){\KonradPSNRCSFtwentyfive \text{ / }\KonradSSIMCSFtwentyfive}
	\end{subfigure}
	\begin{subfigure}[t]{0.32\columnwidth}
		\centering
		\caption{\textbf{Proposed}}
		\includegraphics[width=\columnwidth]{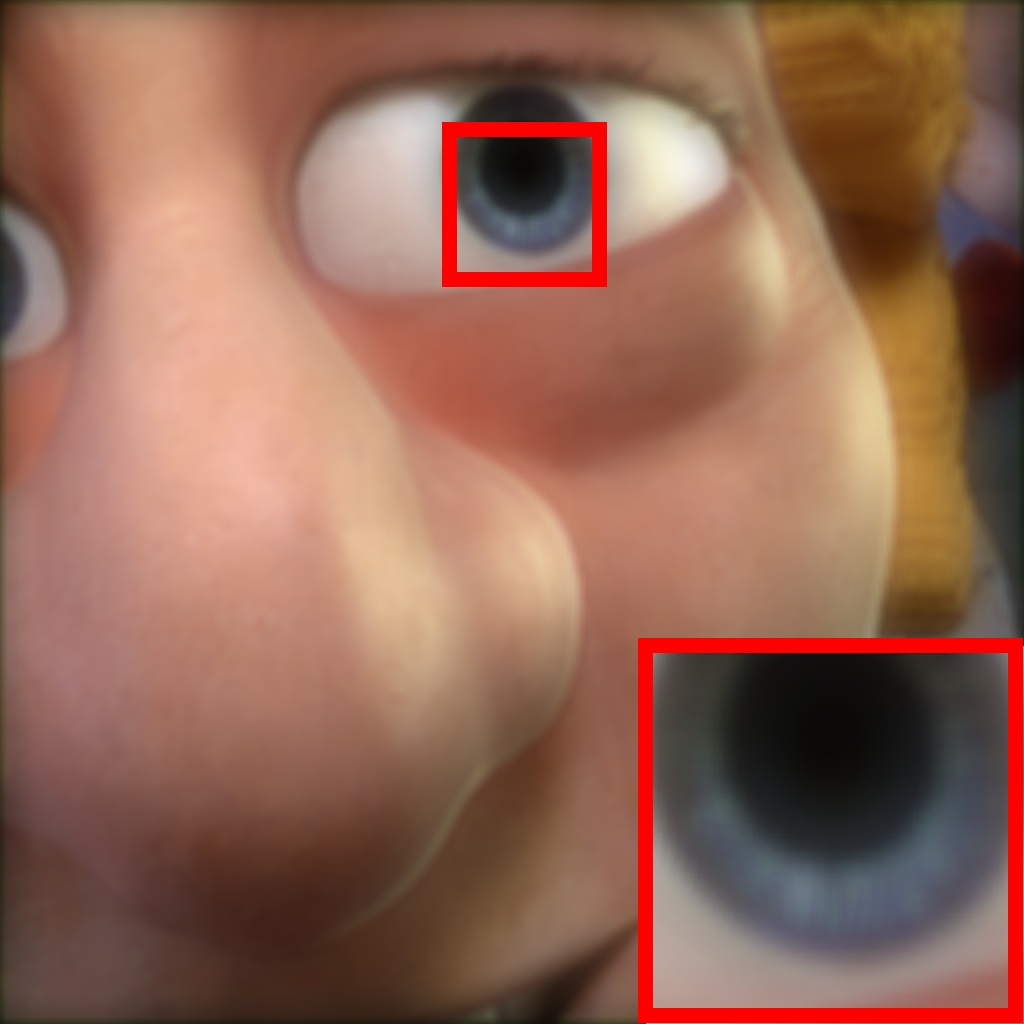}
		\put(-63,-9){\EyePSNRCSFtwentyfive \text{ / }\EyeSSIMCSFtwentyfive}
	\end{subfigure}
	
	\begin{subfigure}[t]{0.32\columnwidth}
		\centering
		\includegraphics[width=\columnwidth]{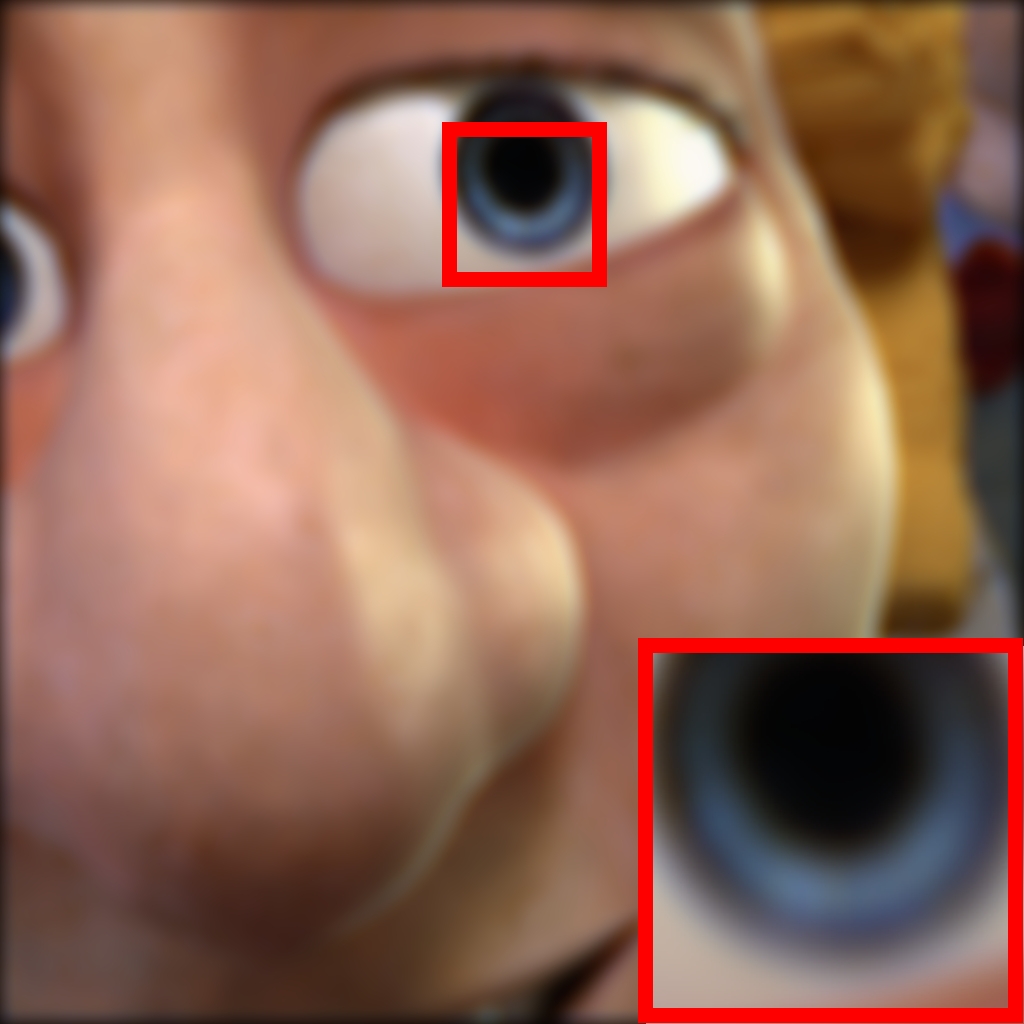}
		\put(-63,-9){\TLPSNRCSFthirtythree \text{ / }\TLSSIMCSFthirtythree}
		\put(-90,25){\rotatebox{90}{3 D}}
	\end{subfigure}
	\begin{subfigure}[t]{0.32\columnwidth}
		\centering
		\includegraphics[width=\columnwidth]{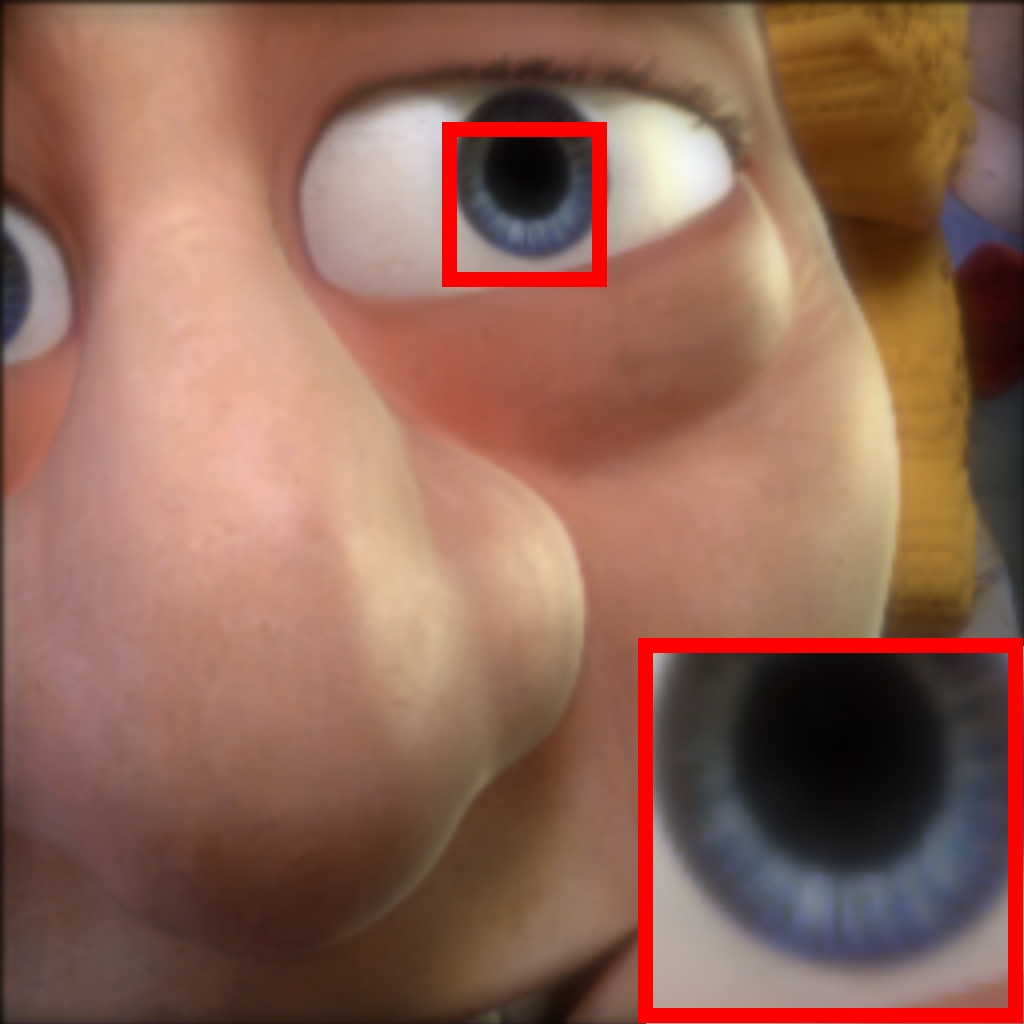}
		\put(-63,-9){\KonradPSNRCSFthirtythree \text{ / }\KonradSSIMCSFthirtythree}
	\end{subfigure}
	\begin{subfigure}[t]{0.32\columnwidth}
		\centering
		\includegraphics[width=\columnwidth]{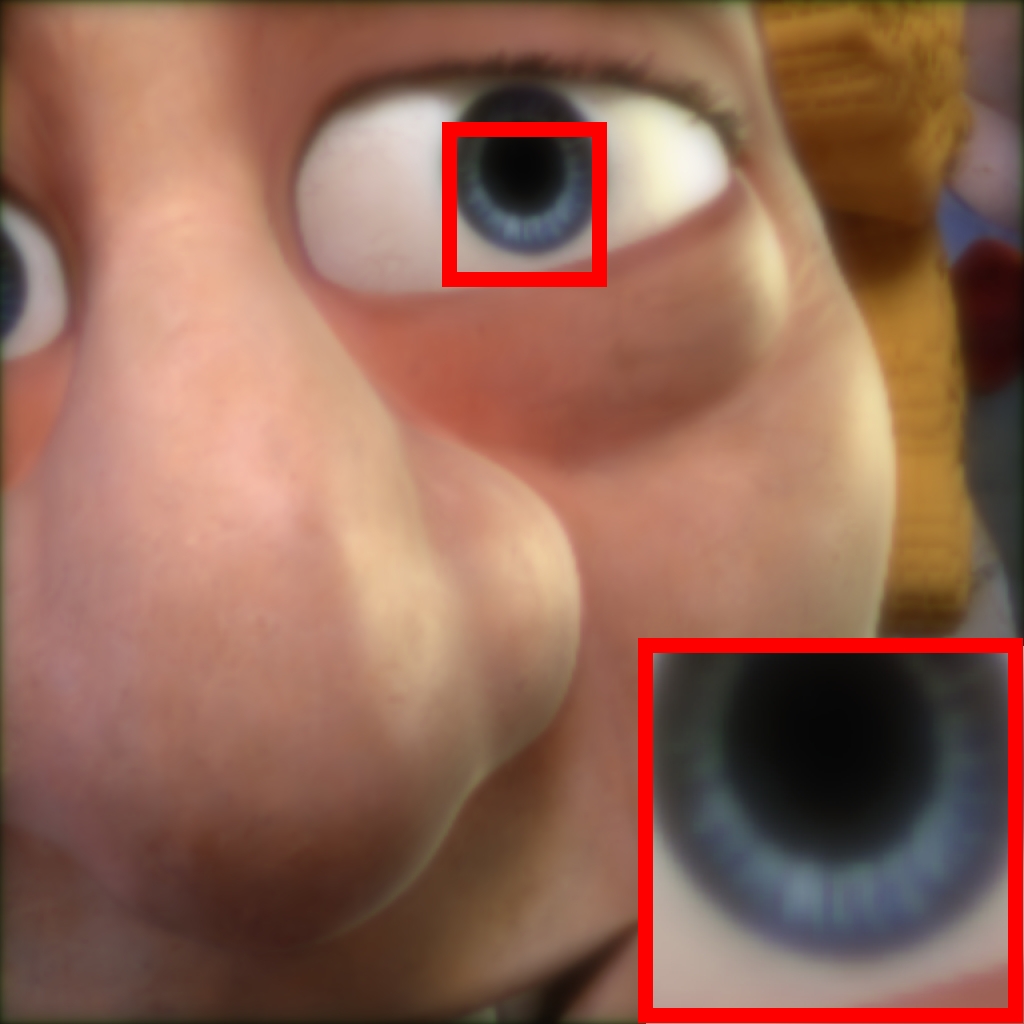}
		\put(-63,-9){\EyePSNRCSFthirtythree \text{ / }\EyeSSIMCSFthirtythree}
	\end{subfigure}
	
	\begin{subfigure}[t]{0.32\columnwidth}
		\centering
		\includegraphics[width=\columnwidth]{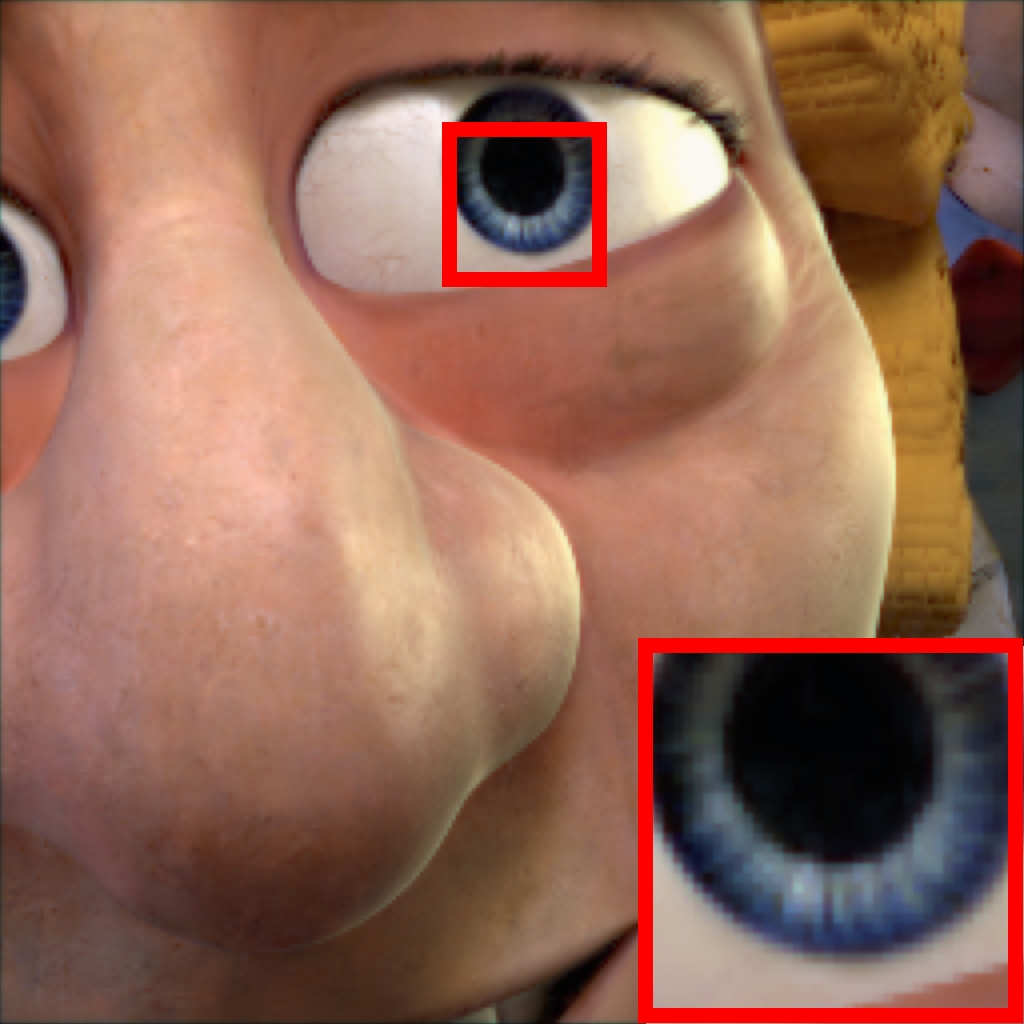}
		\put(-63,-9){\TLPSNRCSFfifty \text{ / }\TLSSIMCSFfifty}
		\put(-90,25){\rotatebox{90}{2 D}}
	\end{subfigure}
	\begin{subfigure}[t]{0.32\columnwidth}
		\centering
		\includegraphics[width=\columnwidth]{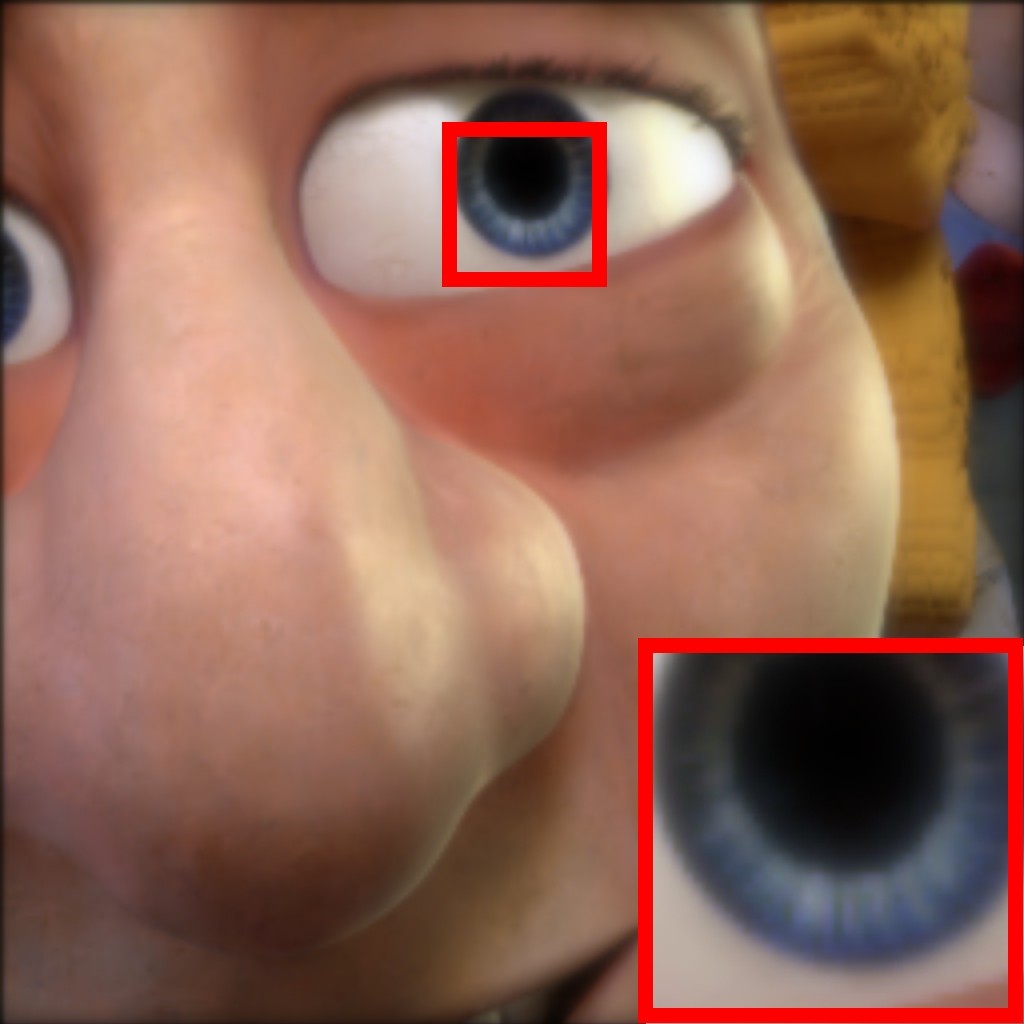}
		\put(-63,-9){\KonradPSNRCSFfifty \text{ / }\KonradSSIMCSFfifty}
	\end{subfigure}
	\begin{subfigure}[t]{0.32\columnwidth}
		\centering
		\includegraphics[width=\columnwidth]{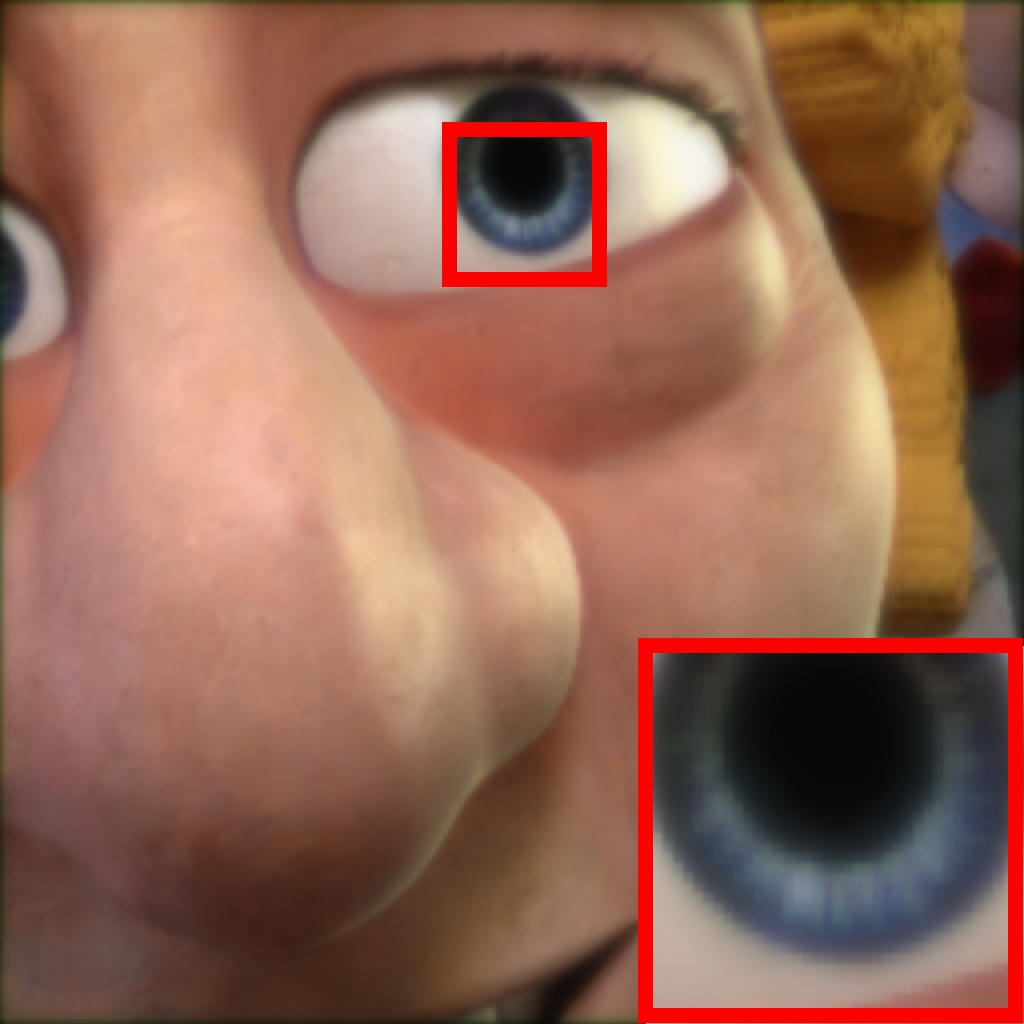}
		\put(-63,-9){\EyePSNRCSFfifty \text{ / }\EyeSSIMCSFfifty}
	\end{subfigure}	
	
	\begin{subfigure}[t]{0.32\columnwidth}
		\centering
		\includegraphics[width=\columnwidth]{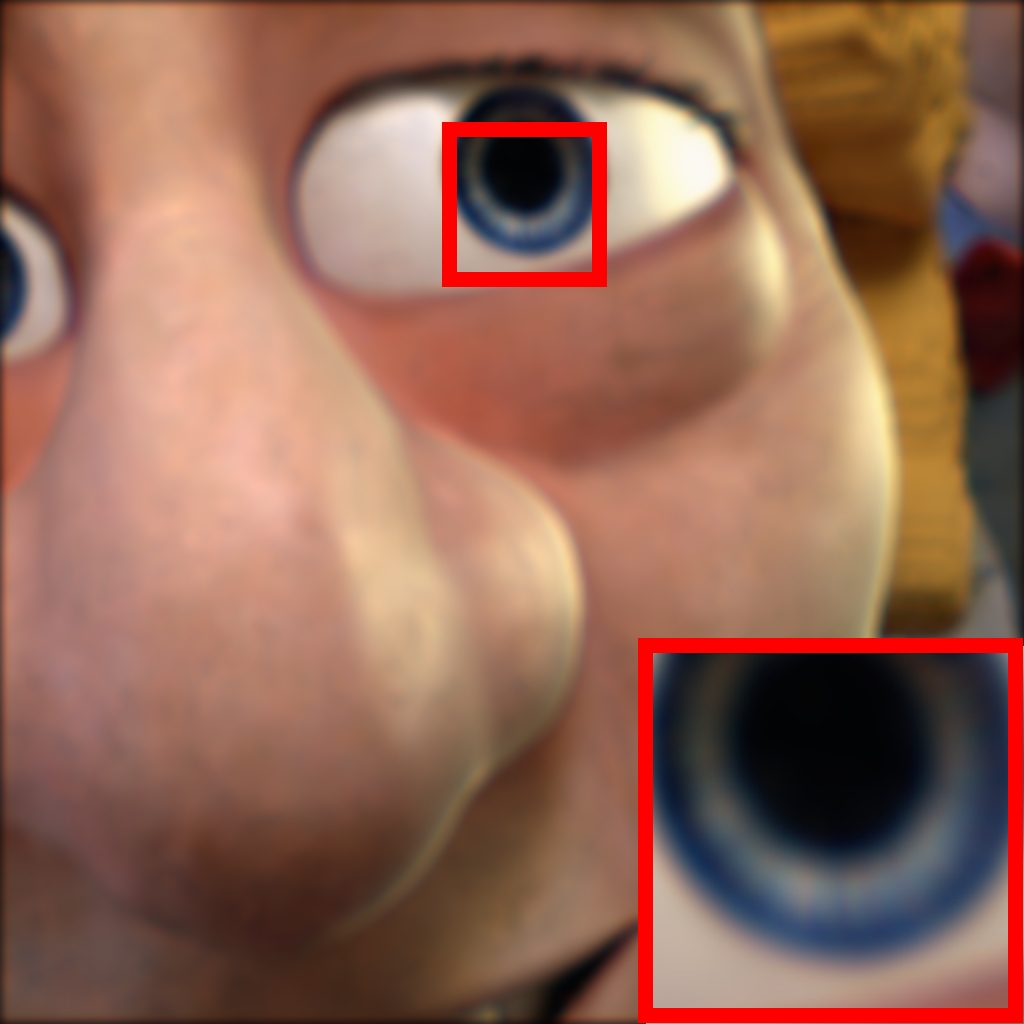}
		\put(-63,-9){\TLPSNRCSFonehundred \text{ / }\TLSSIMCSFonehundred}
		\put(-90,25){\rotatebox{90}{1 D}}
	\end{subfigure}
	\begin{subfigure}[t]{0.32\columnwidth}
		\centering
		\includegraphics[width=\columnwidth]{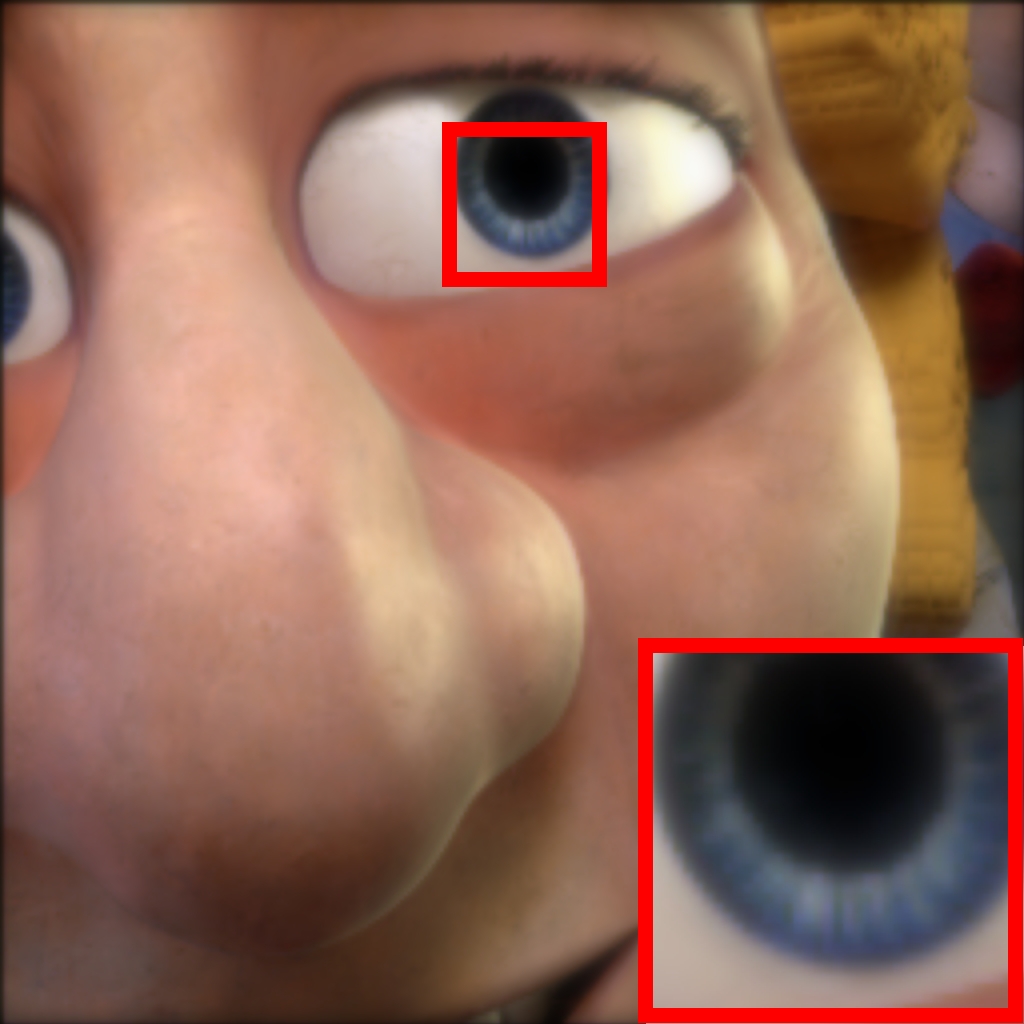}
		\put(-63,-9){\KonradPSNRCSFonehundred \text{ / }\KonradSSIMCSFonehundred}
	\end{subfigure}
	\begin{subfigure}[t]{0.32\columnwidth}
		\centering
		\includegraphics[width=\columnwidth]{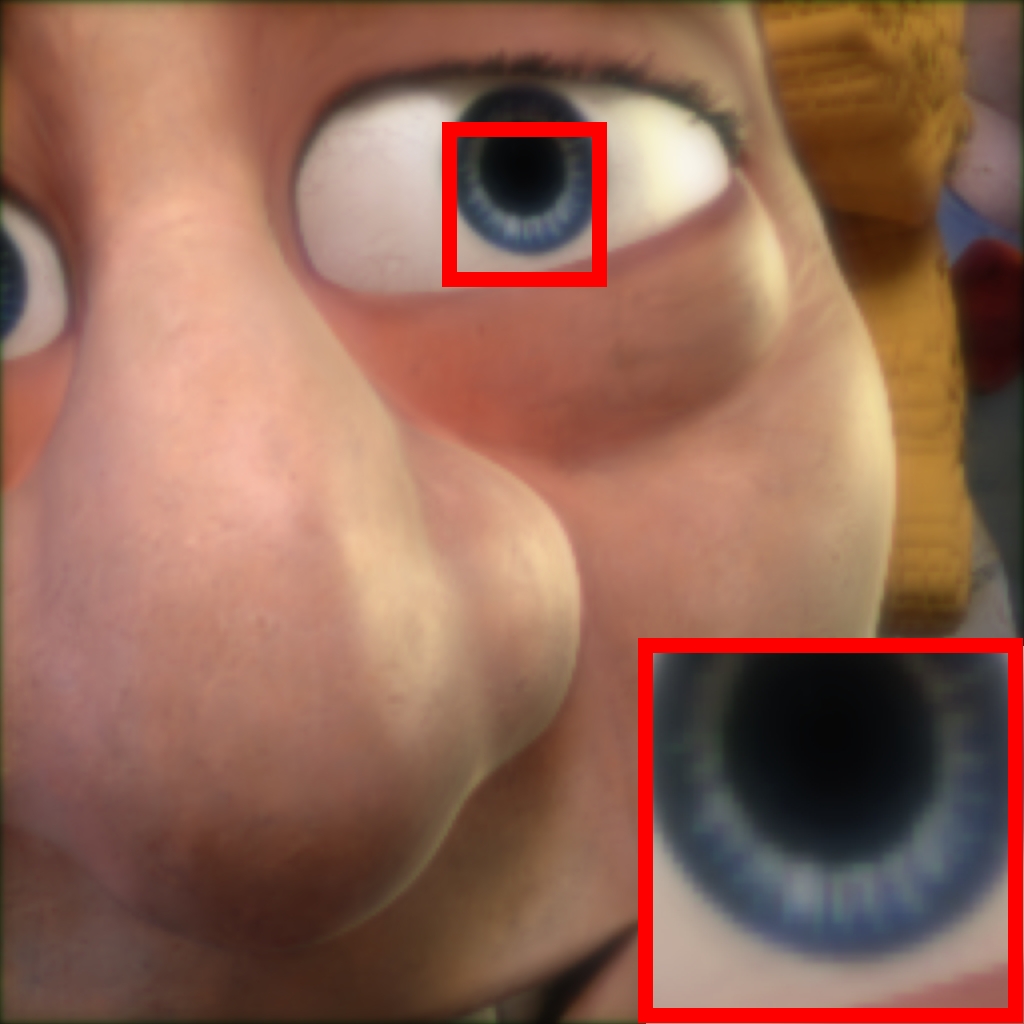}
		\put(-63,-9){\EyePSNRCSFonehundred \text{ / }\EyeSSIMCSFonehundred}
	\end{subfigure}	
	
	\begin{subfigure}[t]{0.32\columnwidth}
		\centering
		\includegraphics[width=\columnwidth]{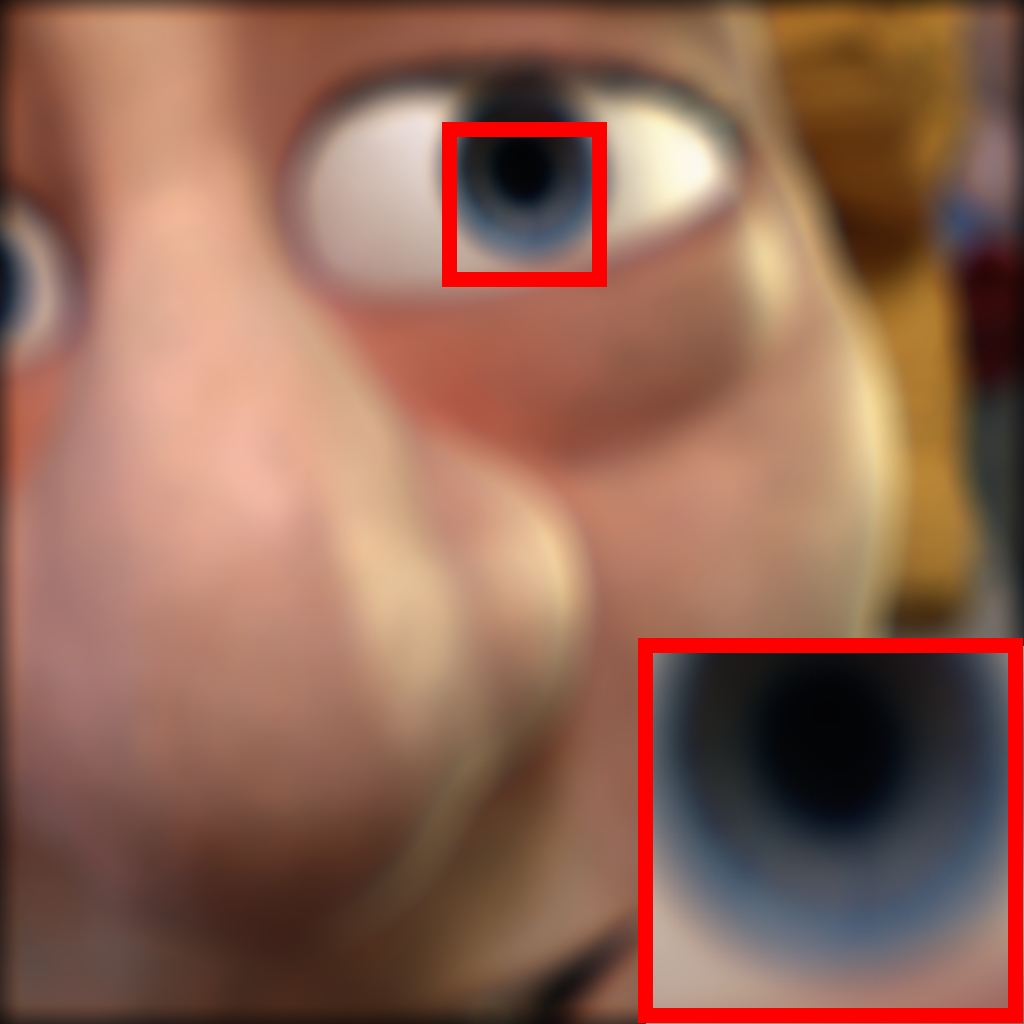}
		\put(-63,-9){\TLPSNRCSFinfinity \text{ / }\TLSSIMCSFinfinity}
		\put(-90,25){\rotatebox{90}{0 D}}
	\end{subfigure}
	\begin{subfigure}[t]{0.32\columnwidth}
		\centering
		\includegraphics[width=\columnwidth]{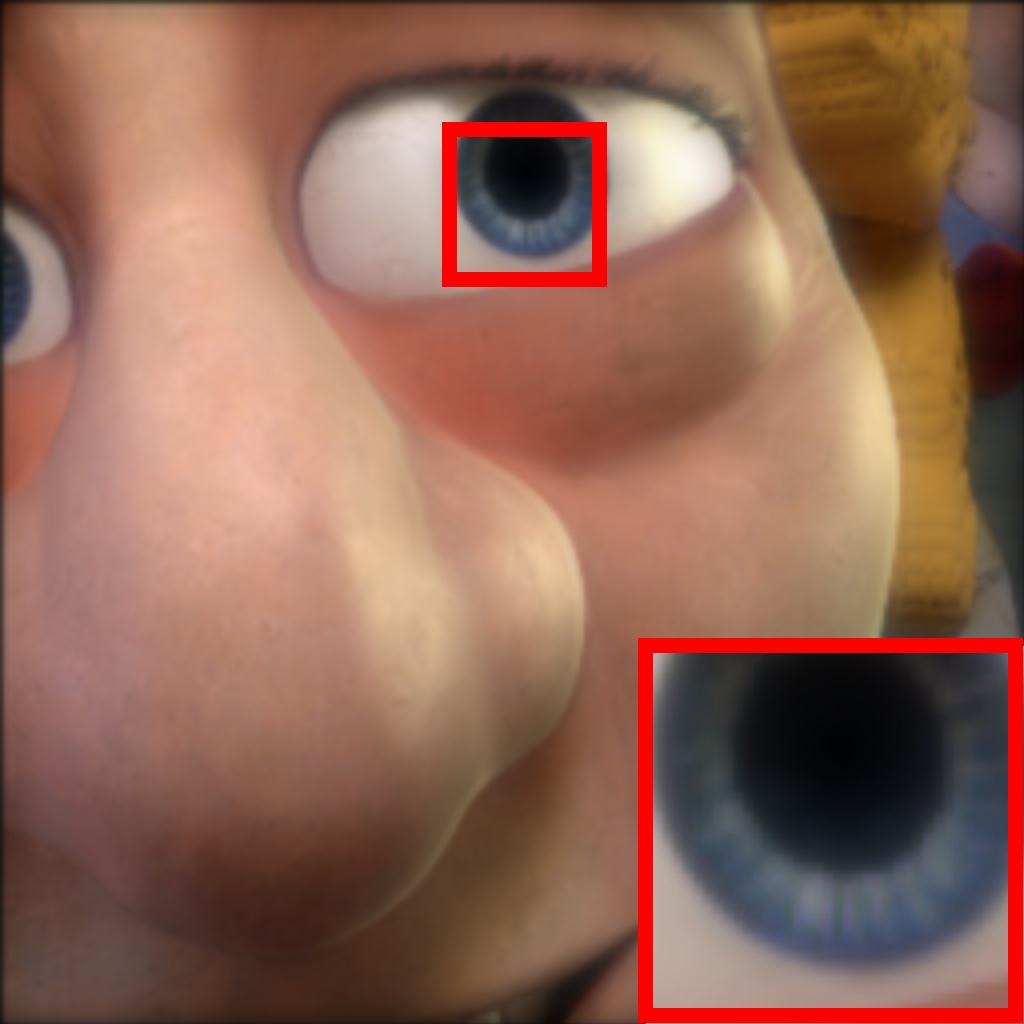}
		\put(-63,-9){\KonradPSNRCSFinfinity \text{ / }\KonradSSIMCSFinfinity}
	\end{subfigure}
	\begin{subfigure}[t]{0.32\columnwidth}
		\centering
		\includegraphics[width=\columnwidth]{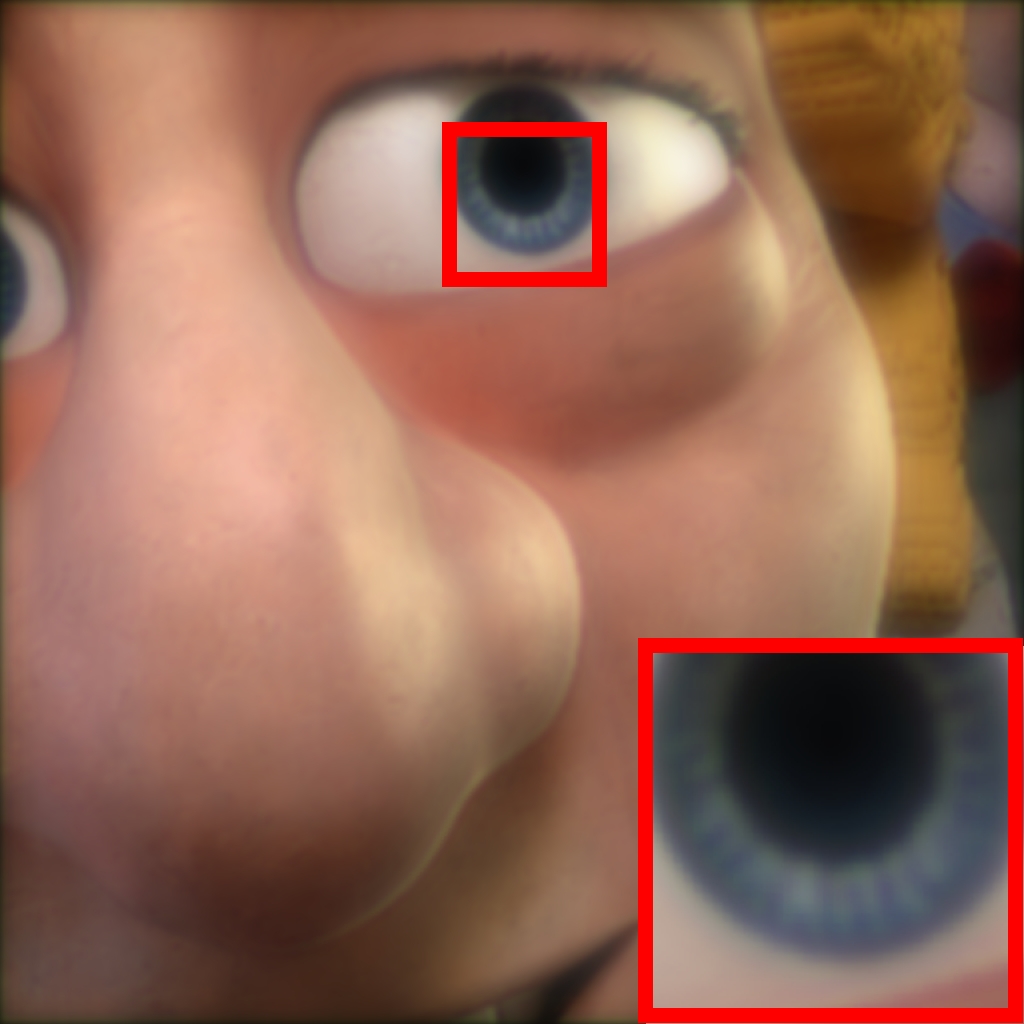}
		\put(-63,-9){\EyePSNRCSFinfinity \text{ / }\EyeSSIMCSFinfinity}
	\end{subfigure}		

	\caption{Comparison of the conventional stereoscopic display with single refractive lens (left), AI NED from Konrad et.al. \cite{KonradAI} (middle), and the proposed (right) displays. The PSNR/SSIM values are given under each image.}
	\label{fig:compare_soa}
	
\end{figure}

%% file: figures/images/DepthIm/X0Y0/Spherical/PSNR.tex
\newcommand\TLPSNRtwentyfive{23.44} % 25 cm, thin lens, Depthim_small image
\newcommand\TLPSNRthirtythree{26.86} % 33 cm, thin lens, Depthim_small image
\newcommand\TLPSNRfifty{35.25} % 50 cm, thin lens, Depthim_small image
\newcommand\TLPSNRonehundred{28.63} % 100 cm, thin lens, Depthim_small image
\newcommand\TLPSNRinfinity{24.20} % Inf cm, thin lens, Depthim_small image

\newcommand\TLPSNRCSFtwentyfive{25.02} % 25 cm, thin lens, Depthim_small image
\newcommand\TLPSNRCSFthirtythree{29.80} % 33 cm, thin lens, Depthim_small image
\newcommand\TLPSNRCSFfifty{37.41} % 50 cm, thin lens, Depthim_small image
\newcommand\TLPSNRCSFonehundred{32.46} % 100 cm, thin lens, Depthim_small image
\newcommand\TLPSNRCSFinfinity{26.02} % Inf cm, thin lens, Depthim_small image

%% file: figures/images/DepthIm/X0Y0/Spherical/SSIM.tex
\newcommand\TLSSIMtwentyfive{0.93} % 25 cm, thin lens, Depthim_small image
\newcommand\TLSSIMthirtythree{0.95} % 33 cm, thin lens, Depthim_small image
\newcommand\TLSSIMfifty{0.99} % 50 cm, thin lens, Depthim_small image
\newcommand\TLSSIMonehundred{0.96} % 100 cm, thin lens, Depthim_small image
\newcommand\TLSSIMinfinity{0.93} % Inf cm, thin lens, Depthim_small image

\newcommand\TLSSIMCSFtwentyfive{0.94} % 25 cm, thin lens, Depthim_small image
\newcommand\TLSSIMCSFthirtythree{0.97} % 33 cm, thin lens, Depthim_small image
\newcommand\TLSSIMCSFfifty{0.99} % 50 cm, thin lens, Depthim_small image
\newcommand\TLSSIMCSFonehundred{0.98} % 100 cm, thin lens, Depthim_small image
\newcommand\TLSSIMCSFinfinity{0.95} % Inf cm, thin lens, Depthim_small image

%% file: figures/images/DepthIm/X0Y0/Konrad5/PSNR.tex
\newcommand\KonradPSNRtwentyfive{23.69} % 25 cm, Konrad, Depthim_small image
\newcommand\KonradPSNRCSFtwentyfive{25.24} % 25 cm, Konrad, CSF, Depthim_small image
\newcommand\KonradPSNRthirtythree{25.28} % 33 cm, Konrad, Depthim_small image
\newcommand\KonradPSNRCSFthirtythree{27.34} % 33 cm, Konrad, CSF, Depthim_small image
\newcommand\KonradPSNRfifty{25.92} % 50 cm, Konrad, Depthim_small image
\newcommand\KonradPSNRCSFfifty{28.23} % 50 cm, Konrad, CSF, Depthim_small image
\newcommand\KonradPSNRonehundred{25.83} % 100 cm, Konrad, Depthim_small image
\newcommand\KonradPSNRCSFonehundred{28.13} % 100 cm, Konrad, CSF, Depthim_small image
\newcommand\KonradPSNRinfinity{25.04} % Inf cm, Konrad, Depthim_small image
\newcommand\KonradPSNRCSFinfinity{27.04} % Inf cm, Konrad, CSF, Depthim_small image

%% file: figures/images/DepthIm/X0Y0/Konrad5/SSIM.tex
\newcommand\KonradSSIMtwentyfive{0.93} % 25 cm, Konrad, Depthim_small image
\newcommand\KonradSSIMCSFtwentyfive{0.95} % 25 cm, Konrad, CSF, Depthim_small image
\newcommand\KonradSSIMthirtythree{0.94} % 33 cm, Konrad, Depthim_small image
\newcommand\KonradSSIMCSFthirtythree{0.96} % 33 cm, Konrad, CSF, Depthim_small image
\newcommand\KonradSSIMfifty{0.95} % 50 cm, Konrad, Depthim_small image
\newcommand\KonradSSIMCSFfifty{0.97} % 50 cm, Konrad, CSF, Depthim_small image
\newcommand\KonradSSIMonehundred{0.95} % 100 cm, Konrad, Depthim_small image
\newcommand\KonradSSIMCSFonehundred{0.97} % 100 cm, Konrad, CSF, Depthim_small image
\newcommand\KonradSSIMinfinity{0.94} % Inf cm, Konrad, Depthim_small image
\newcommand\KonradSSIMCSFinfinity{0.96} % Inf cm, Konrad, CSF, Depthim_small image

%% file: figures/images/DepthIm/X0Y0/Our/PSNR.tex
\newcommand\EyePSNRtwentyfive{25.1} % 25 cm, our method, Eye, Depthim_small image
\newcommand\EyePSNRthirtythree{27.2} % 33 cm, our method, Eye, Depthim_small image
\newcommand\EyePSNRfifty{28.49} % 50 cm, our method, Eye, Depthim_small image
\newcommand\EyePSNRonehundred{27.92} % 100 cm, our method, Eye, Depthim_small image
\newcommand\EyePSNRinfinity{25.35} % Inf cm, our method, Eye, Depthim_small image

\newcommand\EyePSNRCSFtwentyfive{27.06} % 25 cm, our method, Eye, Depthim_small image
\newcommand\EyePSNRCSFthirtythree{30.17} % 33 cm, our method, Eye, Depthim_small image
\newcommand\EyePSNRCSFfifty{32.18} % 50 cm, our method, Eye, Depthim_small image
\newcommand\EyePSNRCSFonehundred{31.10} % 100 cm, our method, Eye, Depthim_small image
\newcommand\EyePSNRCSFinfinity{27.41} % Inf cm, our method, Eye, Depthim_small image

%% file: figures/images/DepthIm/X0Y0/Our/SSIM.tex
\newcommand\EyeSSIMtwentyfive{0.94} % 25 cm, our method, Eye, Depthim_small image
\newcommand\EyeSSIMthirtythree{0.96} % 33 cm, our method, Eye, Depthim_small image
\newcommand\EyeSSIMfifty{0.96} % 50 cm, our method, Eye, Depthim_small image
\newcommand\EyeSSIMonehundred{0.96} % 100 cm, our method, Eye, Depthim_small image
\newcommand\EyeSSIMinfinity{0.94} % Inf cm, our method, Eye, Depthim_small image

\newcommand\EyeSSIMCSFtwentyfive{0.96} % 25 cm, our method, Eye, Depthim_small image
\newcommand\EyeSSIMCSFthirtythree{0.98} % 33 cm, our method, Eye, Depthim_small image
\newcommand\EyeSSIMCSFfifty{0.98} % 50 cm, our method, Eye, Depthim_small image
\newcommand\EyeSSIMCSFonehundred{0.98} % 100 cm, our method, Eye, Depthim_small image
\newcommand\EyeSSIMCSFinfinity{0.96} % Inf cm, our method, Eye, Depthim_small image

%% file: figures/images/PCNN_Output_v2.tex
\begin{figure}[htbp]
    \centering
        \begin{subfigure}[t]{0.17\columnwidth}
		\centering
            \caption{$I$}
            \hfill
		\includegraphics[width=\columnwidth]{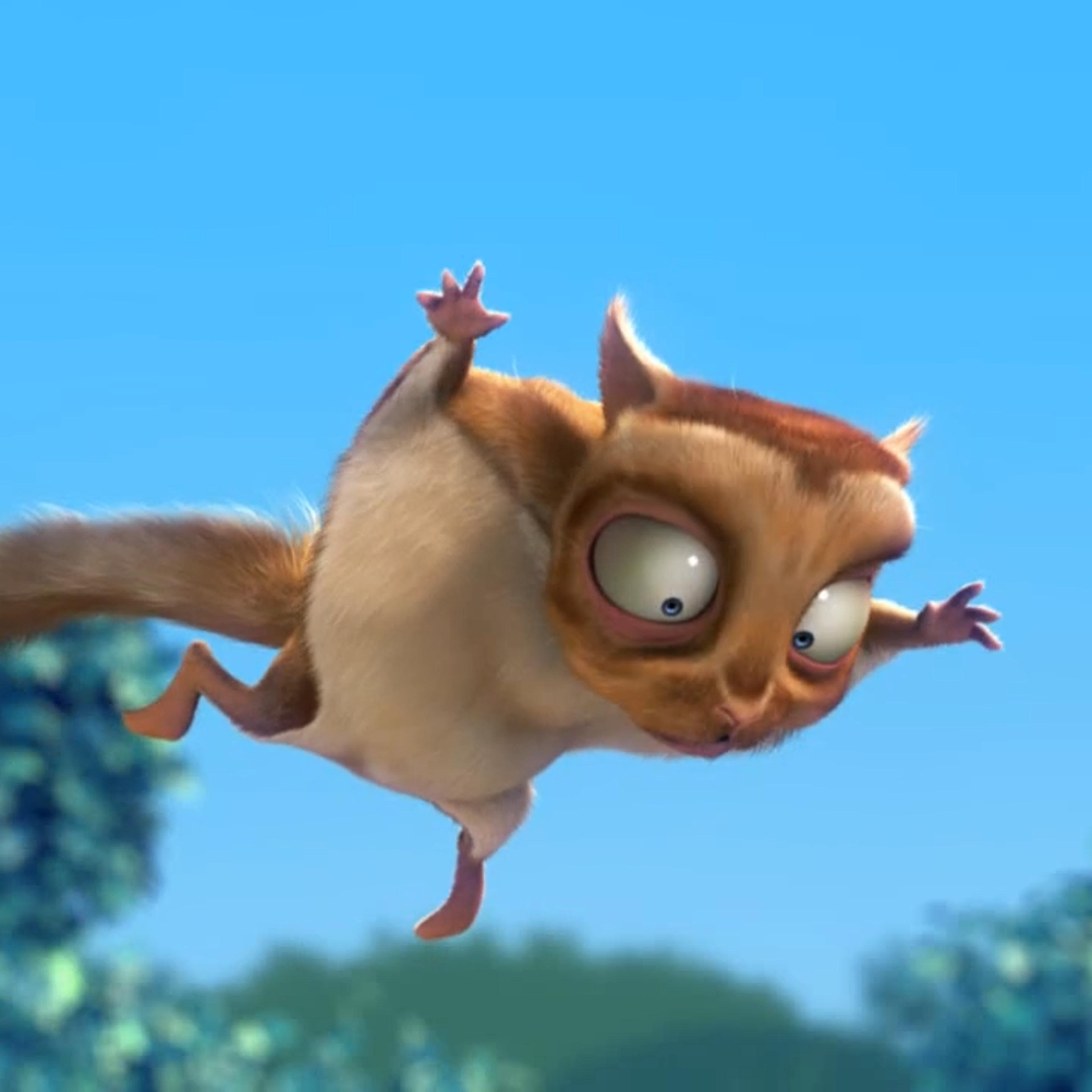}
		%\put(20,-60){$I$}
	\end{subfigure}
        \begin{subfigure}[t]{0.17\columnwidth}
		\centering
		\caption{$I^d$}
		\includegraphics[width=\columnwidth]{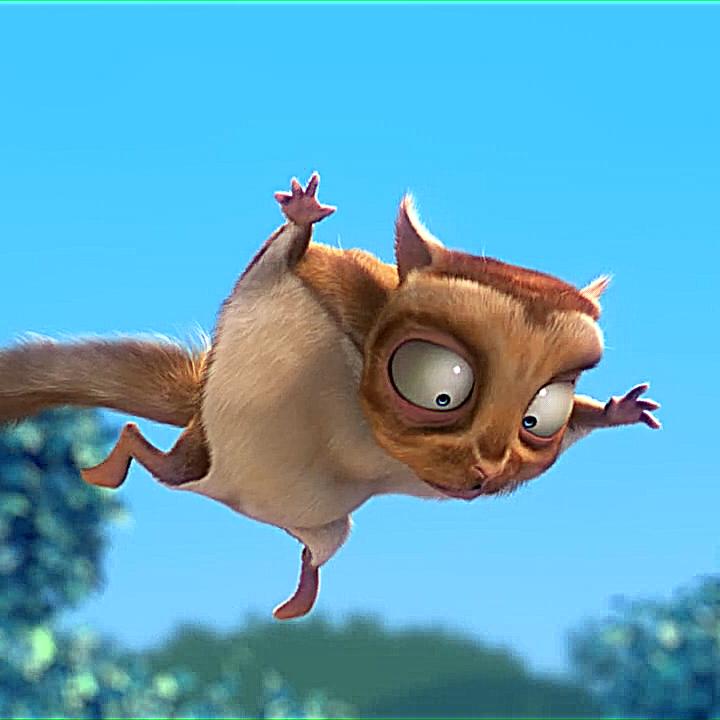}
	\end{subfigure}
        \begin{subfigure}[t]{0.17\columnwidth}
		\centering
		\caption{$I^r$}
		\includegraphics[width=\columnwidth]{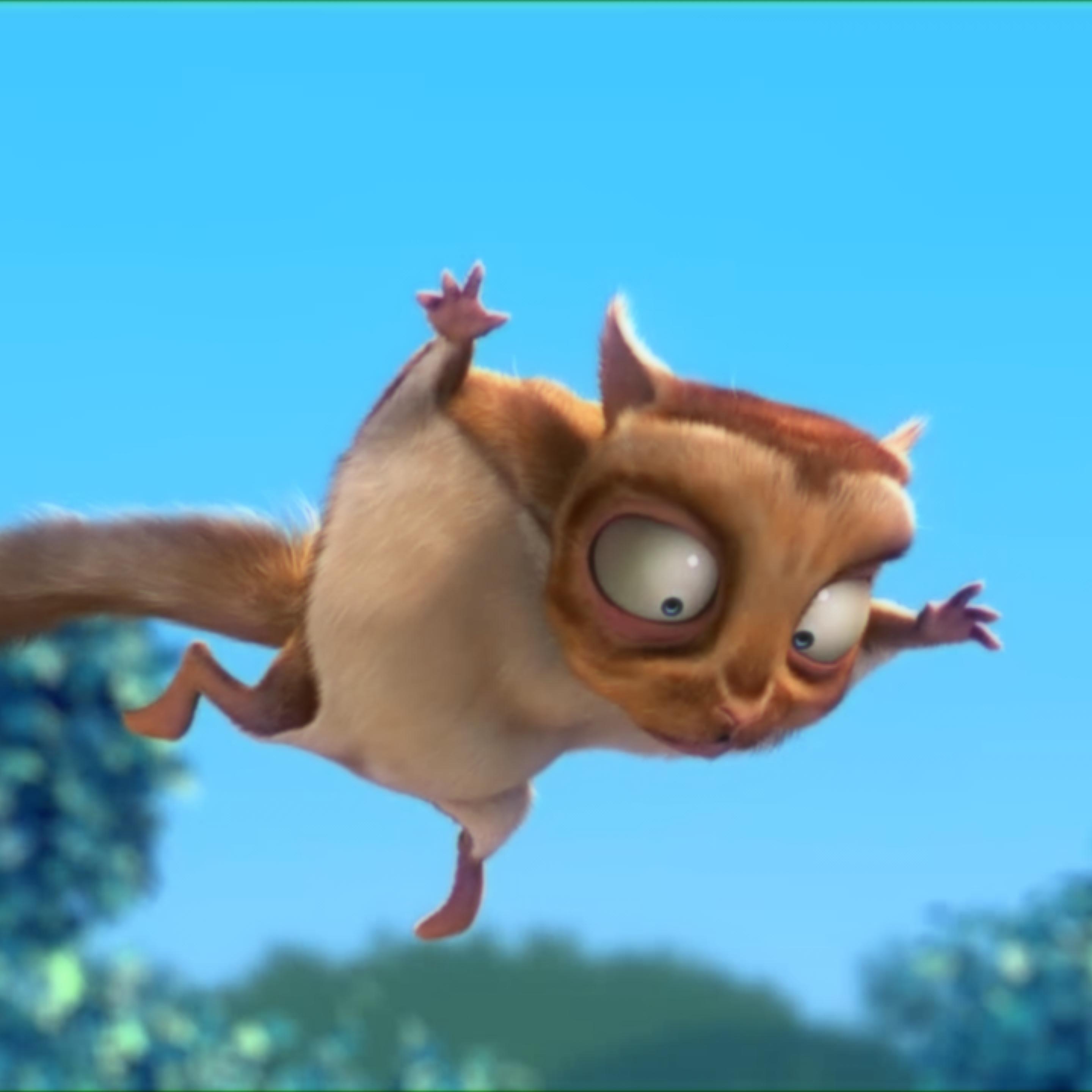}
	\end{subfigure}
        \begin{subfigure}[t]{0.17\columnwidth}
		\centering
		\caption{$\Tilde{I}^r$}
		\includegraphics[width=\columnwidth]{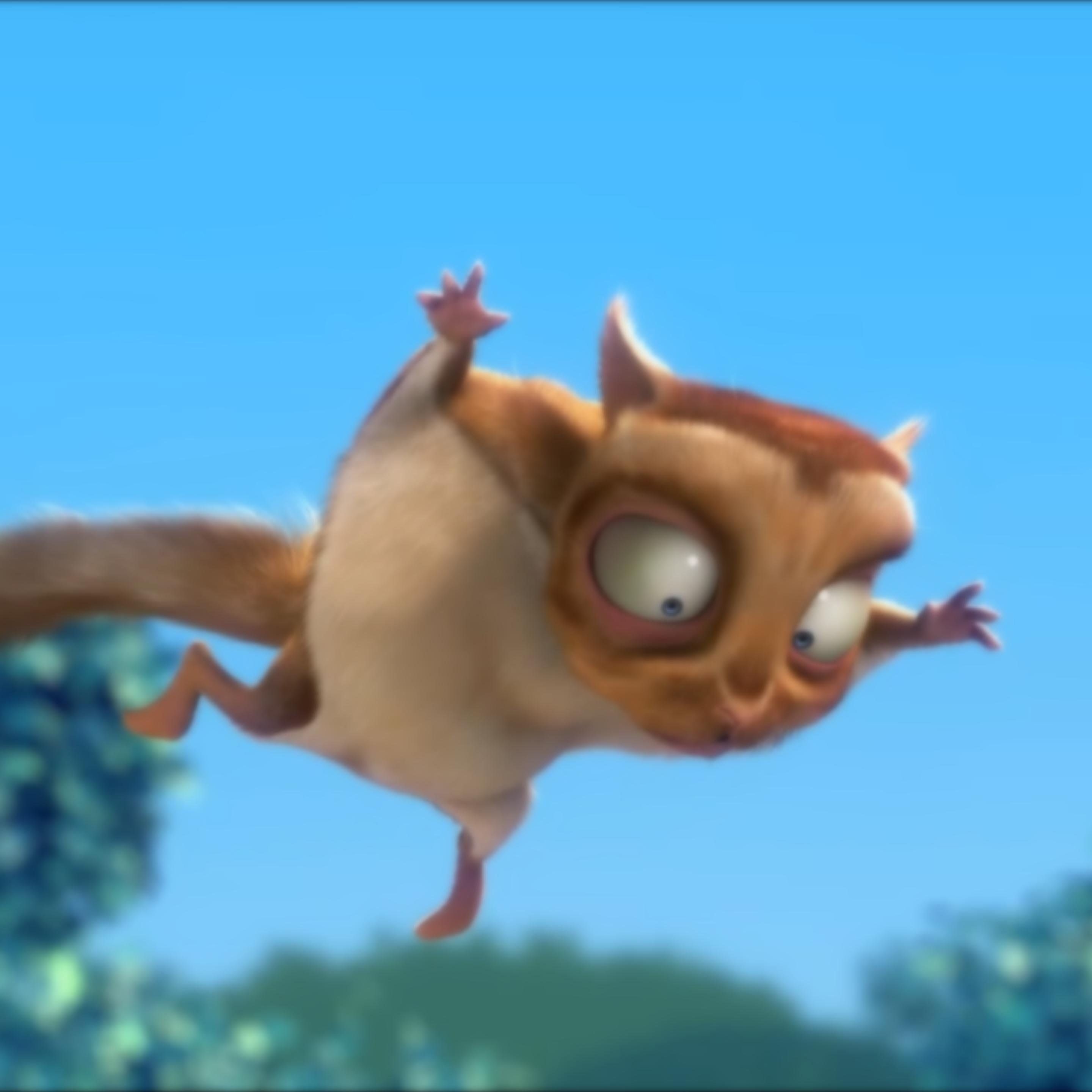}
	\end{subfigure}
 
    \centering
        \begin{subfigure}[t]{0.17\columnwidth}
		\centering
            %\caption{\textbf{3 D}}
		\includegraphics[width=\columnwidth]{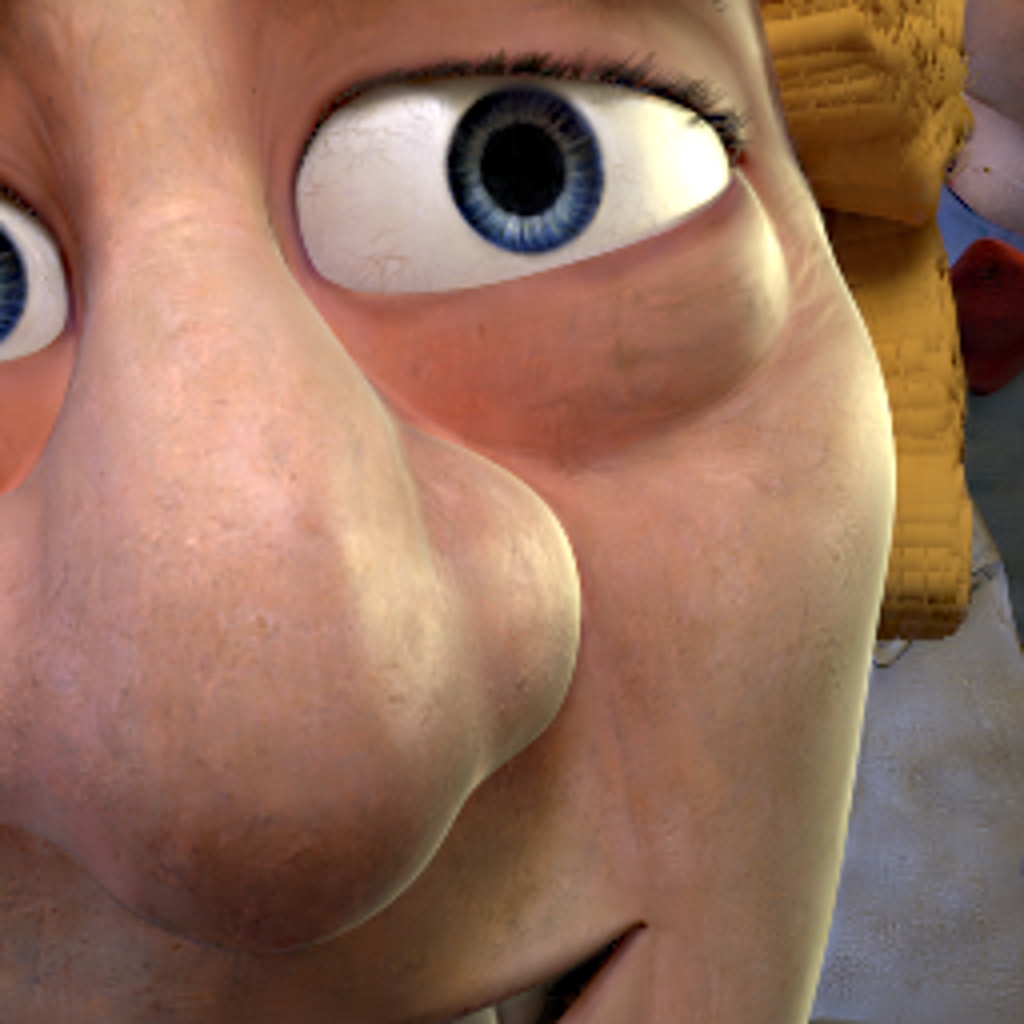}
            \put(-60,20){\rotatebox{90}{2 D}}
	\end{subfigure}
        \begin{subfigure}[t]{0.17\columnwidth}
		\centering
            %\caption{\textbf{3 D}}
		\includegraphics[width=\columnwidth]{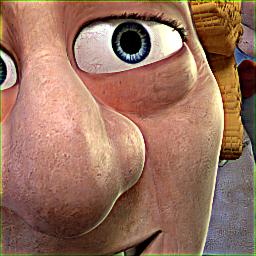}
	\end{subfigure}
        \begin{subfigure}[t]{0.17\columnwidth}
		\centering
            %\caption{\textbf{3 D}}
		\includegraphics[width=\columnwidth]{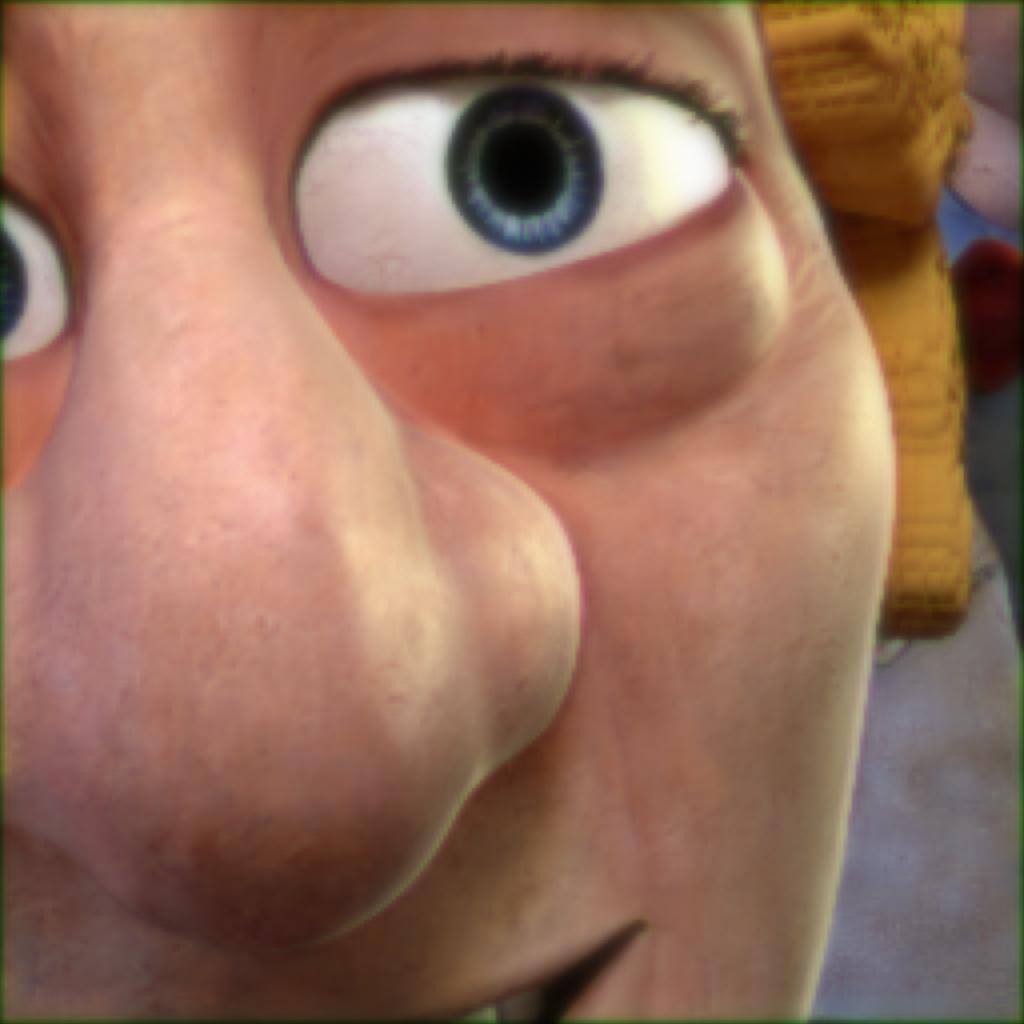}
	\end{subfigure}
        \begin{subfigure}[t]{0.17\columnwidth}
		\centering
            %\caption{\textbf{3 D}}
		\includegraphics[width=\columnwidth]{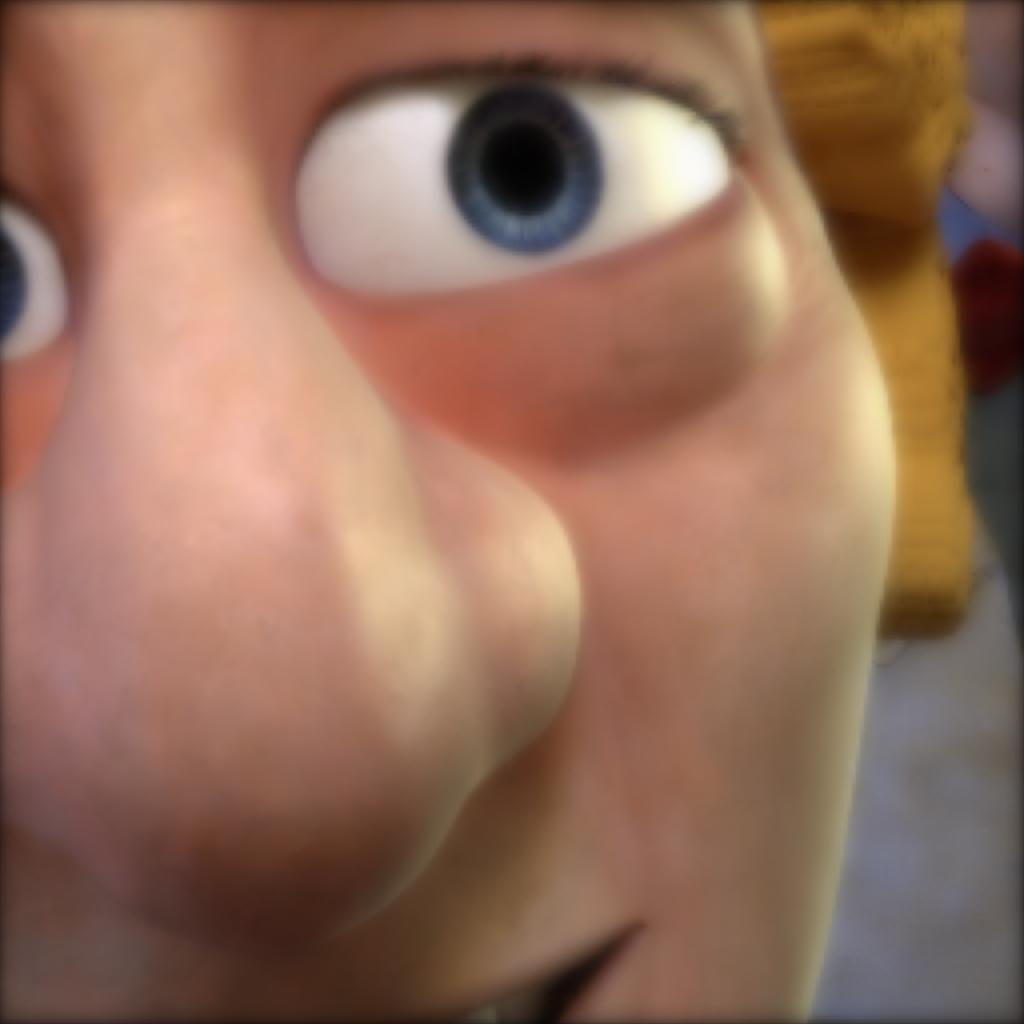}
	\end{subfigure}

    \centering
        \begin{subfigure}[t]{0.17\columnwidth}
		\centering
            %\caption{$I$}
		\includegraphics[width=\columnwidth]{figures/images/Display_Images/BigBuckBunny/gt.jpg}
		%\put(20,-60){$I$}
	\end{subfigure}
        \begin{subfigure}[t]{0.17\columnwidth}
		\centering
		%\caption{$I^d$}
		\includegraphics[width=\columnwidth]{figures/images/Display_Images/BigBuckBunny/Display.jpg}
	\end{subfigure}
        \begin{subfigure}[t]{0.17\columnwidth}
		\centering
		%\caption{$I^r$}
		\includegraphics[width=\columnwidth]{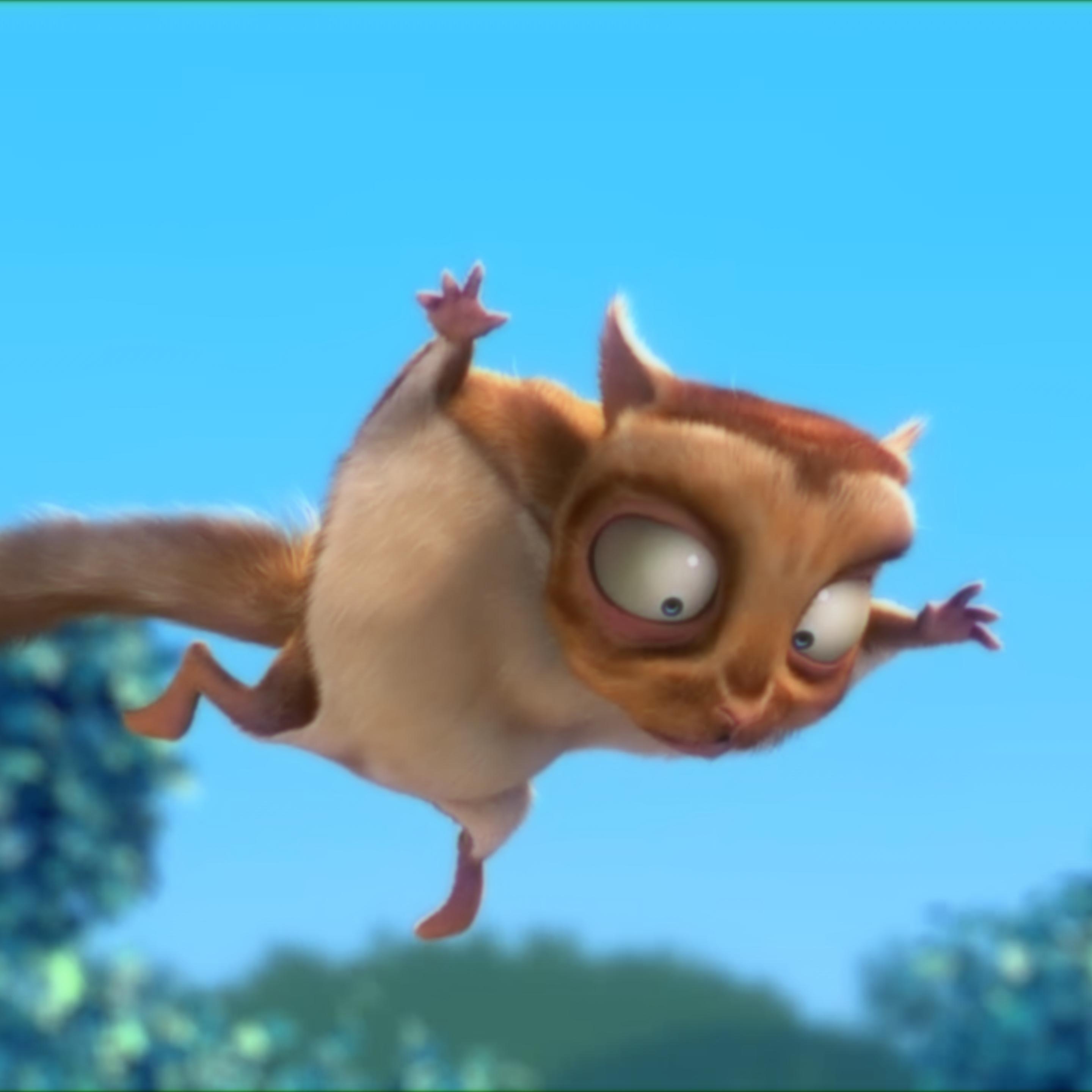}
	\end{subfigure}
        \begin{subfigure}[t]{0.17\columnwidth}
		\centering
		%\caption{$\Tilde{I}^r$}
		\includegraphics[width=\columnwidth]{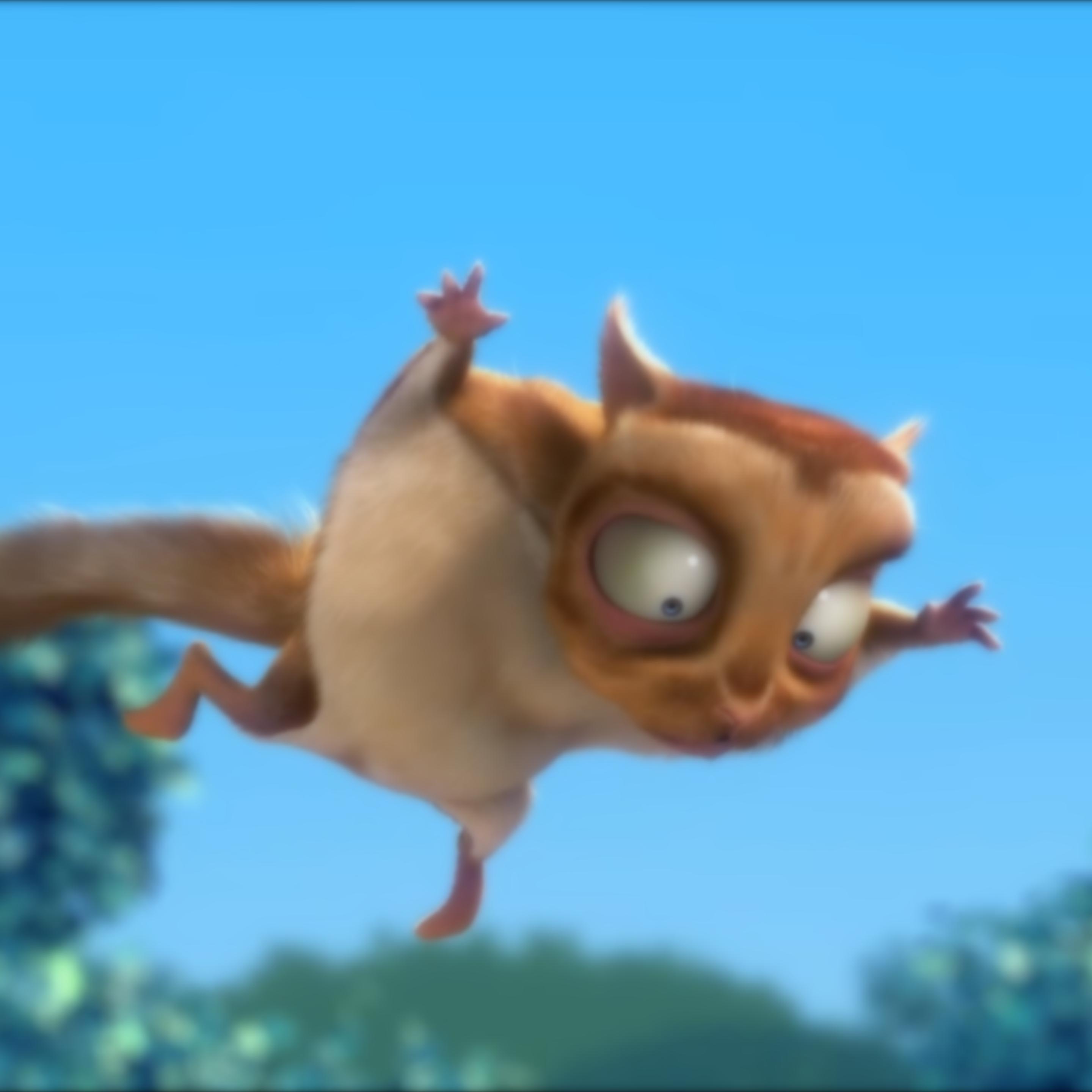}
	\end{subfigure}
 
    \centering
        \begin{subfigure}[t]{0.17\columnwidth}
		\centering
            %\caption{\textbf{3 D}}
		\includegraphics[width=\columnwidth]{figures/images/Display_Images/DepthIm/gt.png}
            \put(-60,20){\rotatebox{90}{3 D}}
	\end{subfigure}
        \begin{subfigure}[t]{0.17\columnwidth}
		\centering
            %\caption{\textbf{3 D}}
		\includegraphics[width=\columnwidth]{figures/images/Display_Images/DepthIm/Display.jpg}
	\end{subfigure}
        \begin{subfigure}[t]{0.17\columnwidth}
		\centering
            %\caption{\textbf{3 D}}
		\includegraphics[width=\columnwidth]{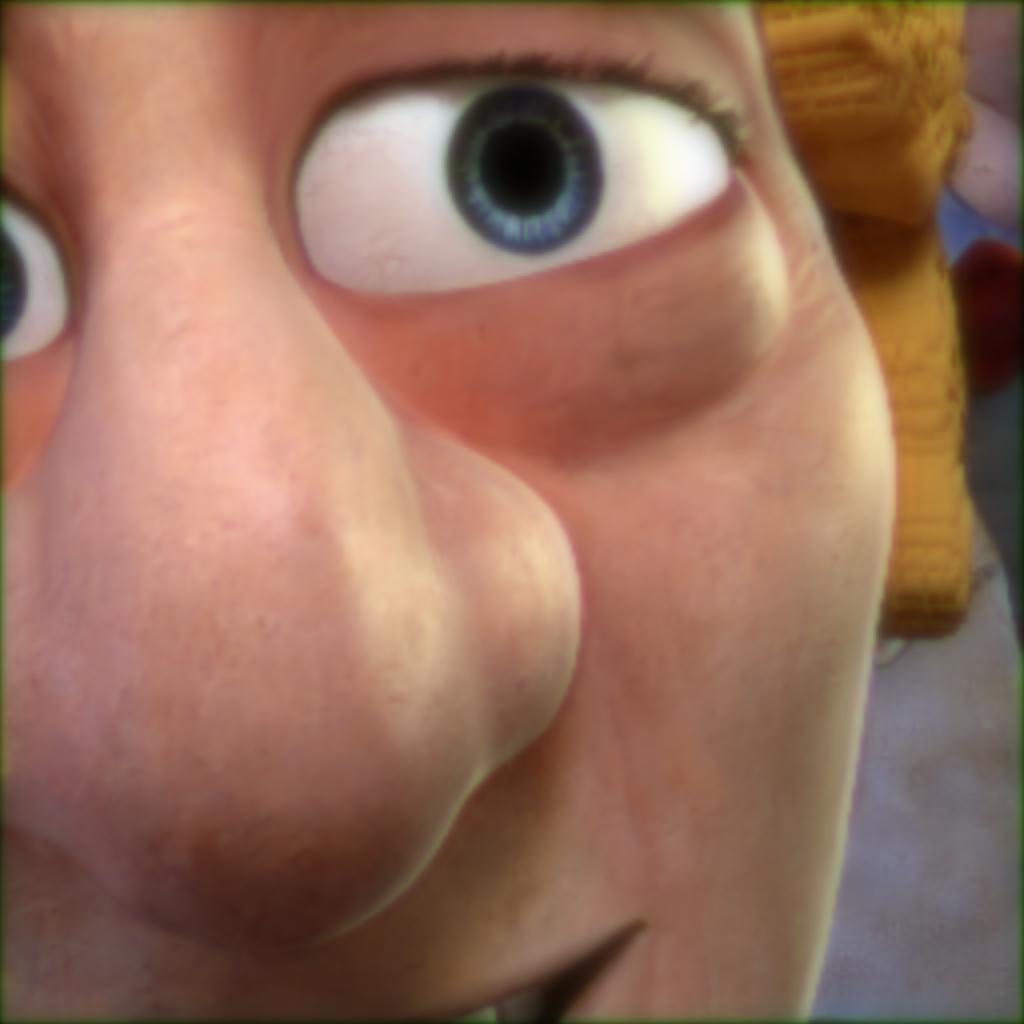}
	\end{subfigure}
        \begin{subfigure}[t]{0.17\columnwidth}
		\centering
            %\caption{\textbf{3 D}}
		\includegraphics[width=\columnwidth]{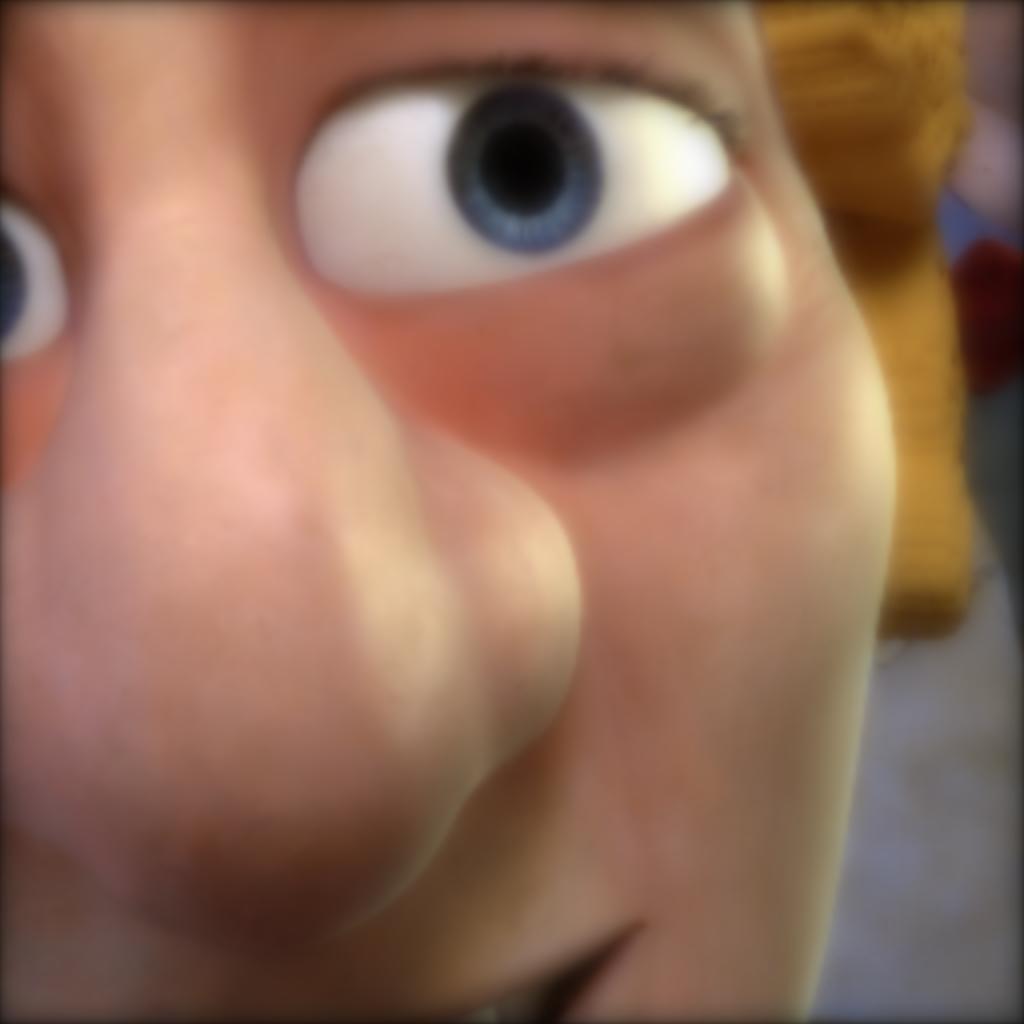}
	\end{subfigure}

\caption{Ablation study to demonstrate P-CNN block's effect in the proposed method, assuming accommodation distances of 2 and 3 diopters. From left to right: Input image, display image after P-CNN, the perceived image at the retina with the P-CNN, the perceived image at the retina when the input image is directly used without P-CNN.}
\label{fig:PCNN_Output}
\end{figure}

%% file: figures/HeightNoise2D/heightnoise.tex
\begin{figure}[ht]
\centering
\begin{subfigure}{0.24\columnwidth}
    \caption{$\sigma_d=\SI{10}{\nm}$}
    \includegraphics[width=\columnwidth]{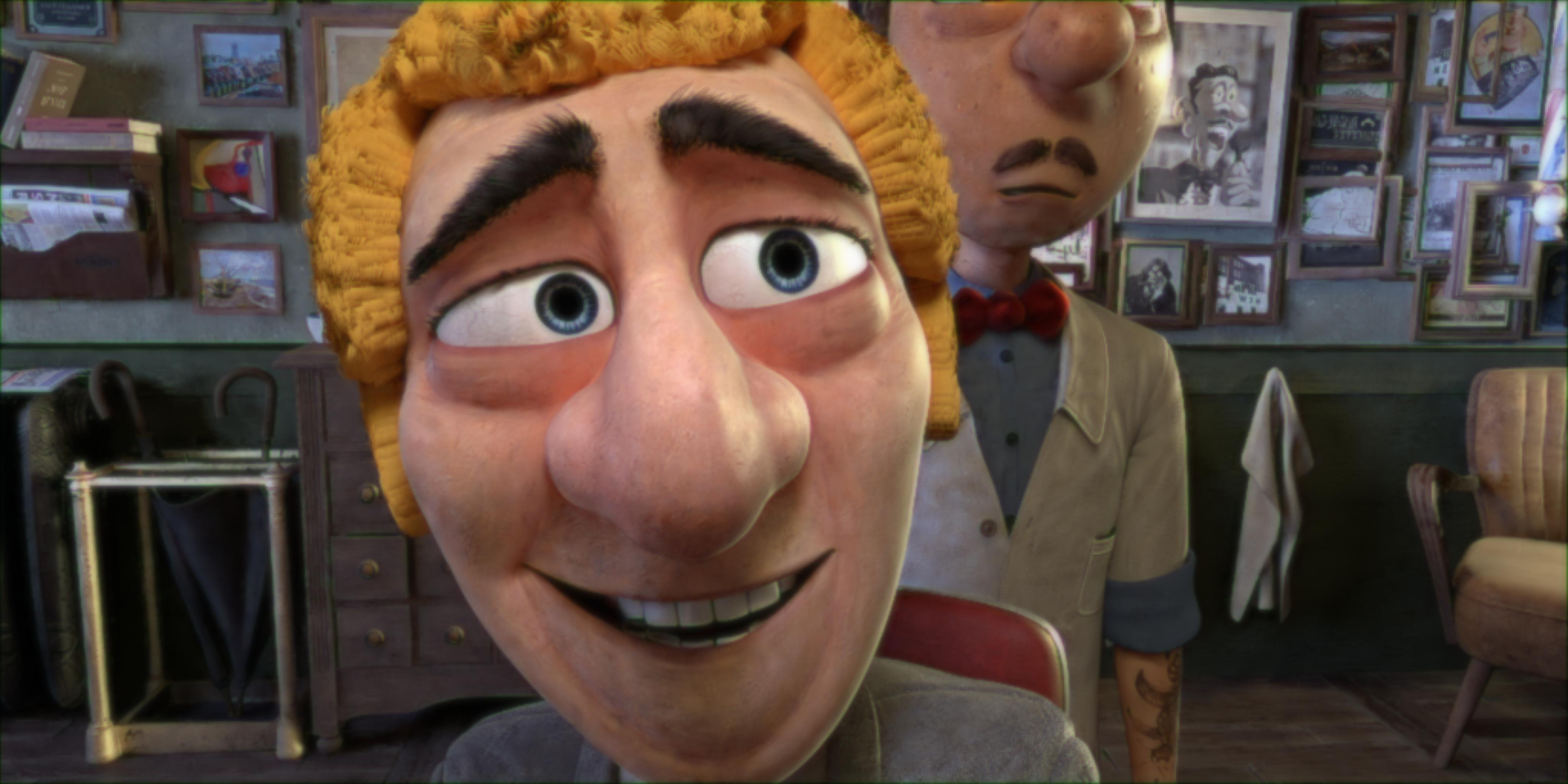}
    \caption{35.59 / 0.95}
    \label{sfig:Hnoise20nm}
\end{subfigure}
\begin{subfigure}{0.24\columnwidth}
    \caption{$\sigma_d=\SI{20}{\nm}$}
    \includegraphics[width=\columnwidth]{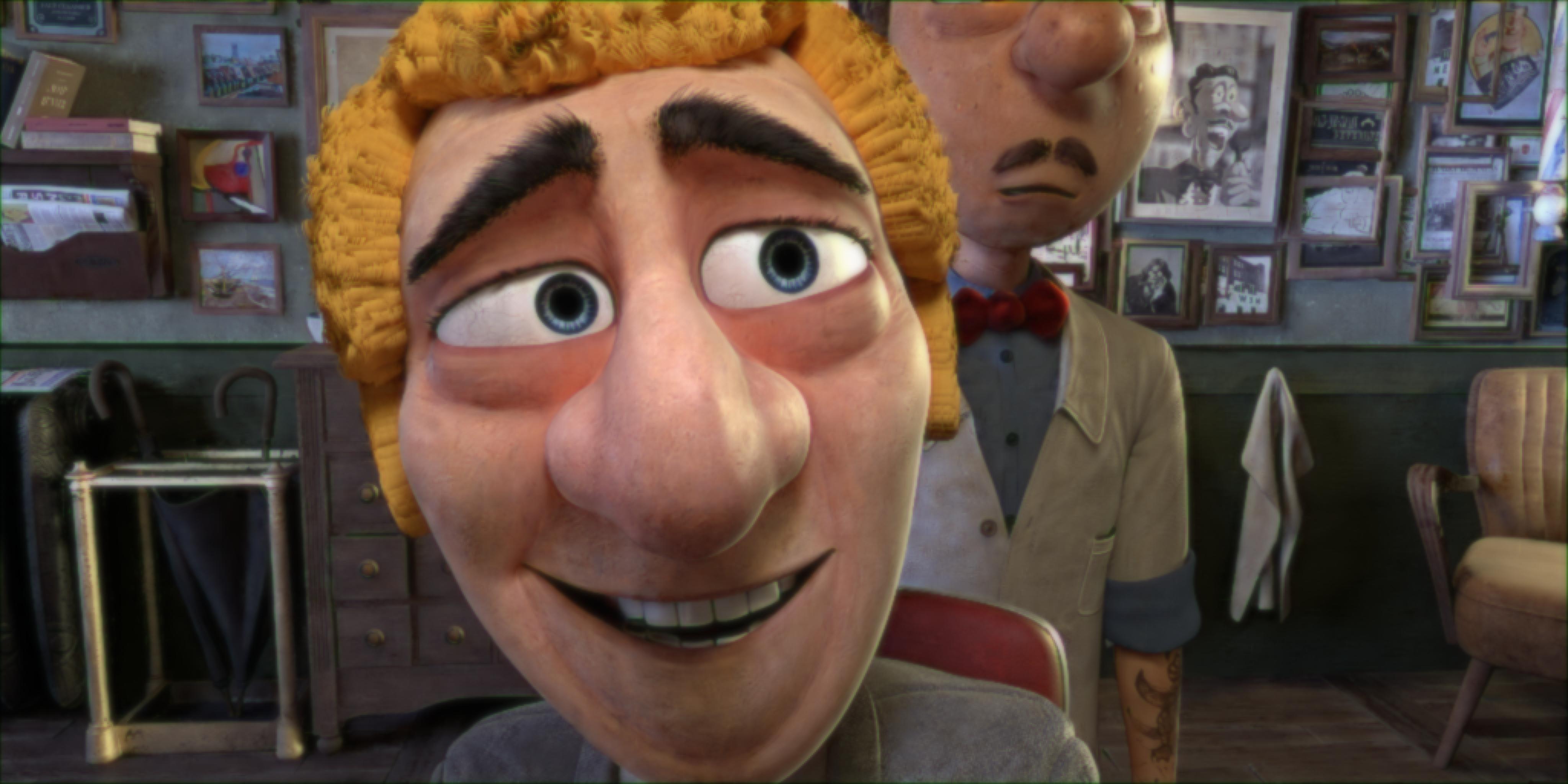}
    \caption{35.40 / 0.95}
    \label{sfig:Hnoise30nm}
\end{subfigure}
\begin{subfigure}{0.24\columnwidth}
    \caption{$\sigma_d=\SI{30}{\nm}$}
    \includegraphics[width=\columnwidth]{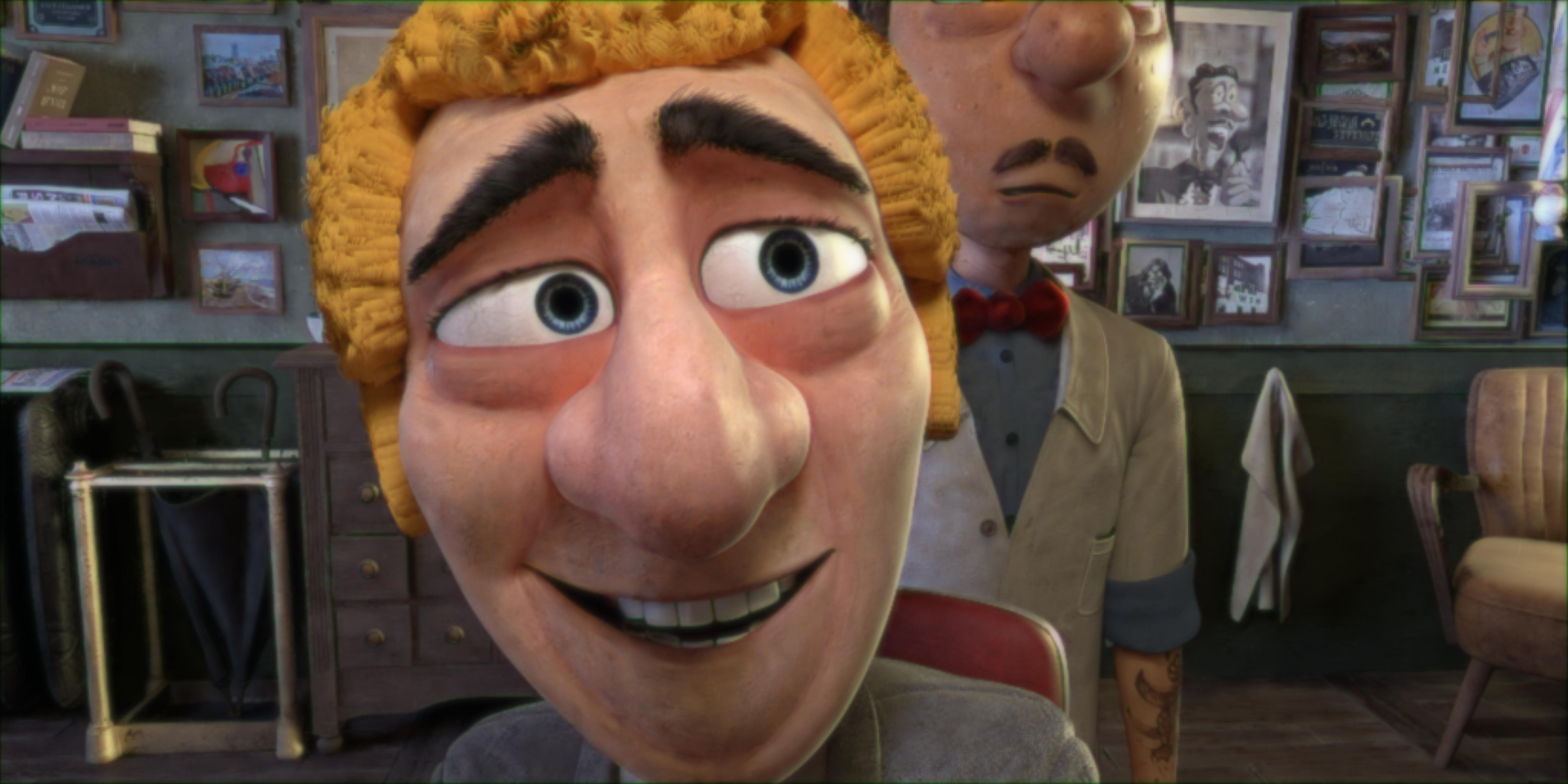}
    \caption{34.88 / 0.95}
    \label{sfig:Hnoise40nm}
\end{subfigure}
\begin{subfigure}{0.24\columnwidth}
    \caption{$\sigma_d=\SI{40}{\nm}$}
    \includegraphics[width=\columnwidth]{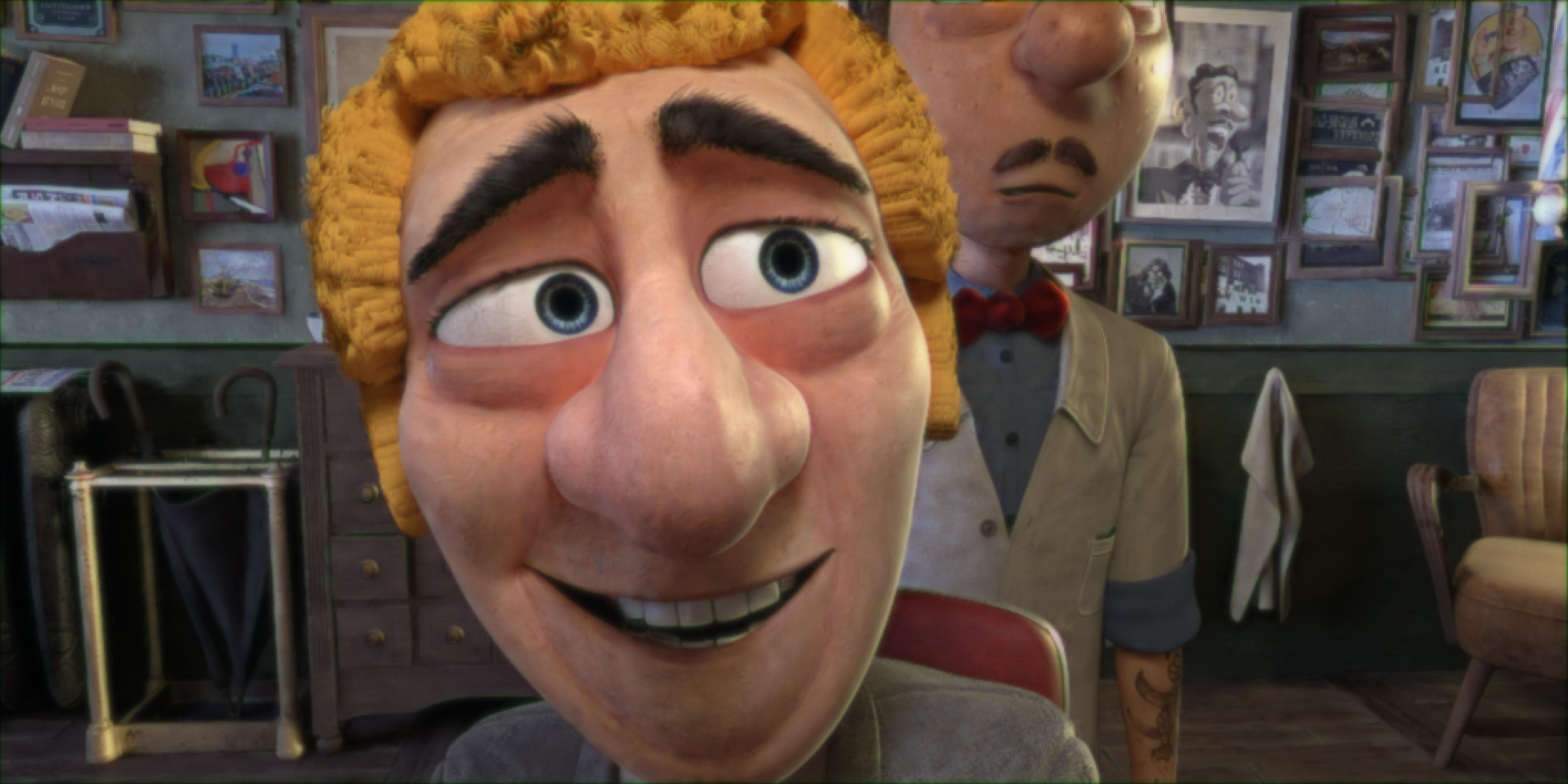}
    \caption{33.89 / 0.94}
    \label{sfig:Hnoise50nm}
\end{subfigure}
\caption{Simulation results with increasing noise in the fabricated height map.}
\label{fig:Hnoise}
\end{figure}

%% file: figures/broadband/figure_broadband.tex
\newcommand\EyeQtwentyfive{6.38} % 25 cm, our method, Eye, first image
\newcommand\EyeQthirtythree{7.10} % 33 cm, our method, Eye, first image
\newcommand\EyeQfifty{7.33} % 50 cm, our method, Eye, first image
\newcommand\EyeQonehundred{7.24} % 100 cm, our method, Eye, first image
\newcommand\EyeQinfinity{6.52} % Inf cm, our method, Eye, first image

\newcommand\TLQtwentyfive{5.35} % 25 cm, thin lens, first image
\newcommand\TLQthirtythree{6.33} % 33 cm, thin lens, first image
\newcommand\TLQfifty{7.31} % 50 cm, thin lens, first image
\newcommand\TLQonehundred{6.70} % 100 cm, thin lens, first image
\newcommand\TLQinfinity{5.56} % Inf cm, thin lens, first image

\begin{figure}[t!]
    \centering
    \begin{subfigure}[t]{0.19\columnwidth}
		\centering
		\caption{4 D}
		\includegraphics[width=\columnwidth]{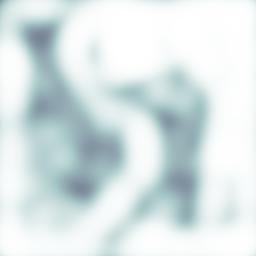}
		\put(-35,-10){\TLQtwentyfive}
		\put(-57,-6){\rotatebox{90}{Conventional}}
	\end{subfigure}
	\vspace{1.3mm}
	\begin{subfigure}[t]{0.19\columnwidth}
		\centering
		\caption{3 D}
		\includegraphics[width=\columnwidth]{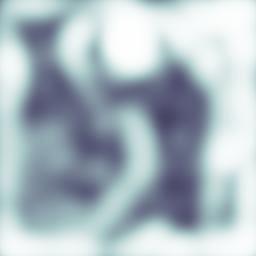}
		\put(-35,-10){\TLQthirtythree}
		%\put(-90,16){\rotatebox{90}{Conventional}}
	\end{subfigure}
	%\vspace{-8mm}
	\begin{subfigure}[t]{0.19\columnwidth}
		\centering
		\caption{2 D}
		\includegraphics[width=\columnwidth]{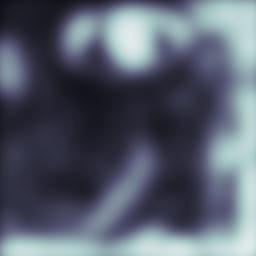}
		\put(-35,-10){\TLQfifty}
	\end{subfigure}
	\begin{subfigure}[t]{0.19\columnwidth}
		\centering
		\caption{1 D}
		\includegraphics[width=\columnwidth]{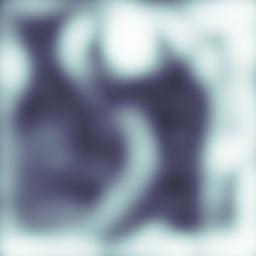}
            \put(-35,-10){\TLQonehundred}
	\end{subfigure}
	\begin{subfigure}[t]{0.19\columnwidth}
		\centering
		\caption{0 D}
		\includegraphics[width=\columnwidth]{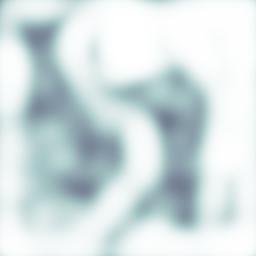}
		\put(-35,-10){\TLQinfinity}
	\end{subfigure}

    \begin{subfigure}[t]{0.19\columnwidth}
		\centering
		\includegraphics[width=\columnwidth]{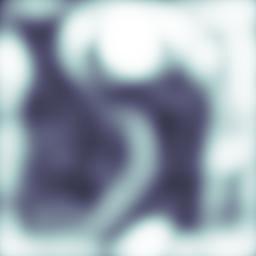}
		\put(-35,-10){\textbf{\EyeQtwentyfive}}
		\put(-57,5){\rotatebox{90}{Proposed}}
	\end{subfigure}
	\begin{subfigure}[t]{0.19\columnwidth}
		\centering
		\includegraphics[width=\columnwidth]{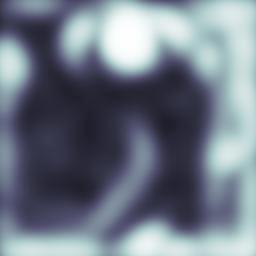}
		\put(-35,-10){\textbf{\EyeQthirtythree}}
		%\put(-90,16){\rotatebox{90}{Conventional}}
	\end{subfigure}
	%\vspace{-8mm}
	\begin{subfigure}[t]{0.19\columnwidth}
		\centering
		\includegraphics[width=\columnwidth]{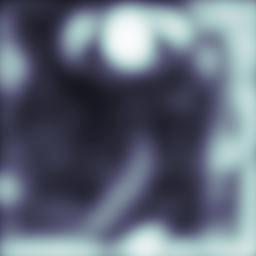}
		\put(-35,-10){\textbf{\EyeQfifty}}
	\end{subfigure}
	\begin{subfigure}[t]{0.19\columnwidth}
		\centering
		\includegraphics[width=\columnwidth]{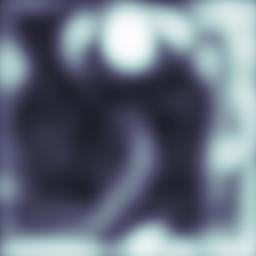}
            \put(-35,-10){\textbf{\EyeQonehundred}}
	\end{subfigure}
	\begin{subfigure}[t]{0.19\columnwidth}
		\centering
		\includegraphics[width=\columnwidth]{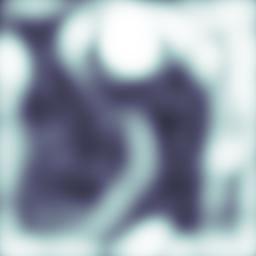}
		\put(-35,-10){\textbf{\EyeQinfinity}}
	\end{subfigure}

	\caption{Predicted visibility maps between the ground truth and the perceived images. Top: conventional stereoscopic display, bottom: proposed continuous-mode AI display. The visibility maps are constructed using HDR-VDP2 \cite{HDRVDP2}. The intensities are scaled between 0 and 1, where brighter intensity means a higher probability that the viewer perceives the artifacts. The resulting quality scores are given under each image.}
	\label{fig:HDRVDP}
\end{figure}

%% file: figures/HeightMapMeasured.tex
\begin{tikzpicture}

\begin{axis}[%
width=0.75\columnwidth,
height=0.75\columnwidth,
at={(0\columnwidth,0\columnwidth)},
scale only axis,
axis on top,
xmin=-5.01,
xmax=5.01,
xtick={-5,  5},
xlabel style={font=\color{white!15!black},yshift=5pt},
xlabel={$s (mm)$},
y dir=reverse,
ymin=-5.01,
ymax=5.01,
ytick={-5,  5},
ylabel style={font=\color{white!15!black},yshift=-14pt},
ylabel={$t (mm)$},
axis background/.style={fill=white},
title style={font=\bfseries},
title={Height map},
% width=0.951\columnwidth,
% height=0.781\columnwidth,
% at={(0\columnwidth,0\columnwidth)},
% scale only axis,
% axis on top,
% xmin=-6.10625,
% xmax=6.10625,
% xlabel style={font=\color{white!15!black}},
% xlabel={s (mm)},
% y dir=reverse,
% ymin=-6.1375,
% ymax=6.1375,
% ylabel style={font=\color{white!15!black}},
% ylabel={t (mm)},
% axis background/.style={fill=white},
% title style={font=\bfseries},
% title={Height Map}
]
\addplot [forget plot] graphics [xmin=-6.10625, xmax=6.10625, ymin=-6.1375, ymax=6.1375] {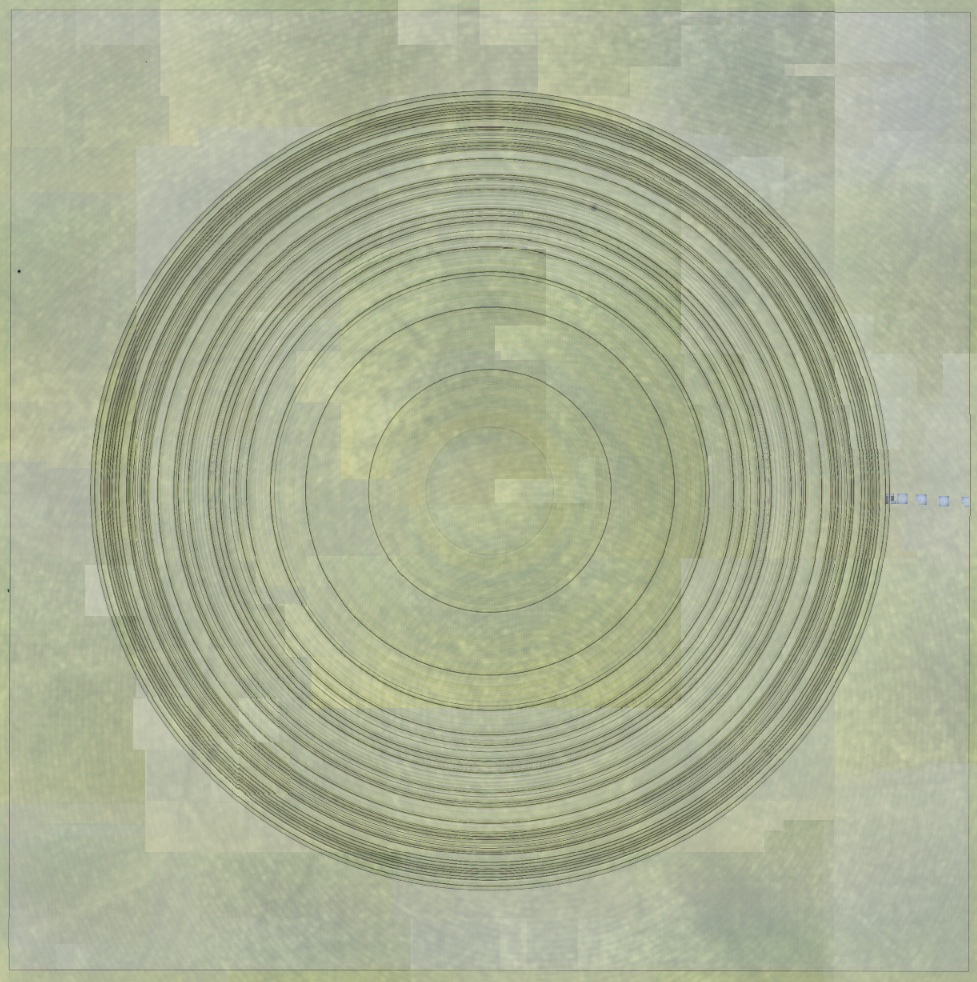};
\end{axis}

% \begin{axis}[%
% width=1.227\columnwidth,
% height=0.958\columnwidth,
% at={(-0.16\columnwidth,-0.105\columnwidth)},
% scale only axis,
% xmin=0,
% xmax=1,
% ymin=0,
% ymax=1,
% axis line style={draw=none},
% ticks=none,
% axis x line*=bottom,
% axis y line*=left
% ]
% \end{axis}

\end{tikzpicture}%

%% file: figures/MTF/Experiments_Slanted_edge/MTF_Lens_SeeYa_v2.tex
\definecolor{mycolor1}{rgb}{0.00000,0.44700,0.74100}%
\definecolor{mycolor2}{rgb}{0.85000,0.32500,0.09800}%
\definecolor{mycolor3}{rgb}{0.92900,0.69400,0.12500}%
\definecolor{mycolor4}{rgb}{0.49400,0.18400,0.55600}%
\definecolor{mycolor5}{rgb}{0.46600,0.67400,0.18800}%
\definecolor{mycolor6}{rgb}{0.30100,0.74500,0.93300}%
\begin{tikzpicture}

\begin{axis}[%
width=\columnwidth,
height=0.75\columnwidth,
at={(0\columnwidth,0\columnwidth)},
scale only axis,
xmin=0,
xmax=16,
xlabel style={font=\color{white!15!black}},
xlabel={Spatial Frequency, cpd},
ymin=0,
ymax=1,
axis background/.style={fill=white},
title style={font=\bfseries},
title={Conventional},
legend style={legend cell align=left, align=left, draw=white!15!black}
]
\addplot [color=mycolor1, line width=1.4pt]
  table[row sep=crcr]{%
0.0111111111111111	1\\
0.797433971645093	0.80027011330471\\
1.59467694476902	0.370453343223935\\
2.3919623559836	0.0601966919848067\\
3.18925837883893	0.202426720667371\\
3.98655864655937	0.235823920252749\\
4.7838610367553	0.178634754258005\\
5.58116463980699	0.0612494866383429\\
6.37846900089813	0.0515380419079528\\
7.17577386735086	0.0837669251491864\\
7.97307908755764	0.0460581177298591\\
8.77038456504057	0.0271490800972479\\
9.56769023548087	0.053126567584563\\
10.3649960543501	0.0458151247120778\\
11.1623019898421	0.0244447228108255\\
11.9596080186324	0.0303266074906556\\
12.7569141232276	0.0479979755087925\\
13.5542202902504	0.0457602476307059\\
14.3515265092963	0.0260247516736422\\
15.148832772151	0.0127200251981494\\
15.9461390722432	0.0257633691017819\\
16.7434454042534	0.0319803985203399\\
17.540751763829	0.0298490911440971\\
18.3380581473746	0.0289868210894167\\
19.1353645518939	0.0237634960998245\\
19.9326709748701	0.0118354395113545\\
20.7299774141736	0.00889044349236959\\
21.5272838679902	0.0275176031796608\\
22.3245903347649	0.0368138375809153\\
23.1218968131572	0.033540909635788\\
23.9192033020054	0.0216420921905563\\
24.7165098002977	0.00815801138298374\\
25.5138163071486	0.00923256799311044\\
26.31112282178	0.0131396134314145\\
27.1084293435055	0.0122814303335639\\
27.9057358717171	0.00917189424935221\\
28.7030424058741	0.0320394800833951\\
29.5003489454945	0.0428737748901516\\
30.2976554901471	0.0340465004104883\\
31.0949620394447	0.00898892000262056\\
31.8922685930389	0.0164413662465905\\
32.6895751506154	0.0175411635553608\\
33.4868817118897	0.0223994479086987\\
34.2841882766039	0.040490124551927\\
35.0814948445233	0.0453168931894095\\
35.8788014154343	0.0332670240115227\\
36.6761079891418	0.0108205692529795\\
37.4734145654674	0.0109304109992104\\
38.2707211442473	0.0192117512505433\\
39.0680277253313	0.0189518702226372\\
39.8653343085812	0.0118920012574472\\
40.6626408938695	0.00761593160827699\\
41.4599474810787	0.0106669153776275\\
42.2572540701	0.0145282419485115\\
43.0545606608328	0.0147011000002715\\
43.8518672531836	0.0149094563051081\\
44.6491738470659	0.0166347169630839\\
45.446480442399	0.00719982347174829\\
46.2437870391079	0.00818362577340934\\
47.0410936371226	0.018119120183313\\
47.8384002363778	0.0169984566955783\\
48.6357068368125	0.0204713621107948\\
49.4330134383697	0.0293760468906263\\
50.2303200409959	0.0256605819447504\\
51.027626644641	0.0100385088748635\\
51.8249332492579	0.0110780059712831\\
52.6222398548025	0.0247162233647432\\
53.4195464612333	0.0248723136681531\\
54.2168530685111	0.0220238707043692\\
55.0141596765991	0.0309021985253209\\
55.8114662854627	0.0249652208562806\\
56.6087728950689	0.0123674799235456\\
57.406079505387	0.0223110506127586\\
58.2033861163876	0.0335614543203566\\
59.000692728043	0.0426320413629273\\
59.7979993403272	0.0388005688483858\\
60.5953059532152	0.0330039868447135\\
61.3926125666834	0.0405138464724423\\
62.1899191807098	0.0402433560945557\\
62.9872257952729	0.0347715894120822\\
63.7845324103527	0.0311714976222055\\
64.5818390259301	0.0264676683723125\\
65.3791456419868	0.0198898276384405\\
66.1764522585055	0.002060090338374\\
66.9737588754697	0.0282045733879484\\
67.7710654928638	0.0466051989384336\\
68.5683721106726	0.0414351187310374\\
69.365678728882	0.0269364227653109\\
70.1629853474782	0.0314937940289873\\
70.9602919664482	0.0394660694566868\\
71.7575985857795	0.0432869789258853\\
72.5549052054603	0.0604192993576338\\
73.3522118254791	0.058080010606668\\
74.149518445825	0.0291167923345819\\
74.9468250664877	0.0166564288359532\\
75.744131687457	0.0532551536064313\\
76.5414383087235	0.064263290344756\\
77.3387449302778	0.055802769204407\\
78.1360515521113	0.0278752958659992\\
78.9333581742154	0.0207613362521909\\
79.7306647965821	0.0340892959327016\\
};
%\addlegendentry{1D}

\addplot [color=mycolor2, line width=1.4pt]
  table[row sep=crcr]{%
0.0111111111111111	1\\
0.820091003322989	0.938064248286917\\
1.63999628492881	0.788837725469854\\
2.45994283248792	0.59671854952458\\
3.27989969848269	0.398931711154998\\
4.09986069204369	0.232826054782586\\
4.91982374942729	0.129406349906247\\
5.73978798614968	0.107939810208891\\
6.55975295996306	0.11818751701651\\
7.37971842517222	0.118972112785302\\
8.19968423435929	0.107483716618909\\
9.01965029371255	0.0910450506721011\\
9.8396165406907	0.0743075022672025\\
10.6595829319958	0.0535508695015485\\
11.4795494367008	0.0338296890197941\\
12.2995160321257	0.0244497663057848\\
13.1194827012606	0.0296769847113068\\
13.9394494310979	0.0376474313434766\\
14.7594162115204	0.0414766528778248\\
15.5793830345412	0.0457126874674781\\
16.3993498937704	0.0399222574259468\\
17.2193167840354	0.0337658373075198\\
18.0392837011041	0.0345742857721643\\
18.8592506414804	0.0384142866011574\\
19.6792176022508	0.0454996322584118\\
20.4991845809679	0.0464990711614337\\
21.3191515755611	0.0406939751783825\\
22.1391185842664	0.0323421266870076\\
22.9590856055716	0.0208285965570253\\
23.7790526381735	0.0178083382038008\\
24.5990196809422	0.027193088940049\\
25.418986732894	0.0343979295715895\\
26.2389537931679	0.0390670033575134\\
27.0589208610073	0.0431850692968485\\
27.8788879357448	0.0460263866834268\\
28.698855016789	0.0412726817164729\\
29.5188221036144	0.0303517896375359\\
30.3387891957523	0.0229396334530581\\
31.1587562927832	0.0181148698357255\\
31.9787233943307	0.01740965694116\\
32.7986905000562	0.021656904128474\\
33.6186576096538	0.0246600566623587\\
34.4386247228471	0.02557286687703\\
35.2585918393851	0.024583645535085\\
36.0785589590399	0.0162800098826084\\
36.8985260816036	0.00335306839297122\\
37.7184932068865	0.00757537938789391\\
38.538460334715	0.0150212593503903\\
39.3584274649301	0.0155425319547742\\
40.1783945973856	0.0125229609973694\\
40.9983617319472	0.011745926919887\\
41.8183288684908	0.00818104945355966\\
42.6382960069022	0.00466638022601974\\
43.4582631470757	0.0070312690397354\\
44.2782302889133	0.0103920811155883\\
45.0981974323243	0.0195064489036871\\
45.9181645772244	0.0225936562672634\\
46.7381317235353	0.02082756937704\\
47.5580988711838	0.0184244065510478\\
48.3780660201022	0.0136900582347963\\
49.1980331702267	0.00892790186452493\\
50.0180003214982	0.010908031084654\\
50.8379674738611	0.0168634421674001\\
51.6579346272634	0.0249851030074025\\
52.4779017816565	0.0372399198981831\\
53.2978689369946	0.0489683211008204\\
54.1178360932348	0.0469234892038962\\
54.9378032503366	0.0416378996848671\\
55.757770408262	0.036158651899189\\
56.5777375669753	0.0304981053214312\\
57.3977047264426	0.0230008811890907\\
58.2176718866321	0.0196833879055996\\
59.0376390475138	0.0240355128656138\\
59.8576062090591	0.0242569891946484\\
60.6775733712412	0.0260930359051501\\
61.4975405340346	0.0356499178832729\\
62.3175076974152	0.039640075617608\\
63.13747486136	0.0323975940677135\\
63.9574420258475	0.02302157140662\\
64.7774091908569	0.0115412081093602\\
65.5973763563687	0.00990195504997484\\
66.4173435223643	0.0133078957676848\\
67.237310688826	0.0212488072626945\\
68.057277855737	0.0257389821431179\\
68.8772450230812	0.0351255000826795\\
69.6972121908433	0.0437846433868156\\
70.5171793590087	0.0488081477733545\\
71.3371465275636	0.0499402439283791\\
72.1571136964946	0.0401279988536352\\
72.9770808657891	0.0240325774505327\\
73.7970480354349	0.0211863841630063\\
74.6170152054206	0.0211675734314764\\
75.4369823757349	0.0189919366667796\\
76.2569495463673	0.0148788703539464\\
77.0769167173077	0.00781742481360636\\
77.8968838885463	0.00638129890187112\\
78.7168510600737	0.00591228279196333\\
79.5368182318812	0.0183981377050479\\
80.35678540396	0.0260324996530346\\
81.176752576302	0.0283930896061464\\
81.9967197488993	0.0305343766333694\\
};
%\addlegendentry{2D}

\addplot [color=mycolor3, line width=1.4pt]
  table[row sep=crcr]{%
0.0111111111111111	1\\
0.843706447506676	0.964921863977619\\
1.68723237166915	0.885504374246029\\
2.53079840704321	0.783764242127832\\
3.3743744720085	0.671771487797869\\
4.21795454898659	0.563022454341438\\
5.06153663200733	0.46592257979988\\
5.90511986134881	0.383416157910288\\
6.74870380714463	0.318138407598513\\
7.59228823057834	0.265283628745849\\
8.43587298835936	0.223013831074447\\
9.27945798930246	0.191050870132828\\
10.1230431726173	0.16414630243186\\
10.9666284962183	0.139192365615102\\
11.8102139300442	0.119851637712152\\
12.65379945205	0.103514827387771\\
13.4973850457021	0.0911243820437552\\
14.340970698357	0.0823323594964861\\
15.184556400181	0.0723335990620852\\
16.0281421434104	0.0653451424789985\\
16.8717279218345	0.0591376849583224\\
17.7153137304255	0.0519210393906062\\
18.5588995650698	0.045945747897552\\
19.402485422369	0.0415665540096317\\
20.2460712994913	0.0364125447710629\\
21.0896571940579	0.0320054550290797\\
21.9332431040561	0.0288922633976037\\
22.7768290277712	0.0285427427457722\\
23.6204149637335	0.030004335326718\\
24.4640009106762	0.0256305348801569\\
25.3075868675011	0.0230163386527359\\
26.1511728332519	0.0256884241456364\\
26.9947588070918	0.0243732280135864\\
27.8383447882855	0.0170739536803077\\
28.681930776184	0.0115148998832147\\
29.5255167702128	0.00675420546253539\\
30.3691027698608	0.00670106321865193\\
31.2126887746726	0.00788975924976217\\
32.0562747842404	0.0101465299933336\\
32.8998607981984	0.0175880106031282\\
33.7434468162173	0.0242280021115697\\
34.587032838	0.0252738262694012\\
35.4306188632777	0.0211756240882232\\
36.2742048918065	0.0177552413332256\\
37.1177909233647	0.0197265324783107\\
37.9613769577504	0.0197390176718639\\
38.8049629947792	0.0163363976179983\\
39.6485490342824	0.0178121758238796\\
40.4921350761053	0.0214966917059231\\
41.3357211201059	0.0261507188401205\\
42.1793071661535	0.0250408279111316\\
43.0228932141278	0.021528010730567\\
43.8664792639175	0.0183181010009009\\
44.71006531542	0.0193815099259084\\
45.55365136854	0.0202217789789226\\
46.3972374231894	0.020468654610022\\
47.2408234792861	0.0236125671643021\\
48.0844095367541	0.0191759121557487\\
48.9279955955224	0.0138197108089414\\
49.7715816555249	0.0148554708195308\\
50.6151677166998	0.0164027442019876\\
51.4587537789896	0.0182962636822364\\
52.3023398423402	0.0170266501669862\\
53.1459259067011	0.0147201965839859\\
53.9895119720251	0.0156169753771706\\
54.8330980382676	0.0152871483950287\\
55.6766841053869	0.0120454358350131\\
56.5202701733437	0.010092316450669\\
57.3638562421011	0.0100619299419101\\
58.2074423116243	0.00993653356195495\\
59.0510283818804	0.0104907709273259\\
59.8946144528385	0.0173225988388693\\
60.7382005244694	0.0228670258619136\\
61.5817865967453	0.021444538312315\\
62.4253726696402	0.0203091739118615\\
63.2689587431293	0.0242249938541891\\
64.1125448171891	0.0216826718111461\\
64.9561308917974	0.0235706748322376\\
65.7997169669331	0.0279984745416519\\
66.6433030425761	0.0251786322652924\\
67.4868891187075	0.0197635290186005\\
68.3304751953091	0.0204505972938413\\
69.1740612723638	0.0225831903883822\\
70.0176473498552	0.019169290771985\\
70.8612334277676	0.0185070298718184\\
71.7048195060863	0.0197391923729664\\
72.548405584797	0.015322061707131\\
73.3919916638862	0.0105263473718435\\
74.2355777433411	0.0115290975493053\\
75.0791638231492	0.0159139491078508\\
75.9227499032989	0.0151149143208126\\
76.7663359837788	0.0169445312236732\\
77.6099220645782	0.0184750347709696\\
78.4535081456868	0.0190891482674402\\
79.2970942270948	0.0180023593023704\\
80.1406803087926	0.0174373335491307\\
80.9842663907712	0.0132434359123384\\
81.8278524730219	0.012882757524463\\
82.6714385555365	0.0150460447443208\\
83.5150246383068	0.0216086016833003\\
84.3586107213253	0.0297777542917937\\
};
%\addlegendentry{3D}

\addplot [color=mycolor4, line width=1.4pt]
  table[row sep=crcr]{%
0.0111111111111111	1\\
0.868805571321384	0.901822075977867\\
1.73743583447889	0.661691585401991\\
2.60610505034732	0.369969901279642\\
3.47478400603126	0.106991984257731\\
4.34346685780287	0.078559293837858\\
5.2121516576515	0.175875909088628\\
6.08083757069676	0.205173797653002\\
6.94952417949347	0.194385617130667\\
7.81821125212601	0.165314373287665\\
8.68689864944433	0.129091168630505\\
9.55558628289814	0.0947869483350452\\
10.4242740934538	0.0708645358406004\\
11.2929620402417	0.0602041383125255\\
12.1616500940693	0.0574670210531237\\
13.0303382335288	0.0531902229889024\\
13.8990264425642	0.050906726653226\\
14.7677147088974	0.0475081559848162\\
15.6364030229787	0.0474288806744143\\
16.5050913772691	0.0469756697730321\\
17.3737797657372	0.0427279256635316\\
18.2424681835004	0.0381264853265397\\
19.1111566265639	0.0341165673591665\\
19.9798450916278	0.0299042680936602\\
20.848533575942	0.0259531689931695\\
21.7172220771964	0.0279929463826068\\
22.5859105934364	0.0351466210868246\\
23.4545991229969	0.0432980863176349\\
24.3232876644508	0.046698197488128\\
25.1919762165677	0.0426508890026398\\
26.0606647782813	0.0331364200595419\\
26.9293533486628	0.0204554223781524\\
27.7980419268997	0.00683050770437904\\
28.6667305122779	0.00947250533508052\\
29.5354191041671	0.0173304418074959\\
30.4041077020095	0.0240072588804708\\
31.2727963053087	0.0256167192730184\\
32.1414849136225	0.0221213622830524\\
33.0101735265548	0.014309752374317\\
33.8788621437505	0.00305114754171713\\
34.7475507648898	0.00886650666173351\\
35.6162393896841	0.0144791009082175\\
36.4849280178723	0.0192891667993476\\
37.3536166492177	0.0194184807464337\\
38.2223052835051	0.0188135561732614\\
39.0909939205382	0.0184348551087201\\
39.959682560138	0.0203059533686252\\
40.8283712021407	0.0199097937649472\\
41.697059846396	0.021049772534769\\
42.5657484927662	0.0232203192261607\\
43.4344371411242	0.0176980765748912\\
44.3031257913532	0.013873462932966\\
45.1718144433452	0.0168367542867527\\
46.0405030970005	0.0170401105865871\\
46.9091917522265	0.0173595464826574\\
47.7778804089377	0.020364270117295\\
48.6465690670544	0.0211976949938739\\
49.5152577265028	0.0183516383030586\\
50.3839463872139	0.0104945358925099\\
51.2526350491235	0.00261388739375519\\
52.1213237121717	0.0074833467756596\\
52.9900123763025	0.0178226694536972\\
53.8587010414635	0.0272254658941479\\
54.7273897076056	0.0281843030905878\\
55.5960783746829	0.0257720128019651\\
56.4647670426523	0.0231736095015935\\
57.333455711473	0.0185058704780885\\
58.2021443811071	0.0222137614216642\\
59.0708330515187	0.0342437901142494\\
59.9395217226739	0.0441936384101251\\
60.8082103945409	0.041025934295844\\
61.6768990670895	0.0375658382965081\\
62.5455877402915	0.0385494176754521\\
63.4142764141199	0.0453711780828281\\
64.2829650885494	0.0437330070935769\\
65.1516537635559	0.0346994760512979\\
66.0203424391166	0.0254629559056316\\
66.88903111521	0.0304236844153555\\
67.7577197918155	0.0324570242106852\\
68.6264084689137	0.0235856621225998\\
69.4950971464861	0.0266926617243378\\
70.3637858245152	0.0309309136608847\\
71.2324745029843	0.0299318580550476\\
72.1011631818774	0.025778914110751\\
72.9698518611794	0.0358301820001221\\
73.838540540876	0.0447591237979628\\
74.7072292209532	0.042889453556621\\
75.575917901398	0.0387104425596405\\
76.4446065821979	0.0360598952200517\\
77.3132952633408	0.0320877189817469\\
78.1819839448155	0.0185378937269291\\
79.0506726266108	0.0185568150157985\\
79.9193613087164	0.0215586190742791\\
80.7880499911223	0.0228301357999485\\
81.6567386738188	0.0200167375312795\\
82.5254273567968	0.00668762450967945\\
83.3941160400475	0.00918975571193112\\
84.2628047235625	0.0224596451861415\\
85.1314934073336	0.02994976288886\\
86.0001820913531	0.0294598244018244\\
86.8688707756136	0.0248867124338388\\
};
%\addlegendentry{4D}

\addplot [color=mycolor5, line width=1.4pt]
  table[row sep=crcr]{%
0.0111111111111111	1\\
0.895266762058885	0.802578933486425\\
1.79036339749632	0.374580744708092\\
2.68549783459185	0.0600883324196322\\
3.58064172359871	0.206041261235884\\
4.47578939351764	0.238986073686411\\
5.37093895392297	0.185163127642796\\
6.26608959461514	0.0732335154694334\\
7.16124091048985	0.0422769903622691\\
8.05639267648763	0.0860085119712202\\
8.95154475757221	0.0569232443118209\\
9.84669706781118	0.0196154179370471\\
10.7418495499161	0.0500595007962451\\
11.6370021642258	0.0518732596020919\\
12.5321548824106	0.0341030930741901\\
13.4273076836956	0.027648770784876\\
14.3224605524996	0.037478187962353\\
15.2176134769074	0.037430040085291\\
16.1127664476517	0.0278401578270518\\
17.0079194574164	0.0161956542131801\\
17.9030725003482	0.0222350768010437\\
18.7982255717091	0.0156547515216211\\
19.6933786676224	0.00821871490575887\\
20.5885317848856	0.0213396051166646\\
21.4836849208299	0.0278781089408253\\
22.3788380732137	0.0225233261789851\\
23.27399124014	0.00676073357954265\\
24.169144419993	0.0113376001702255\\
25.0642976113878	0.0160787453280338\\
25.9594508131303	0.0146217291316177\\
26.8546040241858	0.0124937833559685\\
27.749757243653	0.00845539707843616\\
28.6449104707433	0.0127923936720762\\
29.5400637047637	0.0201009478252509\\
30.4352169451028	0.0193236687500883\\
31.3303701912189	0.00721316728267767\\
32.2255234426306	0.0151952097932188\\
33.1206766989086	0.0312389100822291\\
34.0158299596686	0.0246694770622519\\
34.9109832245659	0.00397227325161597\\
35.8061364932902	0.0261372048643681\\
36.7012897655615	0.0289329500058042\\
37.5964430411264	0.016408546889245\\
38.4915963197551	0.0206625785802294\\
39.3867496012387	0.0321790230258757\\
40.281902885387	0.0326412694784617\\
41.1770561720261	0.0297284471149067\\
42.072209460997	0.0316859119524625\\
42.967362752154	0.0284718315972613\\
43.8625160453632	0.0150467873059953\\
44.7576693405016	0.00132994558413991\\
45.6528226374556	0.00488336285033329\\
46.5479759361205	0.00591141401373059\\
47.4431292363994	0.0166976307250752\\
48.3382825382028	0.0260102737938169\\
49.2334358414473	0.023017978109996\\
50.1285891460559	0.02157509383909\\
51.0237424519567	0.0314129169518489\\
51.9188957590829	0.0365498040244395\\
52.8140490673722	0.0347748811950051\\
53.7092023767664	0.0232531957787398\\
54.6043556872112	0.0159687109597861\\
55.4995089986558	0.00828603447941482\\
56.3946623110525	0.00814284240679547\\
57.2898156243567	0.0134312538702877\\
58.1849689385266	0.0152114857031543\\
59.0801222535227	0.00695102067531523\\
59.9752755693081	0.00583823461308558\\
60.870428885848	0.0127825298649928\\
61.7655822031095	0.0121049547942514\\
62.6607355210618	0.00764380222464245\\
63.5558888396756	0.0157463457928315\\
64.4510421589234	0.0246583153800414\\
65.3461954787792	0.0289288114389543\\
66.2413487992182	0.0149828133340667\\
67.1365021202172	0.00959571036955805\\
68.031655441754	0.0197557973090561\\
68.9268087638077	0.0210523306528711\\
69.8219620863584	0.0166331766533564\\
70.7171154093873	0.0208950854251197\\
71.6122687328763	0.0270395339315374\\
72.5074220568086	0.0216874908933528\\
73.4025753811678	0.0134114579930344\\
74.2977287059385	0.0188612366942013\\
75.192882031106	0.031419467916961\\
76.0880353566563	0.0454481134968218\\
76.9831886825761	0.0463082598554602\\
77.8783420088526	0.0355946622301741\\
78.7734953354737	0.0300519480342922\\
79.6686486624277	0.0310723215087917\\
80.5638019897036	0.0495968766410223\\
81.4589553172907	0.0615454434492573\\
82.3541086451789	0.0611600920903977\\
83.2492619733585	0.0492158170298069\\
84.1444153018201	0.0400328285814865\\
85.0395686305549	0.0351127470439745\\
85.9347219595544	0.0390483803198512\\
86.8298752888103	0.029289544043297\\
87.7250286183148	0.0216431412668283\\
88.6201819480604	0.025432373937692\\
89.5153352780398	0.0415469301636337\\
};
%\addlegendentry{5D}

\addplot [color=gray, line width=1.4pt, forget plot]
  table[row sep=crcr]{%
0.0111111111111111	0.1\\
0.895266762058885	0.1\\
1.79036339749632	0.1\\
2.68549783459185	0.1\\
3.58064172359871	0.1\\
4.47578939351764	0.1\\
5.37093895392297	0.1\\
6.26608959461514	0.1\\
7.16124091048985	0.1\\
8.05639267648763	0.1\\
8.95154475757221	0.1\\
9.84669706781118	0.1\\
10.7418495499161	0.1\\
11.6370021642258	0.1\\
12.5321548824106	0.1\\
13.4273076836956	0.1\\
14.3224605524996	0.1\\
15.2176134769074	0.1\\
16.1127664476517	0.1\\
17.0079194574164	0.1\\
17.9030725003482	0.1\\
18.7982255717091	0.1\\
19.6933786676224	0.1\\
20.5885317848856	0.1\\
21.4836849208299	0.1\\
22.3788380732137	0.1\\
23.27399124014	0.1\\
24.169144419993	0.1\\
25.0642976113878	0.1\\
25.9594508131303	0.1\\
26.8546040241858	0.1\\
27.749757243653	0.1\\
28.6449104707433	0.1\\
29.5400637047637	0.1\\
30.4352169451028	0.1\\
31.3303701912189	0.1\\
32.2255234426306	0.1\\
33.1206766989086	0.1\\
34.0158299596686	0.1\\
34.9109832245659	0.1\\
35.8061364932902	0.1\\
36.7012897655615	0.1\\
37.5964430411264	0.1\\
38.4915963197551	0.1\\
39.3867496012387	0.1\\
40.281902885387	0.1\\
41.1770561720261	0.1\\
42.072209460997	0.1\\
42.967362752154	0.1\\
43.8625160453632	0.1\\
44.7576693405016	0.1\\
45.6528226374556	0.1\\
46.5479759361205	0.1\\
47.4431292363994	0.1\\
48.3382825382028	0.1\\
49.2334358414473	0.1\\
50.1285891460559	0.1\\
51.0237424519567	0.1\\
51.9188957590829	0.1\\
52.8140490673722	0.1\\
53.7092023767664	0.1\\
54.6043556872112	0.1\\
55.4995089986558	0.1\\
56.3946623110525	0.1\\
57.2898156243567	0.1\\
58.1849689385266	0.1\\
59.0801222535227	0.1\\
59.9752755693081	0.1\\
60.870428885848	0.1\\
61.7655822031095	0.1\\
62.6607355210618	0.1\\
63.5558888396756	0.1\\
64.4510421589234	0.1\\
65.3461954787792	0.1\\
66.2413487992182	0.1\\
67.1365021202172	0.1\\
68.031655441754	0.1\\
68.9268087638077	0.1\\
69.8219620863584	0.1\\
70.7171154093873	0.1\\
71.6122687328763	0.1\\
72.5074220568086	0.1\\
73.4025753811678	0.1\\
74.2977287059385	0.1\\
75.192882031106	0.1\\
76.0880353566563	0.1\\
76.9831886825761	0.1\\
77.8783420088526	0.1\\
78.7734953354737	0.1\\
79.6686486624277	0.1\\
80.5638019897036	0.1\\
81.4589553172907	0.1\\
82.3541086451789	0.1\\
83.2492619733585	0.1\\
84.1444153018201	0.1\\
85.0395686305549	0.1\\
85.9347219595544	0.1\\
86.8298752888103	0.1\\
87.7250286183148	0.1\\
88.6201819480604	0.1\\
89.5153352780398	0.1\\
};
\end{axis}
\end{tikzpicture}%

%% file: figures/MTF/Experiments_Slanted_edge/MTF_DOE_PCNN_SeeYa_v2.tex
\definecolor{mycolor1}{rgb}{0.00000,0.44700,0.74100}%
\definecolor{mycolor2}{rgb}{0.85000,0.32500,0.09800}%
\definecolor{mycolor3}{rgb}{0.92900,0.69400,0.12500}%
\definecolor{mycolor4}{rgb}{0.49400,0.18400,0.55600}%
\definecolor{mycolor5}{rgb}{0.46600,0.67400,0.18800}%
\definecolor{mycolor6}{rgb}{0.30100,0.74500,0.93300}%
\begin{tikzpicture}

\begin{axis}[%
width=\columnwidth,
height=0.75\columnwidth,
at={(0\columnwidth,0\columnwidth)},
scale only axis,
xmin=0,
xmax=16,
xlabel style={font=\color{white!15!black}},
xlabel={Spatial Frequency, cpd},
ymin=0,
ymax=1,
axis background/.style={fill=white},
title style={font=\bfseries},
title={Proposed},
legend style={legend cell align=left, align=left, draw=white!15!black,
at={(1,1)},anchor=north east,
nodes={scale=0.75, transform shape}}
]
\addplot [color=mycolor1, line width=1.4pt]
  table[row sep=crcr]{%
0.0111111111111111	1\\
0.797460215831496	0.825972230926005\\
1.59472943942753	0.484483873082962\\
2.392041099718	0.248331397835377\\
3.18936337129994	0.214964598053075\\
3.98668988760726	0.220554600715621\\
4.78401852632021	0.177960493000804\\
5.58134837784899	0.111046221623565\\
6.37867898739228	0.0600892520332017\\
7.17601010228052	0.0489288497444948\\
7.97334157091115	0.0630715096012571\\
8.77067329680947	0.0710288272546742\\
9.56800521565881	0.0617176506709028\\
10.3653372829322	0.0379542498429087\\
11.1626694668245	0.00613517949653274\\
11.960001744012	0.0259526682020314\\
12.757334097002	0.038720853895718\\
13.5546665124175	0.0321594608945821\\
14.3519989798543	0.0177476910172716\\
15.1493314910985	0.0104713538789859\\
15.9466640395791	0.013145368395227\\
16.7439966199765	0.0160110606324641\\
17.5413292279385	0.020473889117151\\
18.3386618598696	0.021455839533952\\
19.1359945127738	0.0233030036211777\\
19.9333271841343	0.0203300040787355\\
20.7306598718215	0.0174695320084975\\
21.5279925740213	0.0300044898834989\\
22.3253252891789	0.0316775470363817\\
23.1226580159536	0.0173747183967545\\
23.9199907531839	0.00896183249959572\\
24.717323499858	0.0183092294812712\\
25.5146562550904	0.0198975181778616\\
26.3119890181031	0.00633535952245595\\
27.1093217882096	0.014109056222541\\
27.9066545648019	0.0220453499072177\\
28.7039873473395	0.0221831954637362\\
29.5013201353404	0.0166896059551178\\
30.2986529283733	0.0186210574071876\\
31.0959857260509	0.0098726732723047\\
31.8933185280251	0.0116464736457054\\
32.6906513339815	0.0197365304321784\\
33.4879841436355	0.00487479183646874\\
34.2853169567292	0.0134896895999827\\
35.0826497730281	0.0144753476765845\\
35.8799825923185	0.0153027399657478\\
36.6773154144054	0.0119350585815453\\
37.4746482391101	0.00750551687653656\\
38.2719810662692	0.0270626962674856\\
39.0693138957322	0.0436379931319496\\
39.8666467273611	0.0392322420469057\\
40.6639795610283	0.0213170806442114\\
41.4613123966164	0.0115071439380058\\
42.2586452340165	0.0116877094024549\\
43.055978073128	0.022494680499049\\
43.8533109138575	0.0128501673325958\\
44.6506437561185	0.0142544650773957\\
45.4479765998302	0.0269659368057186\\
46.2453094449176	0.0230308043577031\\
47.0426422913108	0.00893132536459634\\
47.8399751389445	0.01263966858002\\
48.6373079877576	0.0204267253056988\\
49.4346408376932	0.0205150693385565\\
50.2319736886977	0.0213572828414102\\
51.0293065407211	0.0221009608845864\\
51.8266393937163	0.0109375110355371\\
52.6239722476392	0.0109828417954445\\
53.4213051024482	0.0132028991460592\\
54.2186379581042	0.0237177901743124\\
55.0159708145704	0.0295775011371611\\
55.8133036718121	0.0185687631335586\\
56.6106365297964	0.00339571550250431\\
57.4079693884926	0.00726384659182551\\
58.2053022478713	0.00626092714895603\\
59.0026351079048	0.00853372601072101\\
59.7999679685669	0.00961823268544292\\
60.5973008298329	0.00122919515667312\\
61.3946336916792	0.00770988995658378\\
62.1919665540835	0.0167506010524743\\
62.9892994170246	0.0235813575058372\\
63.7866322804823	0.0199128001741821\\
64.5839651444376	0.00997186740591676\\
65.3812980088722	0.0216337428757725\\
66.1786308737689	0.0253720199642353\\
66.975963739111	0.0146286805505325\\
67.7732966048829	0.0174749961747146\\
68.5706294710696	0.022837113655592\\
69.3679623376567	0.000388468208884499\\
70.1652952046308	0.0233157783204425\\
70.9626280719786	0.0195960607502133\\
71.7599609396877	0.00153308734319477\\
72.5572938077463	0.0191589189773434\\
73.3546266761429	0.0259846724235668\\
74.1519595448666	0.0232600804016562\\
74.949292413907	0.0206480448914844\\
75.7466252832541	0.0202157718716417\\
76.5439581528982	0.010022365600443\\
77.3412910228303	0.00998054744186279\\
78.1386238930415	0.00375797838799818\\
78.9359567635234	0.0156176354011499\\
79.7332896342677	0.0294965934206809\\
};
\addlegendentry{1D}

\addplot [color=mycolor2, line width=1.4pt]
  table[row sep=crcr]{%
0.0111111111111111	1\\
0.819728747900783	0.793977167277376\\
1.63927169201005	0.450732403744705\\
2.45885592030401	0.276112906478603\\
3.27845047159406	0.204386570167225\\
4.09804915227476	0.157245420111132\\
4.91764989769034	0.118826267037141\\
5.73725182296601	0.0827155946725498\\
6.55685448565849	0.0819249855144863\\
7.37645763996398	0.0904905782989455\\
8.19606113839943	0.069195162301576\\
9.01566488711166	0.0598322779896352\\
9.8352688235317	0.0436115317156811\\
10.6548729043425	0.0121016568592409\\
11.4744770986033	0.0257167046665375\\
12.2940813836242	0.0382558441846817\\
13.1136857423876	0.0365506599375168\\
13.9332901618802	0.0299843856278192\\
14.7528946319805	0.023015509394343\\
15.5724991446978	0.0215763704575925\\
16.3921036936395	0.0227446151640742\\
17.2117082736309	0.0252066475971394\\
18.0313128804377	0.0209140597250963\\
18.8509175105624	0.00936821080076563\\
19.6705221610902	0.0112116209509575\\
20.4901268295727	0.00867403384974732\\
21.3097315139383	0.00588175449333671\\
22.1293362124222	0.0055426196055631\\
22.9489409235116	0.00124327428809275\\
23.7685456459026	0.00563485799208047\\
24.5881503784651	0.00676180594992086\\
25.4077551202146	0.0158042396612958\\
26.2273598702899	0.0208941284237003\\
27.0469646279341	0.0190424738331802\\
27.8665693924794	0.0200840571463665\\
28.6861741633342	0.0202437201933122\\
29.5057789399728	0.0243537428338656\\
30.3253837219262	0.0199207042849574\\
31.1449885087747	0.00619583363501023\\
31.9645933001419	0.00591681607856949\\
32.7841980956889	0.000799147631439061\\
33.6038028951098	0.00910725225389324\\
34.4234076981279	0.00997041462978364\\
35.2430125044922	0.0150044246989344\\
36.0626173139746	0.0143863349564257\\
36.8822221263673	0.0258006336342975\\
37.7018269414803	0.0228607410944377\\
38.5214317591402	0.0171107626958002\\
39.3410365791876	0.0135307989924955\\
40.1606414014765	0.0190777003978383\\
40.9802462258722	0.0111039628602576\\
41.7998510522511	0.00364822620771095\\
42.6194558804984	0.0192420806856999\\
43.4390607105086	0.0236776094566712\\
44.2586655421837	0.0202427004947568\\
45.0782703754329	0.014975746360543\\
45.8978752101719	0.0284318545885231\\
46.7174800463221	0.0362008452450546\\
47.5370848838108	0.0294122266927241\\
48.3566897225697	0.0203113289356986\\
49.1762945625354	0.00914515778351194\\
49.9958994036485	0.00898085829133797\\
50.8155042458535	0.0167554876288955\\
51.6351090890985	0.0135134287203079\\
52.4547139333346	0.018324499443488\\
53.2743187785162	0.0289731944982083\\
54.0939236246001	0.0238819412566208\\
54.9135284715461	0.0190130506186912\\
55.7331333193162	0.0223710092490007\\
56.5527381678744	0.0162565975864023\\
57.372343017187	0.011152604824356\\
58.1919478672221	0.0192648353163872\\
59.0115527179497	0.0268796796284856\\
59.8311575693412	0.0220760143919558\\
60.6507624213698	0.0229628309788225\\
61.4703672740099	0.0293408965149719\\
62.2899721272375	0.0287151374213448\\
63.1095769810296	0.0215408565674868\\
63.9291818353645	0.0147773330807427\\
64.7487866902216	0.013796561133622\\
65.5683915455814	0.022484600759435\\
66.3879964014252	0.0367596345610822\\
67.2076012577353	0.0306561990530938\\
68.0272061144948	0.0102191281214586\\
68.8468109716877	0.0122058383910861\\
69.6664158292988	0.0166353211965739\\
70.4860206873133	0.017753023160187\\
71.3056255457175	0.0150395455706655\\
72.1252304044979	0.0253836460866916\\
72.944835263642	0.0189301404312387\\
73.7644401231377	0.0246676778285683\\
74.5840449829732	0.033572101736865\\
75.4036498431376	0.0370508762812114\\
76.2232547036202	0.0342356285460783\\
77.0428595644109	0.0230004354996103\\
77.8624644255	0.022625765255205\\
78.6820692868781	0.0185381143273185\\
79.5016741485363	0.0155748591928246\\
80.3212790104659	0.0116969902940733\\
81.1408838726589	0.00392035224521629\\
81.9604887351072	0.00250069822611485\\
};
\addlegendentry{2D}

\addplot [color=mycolor3, line width=1.4pt]
  table[row sep=crcr]{%
0.0111111111111111	1\\
0.843685315376043	0.861650955247832\\
1.68719010288624	0.598012982361825\\
2.53073500261246	0.388585809150995\\
3.37428993218114	0.278349299983475\\
4.21784887386314	0.238667806471525\\
5.06140982163804	0.223854792643919\\
5.90497191576239	0.189070190486637\\
6.74853472635905	0.138571227446084\\
7.59209801460556	0.10496584028647\\
8.43566163720775	0.0942121741027436\\
9.27922550297811	0.0971285122646666\\
10.1227895511248	0.0943108399276838\\
10.9663537395612	0.0814209500197545\\
11.8099180382252	0.0621726704154341\\
12.6534824250713	0.0475197643587947\\
13.4970468835655	0.0422073129481629\\
14.340611401064	0.0410761928964368\\
15.1841759677328	0.0401560146311423\\
16.0277405758081	0.0327424122917661\\
16.8713052190789	0.0156556943354417\\
17.7148698925175	0.00714426177184196\\
18.5584345920098	0.00737417649340753\\
19.4019993141578	0.011036990093475\\
20.2455640561293	0.0180595420370001\\
21.0891288155456	0.0140380660182616\\
21.9326935903938	0.00979843441752363\\
22.7762583789593	0.011644034457675\\
23.6198231797723	0.011887377664914\\
24.4633879915659	0.00254403556634733\\
25.3069528132421	0.00735523633032736\\
26.1505176438443	0.0163708456061247\\
26.9940824825359	0.0288285554508889\\
27.8376473285814	0.0371547467399797\\
28.6812121813319	0.0402737224096411\\
29.5247770402128	0.0305514474378041\\
30.3683419047131	0.0189299767724105\\
31.2119067743773	0.0139946513030935\\
32.0554716487976	0.0202629054260476\\
32.8990365276083	0.0158017347356545\\
33.7426014104799	0.0194077665323217\\
34.5861662971155	0.0294202837922733\\
35.4297311872461	0.0319110068407472\\
36.2732960806279	0.0231291254832181\\
37.1168609770392	0.0144468656761042\\
37.9604258762781	0.0123336627788405\\
38.8039907781601	0.00552154652834258\\
39.6475556825166	0.0106839671392409\\
40.4911205891928	0.0225025095857175\\
41.3346854980468	0.023107237561635\\
42.1782504089479	0.0227020326189568\\
43.0218153217757	0.019503872393663\\
43.865380236419	0.0155499030136701\\
44.708945152775	0.0120498220830037\\
45.5525100707487	0.0122823011928879\\
46.3960749902517	0.0102718343231195\\
47.2396399112022	0.0078248273766675\\
48.0832048335239	0.00425866734836456\\
48.9267697571459	0.00759551602153052\\
49.7703346820022	0.00917029046061249\\
50.613899608031	0.0136831834443924\\
51.4574645351746	0.0185498486170242\\
52.3010294633791	0.0243090561140155\\
53.144594392594	0.0199617459150565\\
53.9881593227718	0.0158187256436224\\
54.8317242538683	0.0174377615352574\\
55.6752891858416	0.0159271764650563\\
56.5188541186524	0.00764073469670429\\
57.3624190522638	0.00119496879300369\\
58.205983986641	0.00286657278699164\\
59.0495489217512	0.00922495264695178\\
59.8931138575634	0.00991695552346327\\
60.7366787940484	0.00098311711413321\\
61.5802437311784	0.00568413759987873\\
62.4238086689274	0.0114563584675102\\
63.2673736072706	0.0233537874672514\\
64.1109385461846	0.0315154238086243\\
64.9545034856471	0.0343656113528686\\
65.7980684256369	0.0369894187046327\\
66.6416333661341	0.0265074125199543\\
67.4851983071197	0.0150623462267915\\
68.3287632485756	0.0162743723128607\\
69.1723281904845	0.0154833974916925\\
70.0158931328301	0.0154752616561309\\
70.8594580755968	0.0123665595590964\\
71.7030230187697	0.0114518680703737\\
72.5465879623347	0.0035674987675765\\
73.3901529062782	0.00692007478534318\\
74.2337178505873	0.0177229498275234\\
75.0772827952498	0.0198792575188535\\
75.9208477402537	0.00364543905102831\\
76.764412685588	0.019018136372669\\
77.6079776312417	0.0235735082722996\\
78.4515425772047	0.0223088846012072\\
79.295107523467	0.0216794447820142\\
80.1386724700191	0.0238782214794037\\
80.9822374168521	0.0289117763763318\\
81.8258023639572	0.0403578877488352\\
82.6693673113261	0.0247488759149551\\
83.5129322589508	0.00766859607099838\\
84.3564972068236	0.0208095506680131\\
};
\addlegendentry{3D}

\addplot [color=mycolor4, line width=1.4pt]
  table[row sep=crcr]{%
0.0111111111111111	1\\
0.868806249251798	0.87674613970168\\
1.73743719047651	0.627644307407093\\
2.60610708438176	0.402425079373053\\
3.47478671809492	0.263811853403974\\
4.3434702478927	0.206211736987394\\
5.21215572576599	0.188090656185816\\
6.08084231683504	0.181529922228017\\
6.949529603655	0.166623355508953\\
7.81821735431042	0.134576172232422\\
8.68690542965137	0.0937094004598274\\
9.55559374112762	0.0732807994663598\\
10.4242822297056	0.0716064171516039\\
11.2929708545157	0.0683510026927914\\
12.1616595863655	0.0599533227870381\\
13.030348403847	0.0457507154886169\\
13.8990372909043	0.0342426578653141\\
14.7677262352595	0.0292913572710571\\
15.6364152273628	0.0232038671002066\\
16.505104259675	0.0172407781956645\\
17.3737933261649	0.017973800870603\\
18.24248242195	0.0194226652832045\\
19.1111715430353	0.0218204867441152\\
19.979860686121	0.0184690690661883\\
20.848549848457	0.011082751906047\\
21.7172390277331	0.008253102573576\\
22.5859282219949	0.00590859109166879\\
23.4546174295772	0.00151438483661555\\
24.3233066490528	0.00662132287503811\\
25.1919958791914	0.00996245178922498\\
26.0606851189267	0.0178482154590368\\
26.92937436733	0.0213068005335761\\
27.7980636235886	0.0137970084063274\\
28.6667528869884	0.00998458030885572\\
29.5354421568994	0.010368412351213\\
30.4041314327634	0.0122123860175016\\
31.2728207140844	0.0146350066625611\\
32.1415100004198	0.00986835674714824\\
33.0101992913738	0.0137988507596379\\
33.8788885865912	0.0150681212849405\\
34.7475778857521	0.0179907662792247\\
35.6162671885681	0.018669023393964\\
36.484956494778	0.0165641219359004\\
37.3536458041451	0.0125410348662395\\
38.2223351164541	0.0272448380485332\\
39.0910244315089	0.0501740307594039\\
39.9597137491304	0.0554032648879663\\
40.8284030691547	0.0419522011514808\\
41.6970923914317	0.030073870470212\\
42.5657817158235	0.0231806357583602\\
43.4344710422032	0.0124269287106259\\
44.3031603704539	0.00253512783843761\\
45.1718497004675	0.0131920920502412\\
46.0405390321444	0.0212049538349117\\
46.9092283653921	0.0277718405040159\\
47.7779177001249	0.0280428411594759\\
48.6466070362633	0.0164262654579727\\
49.5152963737333	0.014547451308459\\
50.3839857124661	0.0136834208818206\\
51.2526750523973	0.0218466945552432\\
52.1213643934672	0.0141603579688817\\
52.9900537356196	0.0111036521237824\\
53.8587430788022	0.014280311219181\\
54.727432422966	0.0132716764217105\\
55.596121768065	0.00801170993586215\\
56.4648111140559	0.00569574607921951\\
57.3335004608983	0.0129512188679401\\
58.2021898085541	0.0115795061463184\\
59.0708791569873	0.010102574560238\\
59.9395685061641	0.0203932078809674\\
60.8082578560527	0.0222253964727477\\
61.676947206623	0.0165328145156908\\
62.5456365578466	0.00989130056091696\\
63.4143259096967	0.010651178287316\\
64.2830152621478	0.00824105117501825\\
65.1517046151759	0.0205504454936462\\
66.0203939687583	0.0220681163911732\\
66.8890833228733	0.0165160140621176\\
67.7577726775004	0.00883594865172389\\
68.6264620326202	0.00457837021375331\\
69.4951513882143	0.0127468825464369\\
70.363840744265	0.0284485417870674\\
71.2325301007557	0.0365299973760334\\
72.1012194576705	0.0244418593250386\\
72.9699088149942	0.00820322339867703\\
73.8385981727123	0.014801577911352\\
74.7072875308111	0.0163732544548366\\
75.5759768892776	0.0150683816580484\\
76.4446662480991	0.0127054656758595\\
77.3133556072637	0.00445833420083055\\
78.1820449667599	0.0121617142645882\\
79.0507343265769	0.0213225998866753\\
79.9194236867041	0.0293190955065371\\
80.7881130471316	0.0313656490091448\\
81.6568024078497	0.0184006502325085\\
82.5254917688494	0.0178821681908764\\
83.3941811301217	0.0279417012747923\\
84.2628704916583	0.03129360977658\\
85.131559853451	0.0174043667625771\\
86.0002492154922	0.0132471011749481\\
86.8689385777744	0.0306808721626046\\
};
\addlegendentry{4D}

\addplot [color=mycolor5, line width=1.4pt]
  table[row sep=crcr]{%
0.0111111111111111	1\\
0.895247433755812	0.829527153400937\\
1.79032473721716	0.49414647875089\\
2.68543984315255	0.257519562934743\\
3.58056440120335	0.214905272630613\\
4.47569274024788	0.220850693248263\\
5.37082296981961	0.184186327259125\\
6.26595427970152	0.12025351336925\\
7.16108626478055	0.0694154839790249\\
8.05621869999237	0.0559073320210651\\
8.95135145029781	0.066319188095031\\
9.84648442976257	0.0745292676918658\\
10.741617581097	0.0663112783324174\\
11.636750864639	0.0439012604971797\\
12.5318842520585	0.010698014410377\\
13.4270177225799	0.0245762054118241\\
14.3221512606217	0.0418337191199687\\
15.2172848542686	0.0449939442183362\\
16.112418494253	0.0341928963077936\\
17.0075521732585	0.0236898606049293\\
17.902685885432	0.0178322562651954\\
18.7978196260351	0.017647921073848\\
19.6929533911911	0.0141189845058904\\
20.5880871776975	0.00596298439637533\\
21.4832209828855	0.00685227835184741\\
22.3783548045132	0.0103736932021906\\
23.2734886406838	0.0233307004605702\\
24.1686224897815	0.0357371856418466\\
25.063756350421	0.0402587707487268\\
25.9588902214086	0.0312535047658911\\
26.8540241017093	0.0196727389399231\\
27.749157990422	0.00417623243878906\\
28.6442918867579	0.0101809509627485\\
29.5394257900241	0.0130033655583031\\
30.434559699609	0.0115648331098957\\
31.3296936149711	0.00617040297827514\\
32.224827535629	0.01365132465751\\
33.1199614611531	0.00487765978762974\\
34.0150953911595	0.0164170038735702\\
34.9102293253032	0.0216296473643642\\
35.805363263274	0.0162628814791749\\
36.7004972047918	0.0168825959257158\\
37.5956311496034	0.00426227536467148\\
38.4907650974788	0.00504457831446197\\
39.3858990482092	0.0199744398515393\\
40.2810330016043	0.0223185691527138\\
41.1761669574902	0.0232399679763996\\
42.0713009157081	0.0194874066109381\\
42.966434876112	0.0170341110637696\\
43.8615688385683	0.023266457257568\\
44.7567028029537	0.0235976247800764\\
45.6518367691548	0.0116084798217816\\
46.5469707370669	0.00792019211078716\\
47.442104706593	0.0135826610337316\\
48.3372386776435	0.0194956332652776\\
49.2323726501353	0.012017369891243\\
50.1275066239912	0.0107001246725462\\
51.0226405991393	0.0218655563954379\\
51.9177745755128	0.0149437474528286\\
52.8129085530495	0.0138769872959625\\
53.708042531691	0.0126159196534071\\
54.6031765113832	0.00799076684607162\\
55.4983104920752	0.0123802407167947\\
56.3934444737194	0.0101513132757898\\
57.288578456271	0.00965518418802567\\
58.1837124396884	0.0165558155144138\\
59.078846423932	0.0124947165377911\\
59.9739804089649	0.0102985797104155\\
60.8691143947523	0.00725482913703672\\
61.7642483812614	0.010581942036551\\
62.6593823684612	0.0242526879437655\\
63.5545163563227	0.0274636838497448\\
64.449650344818	0.0170308037548284\\
65.3447843339214	0.0141831439313543\\
66.239918323608	0.0145423215217537\\
67.1350523138546	0.0281473535129883\\
68.030186304639	0.0363073555916768\\
68.9253202959404	0.0324595242233269\\
69.8204542877388	0.0172570106606224\\
70.7155882800153	0.00986995113969667\\
71.610722272752	0.015572888452601\\
72.5058562659319	0.0258403259220646\\
73.4009902595388	0.0282216514379086\\
74.2961242535572	0.0201491287148634\\
75.1912582479725	0.0166665366200539\\
76.0863922427705	0.0197875236412837\\
76.981526237938	0.0143444721062633\\
77.8766602334623	0.0138801987458155\\
78.7717942293311	0.0207849705393367\\
79.6669282255329	0.0294044152148691\\
80.5620622220565	0.0252553267051852\\
81.4571962188914	0.0193590023809217\\
82.3523302160274	0.0210366402932451\\
83.2474642134547	0.0243652227264967\\
84.1425982111642	0.0159012112669893\\
85.0377322091468	0.0140021944057325\\
85.932866207394	0.0183735827470806\\
86.8280002058977	0.0208674145945452\\
87.72313420465	0.0320752971488706\\
88.6182682036434	0.0416054582124684\\
89.5134022028706	0.0360921133566622\\
};
\addlegendentry{5D}

\addplot [color=gray, line width=1.4pt, forget plot]
  table[row sep=crcr]{%
0.0111111111111111	0.1\\
0.895247433755812	0.1\\
1.79032473721716	0.1\\
2.68543984315255	0.1\\
3.58056440120335	0.1\\
4.47569274024788	0.1\\
5.37082296981961	0.1\\
6.26595427970152	0.1\\
7.16108626478055	0.1\\
8.05621869999237	0.1\\
8.95135145029781	0.1\\
9.84648442976257	0.1\\
10.741617581097	0.1\\
11.636750864639	0.1\\
12.5318842520585	0.1\\
13.4270177225799	0.1\\
14.3221512606217	0.1\\
15.2172848542686	0.1\\
16.112418494253	0.1\\
17.0075521732585	0.1\\
17.902685885432	0.1\\
18.7978196260351	0.1\\
19.6929533911911	0.1\\
20.5880871776975	0.1\\
21.4832209828855	0.1\\
22.3783548045132	0.1\\
23.2734886406838	0.1\\
24.1686224897815	0.1\\
25.063756350421	0.1\\
25.9588902214086	0.1\\
26.8540241017093	0.1\\
27.749157990422	0.1\\
28.6442918867579	0.1\\
29.5394257900241	0.1\\
30.434559699609	0.1\\
31.3296936149711	0.1\\
32.224827535629	0.1\\
33.1199614611531	0.1\\
34.0150953911595	0.1\\
34.9102293253032	0.1\\
35.805363263274	0.1\\
36.7004972047918	0.1\\
37.5956311496034	0.1\\
38.4907650974788	0.1\\
39.3858990482092	0.1\\
40.2810330016043	0.1\\
41.1761669574902	0.1\\
42.0713009157081	0.1\\
42.966434876112	0.1\\
43.8615688385683	0.1\\
44.7567028029537	0.1\\
45.6518367691548	0.1\\
46.5469707370669	0.1\\
47.442104706593	0.1\\
48.3372386776435	0.1\\
49.2323726501353	0.1\\
50.1275066239912	0.1\\
51.0226405991393	0.1\\
51.9177745755128	0.1\\
52.8129085530495	0.1\\
53.708042531691	0.1\\
54.6031765113832	0.1\\
55.4983104920752	0.1\\
56.3934444737194	0.1\\
57.288578456271	0.1\\
58.1837124396884	0.1\\
59.078846423932	0.1\\
59.9739804089649	0.1\\
60.8691143947523	0.1\\
61.7642483812614	0.1\\
62.6593823684612	0.1\\
63.5545163563227	0.1\\
64.449650344818	0.1\\
65.3447843339214	0.1\\
66.239918323608	0.1\\
67.1350523138546	0.1\\
68.030186304639	0.1\\
68.9253202959404	0.1\\
69.8204542877388	0.1\\
70.7155882800153	0.1\\
71.610722272752	0.1\\
72.5058562659319	0.1\\
73.4009902595388	0.1\\
74.2961242535572	0.1\\
75.1912582479725	0.1\\
76.0863922427705	0.1\\
76.981526237938	0.1\\
77.8766602334623	0.1\\
78.7717942293311	0.1\\
79.6669282255329	0.1\\
80.5620622220565	0.1\\
81.4571962188914	0.1\\
82.3523302160274	0.1\\
83.2474642134547	0.1\\
84.1425982111642	0.1\\
85.0377322091468	0.1\\
85.932866207394	0.1\\
86.8280002058977	0.1\\
87.72313420465	0.1\\
88.6182682036434	0.1\\
89.5134022028706	0.1\\
};
\end{axis}
\end{tikzpicture}%

%% file: figures/Exp_figure_compare_soa_SeeYa_v2.tex
\begin{figure*}[t!]
    \centering

    \begin{subfigure}[t]{0.32\columnwidth}
		\centering
		\caption{\textbf{5 D}}
		\includegraphics[width=\columnwidth]{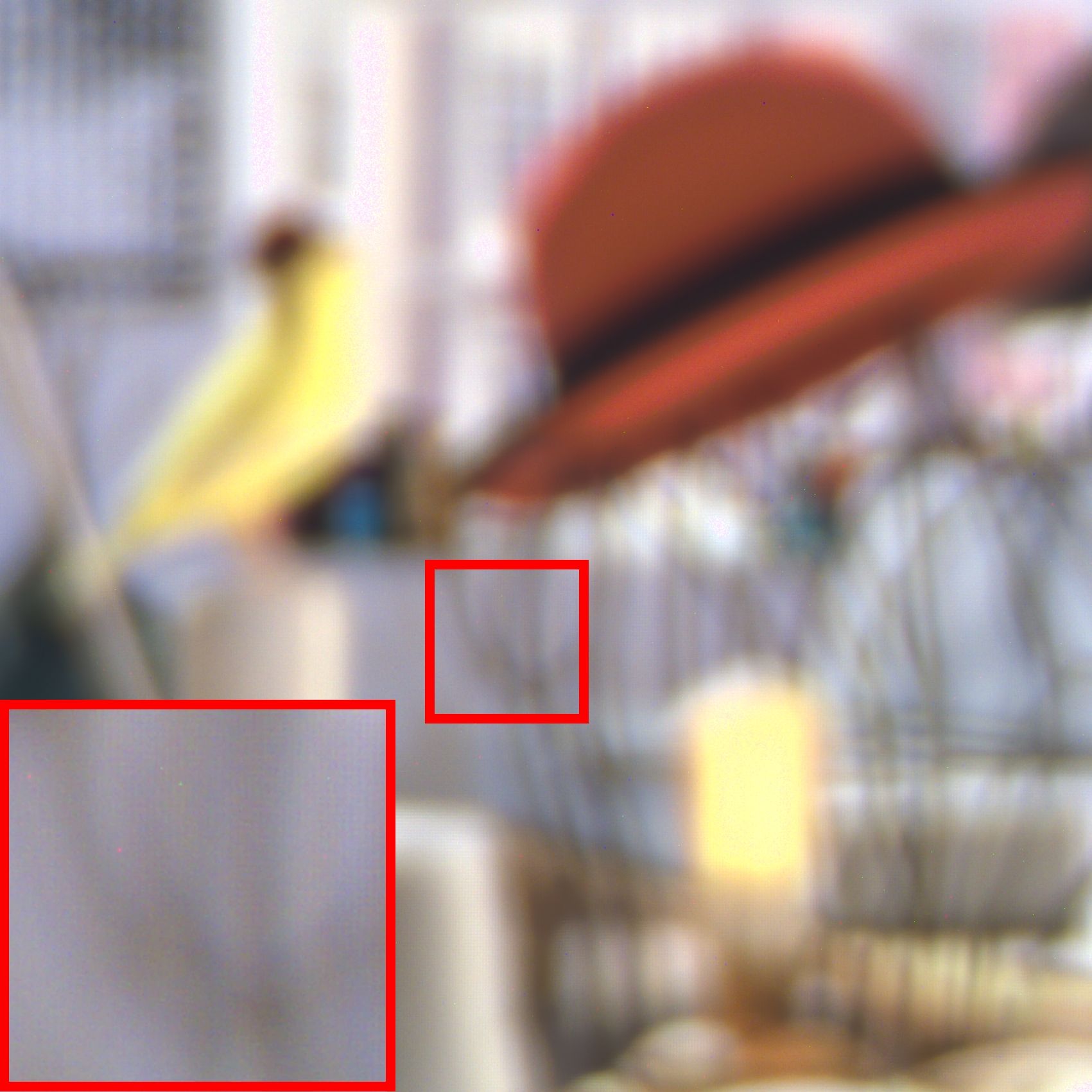}
            \put(-92,12){\rotatebox{90}{Conventional}}
    \end{subfigure}
     \vspace{1mm}
	%\put(-112,-74){\rotatebox{90}{\textbf{(0, 0)}}}
	\begin{subfigure}[t]{0.32\columnwidth}
		\centering
		\caption{\textbf{4 D}}
		\includegraphics[width=\columnwidth]{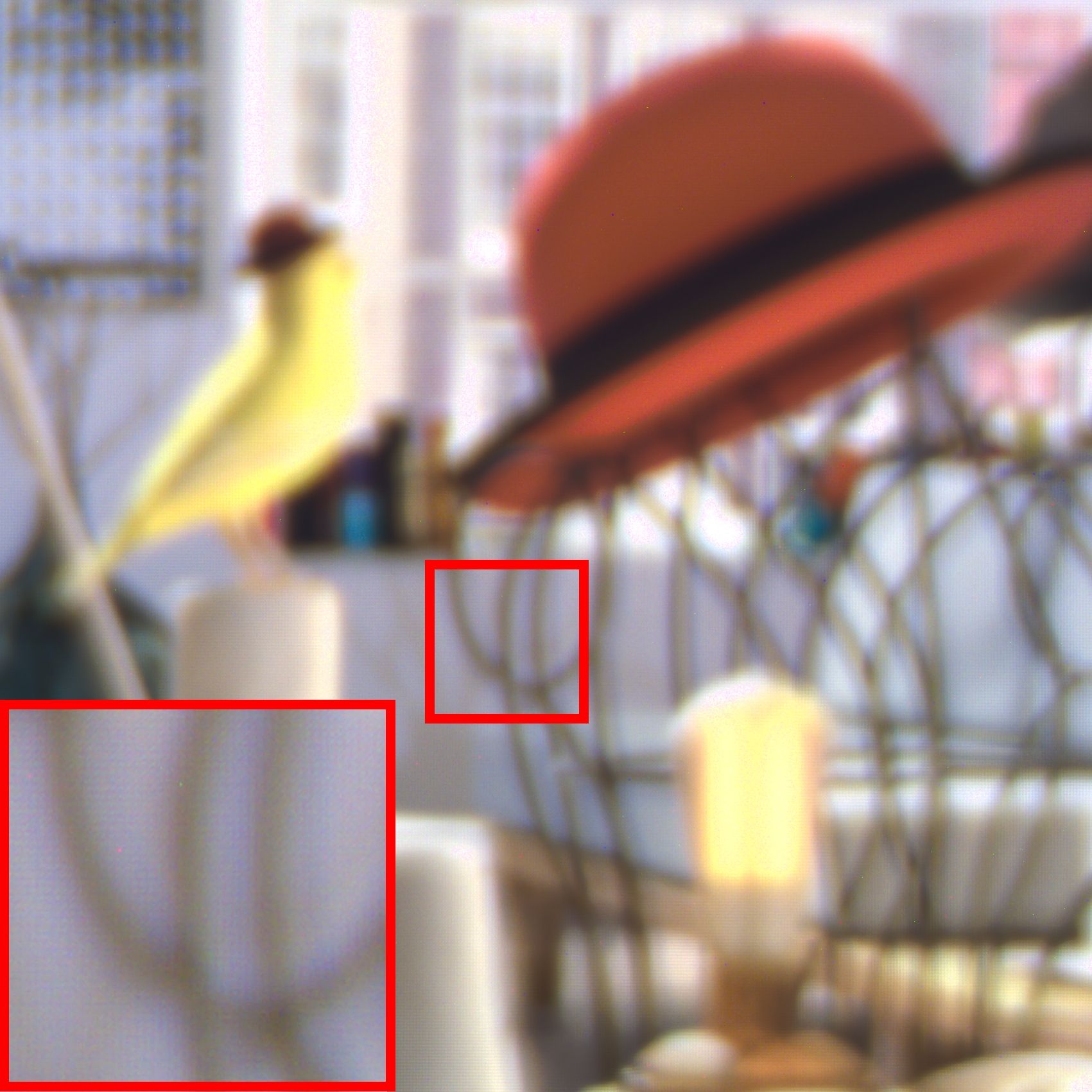}
	\end{subfigure}
	\begin{subfigure}[t]{0.32\columnwidth}
		\centering
		\caption{\textbf{3 D}}
		\includegraphics[width=\columnwidth]{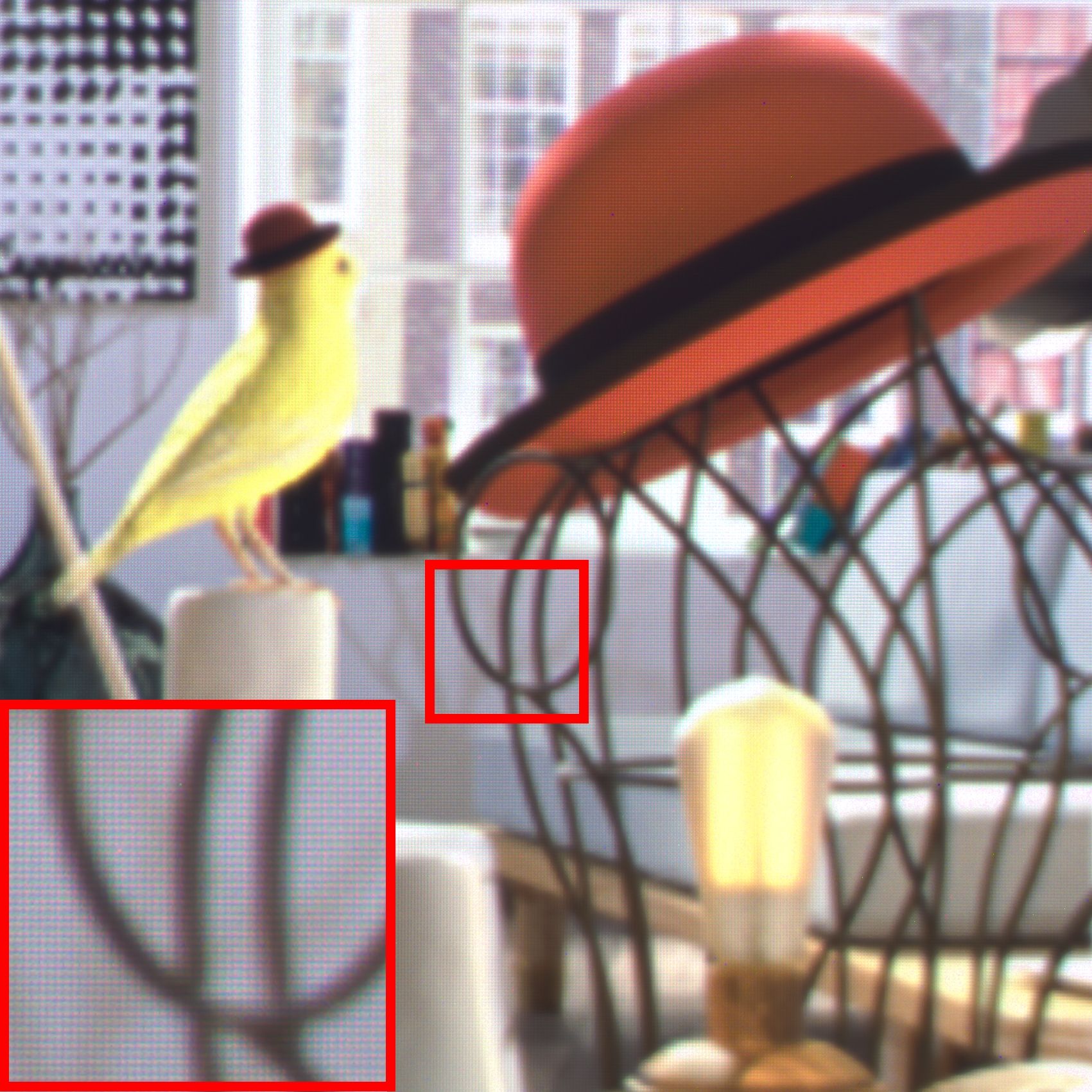}
	\end{subfigure}
	\begin{subfigure}[t]{0.32\columnwidth}
		\centering
		\caption{\textbf{2 D}}
		\includegraphics[width=\columnwidth]{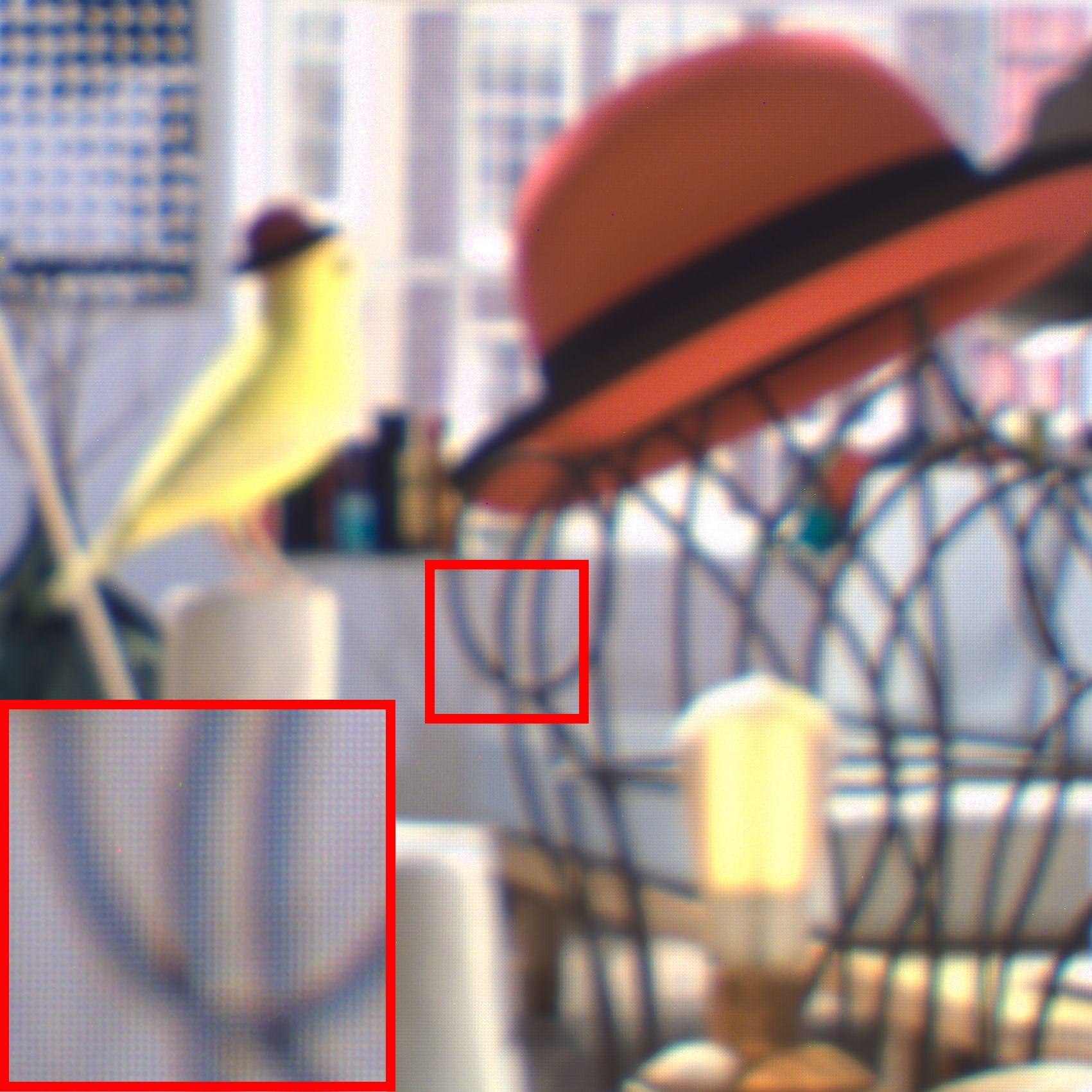}
	\end{subfigure}
	\begin{subfigure}[t]{0.32\columnwidth}
		\centering
		\caption{\textbf{1 D}}
		\includegraphics[width=\columnwidth]{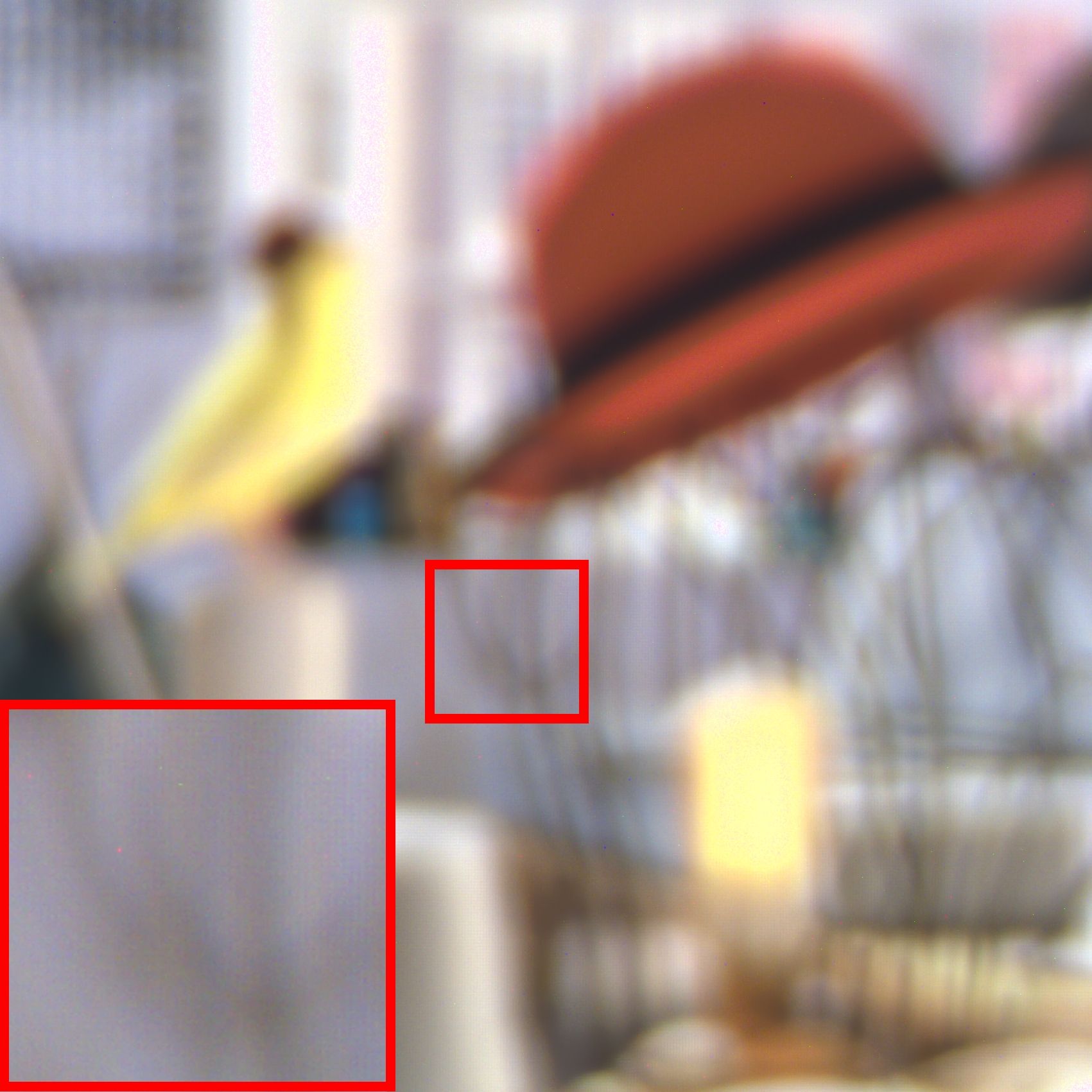}
	\end{subfigure}

    \begin{subfigure}[t]{0.32\columnwidth}
		\centering
		\includegraphics[width=\columnwidth]{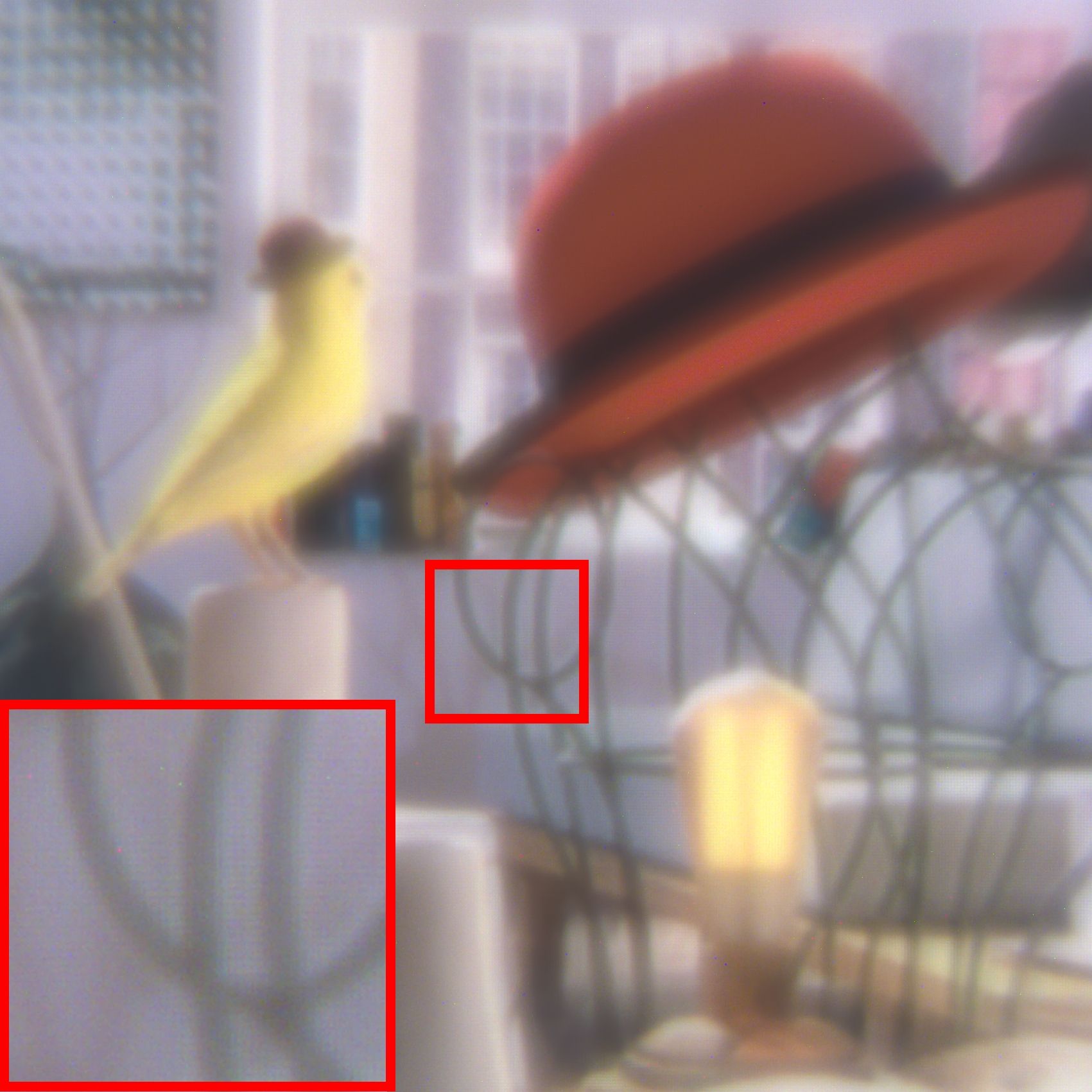}
            \put(-92,20){\rotatebox{90}{Proposed}}
	\end{subfigure}
        \vspace{1mm}
	\begin{subfigure}[t]{0.32\columnwidth}
		\centering
		\includegraphics[width=\columnwidth]{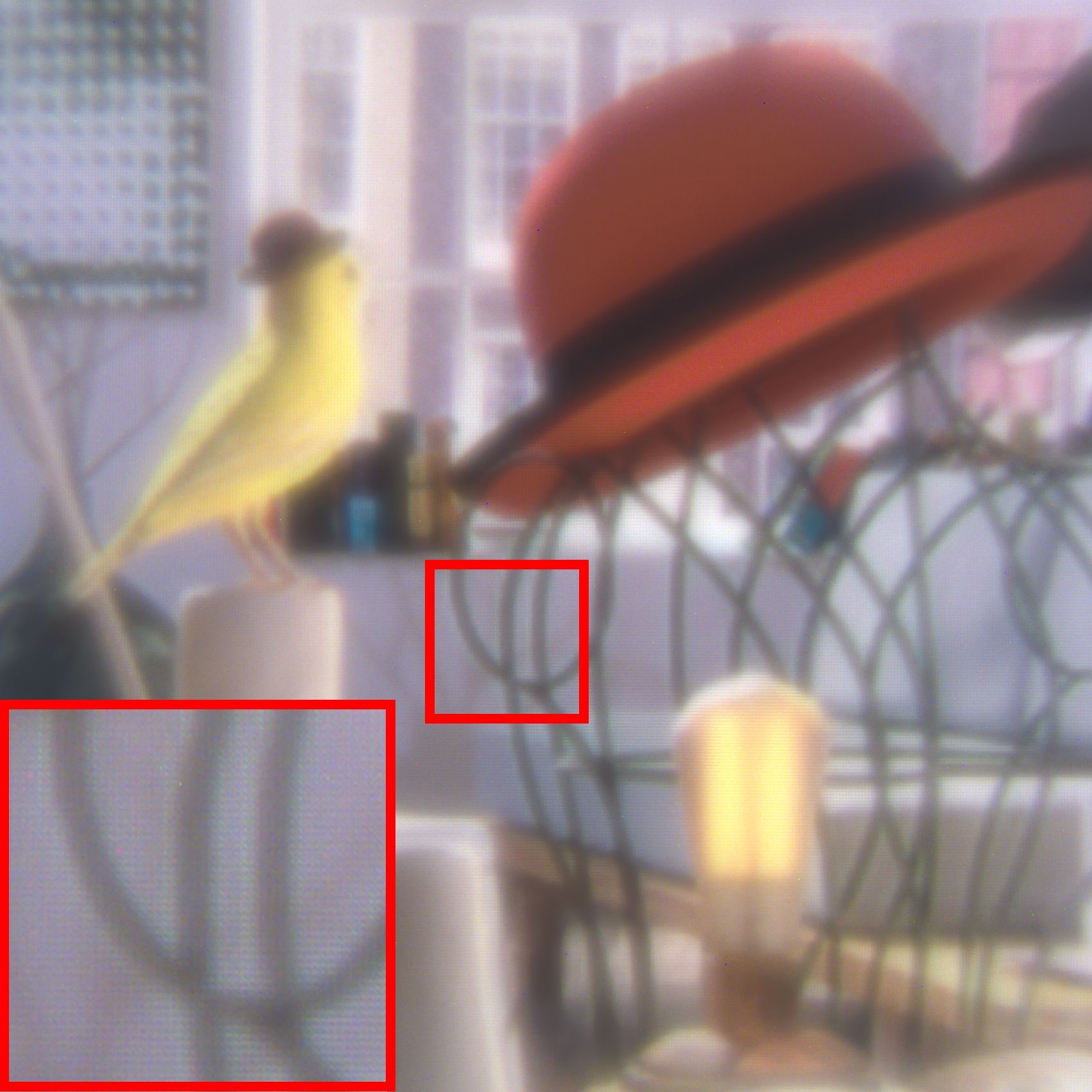}
	\end{subfigure}
	\begin{subfigure}[t]{0.32\columnwidth}
		\centering
		\includegraphics[width=\columnwidth]{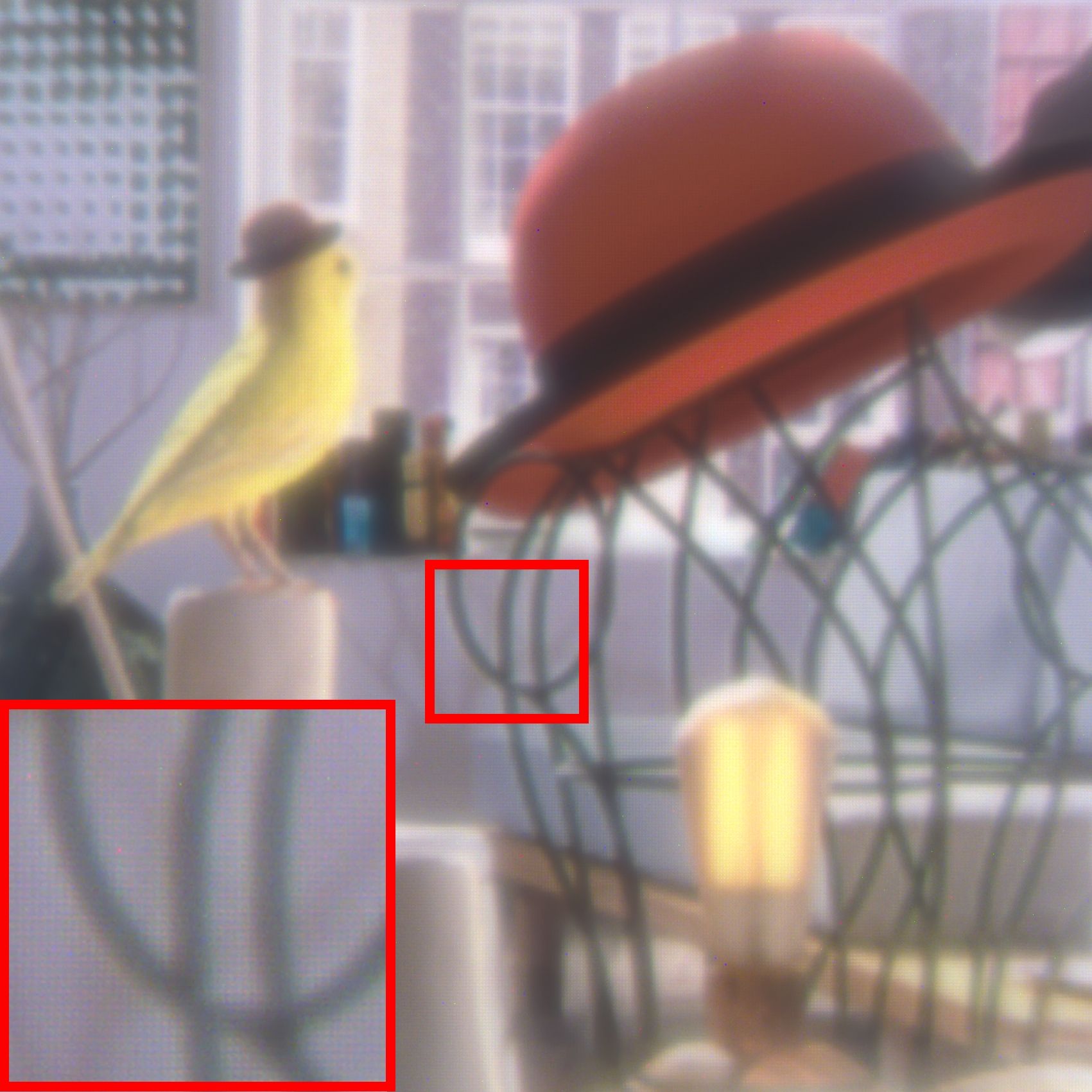}
	\end{subfigure}
	\begin{subfigure}[t]{0.32\columnwidth}
		\centering
		\includegraphics[width=\columnwidth]{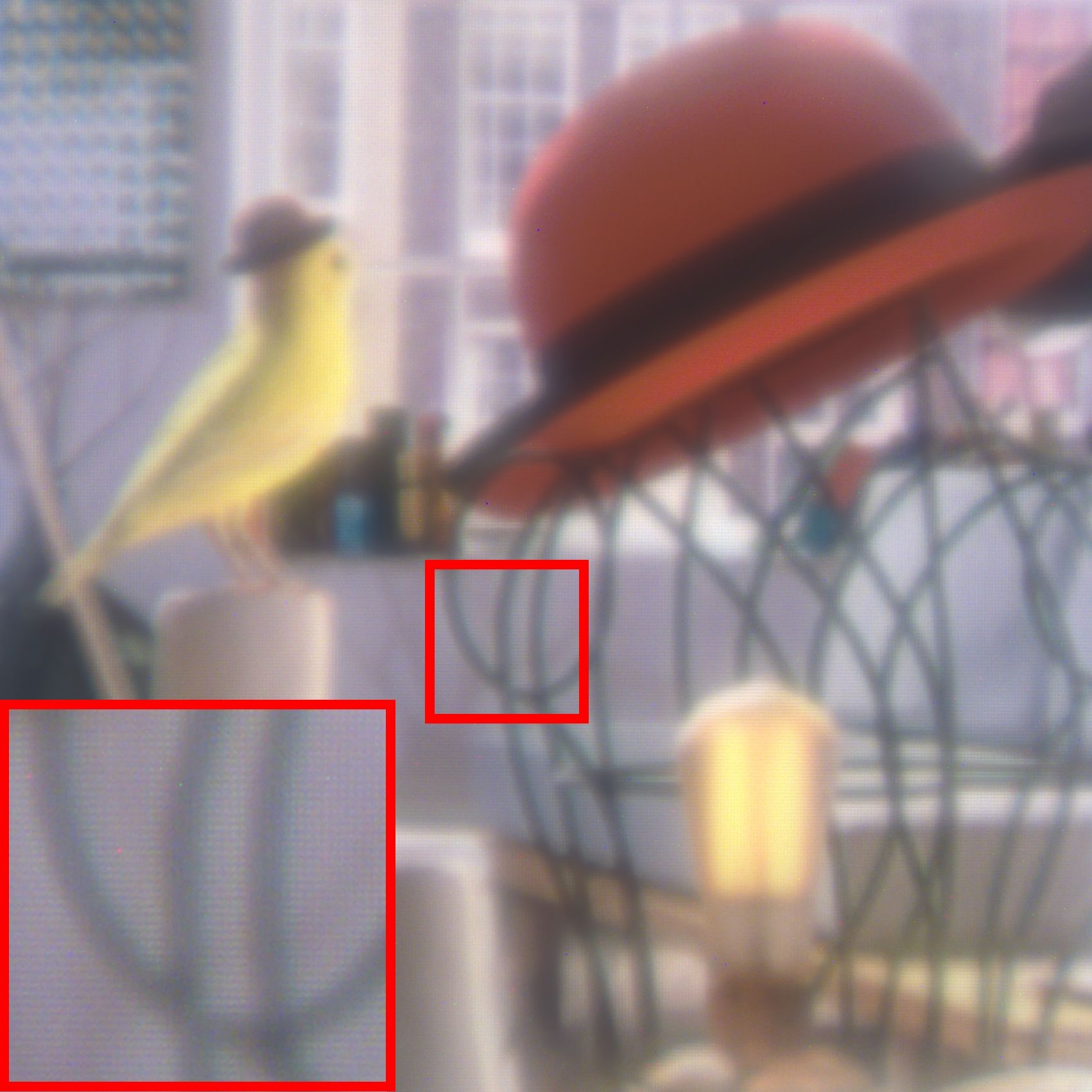}
	\end{subfigure}
	\begin{subfigure}[t]{0.32\columnwidth}
		\centering
		\includegraphics[width=\columnwidth]{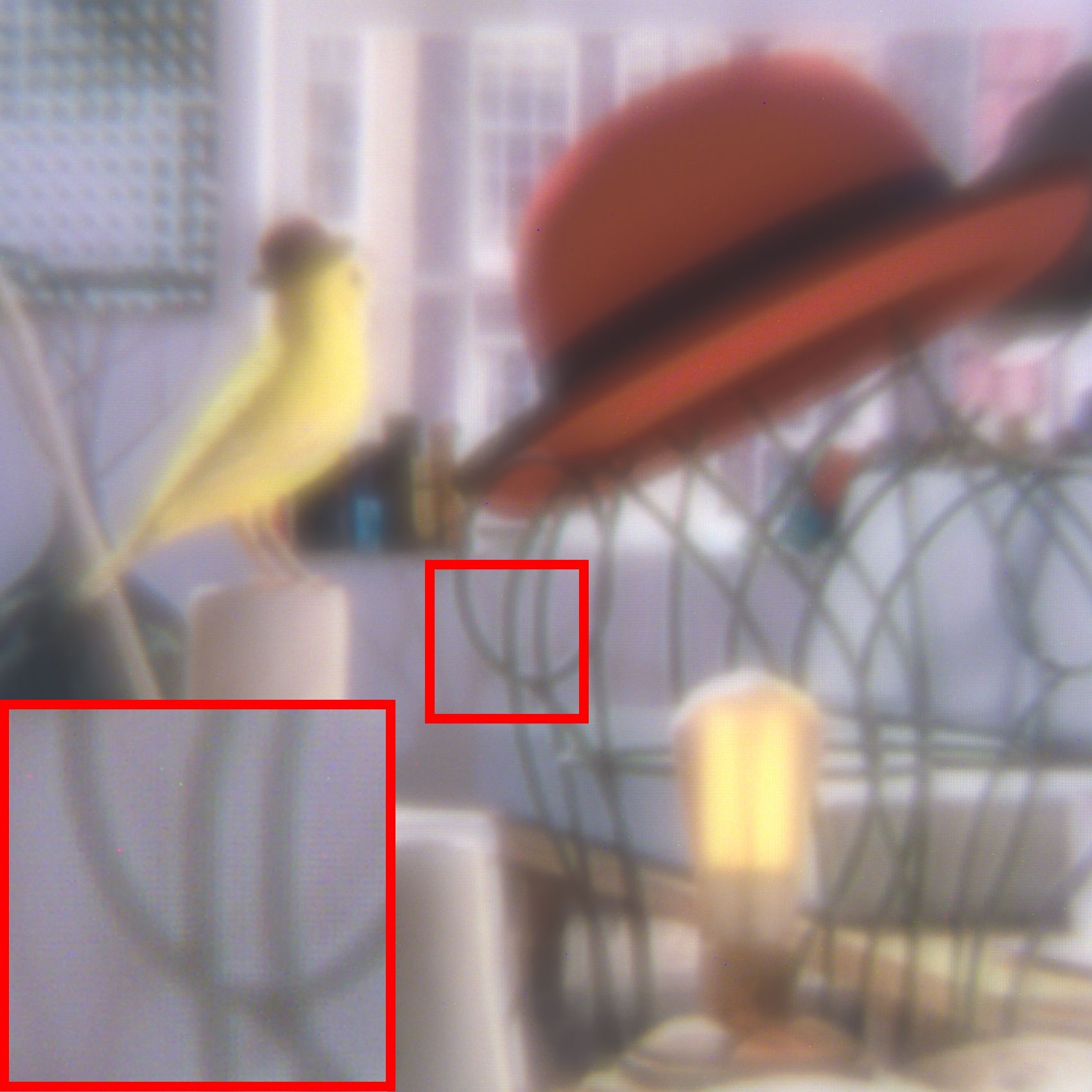}
	\end{subfigure}

        \begin{subfigure}[t]{0.32\columnwidth}
		\centering
		\includegraphics[width=\columnwidth]{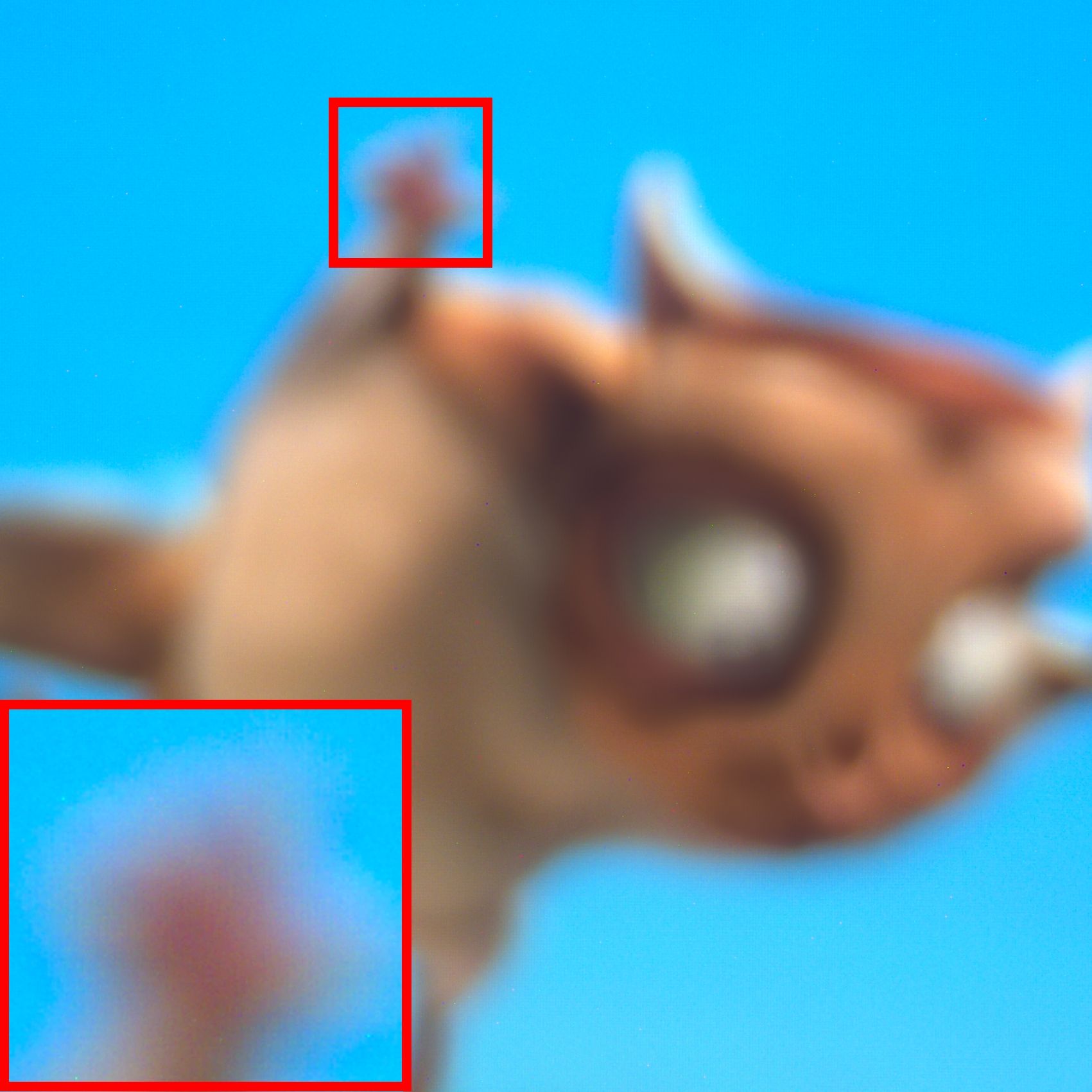}
            \put(-92,12){\rotatebox{90}{Conventional}}
    \end{subfigure}
     \vspace{1mm}
	%\put(-112,-74){\rotatebox{90}{\textbf{(0, 0)}}}
	\begin{subfigure}[t]{0.32\columnwidth}
		\centering
		\includegraphics[width=\columnwidth]{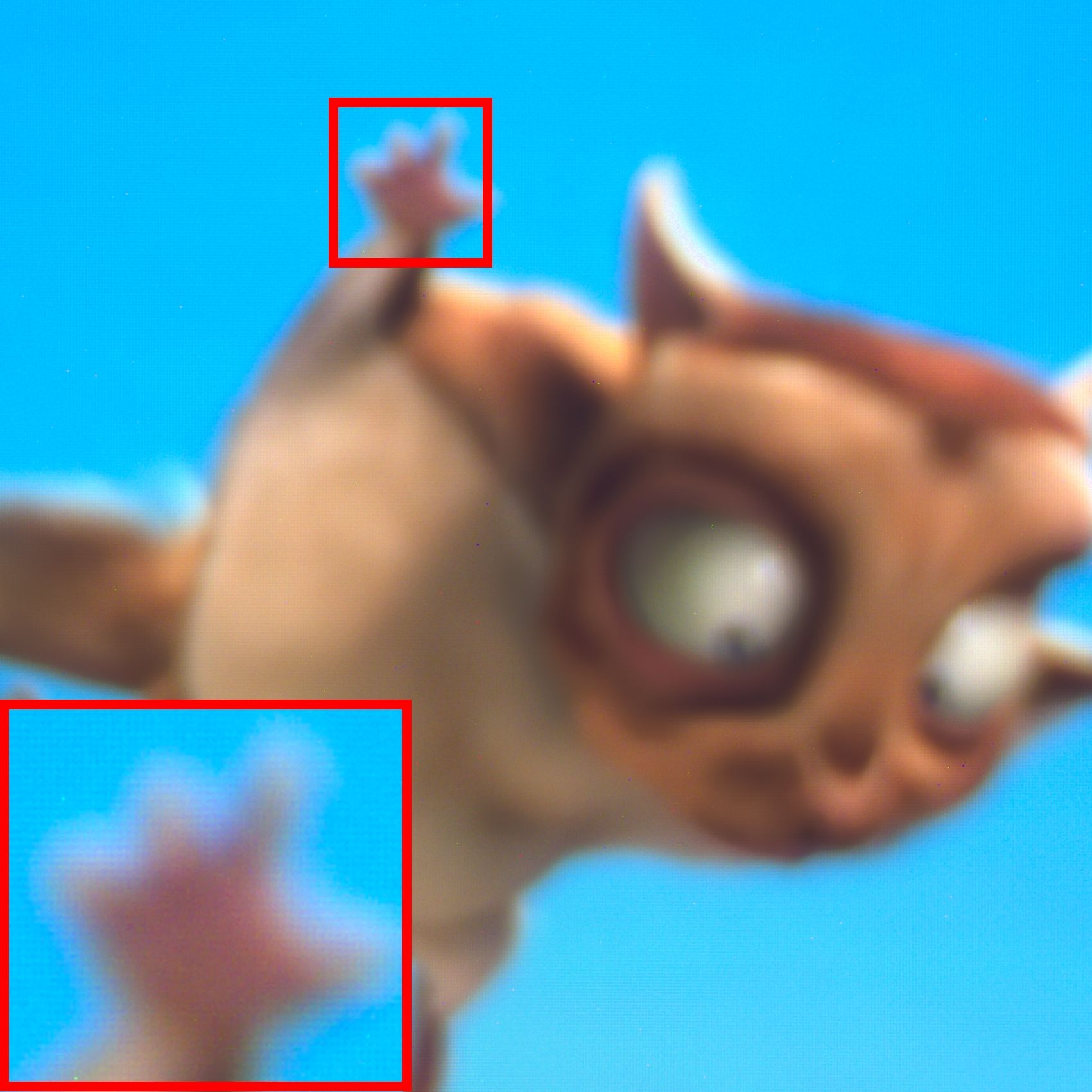}
	\end{subfigure}
	\begin{subfigure}[t]{0.32\columnwidth}
		\centering
		\includegraphics[width=\columnwidth]{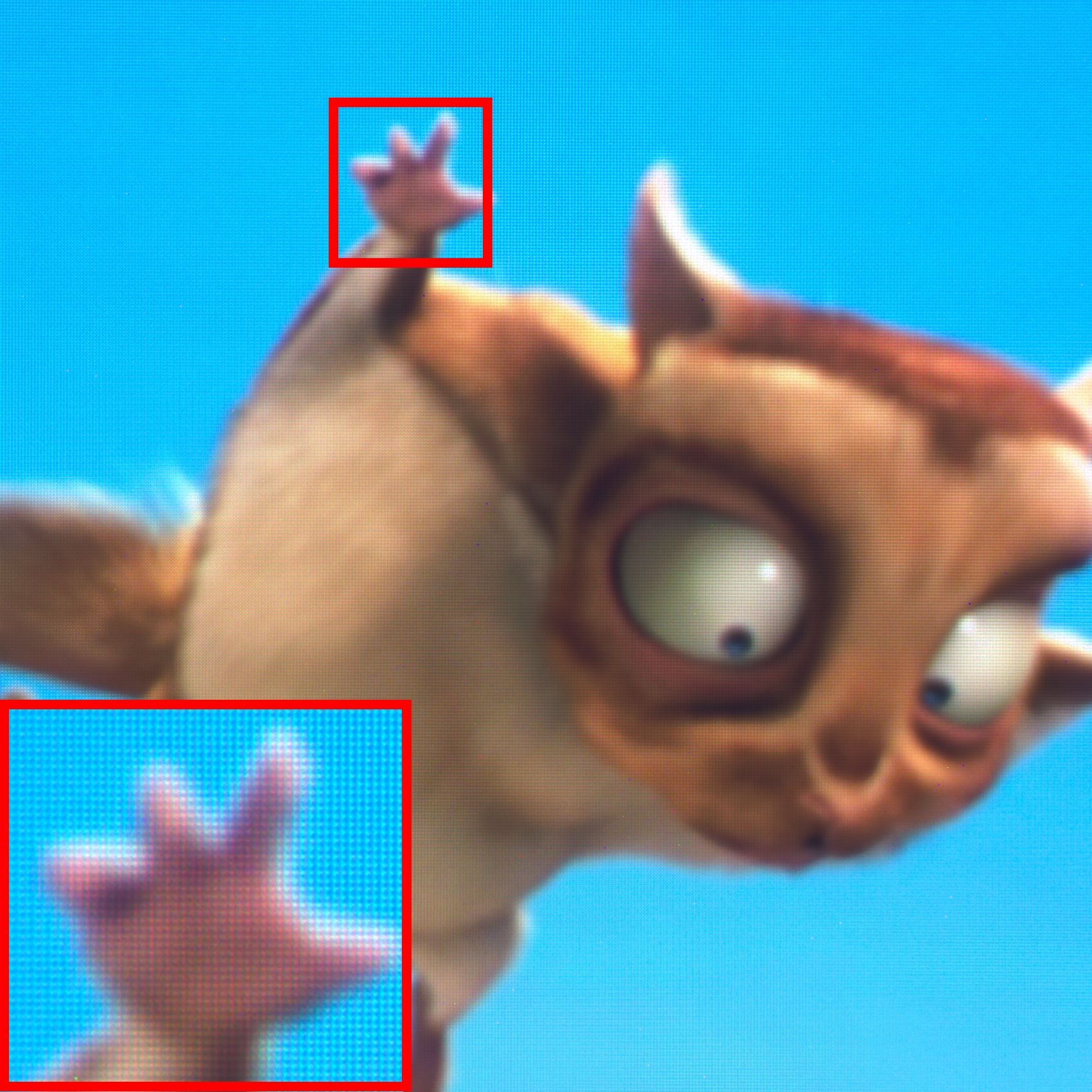}
	\end{subfigure}
	\begin{subfigure}[t]{0.32\columnwidth}
		\centering
		\includegraphics[width=\columnwidth]{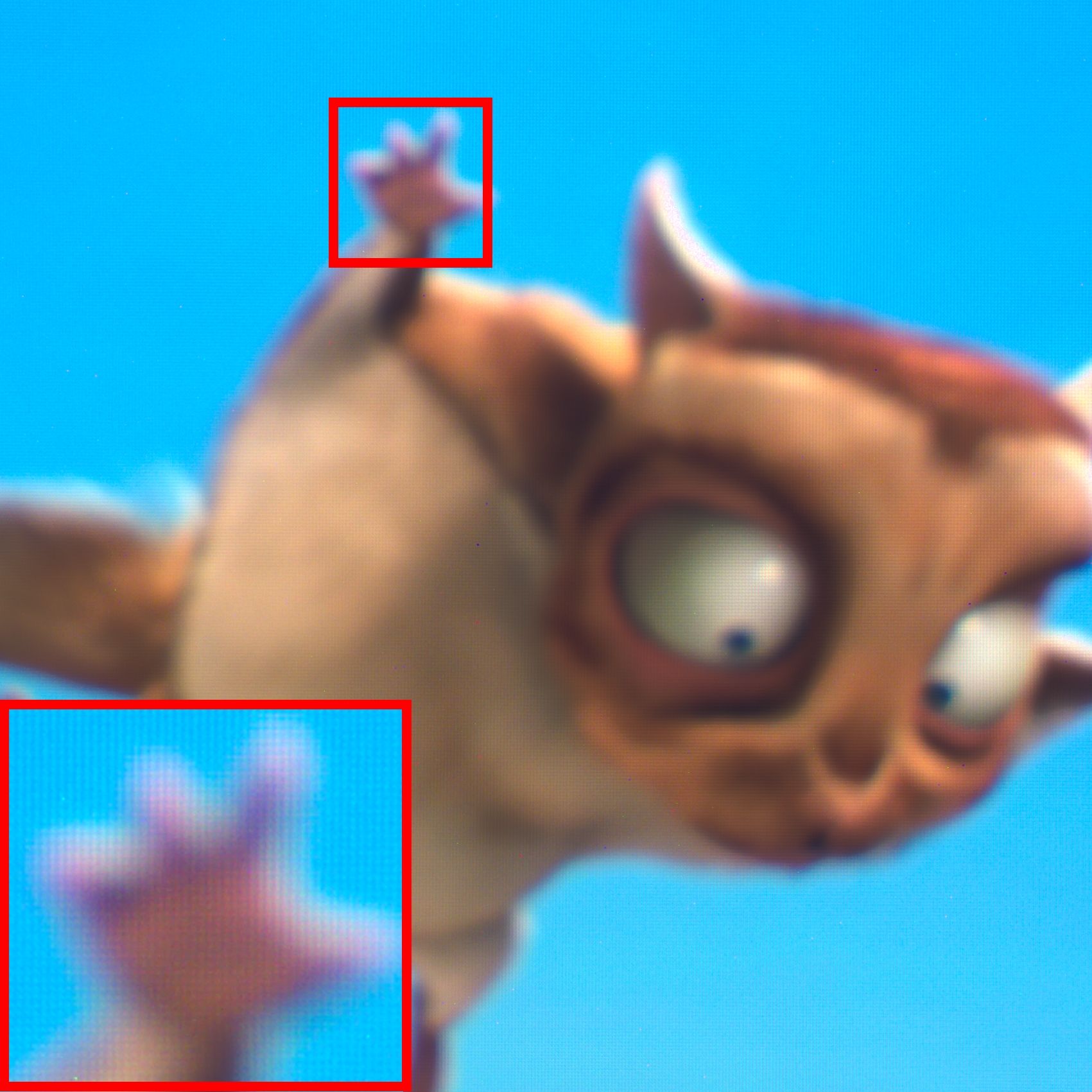}
	\end{subfigure}
	\begin{subfigure}[t]{0.32\columnwidth}
		\centering
		\includegraphics[width=\columnwidth]{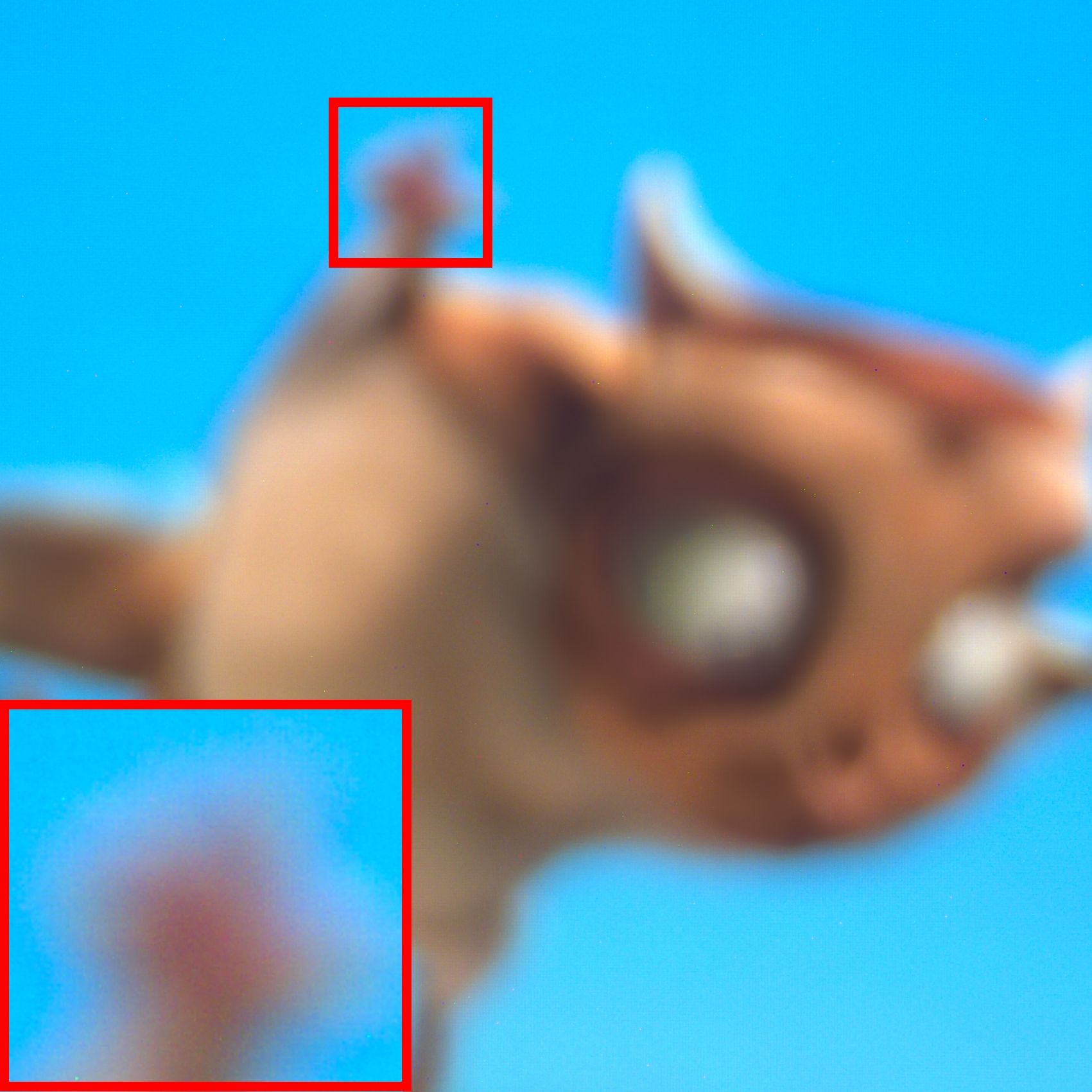}
	\end{subfigure}
	
	\begin{subfigure}[t]{0.32\columnwidth}
		\centering
		\includegraphics[width=\columnwidth]{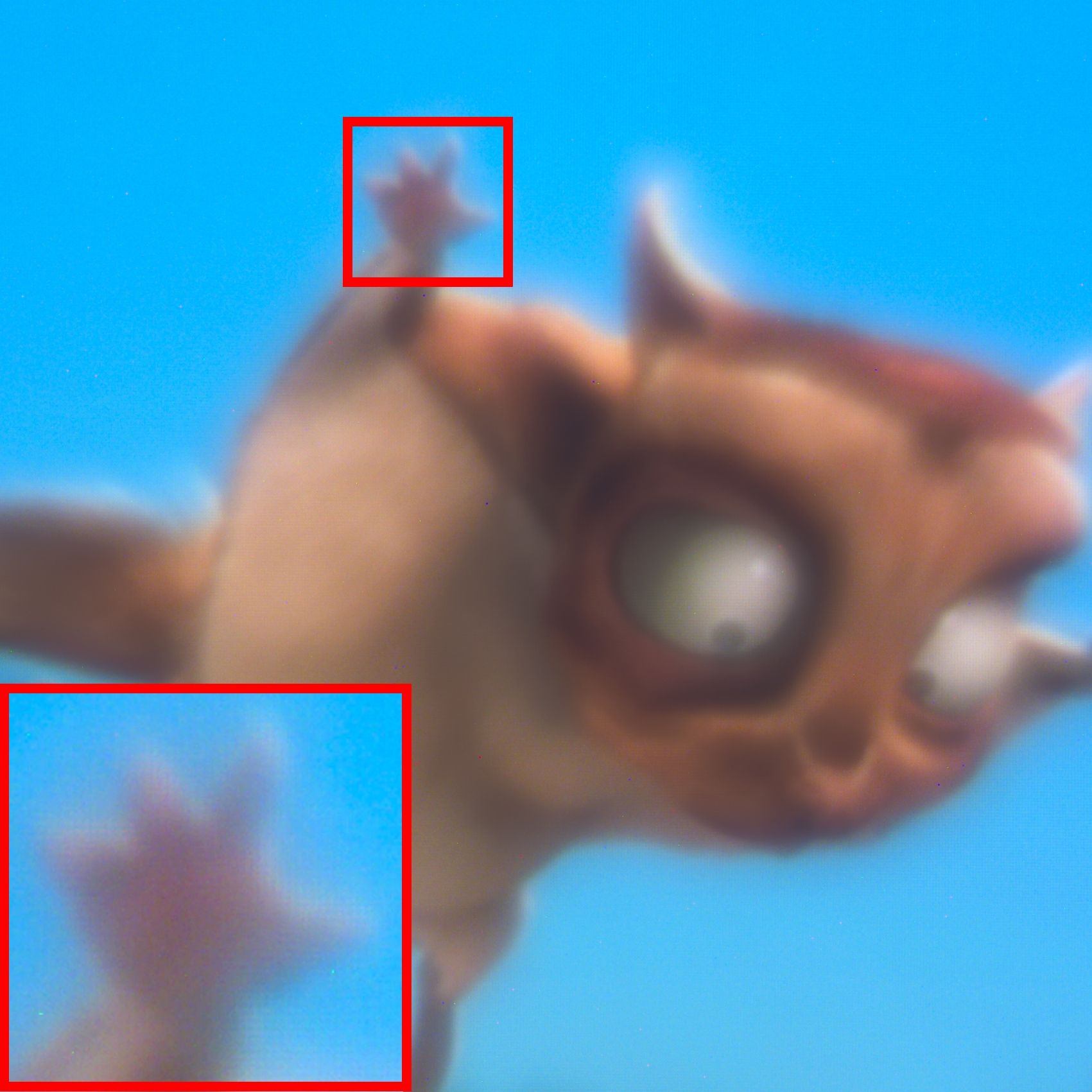}
            \put(-92,20){\rotatebox{90}{Proposed}}
	\end{subfigure}
        \vspace{1mm}
	\begin{subfigure}[t]{0.32\columnwidth}
		\centering
		\includegraphics[width=\columnwidth]{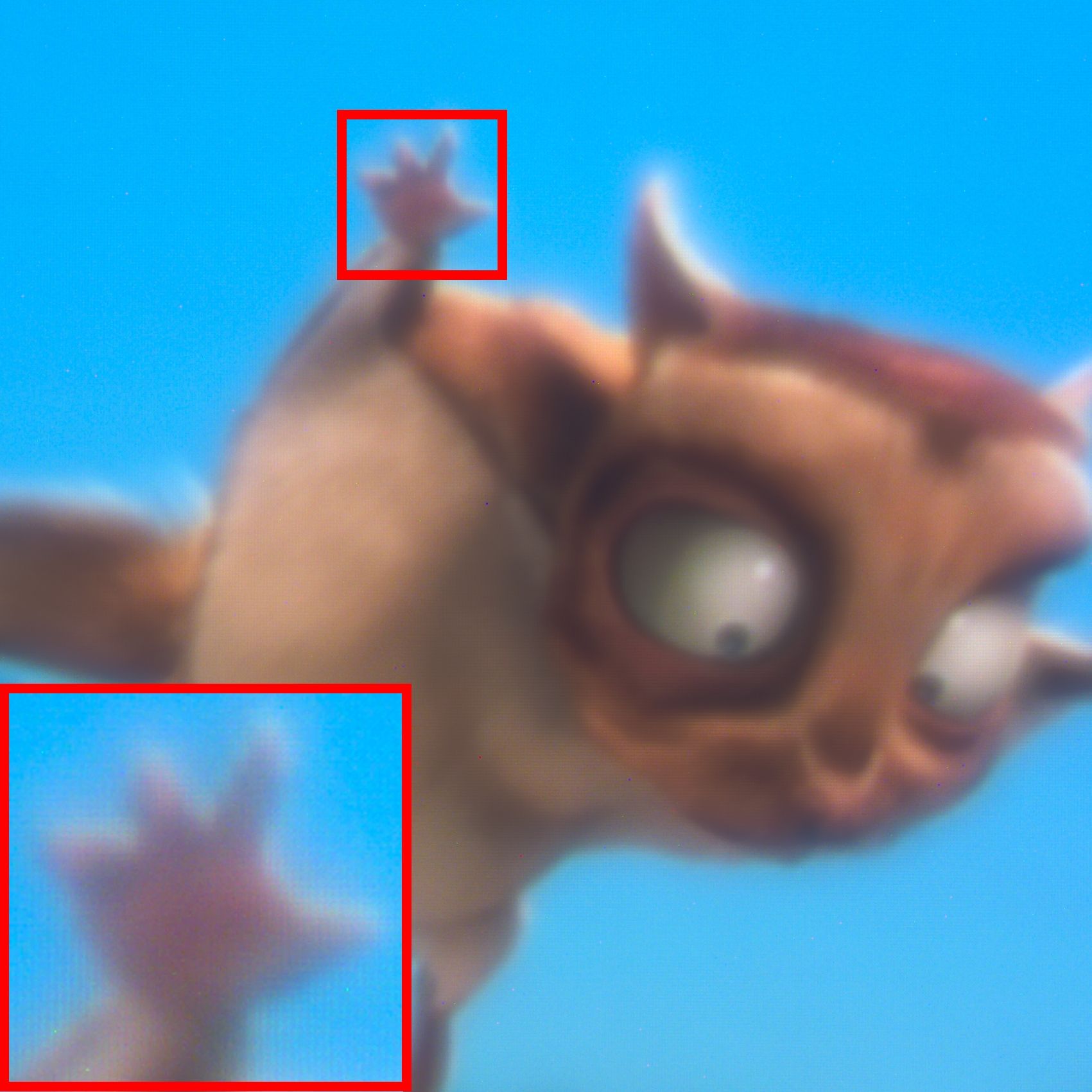}
	\end{subfigure}
	\begin{subfigure}[t]{0.32\columnwidth}
		\centering
		\includegraphics[width=\columnwidth]{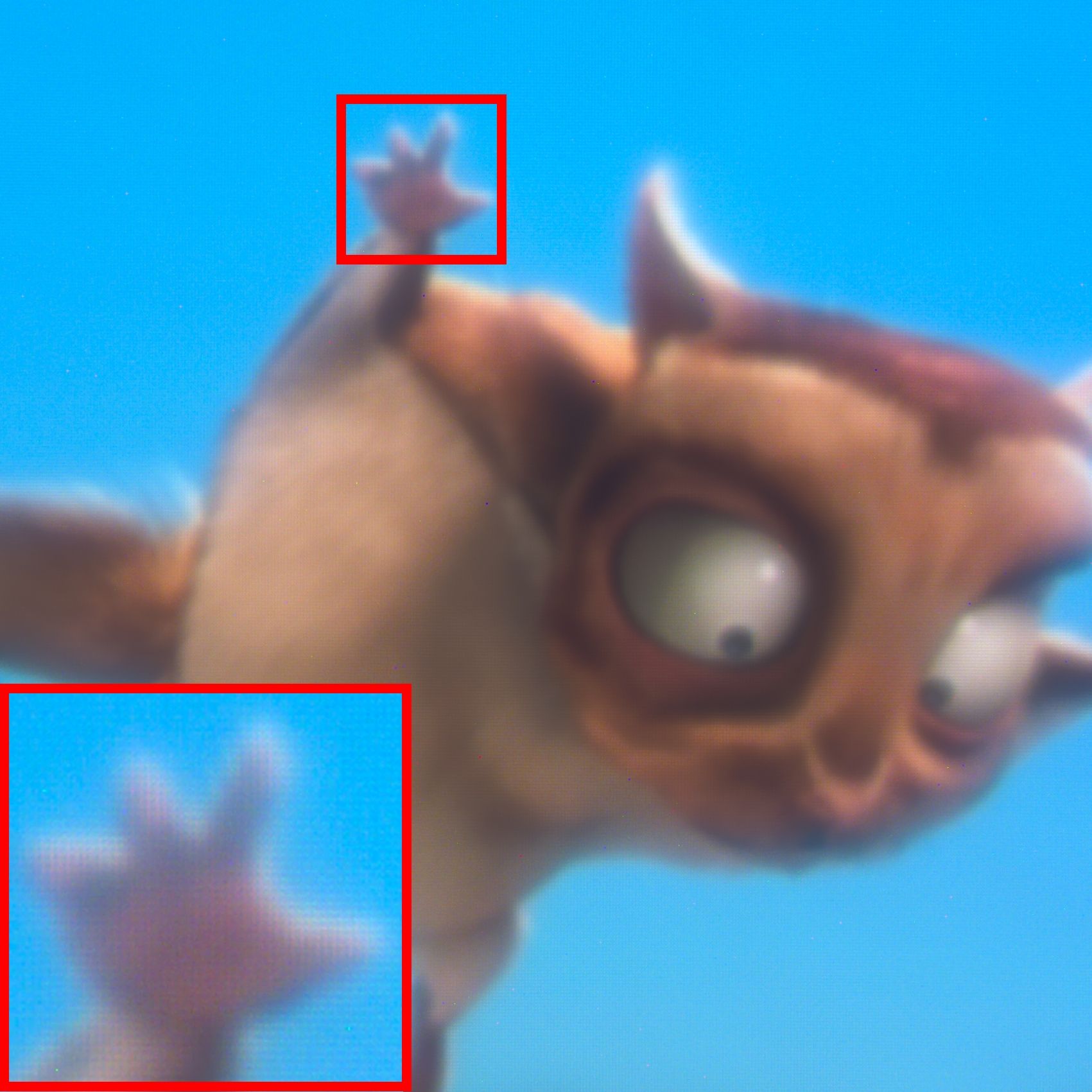}
	\end{subfigure}
	\begin{subfigure}[t]{0.32\columnwidth}
		\centering
		\includegraphics[width=\columnwidth]{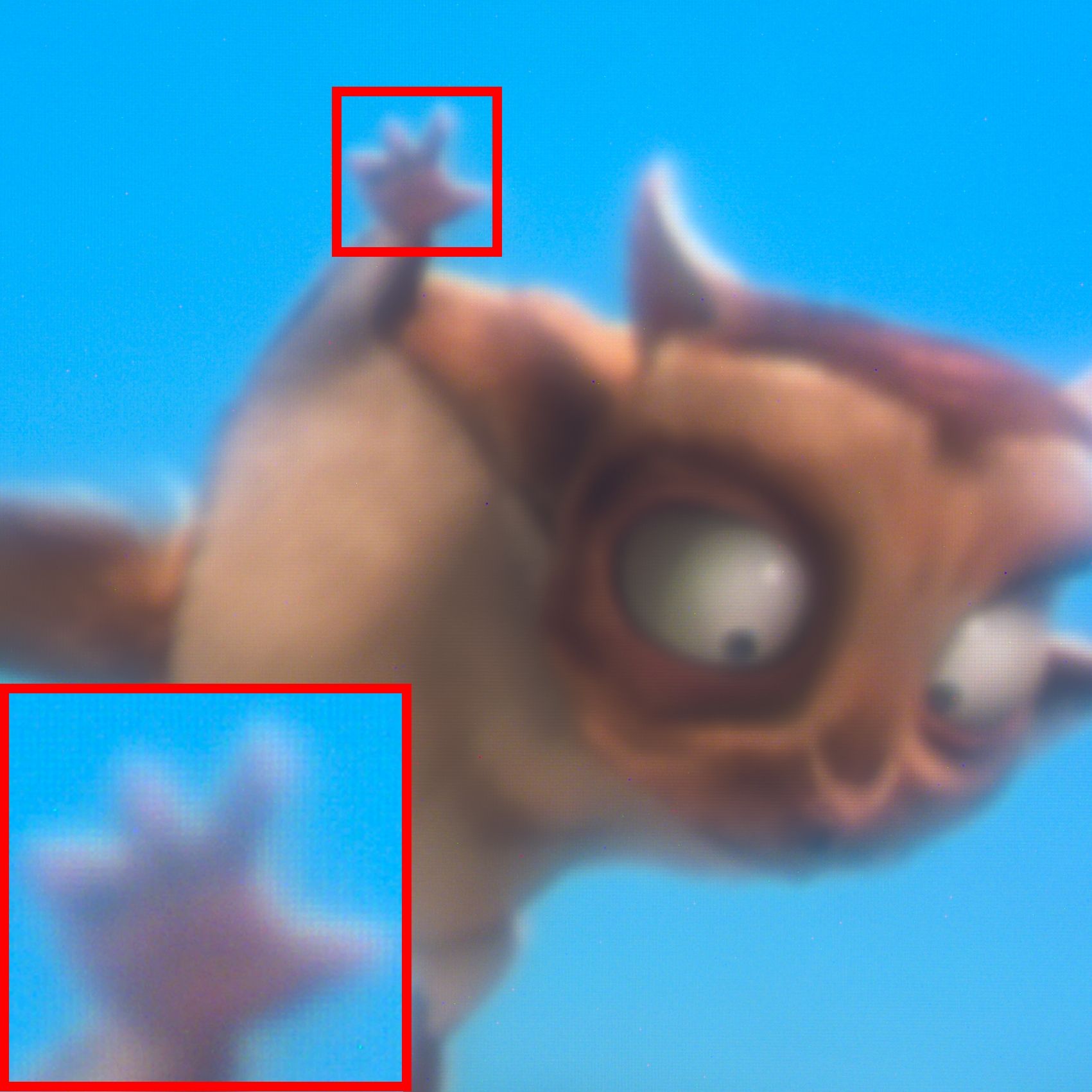}
	\end{subfigure}
	\begin{subfigure}[t]{0.32\columnwidth}
		\centering
		\includegraphics[width=\columnwidth]{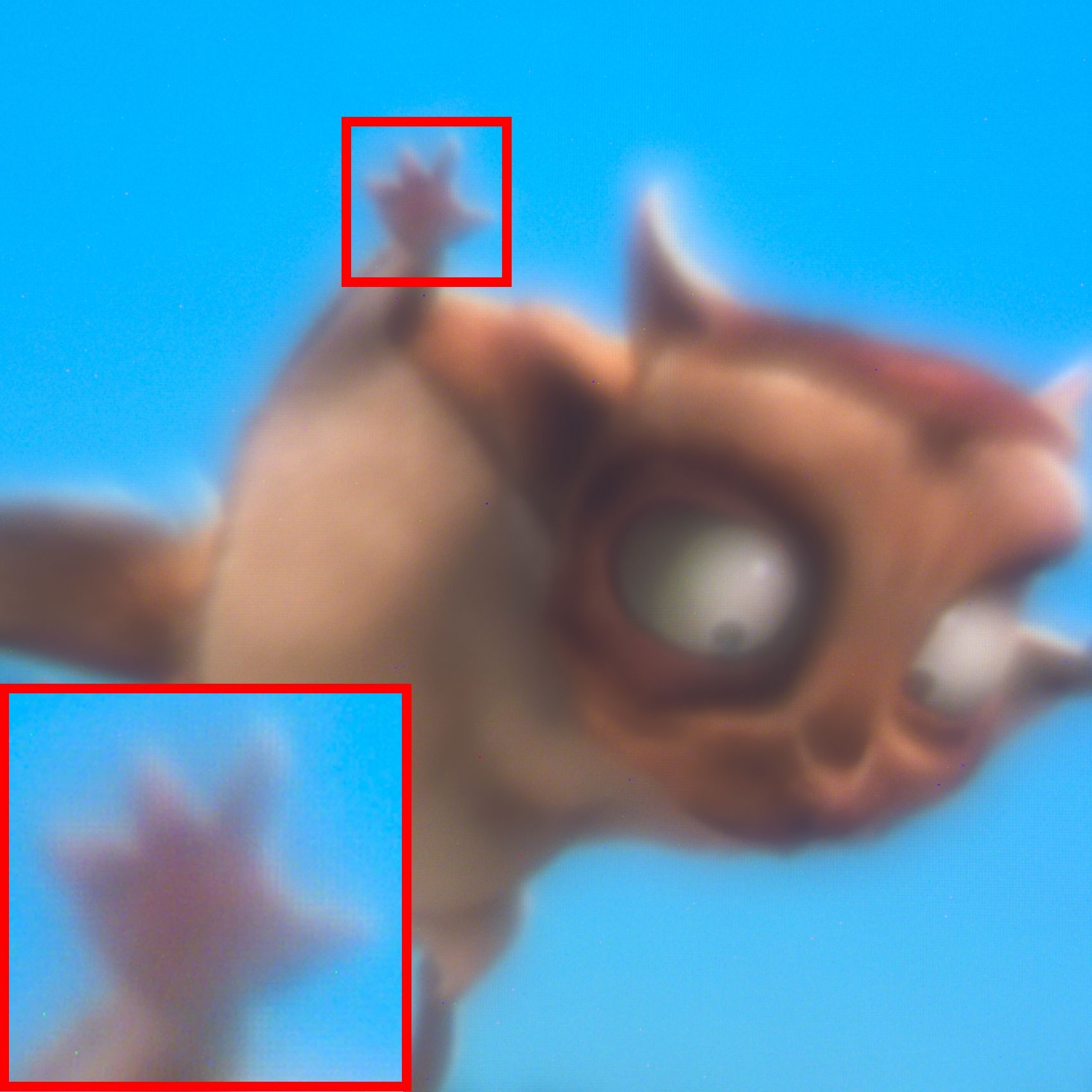}
	\end{subfigure}

    \begin{subfigure}[t]{0.32\columnwidth}
		\centering
		\includegraphics[width=\columnwidth]{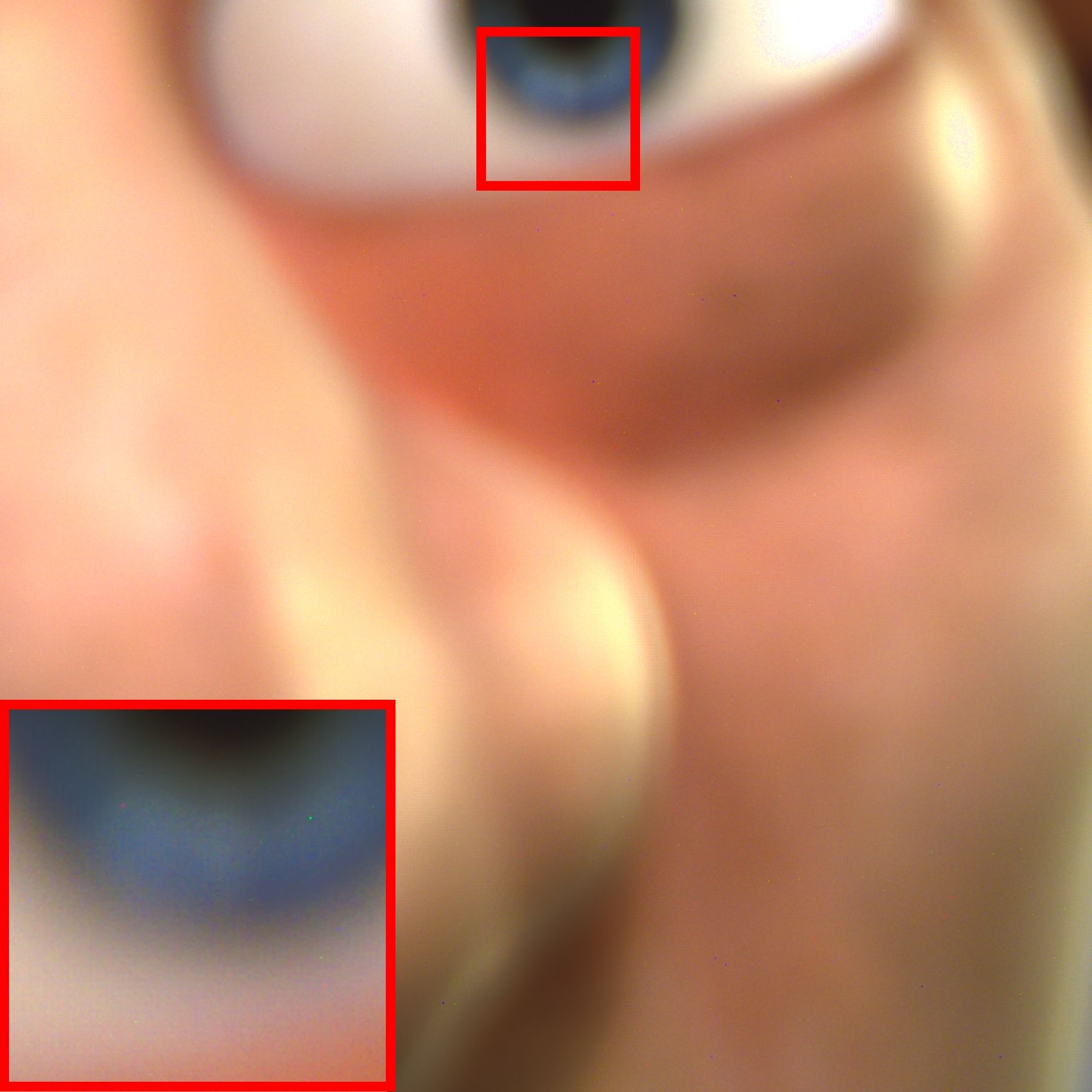}
            \put(-92,12){\rotatebox{90}{Conventional}}
    \end{subfigure}
     \vspace{1mm}
	%\put(-112,-74){\rotatebox{90}{\textbf{(0, 0)}}}
	\begin{subfigure}[t]{0.32\columnwidth}
		\centering
		\includegraphics[width=\columnwidth]{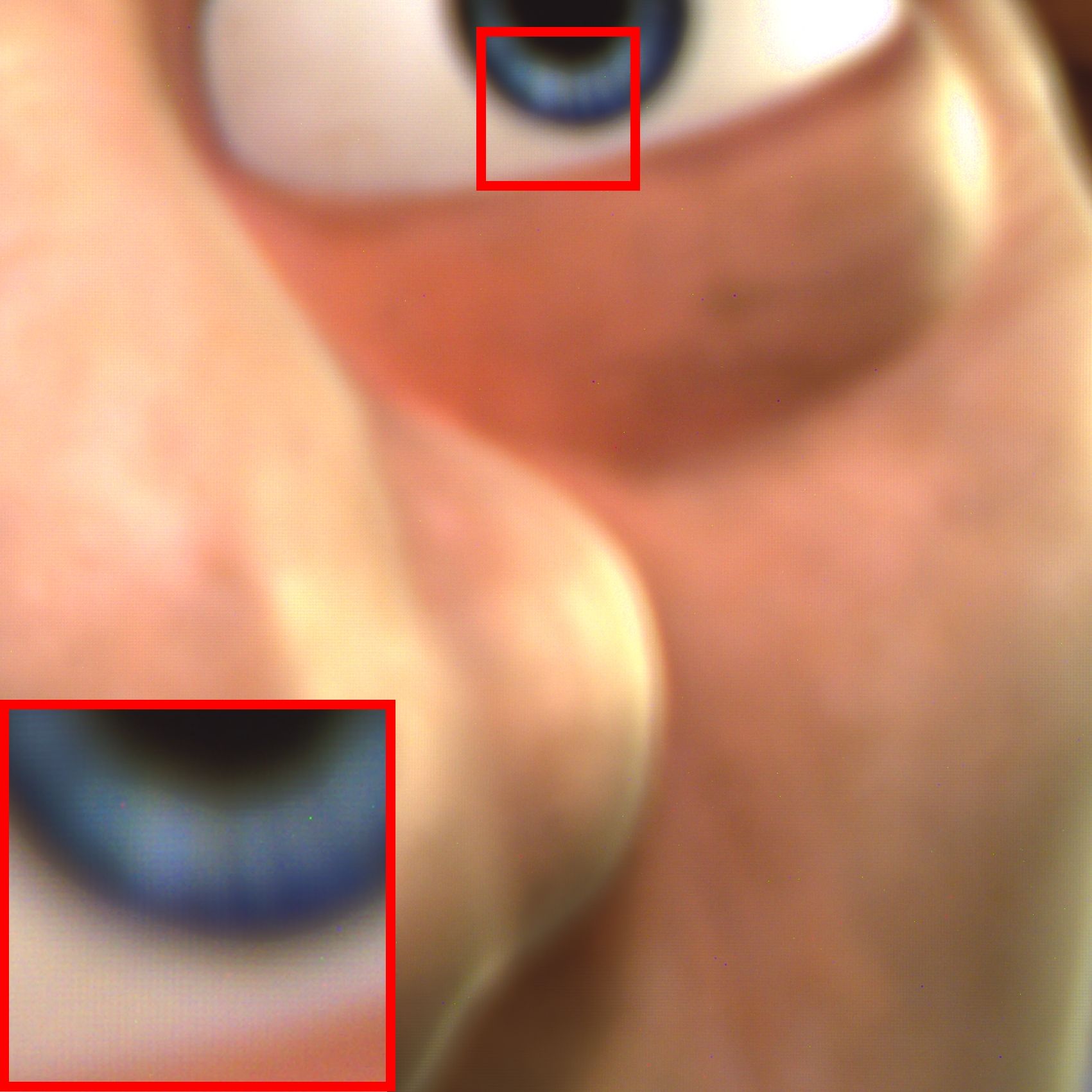}
	\end{subfigure}
	\begin{subfigure}[t]{0.32\columnwidth}
		\centering
		\includegraphics[width=\columnwidth]{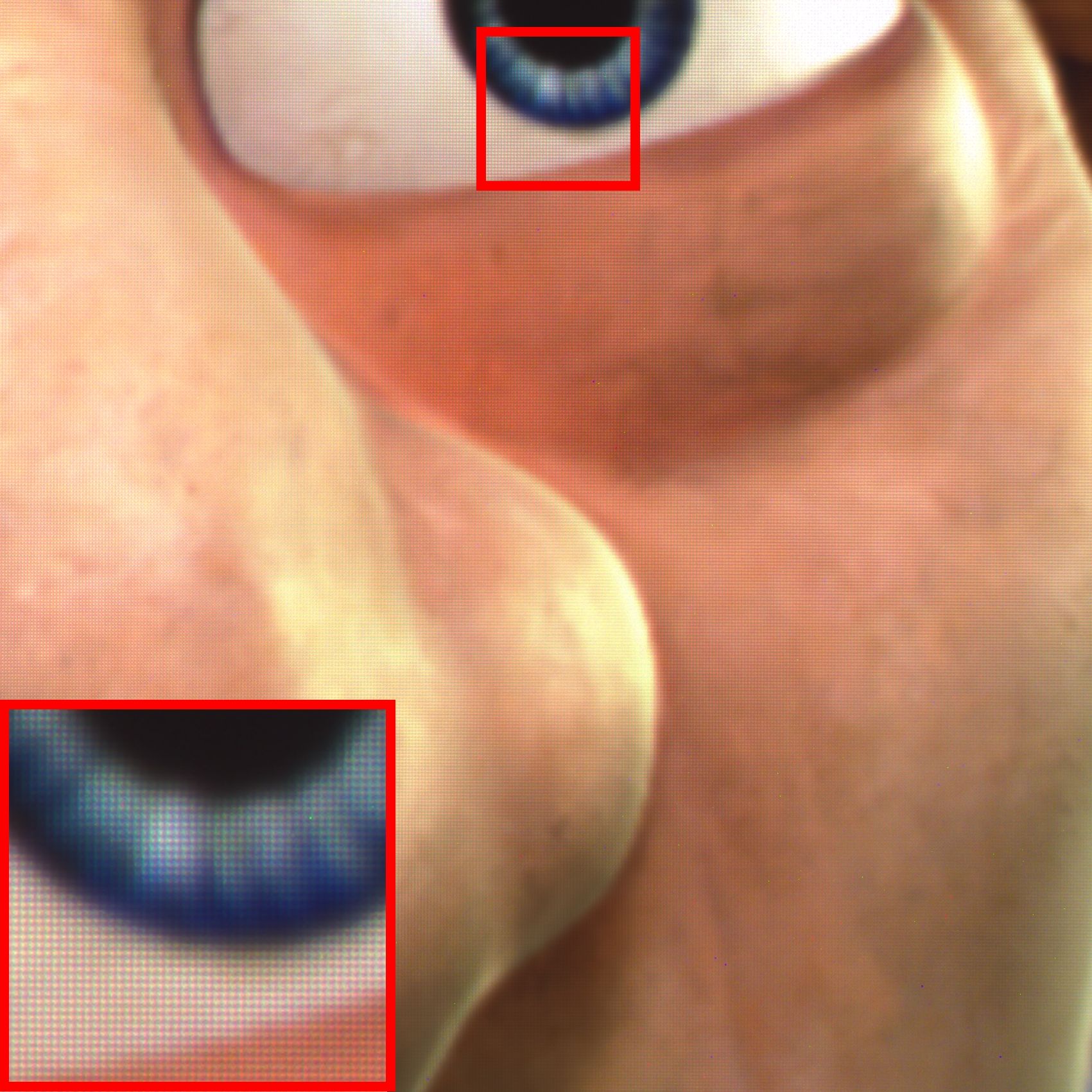}
	\end{subfigure}
	\begin{subfigure}[t]{0.32\columnwidth}
		\centering
		\includegraphics[width=\columnwidth]{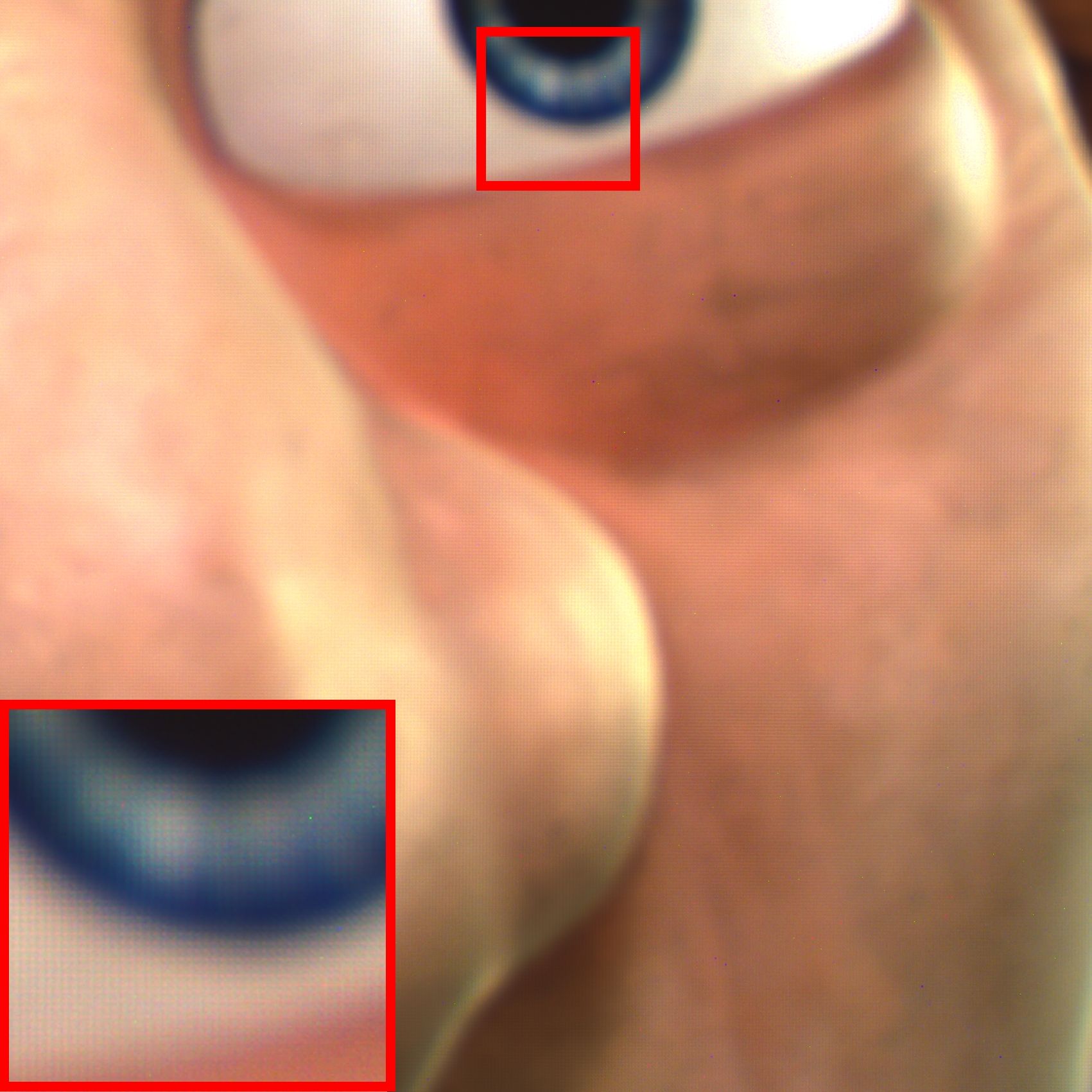}
	\end{subfigure}
	\begin{subfigure}[t]{0.32\columnwidth}
		\centering
		\includegraphics[width=\columnwidth]{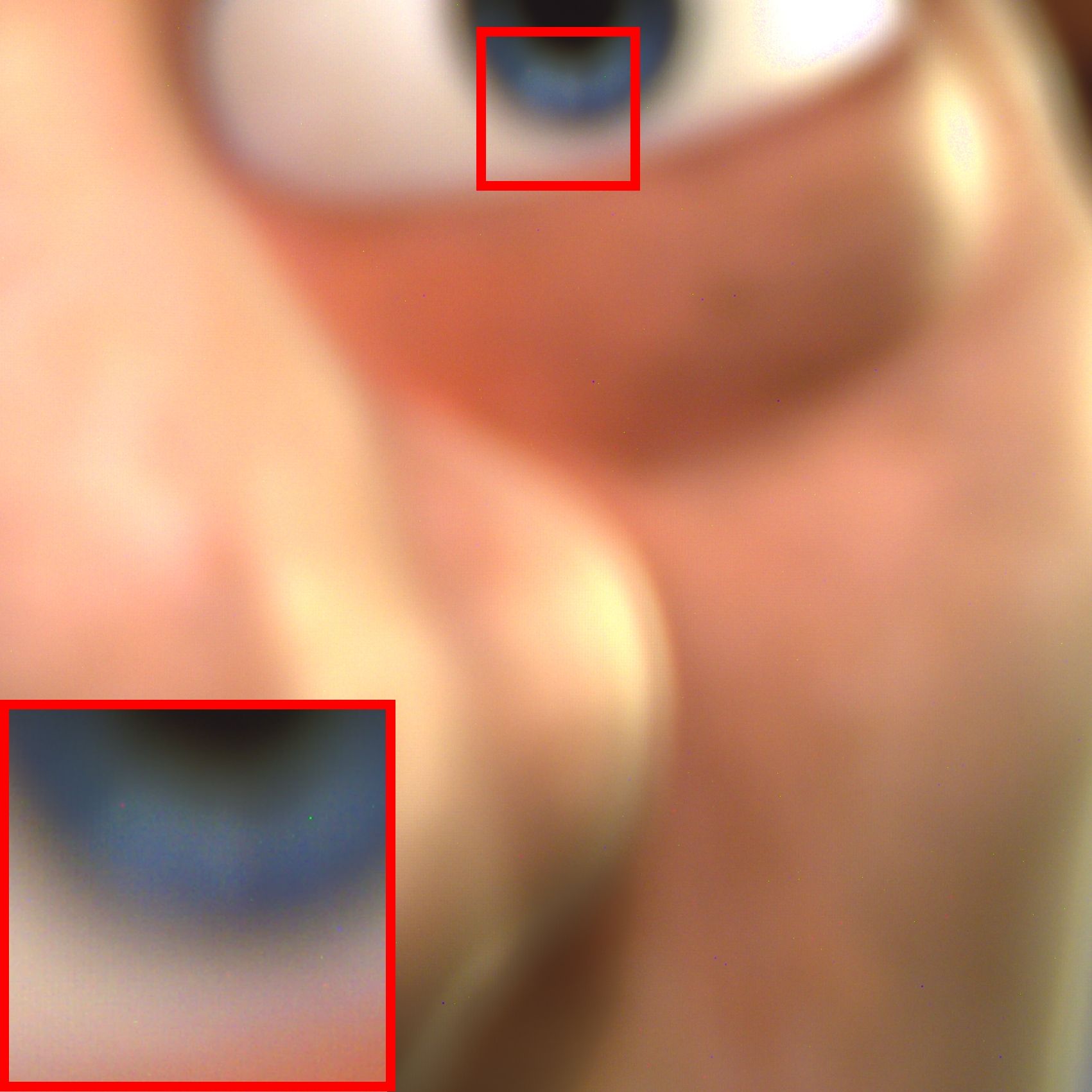}
	\end{subfigure}
	
	\begin{subfigure}[t]{0.32\columnwidth}
		\centering
		\includegraphics[width=\columnwidth]{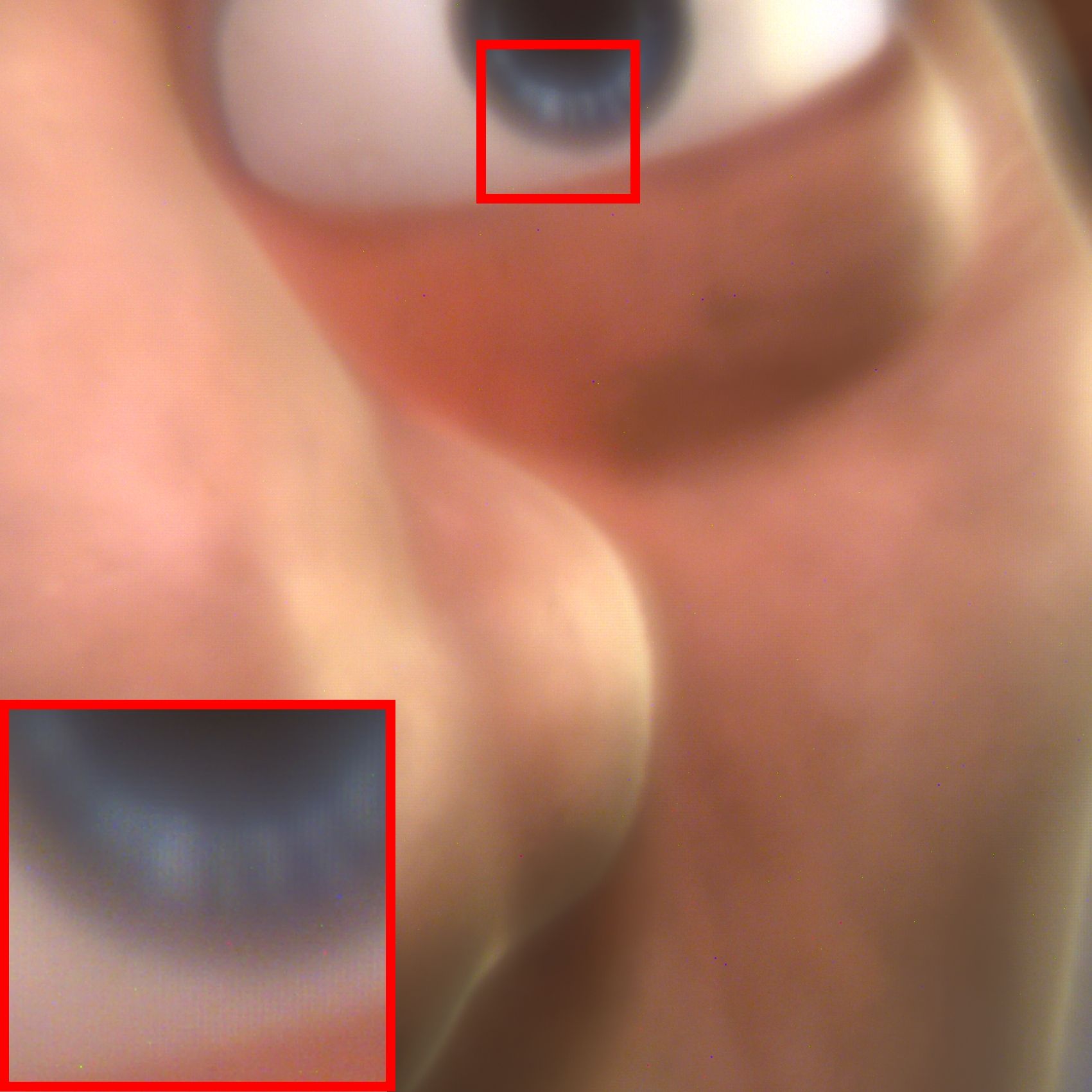}
            \put(-92,20){\rotatebox{90}{Proposed}}
	\end{subfigure}
        \vspace{1mm}
	\begin{subfigure}[t]{0.32\columnwidth}
		\centering
		\includegraphics[width=\columnwidth]{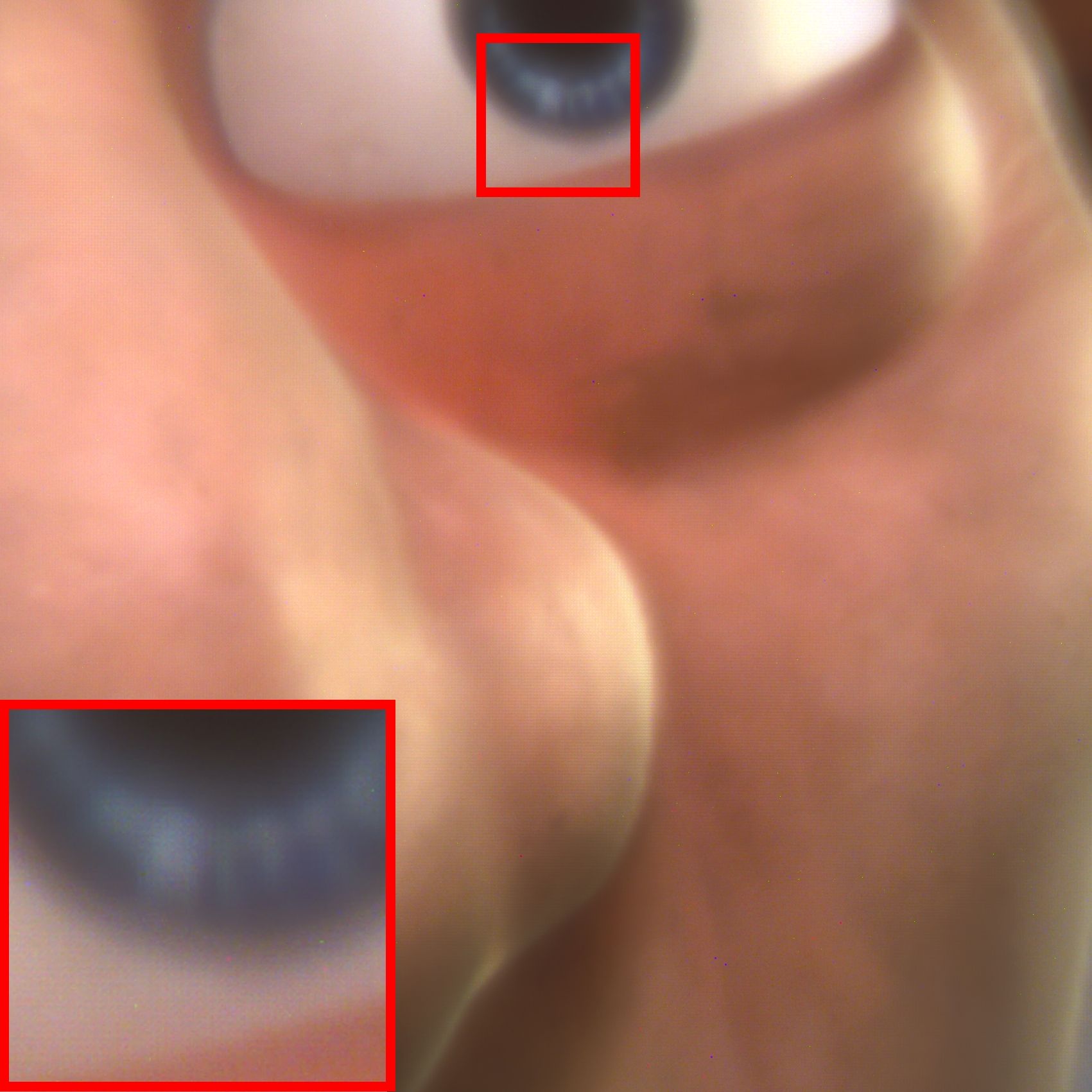}
	\end{subfigure}
	\begin{subfigure}[t]{0.32\columnwidth}
		\centering
		\includegraphics[width=\columnwidth]{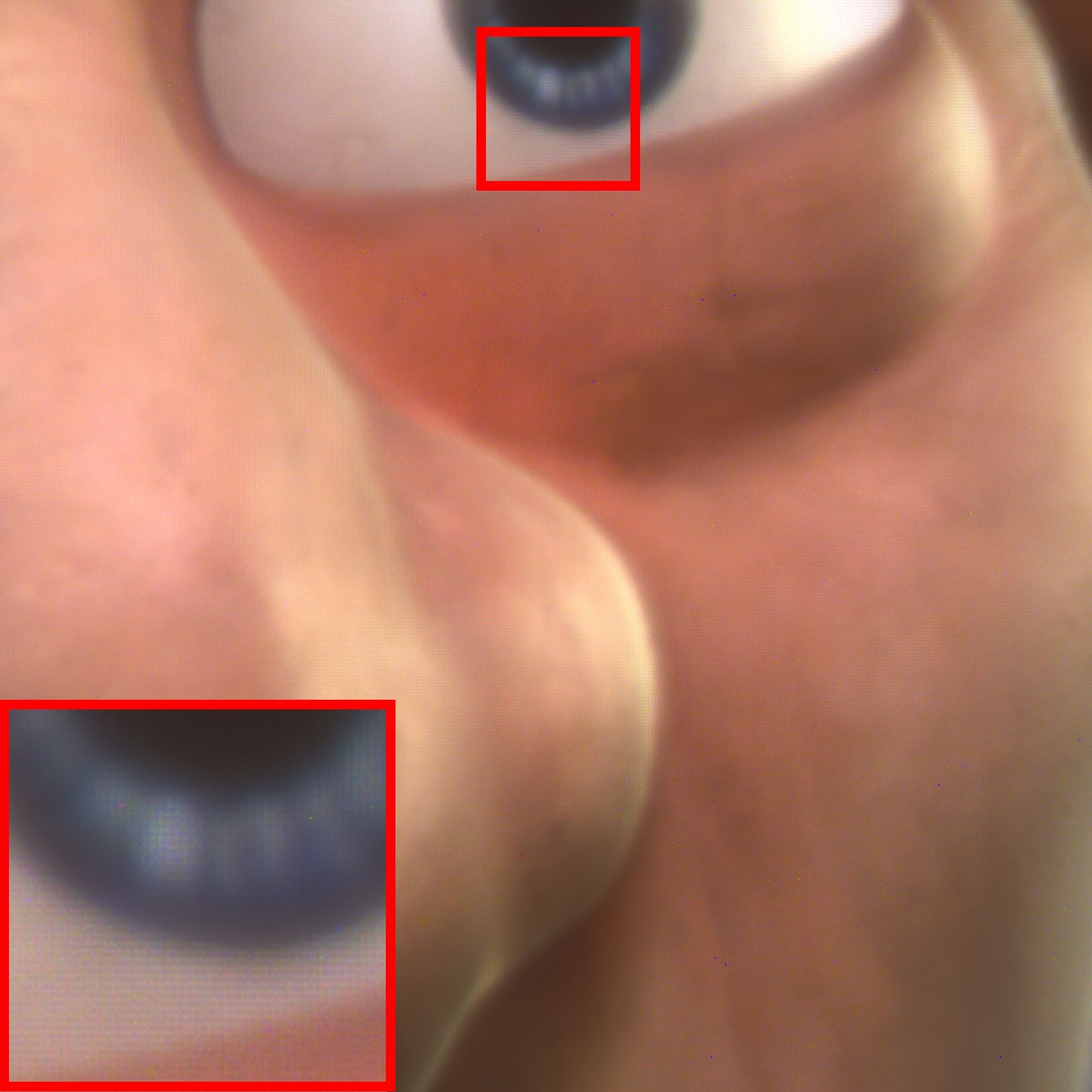}
	\end{subfigure}
	\begin{subfigure}[t]{0.32\columnwidth}
		\centering
		\includegraphics[width=\columnwidth]{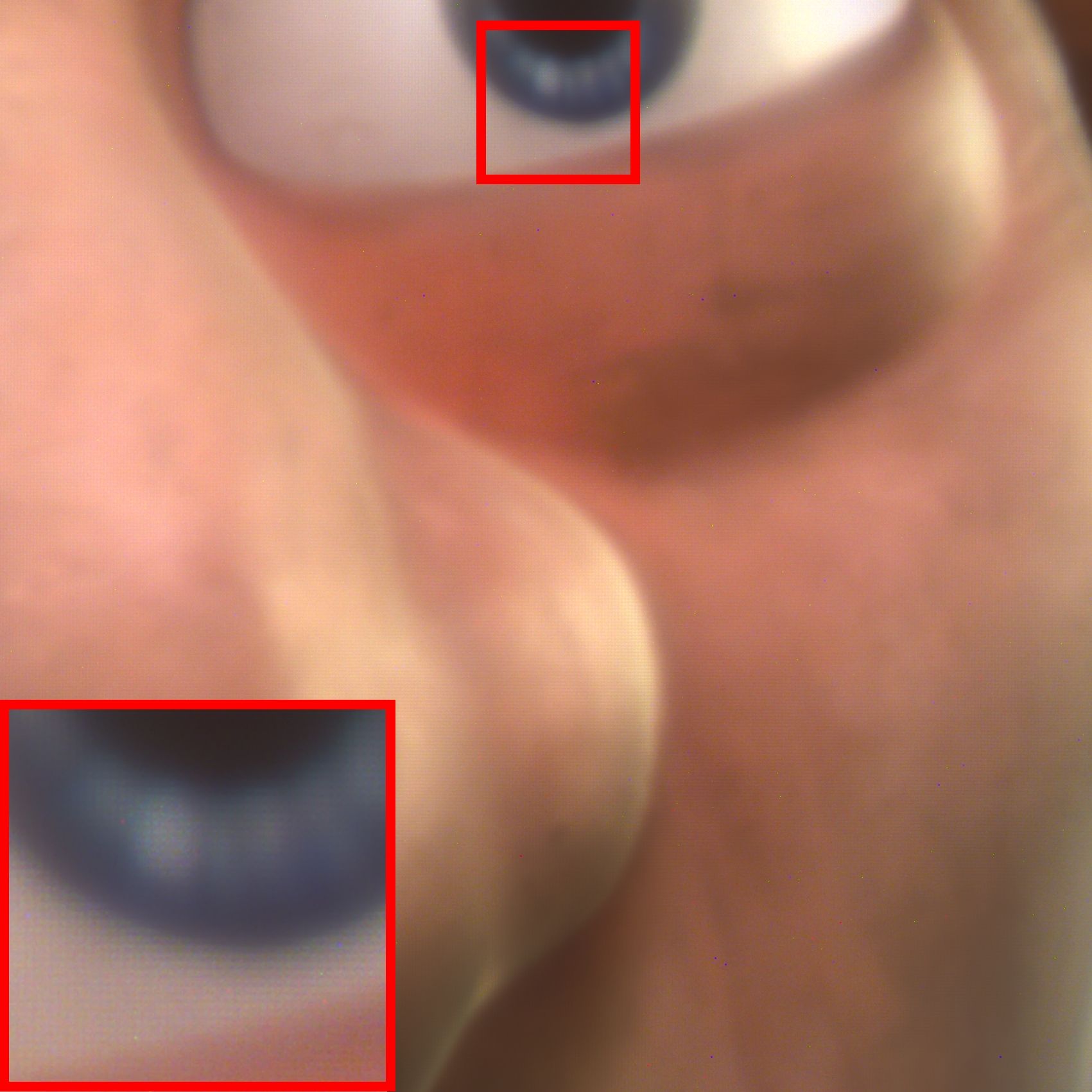}
	\end{subfigure}
	\begin{subfigure}[t]{0.32\columnwidth}
		\centering
		\includegraphics[width=\columnwidth]{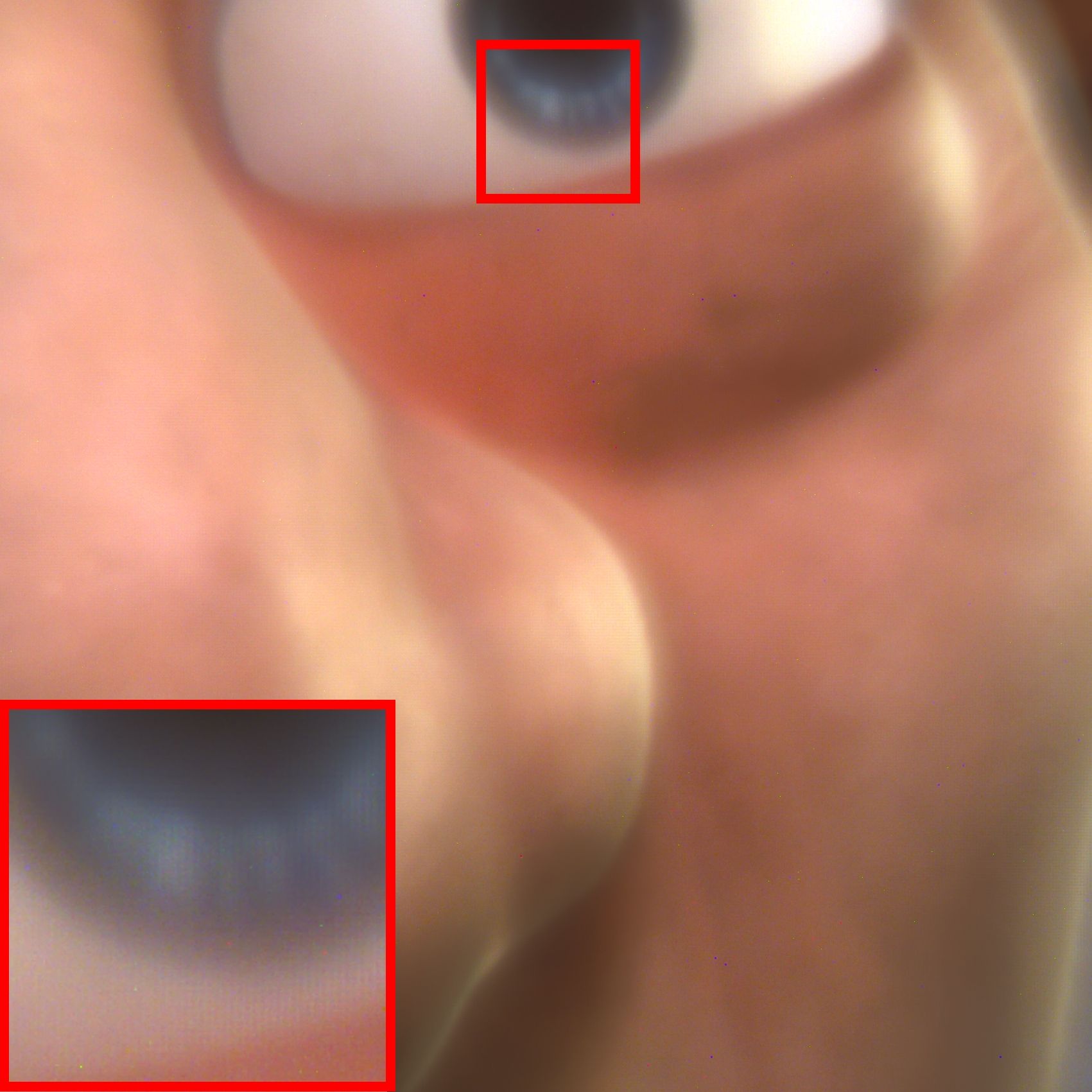}
	\end{subfigure}

	\caption{Experimental verification of the proposed method, compared with the conventional approach (First and second rows, source image courtesy: “Interior Scene”, \url{www.cgtrader.com}). %Preliminary experimental results taken with the wearable prototype.
 }
	\label{fig:Exp_compare_soa}
	
\end{figure*}